\documentclass[12pt,PhD]{phdthesis}
\usepackage {color, amsfonts, amsmath, amsthm, amssymb}
\usepackage {hyperref}
\usepackage {epsf, epsfig}
\usepackage {graphicx}
\usepackage {verbatim}
\usepackage{latexsym}
\usepackage{bm}
\usepackage[english]{babel}

\newcommand{\formula}[2]{\begin{equation}\label{#1} #2 \end{equation}}

\newcommand{\stac}[2]{\stackrel{\scriptscriptstyle {#1}}{#2}}
\newcommand*{\py}{\partial_y}
\newcommand*{\cD}{{\cal D}}
\newcommand*{\cF}{{\cal F}}
\newcommand*{\cJ}{{\cal J}}
\newcommand*{\cK}{{\cal K}}
\newcommand*{\e}{{\rm e}}

\newcommand{\B}[1]{\mathbf{#1}}

\begin{document}

\title{On the four-dimensional effective theories in brane-worlds}

\author{Frederico M. A. Arroja}

\dept {Institute of Cosmology and Gravitation}

\principaladvisor{Dr. Kazuya Koyama}

\secondsupervisor {Prof. Roy Maartens}

\secondreader {Dr. Robert Crittenden} 



\copyrightfalse


\tablespagefalse

\beforeabstract

\prefacesection{Abstract}

Models with extra dimensions have attracted much interest
recently. The reason is that string theory, a serious candidate
for the theory of unification of all interactions, predicts the
existence of up to seven extra dimensions. These models have many
attractive features and may turn out to provide the solution for
many long standing problems in physics. One interesting and very
attractive idea is that our visible universe is confined to a
four-dimensional hypersurface in a higher-dimensional spacetime.
This membrane like universe was dubbed brane-world.

The main goal of this thesis is the study of the four-dimensional
(4D) effective theories and their observational consequences in
the brane-world universe.

After introducing the brane-world idea with more detail we shall
use the gradient expansion method to obtain the 4D effective
theories of gravity for several higher-dimensional theories with
different numbers of extra-dimensions.

In chapter \ref{chapter:5D}, we shall consider a five-dimensional
two brane model with a scalar field living in the extra-dimension.
This model reduces to the Randall and Sundrum model and to the
Ho\v{r}ava-Witten theory for a particular choice of our
parameters. We will show that the 4D effective theory can
reproduce exact dynamically unstable solutions present in the
literature and we will identify the origin of this instability. We
will argue against a claim that the 4D effective theory allows a
wider class of solutions than the 5D theory.

In the following chapter, we shall derive the low energy effective
theory on a brane in 6D supergravity. In order to consider matter
other than tension we will regularize the singular brane by a
codimension one brane. We will show that the 4D effective theory
is a Brans-Dicke (BD) theory with the BD parameter
$\omega_{BD}=1/2$. We will prove that time-dependent solutions in
the 4D theory can be lifted up to exact 6D solutions found in the
literature. We will also discuss the possibility of stabilizing
the BD field in order to recover standard cosmology. At the end of
chapter \ref{chapter:6D} we comment on the implications of these
models for the cosmological constant problem.

In chapter \ref{chapter:10D}, we will present a systematic way to
derive the 4D effective theory for warped compactifications with
fluxes and branes in the 10D type IIB supergravity. We will obtain
the 4D effective potential for the universal K\"{a}hler modulus.

In the second half of the thesis, after introducing the concept of
brane-inflation we will focus on some observational consequences
of these low energy effective theories. In particular, chapter
\ref{chapter:trispectrum} is devoted to the study of
non-Gaussianities in general single field models of inflation. We
will compute the fourth order action for scalar and second order
tensor perturbations for a fairly general model that includes the
brane-inflation scenarios as particular cases. This will enable us
to calculate the trispectrum at leading order in the slow-roll
expansion. We will point out that in order to obtain the correct
leading order result using the comoving gauge action, one has to
include the second order tensor perturbations. This fact has been
wrongly ignored in the literature. We will also provide the
necessary formalism to calculate the next-to-leading order
contribution that might become observationally important in the
future.

The final chapter before the conclusion studies the bispectrum in
general multiple field models of inflation. These general models
include multi-field $K$-inflation and multi-field
Dirac-Born-Infeld inflation as particular cases. We shall derive
the exact second and third order actions including the metric
perturbations. In the small sound speed limit and at leading order
in the slow-roll expansion, we shall derive the three point
function for the curvature perturbation which depends on both
adiabatic and entropy perturbations. We will show that the
contribution from the entropy perturbation has a different shape
dependence from the adiabatic perturbation contribution if the
sound speeds of these perturbations are different. This provides a
way to distinguish models.

The final chapter is for conclusions and discussions.


\afterabstract



\prefacesection{Preface}

The work of this thesis was carried out at the Institute of
Cosmology and Gravitation, University of Portsmouth, United
Kingdom. The author was supported by ``Funda\c{c}\~{a}o para a
Ci\^{encia} e a Tecnologia (Portugal)", with the fellowship's
reference's number: SFRH/BD/18116/2004.
\\

The following chapters are based on published work:
\begin{itemize}
\item Chapter \ref{chapter:5D} - F. Arroja and K. Koyama,
``Moduli instability in warped compactification: 4D effective
theory approach,'' Class.\ Quant.\ Grav.\  {\bf 23} (2006) 4249
[arXiv:hep-th/0602068].

\item Chapter \ref{chapter:6D} - F. Arroja, T. Kobayashi, K. Koyama and T. Shiromizu,
``Low energy effective theory on a regularized brane in 6D gauged
chiral supergravity,'' JCAP {\bf 0712} (2007) 006
[arXiv:0710.2539[hep-th]].

\item Chapter \ref{chapter:10D} - K. Koyama, K. Koyama and F. Arroja,
``On the 4D effective theory in warped compactifications with
fluxes and branes,'' Phys.\ Lett.\  B {\bf 641} (2006) 81
[arXiv:hep-th/0607145].

\item Chapter \ref{chapter:trispectrum} - F. Arroja and K. Koyama,
``Non-gaussianity from the trispectrum in general single field
inflation,'' Phys.\ Rev.\  D {\bf 77} (2008) 083517
[arXiv:0802.1167 [hep-th]].

\item Chapter \ref{chapter:bispectrum} - F. Arroja, S. Mizuno and K. Koyama,
``Non-gaussianity from the bispectrum in general multiple field
inflation,'' arXiv:0806.0619 [astro-ph], accepted for publication
in JCAP.

\end{itemize}

\afterpreface

\prefacesection{Acknowledgements}

First of all I would like to thank Kazuya Koyama for supervising
my Ph.D and for teaching me a lot of physics. I also would like to
thank my collaborators Kayoko Koyama, Tsutomu Kobayashi, Tetsuya
Shiromizu and Shuntaro Mizuno. I thank Ben Hoyle for correcting
the English of several sections of this thesis. I thank
``Funda\c{c}\~{a}o para a Ci\^{e}ncia e a Tecnologia" and the
Portuguese government for providing the funding for my Ph.D.

Special thanks to Roy Maartens, the Institute of Cosmology and
Gravitation of the University of Portsmouth and its past and
present members for the items listed below.

Many thanks for:
\begin{itemize}
\item providing a friendly and welcoming environment that
can (almost) substitute our home;
\item providing an stimulating research group where it is possible
to discuss and share ideas about many areas of physics with many
researchers from all over the world;
\item teaching me so much about physics;
\item giving me the opportunity to attend so many national and
international conferences and workshops;
\item playing sport (football, volleyball, squash, tennis, badminton, ...) that
helped me to keep a healthy body and mind;
\item the almost countless parties and nights out during my stay.
\end{itemize}
To all past, present members and visitors of the ICG, thank you
very much for everything!


\afterpreface

\chapter{Introduction to the brane-world}\label{chapter:braneintro}

Superstring theory, or better, the union of the five different
string theories known as M-Theory is regarded as a promising
candidate for the ``Theory of Everything". Scientists hope that,
once completely formulated, this theory will not only be able to
reproduce, in certain limits, the two theories that are the
pillars of the $20^{th}$ century physics, quantum mechanics and
general relativity, but it will also be able to explain new
phenomena. One of the most interesting and surprising aspects of
this theory is the fact that it can only be correctly formulated
in a higher-dimensional spacetime, to be more specific a $10+1$
dimensional spacetime. But it is also true that almost all of our
daily experiences can be perfectly explained using just a
four-dimensional spacetime, $3+1$ dimensions. Therefore we need a
mechanism of compactification of these seven extra dimensions, so
that they become ``invisible" at least at low energy scales.

In 1996, Ho\v{r}ava and Witten \cite{Horava:1995qa} came up with a
model where this M-Theory compactified in a certain way would
reduce to a known string theory ($E_8\times E_8$ \emph{Heterotic
String}). Basically, this model can be realized, from an effective
point of view \cite{Lukas:1998yy}, as a five-dimensional (5D)
spacetime, the bulk, with two boundary hyper surfaces, the branes.
All the Standard Model particles and fields are confined to these
hyper surfaces while gravity is the only interaction that can
occur in the bulk.

In 1999, motivated by these ideas, Randall and Sundrum
\cite{Randall:1999ee} proved that in this scenario one of the
biggest problems in physics, the hierarchy problem could be
explained in a simple way. I shall describe their model in section
\ref{sec:RES}. In the same year, Randall and Sundrum published
another paper \cite{Randall:1999vf} where they proved that
Newton's law of gravity could be retrieved in these models with
extra dimensions. Although this was a very important paper, it was
somehow incomplete because the recovery of Newton's theory is not
enough when we know that the correct theory is general relativity.
The gap was filled with the work of Garriga and Tanaka
\cite{Garriga:1999yh} (see also \cite{Giddings:2000mu}) and the
subject of linearized gravity will be discussed is section
\ref{sec:LG}.

The effective non-linear Einstein equations of the 3-brane world
were derived in \cite{Shiromizu:1999wj} and they will be under
consideration in section \ref{sec:EE}. I shall say a few words
about the cosmological implications
\cite{Binetruy:1999ut,Binetruy:1999hy} of this new set of
equations in subsection \ref{subsec:CS}. The effective Einstein
equations are very difficult to solve analytically. In fact, it
was shown that in general one needs to solve the full 5D bulk
equations in order to find the 4D geometry, so it is crucial to
develop tools to tackle this problem. Kanno and Soda
\cite{Kanno:2002iaa,Kanno:2002ia} developed such a tool and
section \ref{sec:LE} will be dedicated to the study of their
method, the low energy expansion scheme. This perturbative method
solves the full 5D equations of motion using an approximation and
after imposing the junction conditions, one obtains the 4D
effective equations of motion.

In the next three chapters of this thesis we will use the low
energy expansion scheme extensively in order to obtain the
four-dimensional (4D) effective theories of gravity for three
different models with several extra-dimensions.

The first model under consideration will be a 5D dilatonic two
brane model that interpolates between the Randall and Sundrum
model and the Ho\v{r}ava-Witten theory, which we shall introduce
in section \ref{sec:introHoravaWitten}.

The second model that we will study is the 6D supergravity
scenario. As an introduction, in section \ref{sec:intro6D}, we
will present the idea of self-tuning in a six-dimensional
brane-world model (co-dimension two brane-world). This self-tuning
property is remarkable and it might prove to be the solution to
another one of the biggest problems in physics, the cosmological
constant problem. We will obtain the 4D effective theory and we
will show that it is a Brans-Dicke (BD) theory with the BD
parameter equals to one half.

It would be desirable to realize the Randall and Sundrum mechanism
of generating large hierarchies in a more fundamental theory like
string theory for example. This was achieved in
\cite{Giddings:2001yu} for the type IIB string theory. In section
\ref{sec:intro10D} we will present the 10D type IIB supergravity
model that is a low energy approximation to the type IIB string
theory. Section \ref{sec:intro10D} is of particular relevance for
the understanding of Chapter \ref{chapter:10D}, where we will
obtain the 4D effective theory in warped compactifications of the
10D supergravity model. This constitutes the last of the three
higher-dimensional theories that we will discuss.

In this thesis, we shall follow the sign conventions and
definitions of the Riemann and Ricci tensors of Wald
\cite{Wald:1984rg}. In particular, we will use the metric
signature $-+++\cdots$ We also choose to work in an unit system
where the speed of light and the reduced Planck constant are equal
to one.

\section{\label{sec:RES}Randall and Sundrum two brane model}

This model \cite{Randall:1999ee} was proposed as an explanation to
the hierarchy problem. So, first I shall clarify what exactly is
this problem and why it is one of the most important problems in
physics.

The hierarchy problem highlights the enormous difference in
strength of gravity and the electroweak interaction. The
electroweak energy scale is $m_{EW}\sim 10^3 GeV$ while the Planck
energy scale (the scale at which general relativity breaks down
and a theory of quantum gravity must be used) is
$M_{PL}=G_N^{-1/2}\sim 10^{19} GeV$. $G_N$ denotes the 4D Newton's
constant. It is problematic to have two (so different) fundamental
energy scales when searching for the theory of unification of all
interactions.

However, there is a crucial difference between the electroweak
scale and the Planck scale. While we have experiments that probe
the electroweak interaction down to $m_{EW}^{-1}\sim 10^{-17} cm$
distances, we do not have experiments to test gravity at distances
of order of $M_{PL}^{-1}\sim 10^{-33} cm$. The smallest scale that
we have access through gravity is of order $10^{-2} cm$
\cite{Kapner:2006si}, so our assumption that $M_{PL}$ is a
fundamental energy scale relies on a huge extrapolation.

This reason led Randall and Sundrum (RS) and others before them to
infer that $M_{PL}$ might not be a fundamental scale and to
proposed mechanisms to generate it. Before RS, the
higher-dimensional mechanisms \cite{ArkaniHamed:1998rs} which
existed solved the hierarchy problem relying on the formula
\formula{MPL}{M_{PL}^2=M^{n+2}V_n.} This formula is valid for a
$4+n$ dimensional spacetime, where the $n$ extra dimensions are
compactified, i.e., the full spacetime is equal to a $4D$
spacetime times a $nD$ compact space, being $V_n$ its volume. $M$
is the $(4+n)D$ Planck mass and we suppose $M\sim M_{EW}$. If
$n=2$ then $V_2^\frac{1}{2}\sim 1mm$.

The ingenious step in RS work \cite{Randall:1999ee} was to
consider a non-factorizable metric (\ref{NFM}) instead of a
factorizable one:
\formula{NFM}{ds^2=g_{AB}dx^Adx^B=e^{-2|y|/l}\eta_{\mu\nu}dx^\mu
dx^\nu+dy^2,} where $l^{-1}$ is of order of $M$ and $y$ is the
coordinate for the extra dimension, $0\leq y\leq L$.
$\eta_{\mu\nu}$ denotes the Minkowski metric.

\begin{figure}[t]
\centering
 \scalebox{.5}
 {\rotatebox{0}{
    \includegraphics*{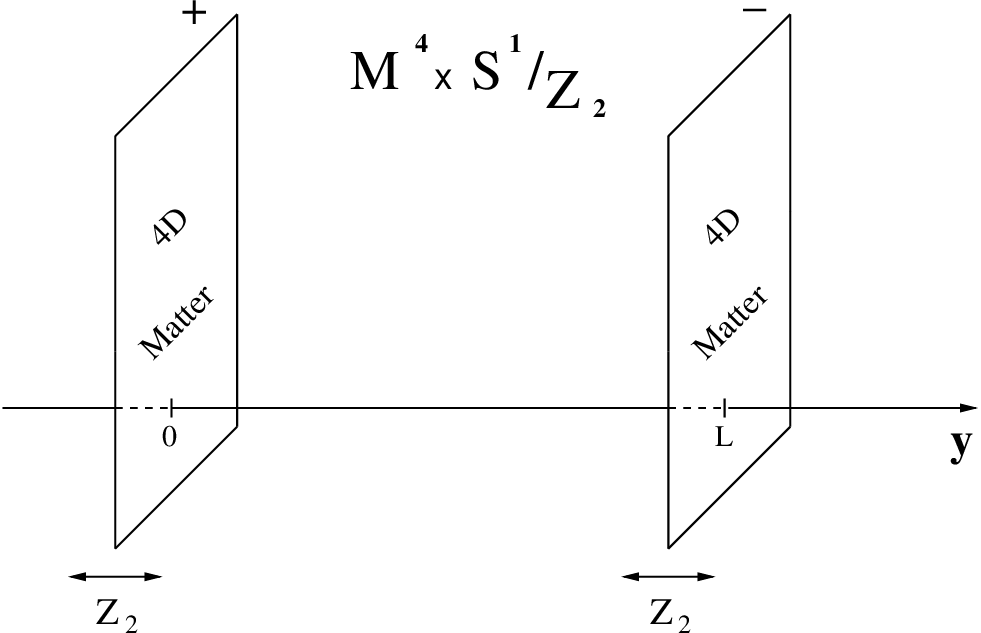}
                 }
 }
\caption{The brane-world universe}\label{fig:fig1}
\end{figure}

Their set-up is represented in figure \ref{fig:fig1}.
It consists of two 3-branes at the fixed points, $y=0$ (the hidden
brane, A) and $y=L$ (the visible brane, B), of the orbifold
$S^1/Z_2$. Effectively the branes are the boundaries of the 5D
spacetime (bulk).

The action is
\begin{eqnarray}
S&=&\frac{1}{2\kappa_G^2}\int d^5 x \sqrt{-g} \left( {}^{(5)}\!\!R
+ \frac{12}{l^2} \right) \nonumber\\&&- \sum_{i=A,B}\sigma_i \int
d^4 x \sqrt{-g^{i\:brane}}+\sum_{i=A,B}\int d^4 x
\sqrt{-g^{i\:brane}}L_{matter}^{i},\label{RSA}
\end{eqnarray}
where $l$ is the curvature radius of $AdS_5$, $\sigma_{vis}$,
$\sigma_{hid}$ are the brane tensions and $\kappa_G$ is
$\kappa_G^2=8\pi{}^{(5)}\!G_N=M^{-3}$. ${}^{(5)}\!\!R$ denotes the
5D Ricci scalar, $g^{i\:brane}$ denotes the induced metric on the
brane.

They have proved that (\ref{NFM}) is a solution of the Einstein's
equations that result from the action (\ref{RSA}), without the
matter terms, and obtained the following relations for the brane
tensions:\formula{RSR}{\sigma_{hid}=-\sigma_{vis}=\frac{6}{\kappa_G^2l}.}
It can also be shown that for a negative tension brane observer
\formula{MPL2}{M_{PL}^2=M^3l\left[e^{2L/l}-1\right].} So if we
take $M\sim m_{EW}$ as a fundamental scale of the ``hidden" brane
(positive tension) we can easily (without introducing other large
hierarchies) obtain a scale for gravity in the``visible" brane
(negative tension) of the order $10^{19}$GeV. This is achieved
thanks to the exponential ``warp" factor $a(y)\equiv e^{-y/l}$.

\section{\label{sec:LG}Linearized gravity}

Once we have seen this new solution for the hierarchy problem it
is relevant to discuss the characteristics of gravity in the RS
model. RS also realized that their model could provide us with an
alternative to compactification, i.e., we can reproduce Newtonian
gravity even with an infinite extra-dimension.

Let us begin with the same set-up of the previous section but with
the negative tension brane ($y=L$) removed. We can do this simply
by allowing $L\rightarrow+\infty$. Now, an observer living on the
brane will calculate the Planck mass as
\formula{MPL3}{M_{PL}^2=M^3l.} As we can see the Planck mass is
finite in spite of an infinite extra-dimension. This is in
contrast with the factorizable metric's models where the Planck
mass would be infinite. As mentioned before, it is the fact that
the extra-dimension is curved that acts as an effective
compactification. It could be said that the 5D graviton only
``sees" an extra-dimension of size $l$.

In order to determine if we recover the Newtonian potential we
need to study the behaviour of tensor perturbation $h_{\mu\nu}$
around the metric (\ref{NFM}),
\formula{PM}{ds^2=\left(g_{AB}+h_{AB}\right)dx^Adx^B=\left(e^{-2|y|/l}
\eta_{\mu\nu}+h_{\mu\nu}\right)dx^\mu dx^\nu+dy^2.} With the gauge
choice $h_{yy}=h_{\mu y}=0$, $h_{\mu,\nu}^\nu=0$ and
$h_\mu^\mu=0$, the Einstein's equations at first order in the
perturbation are \cite{Randall:1999vf,Garriga:1999yh}:
\formula{PE}{\left[
e^{2|y|/l}\eta^{\mu\nu}\partial_\mu\partial_\nu+\partial_y^2
-\frac{4}{l^2}+\frac{4}{l}\delta(y) \right]h_{\alpha\beta}=0.}

We may look for a solution of the form
$h(x^\mu,y)=\psi_m(y)e^{ip^\mu x_\mu}$ (dropping tensor indices
for simplicity), with $p^2=-m^2$. The resulting equation for
$\psi_m(y)$ can be written in the form
\formula{SCH}{\frac{d^2\hat{\psi}_m}{dz^2}-V(z)\hat{\psi}_m=-m^2\hat{\psi}_m,
\;\;\;\;\;\;\;\;\;\;V(z)=\frac{15}{4\left(|z|+l\right)^2}-\frac{3\delta(z)}{l},}
where the new variable is $z=\int a^{-1}(y)\;dy$ and with the
definition $\hat{\psi}_m=a^{-1/2}\psi_m$.

The problem now reduces to a trivial quantum mechanics' exercise.
A particle (the 5D graviton) governed by a Schr\"{o}dinger like
equation in a``volcano" shaped potential. The eigenvalue $m=0$
corresponds the eigenfunction
$\psi_0(y)=\frac{e^{-2|y|/l}}{\sqrt{l}}$. This is the ``wave
function" of the massless graviton of the 4D theory and we will
see that it reproduces Newton's potential energy between two
masses on the brane. The other eigenmodes (continuum) are plane
waves, asymptotically away from the brane, and near the brane
their amplitudes are suppressed. Because of this they are weakly
coupled to the matter on the brane but they will introduce small
corrections to the standard 4D physics.

Outside a spherically symmetric source of mass $M_0$ on the brane,
the gravitational potential is given by
\formula{NEWC}{V(r)\approx\frac{G_NM_0}{r}\left(1+\frac{2l^2}{3r^2}\right).}
This formula is valid in the limit $r\gg l$ and the second term is
the first correction due to the continuum of Kaluza-Klein (KK)
modes.

When $r\ll l$, the small scale limit, the potential reveals its
pure 5D character \formula{POT}{V(r)\approx\frac{G_NM_0l}{r^2}.}

The best gravitational experiments available today place an upper
bound of $10^{-1}mm$ for $l$. Formulas (\ref{RSR}) and
(\ref{MPL3}) imply the inequalities
\formula{BOU}{\sigma>10^{13}GeV^4,\;\;\;\;\;\;\;M>10^8GeV.} So, we
should not expect to find KK massive modes of the graviton in
future collider experiments.

In \cite{Garriga:1999yh}, it was proved, for the one brane case,
that we do recover the linearized Einstein's equations and not
only Newton's theory. When we introduce a second brane (negative
tension), the KK modes no longer form a continuum but rather have
a discrete set of values. At linear order, we get a scalar-tensor
theory as an effective theory on the brane. The physics on both
branes is quite different, for example, the Newton's constant is
\formula{Gpm}{G^{(\pm)}=\frac{{}^{(5)}\!G_N\;l^{-1}e^{\pm
L/l}}{2\sinh(L/l)},} where $(\pm)$ refers to the sign of the
branes' tension.

This scalar-tensor theory is a Brans-Dicke (BD) theory with the BD
parameter given by
\begin{equation}
\omega_{BD}^{(\pm)}=\frac{3}{2}\left(e^{\pm 2L/l}-1\right).
\end{equation}

Observations require $\omega_{BD}\gg 1$ \cite{Will:2005va}. This
is easily achieved on the positive tension brane and we have a
perfectly acceptable theory of gravity even without stabilizing
the BD field. In the negative tension brane (where we have an
explanation for the hierarchy problem) the BD parameter is always
negative so this cannot be our universe unless we stabilize the
distance between the two branes.

\section{\label{sec:EE}The Einstein's equations of the
brane-world}

In the previous section, we have discussed linearized gravity in
the brane-world. However for example for cosmology, we need to
know the non-linear/strong behaviour of gravity. Here we shall
obtain the full non-linear projected equations for gravity in the
brane-world. In the preceding sections, we have used Einstein's
equations on the brane-world context but we have always assumed
particular symmetries, in the metric or in the energy momentum
tensor. In this section, following \cite{Shiromizu:1999wj}, we
shall derive the equations that govern the brane-world dynamics
without assuming any symmetries. We will proceed in this manner as
far as possible and only then we will use some quite general
brane-world's hypotheses.

For any given 5-dimensional spacetime $\left(V,g_{\mu\nu}\right)$
and any (well behaved) 4-dimensional hypersurface
$\left(M,q_{\mu\nu}\right)$ embedded in this spacetime, it is
possible to write the components of the Riemann tensor of this 4D
hypersurface as (Gauss' equation)
\formula{GAUSS}{{}^{(4)}R^\alpha_{\;\beta\gamma\delta}=
{}^{(5)}R^\mu_{\;\nu\rho\sigma}q_\mu^\alpha q_\beta^\nu
q_\gamma^\rho q_\delta^\sigma+K^\alpha_\gamma
K_{\beta\delta}-K^\alpha_\delta K_{\beta\gamma},} where
$q_{\mu\nu}=g_{\mu\nu}-n_\mu n_\nu$ is the projected metric on the
surface, $n^\mu$ is the unit normal vector to $M$ and $K_{\mu\nu}$
is the extrinsic curvature tensor given by
$K_{\mu\nu}=q_\mu^\alpha q_\nu^\beta \nabla_\alpha n_\beta$. The
Codacci's equation reads \formula{CODACCI}{D_\nu K_\mu^\nu-D_\mu
K={}^{(5)}R_{\rho\sigma}n^\sigma q_\mu^\rho,} where $D_\mu$ is the
covariant derivative with respect to $q_{\mu\nu}$.

Gauss' equation implies
\begin{eqnarray}
{}^{(4)}G_{\mu\nu}&=&{}^{(4)}R_{\mu\nu}-\frac{1}{2}q_{\mu\nu}{}^{(4)}R
\nonumber \\
&=&\left({}^{(5)}R_{\rho\sigma}-\frac{1}{2}g_{\rho\sigma}{}^{(5)}R\right)q_\mu^\rho
q_\nu^\sigma + {}^{(5)}R_{\rho\sigma}n^\rho n^\sigma
q_{\mu\nu}\nonumber
\\&& \mbox{}+ KK_{\mu\nu}-K_\nu^\rho
K_{\nu\sigma}-\frac{1}{2}q_{\mu\nu}\left(K^2-K^{\alpha\beta}K_{\alpha\beta}\right)-\tilde{E}_{\mu\nu},
\label{4DEE}
\end{eqnarray}
where
\formula{ET}{\tilde{E}_{\mu\nu}\equiv{}^{(5)}R^\alpha_{\;\beta\rho\sigma}n_\alpha
n^\rho q_\mu^\beta q_\nu^\sigma.}

The 5-dimensional Einstein's equations are
\formula{5EIN}{{}^{(5)}R_{\alpha\beta}-\frac{1}{2}g_{\alpha\beta{}}^{(5)}R=\kappa_G^2T_{\alpha\beta},}
where $T_{\alpha\beta}$ is the 5D energy momentum tensor. Using
the decomposition of the Riemann's tensor
\formula{DECO}{{}^{(5)}R_{\mu\alpha\nu\beta}=\frac{2}{3}
\left(g_{\mu[\nu}{}^{(5)}R_{\beta]\alpha}-g_{\alpha[\nu}{}^{(5)}R_{\beta]\mu}\right)
-\frac{1}{6}g_{\mu[\nu}g_{\beta]\alpha}{}^{(5)}R+{}^{(5)}C_{\mu\alpha\nu\beta},}
where $C_{\mu\alpha\nu\beta}$ is the Weyl's tensor, we obtain
\begin{eqnarray}
{}^{(4)}G_{\mu\nu}&=&\frac{2\kappa_G^2}{3}\left[T_{\rho\sigma}q_\mu^\rho
 q_\nu^\sigma+\left(T_{\rho\sigma}n^\rho n^\sigma-\frac{1}{4}T\right)q_{\mu\nu}\right]+
KK_{\mu\nu}-K_\mu^\sigma
K_{\nu\sigma}\nonumber\\&&-\frac{1}{2}q_{\mu\nu}\left(K^2-K^{\alpha\beta}K_{\alpha\beta}\right)-E_{\mu\nu},\label{4EEE}
\end{eqnarray}
where
\formula{WEYL}{E_{\mu\nu}\equiv{}^{(5)}C^\alpha_{\;\beta\rho\sigma}n_\alpha
n^\rho q_\mu^\beta q_\nu^\sigma,} is traceless. The Codacci's
equation gives \formula{ECON}{D_\nu K_\mu^\nu-D_\mu
K=\kappa_G^2T_{\rho\sigma}n^\sigma q_\mu^\rho.} Equation
(\ref{4EEE}) and (\ref{ECON}) are completely general and can be
used in all cases.

In order to proceed, we shall assume the following brane-world's
symmetries:
\begin{itemize}
 \item Gaussian normal coordinates,
 $ds^2=dy^2+q_{\mu\nu}dx^\mu dx^\nu,$ where the brane is at y=0,
 \item $T_{\mu\nu}=-\Lambda g_{\mu\nu}+S_{\mu\nu}\delta(y),$ where
 $\Lambda$ is the cosmological constant in the bulk and
 $S_{\mu\nu}$ is the distributional energy momentum tensor on the
 brane,
 \item $S_{\mu\nu}=-\lambda q_{\mu\nu}+\tau_{\mu\nu},$ where
 $\lambda$ is the brane's tension and $\tau_{\mu\nu}$ is the
 energy momentum tensor with the property $\tau_{\mu\nu}n^\nu=0$.
\end{itemize}

We then impose Israel's junction conditions \cite{Israel:1966rt}
\begin{subequations}
\label{JCintro}
 \begin{eqnarray}
  \left[q_{\mu\nu}\right]{\!}_|{}_{y=0}&=&0,\label{JCG}
  \\
  \left[K_{\mu\nu}\right]{\!}_|{}_{y=0}&=&-\kappa_G^2\left(S_{\mu\nu}
  -\frac{1}{3}q_{\mu\nu}S\right).\label{JCDG}
 \end{eqnarray}
\end{subequations}
The previous equations together with (\ref{4EEE}), assuming $Z_2$
symmetry around the brane, give us the effective 4D equations that
replace the standard Einstein's equations in the brane-world
context,
\formula{EEE}{{}^{(4)}G_{\mu\nu}=-\Lambda_4q_{\mu\nu}+8\pi
G_N\tau_{\mu\nu}+\kappa_G^4\pi_{\mu\nu}-E_{\mu\nu},} where
\begin{subequations}
\label{DEFS}
 \begin{eqnarray}
  \Lambda_4&=&\frac{1}{2}\kappa_G^2\left(\Lambda+\frac{1}{6}\kappa_G^2\lambda^2\right),\label{EBCC}
  \\
  G_N&=&\frac{\kappa_G^4\lambda}{48\pi},\label{NC}
  \\
  \pi_{\mu\nu}&=&-\frac{1}{4}\tau_{\mu\alpha}\tau_\nu^\alpha+
  \frac{1}{12}\tau\tau_{\mu\nu} +\frac{1}{8}q_{\mu\nu}\tau_{\alpha\beta}\tau^{\alpha\beta}
  -\frac{1}{24}q_{\mu\nu}\tau^2.\label{QT}
 \end{eqnarray}
\end{subequations}

Important comments about these equations:
\begin{itemize}
 \item When the effective cosmological constant of the brane $\Lambda_4$ is zero, it
corresponds to the RS fine-tuning between the brane's tension and
the bulk cosmological constant,
 \item From Eq. (\ref{NC}) it is possible to see that $\lambda$ has to be positive,
 \item The $\pi_{\mu\nu}$ tensor is quadratic in $\tau$ and it will
 be important at high energies,
 \item There is the new $E_{\mu\nu}$ term. It is zero for an $AdS$ bulk but not
  zero otherwise. Some, but not all, of its degrees of freedom are constrained by the brane's matter distribution as $D^\mu E_{\mu\nu}=\kappa_G^2D^\mu\pi_{\mu\nu}$. These
non-determined degrees of freedom are what transforms the system
in a non-closed system, i.e., we need to solve the full 5-D
Einstein's equations and not only the 4-D effective ones in order
to evaluate $E_{\mu\nu}$,
 \item A dimensional analysis of (\ref{EEE}) suggests that in the low energy limit, i.e. when the brane's tension is much larger then the characteristic energy scale of the matter, it simplifies
to ${}^{(4)}G_{\mu\nu}\approx-\Lambda_4q_{\mu\nu}+8\pi
G_N\tau_{\mu\nu}$ (Einstein's equation),
 \item From (\ref{ECON}) and (\ref{JCDG}) it is possible to show that
 energy conservation on the brane still holds $D_\mu\tau_\nu^\mu=0$.
\end{itemize}

\subsection{\label{subsec:CS}Cosmological scenarios}

As natural, there is a myriad of cosmological models that have
been suggested as models for our universe. For a good overview of
these models see
\cite{Langlois:2002bb,Maartens:2003tw,Langlois:2004gi} and
references therein.

Because we are interested in cosmological solutions (isotropic and
homogeneous) we will consider the Friedman-Robertson-Walker metric
\formula{FRWintro}{dS_{(4)}^2=q_{\mu\nu}dx^\mu
dx^\nu=-dt^2+a^2(t)\left[\frac{dr^2}{1-kr^2}+r^2\left(d\theta^2+\sin^2\theta
d\phi^2\right)\right],} where $a(t)$ is the scale factor of the
3-brane. We assume the perfect fluid energy momentum tensor form
for $\tau_{\mu\nu}=\rho t_\mu t_\nu+Ph^{\mu\nu}$, where $t_\mu$ is
the fluid's 4-velocity and $h^{\mu\nu}=q^{\mu\nu}+t^\mu t^\nu$ is
the metric of the spatial surfaces orthogonal to $t^\mu$. $\rho$
and $P$ are the energy density and pressure respectively. This
implies that $\pi^{\mu\nu}$ is
\formula{PIP}{\pi^{\mu\nu}=\frac{1}{12}\rho\left[\rho t^\mu
t^\nu+\left(\rho+2P\right)h^{\mu\nu}\right],} and its trace is
$\pi=\frac{\rho}{6}\left(\rho+3P\right)$. A general form for
$E_{\mu\nu}$ respecting isotropy and homogeneity is
\formula{EUV}{E^\mu_\nu=-diag\left(-\rho_E,P_E,P_E,P_E\right).}
$E_{\mu\nu}$ being traceless implies
\formula{RADIATION}{P_E=\frac{1}{3}\rho_E.} It is possible to show
$D^\mu\pi_{\mu\nu}=0$, hence \formula{DEUV}{D_\mu E^\mu_\nu=0,}
for this cosmological solution. So (\ref{RADIATION}) and
(\ref{DEUV}) give us
\formula{CONS}{3H\left(\rho_E+P_E\right)+\dot{\rho}_E=0,} where
$H\equiv\frac{\dot{a}}{a}$. Eq. (\ref{CONS}) can be integrated to
obtain \formula{RHOE}{\rho_E=Ca^{-4},} where $C$ is just an
integration constant. With the previous results and after a few
algebraic manipulations of (\ref{EEE}) we obtain the modified
Friedman's equation
\formula{NFRD}{H^2=-\frac{k}{a^2}+\frac{\Lambda_4}{3}+\frac{\rho_E}{3}+\frac{8\pi
G_N\rho}{3}+\frac{\kappa_G^4\rho^2}{36},} with
$\rho_E=\frac{C}{a^4}$.

The $\rho^2$ term (comes from the quadratic term in
$\tau_{\mu\nu}$) will have important implications in the very
early universe, when the energy density is high. The $\rho_E$ term
is called ``dark radiation" because of its $a^{-4}$ behaviour, and
it is related with the mass of a black hole that exists in the
bulk.

\section{\label{sec:LE}Low energy expansion method}

As we have seen in the previous section, the brane-world Einstein
equations are not easy to solve and only a few exact solutions are
known. This is because of the term $E_{\mu\nu}$. This term
$E_{\mu\nu}$ in the projected 4D Einstein equations makes the
system non-closed. As a consequence if one wants to find the 4D
dynamics one has to determine $E_{\mu\nu}$ by solving the full 5D
non-linear differential equations \cite{Shiromizu:1999wj}.
Motivated by this necessity, Kanno and Soda (KS) developed a
perturbative method to solve the 5D equations and to find
$E_{\mu\nu}$. I shall now describe their method which was applied
to RS 2 branes model in \cite{Kanno:2002iaa,Kanno:2002ia}. Because
we will consider the matter terms in (\ref{RSA}), the branes will
not in general be flat. Consequently we cannot put both branes at
$y=0$ and $y=l$ and use Gaussian normal coordinates. The line
element that captures the problem's symmetries is
\formula{KSM}{ds^2=e^{2\phi(y,x^\mu)}dy^2+g_{\mu\nu}(y,x^\mu)dx^\mu
dx^\nu.} And so, the physical distance between the A-brane ($y=0$)
and the B-brane ($y=l$) is
\formula{DIS}{d(x)=\int_0^le^{\phi(y,x^\mu)}dy,} $\phi(y,x^\mu)$
is the so-called radion. The 5D Einstein's field equations are
given by
\begin{eqnarray}
&&e^{-\phi}\left(e^{-\phi}K^\mu_\nu\right)\!\!_{,y}-\left(e^{-\phi}K\right)\left(e^{-\phi}K^\mu_\nu\right)
 +{}^{(4)}\!R^\mu_\nu-\nabla^\mu\nabla_\nu\phi-\nabla^\mu\phi\nabla_\nu\phi
 \nonumber\\
&&\quad\quad=-\frac{4}{l^2}\delta^\mu_\nu+\kappa_G^2\left(\frac{1}{3}\delta^\mu_\nu\sigma_A+\stackrel{A}{T^\mu_\nu}-
 \frac{1}{3}\delta^\mu_\nu\stackrel{A}{T}\right) e^{-\phi}\delta(y)
 \nonumber\\&&\quad\quad\quad+ \kappa_G^2\left(\frac{1}{3}\delta^\mu_\nu\sigma_B+\stackrel{B}{\tilde{T}^\mu_\nu}-
 \frac{1}{3}\delta^\mu_\nu\stackrel{B}{\tilde{T}}\right)
 e^{-\phi}\delta(y-l),\label{KSEEF}
\end{eqnarray}
\formula{KSEET}{
 \begin{array}{l}
  e^{-\phi}\left(e^{-\phi}K\right)\!\!_{,y}-\left(e^{-\phi}K^{\alpha\beta}\right)
 \left(e^{-\phi}K_{\alpha\beta}\right)-\nabla^\alpha\nabla_\alpha\phi-
 \nabla^\alpha\phi\nabla_\alpha\phi
 \\
 \quad=-\frac{4}{l^2}-\frac{\kappa_G^2}{3}\left(-4\sigma_A+\stackrel{A}{T}\right)e^{-\phi}\delta(y)-
 \frac{\kappa_G^2}{3}\left(-4\sigma_B+\stackrel{B}{\tilde{T}}\right)e^{-\phi}\delta(y-l),
 \end{array}
}
\formula{KSEEY}{\nabla_\nu\left(e^{-\phi}K^\nu_\mu\right)-\nabla_\mu\left(e^{-\phi}K\right)=0,}
where $\nabla_\mu$ denotes the covariant derivative with respect
to $g_{\mu\nu}$ and the extrinsic curvature is
$K_{\mu\nu}=-\frac{1}{2}g_{\mu\nu,y}$. From the singular behaviour
of the equations (\ref{KSEEF}-\ref{KSEEY}) we find the junction
conditions \formula{JCA}{e^{-\phi}\left[K_\nu^\mu-\delta_\nu^\mu
K\right]{\!}_|{}_{y=0^+}=\frac{\kappa_G^2}{2}\left(-\sigma_A\delta^\mu_\nu
+\stackrel{A}{T^\mu_\nu}\right),}
\formula{JCB}{e^{-\phi}\left[K_\nu^\mu-\delta_\nu^\mu
K\right]{\!}_|{}_{y=l^-}=-\frac{\kappa_G^2}{2}\left(-\sigma_B\delta^\mu_\nu
+\stackrel{B}{\tilde{T}^\mu_\nu}\right).}

Using the definitions
\formula{DEFE}{e^{-\phi}K_{\mu\nu}=\Sigma_{\mu\nu}+\frac{1}{4}g_{\mu\nu}Q,\;\;\;\;\;\;Q=
-e^{-\phi}\frac{\partial}{\partial y}\ln\sqrt{-g}.}

The system of equations that we will solve is:
\begin{eqnarray}
e^{-\phi}\Sigma^\mu_{\nu,y}-Q\Sigma^\mu_\nu=
-\Big[&&\!\!\!\!\!\!\!\!\!\!{}^{(4)}\!R^\mu_\nu-\frac{1}{4}\delta^\mu_\nu{}^{(4)}R-
\nabla^\mu\nabla_\nu\phi-\nabla^\mu\phi\nabla_\nu\phi
\nonumber\\
&&\!\!\!\!\!\!+\frac{1}{4}\delta^\mu_\nu\left(\nabla^\alpha\nabla_\alpha\phi+
\nabla^\alpha\phi\nabla_\alpha\phi\right)\Big],\label{EEB1}
\end{eqnarray}
\formula{EEB2}{\frac{3}{4}Q^2-\Sigma^\alpha_\beta\Sigma^\beta_\alpha={}^{(4)}\!R+\frac{12}{l^2},}
\formula{EEB3}{e^{-\phi}Q_{,y}-\frac{1}{4}Q^2-\Sigma^{\alpha\beta}\Sigma_{\alpha\beta}
=\nabla^\alpha\nabla_\alpha\phi+\nabla^\alpha\phi\nabla_\alpha\phi-\frac{4}{l^2},}
\formula{EEB4}{\nabla_\lambda\Sigma_\mu^\lambda-\frac{3}{4}\nabla_\mu
Q=0,} for the bulk equations (obtained from (\ref{KSEEF},
\ref{KSEET}, \ref{KSEEY})), and
\formula{JCAN}{\left[\Sigma_\nu^\mu-\frac{3}{4}\delta_\nu^\mu
Q\right]{\!}\Bigg|_{y=0^+}=\frac{\kappa_G^2}{2}\left(-\sigma_A\delta^\mu_\nu
+\stackrel{A}{T^\mu_\nu}\right),}
\formula{JCBN}{\left[\Sigma_\nu^\mu-\frac{3}{4}\delta_\nu^\mu
Q\right]{\!}\Bigg|_{y=l^-}=-\frac{\kappa_G^2}{2}\left(-\sigma_B\delta^\mu_\nu
+\stackrel{B}{\tilde{T}^\mu_\nu}\right),} for the junction
conditions.

In order to solve a differentiable equation using perturbation
theory we need to identify a suitable small parameter of our
equations. The quantities involved in the equations have two very
distinct characteristic scales. The characteristic length scale of
the extra-dimension is $l\sim10^{-2}cm$ and
$g_{\mu\nu,y}\sim\frac{g_{\mu\nu}}{l}$. The characteristic brane's
curvature length scale is $D$ and is given by
$R\sim\frac{g_{\mu\nu}}{D^2}$. Then from the 4D Einstein's
equations we see $\rho_i\sim\frac{1}{{}^{(4)}\!G_ND^2}$ and
equations (\ref{RSR}, \ref{MPL3}) imply
$\sigma\sim\frac{1}{{}^{(4)}\!G_Nl^2}$, so
$\frac{\rho_i}{\sigma}\sim\left(\frac{l}{D}\right)^2$.
Astrophysical (or cosmological) events have length scales $D\gg
l$, so the natural small parameter is
$\epsilon=\frac{\rho_i}{\sigma}\sim\left(\frac{l}{D}\right)^2$.
This means that the energy density of the matter is small compared
to the brane's tension but it does not necessarily imply weak
gravity.

The next steps in the derivation are completely analogous to any
perturbative scheme. We expand the unknown functions in a series
\formula{SERIES}{g_{\mu\nu}(y,x^\mu)=a^2(y,x)\left[h_{\mu\nu}(x^\mu)+
\stackrel{(1)}{g_{\mu\nu}}(y,x^\mu)+\stackrel{(2)}{g_{\mu\nu}}(y,x^\mu)+\ldots\right],}
with the boundary conditions on the A-brane reading
\begin{subequations} \label{BC}
 \begin{eqnarray}
  g_{\mu\nu}(y=0,x^\mu)&=&h_{\mu\nu}(x^\mu),\label{BC1}
  \\
  \stackrel{(n)}{g_{\mu\nu}}(y=0,x^\mu)&=&0, \;\;\mbox{with
}  n=1,2,\cdots. \label{BC2}
 \end{eqnarray}
\end{subequations} Then we plug them back into the initial system and we will obtain successive
systems for the different orders in $\epsilon$. The $0^{th}$ order
system has the solution
\formula{dS20}{ds^2=e^{2\phi(y,x)}dy^2+a^2(y,x)h_{\mu\nu}(x)dx^\mu
dx^\nu,} with $a(y,x)=exp\left[-\frac{1}{l}\int_0^ydy
e^{\phi(y,x)}\right].$ To proceed we will assume
$\phi(y,x)\equiv\phi(x)$ thus
$a(y,x)=exp\left[-\frac{y}{l}e^\phi\right].$ In the $1^{st}$ order
system, the curvature term that has been ignored in the $0^{th}$
order calculation comes into play. After solving the set of
equations, it is possible to show that
$\stackrel{(1)}{g}_{\mu\nu}$ is written in terms of
$h_{\mu\nu}(x)$, $\phi(x)$ and an integration constant
$\chi_{\mu\nu}(x)$ (traceless and transverse) that basically is
the Weyl's tensor contribution. The junction condition for the
A-brane gives
\formula{EIE}{\frac{l}{2}G^\mu_\nu(h)+\chi^\mu_\nu=\frac{\kappa_G^2}{2}\stackrel{A}{T^\mu_\nu},}
where $\chi^\mu_\nu$ is called generalized dark radiation tensor
and contains the effect of the bulk and the second brane.
Combining (\ref{EIE}) and its analogy for the B-brane one obtains
\begin{eqnarray}
G^\mu_\nu(h)&=&\frac{\kappa_G^2}{l\Psi}\stackrel{A}{T^\mu_\nu}+
\frac{\kappa_G^2\left(1-\Psi\right)}{l\Psi}\stackrel{B}{T^\mu_\nu}+
\frac{1}{\Psi}\left(\Psi^{|\mu}_{|\nu}-\delta^\mu_\nu\Psi^{|\alpha}_{|\alpha}\right)
\nonumber\\
&&+\frac{\omega(\Psi)}{\Psi^2}\left(\Psi^{|\mu}\Psi_{|\nu}-
\frac{1}{2}\delta^\mu_\nu\Psi^{|\alpha}\Psi_{|\alpha}\right),\label{4KSEE}
\end{eqnarray}
where $\Psi\equiv 1-\Omega^2\equiv 1-e^{-2e^\phi}$ and
$\omega(\Psi)=\frac{3}{2}\frac{\Psi}{1-\Psi}$. The radion field
obeys
\formula{RADE}{\Box\Psi=\frac{\kappa_G^2}{l}\frac{\stackrel{A}{T}+
\stackrel{B}{T}}{2\omega+3}-\frac{1}{2\omega+3}\frac{d\omega}{d\Psi}\Psi^{|\mu}\Psi_{|\mu}.}
The important point about equations (\ref{4KSEE}) and (\ref{RADE})
is that they do not include the Weyl's tensor term $\chi^\mu_\nu$,
although, they include the energy momentum tensor of the B-brane,
KS called this theory ``quasi-scalar-tensor" gravity.

Equations (\ref{4KSEE}) and (\ref{RADE}) are the only equations we
need to solve to determine the full spacetime dynamics. Let us see
what these equations tell us about the cosmology.

Because we are interested in isotropic and homogeneous
cosmological models we will choose $h$ to be the FRW metric. We
will assume that the matter on the branes has a perfect fluid form
and so the energy momentum tensor of the A-brane and B-brane will
be respectively
\begin{subequations} \label{TS}
 \begin{eqnarray}
  \stackrel{A}{T_\nu^\mu}&=&diag(-\rho_A,P_A,P_A,P_A),\label{TA}
  \\
  \stackrel{B}{T_\nu^\mu}&=&\Omega^2diag(-\rho_B,P_B,P_B,P_B),\label{TB}
 \end{eqnarray}
\end{subequations}
where the $\Omega^2$ factor results from the fact that the
B-brane's metric is $\Omega^2h_{\mu\nu}$. The symmetries imply
that $\Psi$ only depends on time. Using the physical distance
between branes instead of $\Psi$ (related through
$\Psi=1-e^{-2d/l}$) we can write (\ref{4KSEE}) as
\begin{subequations} \label{SM}
 \begin{equation}
  3\left(1-e^{-2d/l}\right)\left(H^2+\frac{k}{a^2}\right)=
  \frac{\kappa_G^2}{l}\left(\rho_A+\rho_Be^{-4d/l}\right)
  +3\frac{\dot{d}}{l}\left(\frac{\dot{d}}{l}-2H\right)e^{-2d/l},\label{SMT}
 \end{equation}
 \begin{eqnarray}
  -\left(1-e^{-2d/l}\right)\left(2\frac{\ddot{a}}{a}+H^2+\frac{k}{a^2}\right)\!\!\!&=&\!\!\!
  \frac{\kappa_G^2}{l}\left(P_A+P_Be^{-4d/l}\right)\nonumber\\&&+
  e^{-2d/l}\left(\frac{2\ddot{d}}{l}-\frac{\dot{d}^2}{l^2}+\frac{4H\dot{d}}{l}\right),\label{SMX}
 \end{eqnarray}
\end{subequations}
and equation (\ref{RADE}) as
\formula{RADED}{\ddot{d}-\frac{\dot{d}^2}{l}+3H\dot{d}=
\kappa_G^2\frac{\left(\rho_A-3P_A\right)+\left(\rho_B-3P_B\right)e^{-2d/l}}{6}.}
Combining these three last equations we can find
\formula{H2}{\frac{\ddot{a}}{a}+H^2+\frac{k}{a^2}=\frac{\kappa_G^2}{6l}\left(\rho_A-3P_A\right).}
If we assume the equation of state $P_A=\omega\rho_A,$ the usual
energy conservation laws in the A-brane imply $\rho_A\sim
a^{-3(1+\omega)}$. Now (\ref{H2}) can be integrated once to give
\formula{FRIEDMANN}{H^2=\frac{\kappa_G^2}{3l}\rho_A-\frac{k}{a^2}+\frac{C}{a^4}.}
This is the well known low energy limit of the modified Friedman's
equation that we have already encountered before in subsection
\ref{subsec:CS}. Once again $C$ appears just as an integration
constant.

As a last remark, I should point out that the effective equations
of motion for gravity on the brane can be consistently derived
from the action
\begin{eqnarray}
S&=&\frac{l}{2\kappa_G^2}\int d^4x\sqrt{-h}\left(\Psi
R-\frac{\omega(\Psi)}{\Psi}\Psi^{|\alpha}\Psi_{|\alpha}\right)
\nonumber\\&&
+\int d^4x\sqrt{-h}L^A+\int
d^4x\sqrt{-h}\left(1-\Psi\right)^2L^B.\label{KSA}
\end{eqnarray}
The solutions at higher orders in the method will be left out of
this introductory chapter, see \cite{Kanno:2002iaa,Kanno:2002ia}
for further details.

In the next three chapters of this thesis we will apply the
gradient expansion method that we have just presented to obtain
the 4D effective theory of gravity for three models with different
number of extra-dimensions and different physical properties. The
models that we will study are: a 5D dilatonic two brane model that
reduces to the Ho\v{r}ava-Witten theory and to the Randall-Sundrum
model for a particular choice of parameters, a 6D supergravity
model and the 10D type IIB supergravity. The next three sections
of this introduction shall be devoted to the presentation of the
essential features of these three models.

\section{\label{sec:introHoravaWitten}The Ho\v{r}ava-Witten theory}

\begin{figure}[t]
\centering
 \scalebox{0.7}
 {\rotatebox{0}{
    \includegraphics{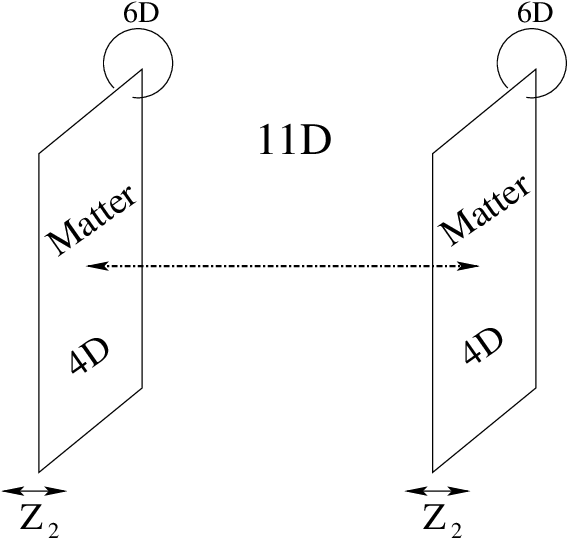}
                 }
 }
\caption{The 11D Ho\v{r}ava-Witten theory compactified on a
$S^1/\mathbb{Z}_2$ orbifold with two end of the world branes. The
six small extra dimensions are compactified on a Calabi-Yau.
}\label{fig:HoravaWitten}
\end{figure}

In this section we will introduce the general features of the
Ho\v{r}ava-Witten theory \cite{Horava:1995qa,Horava:1996ma}.
Ho\v{r}ava and Witten showed that the eleven-dimensional limit of
M-theory compactified on $S^1/\mathbb{Z}_2$ orbifold can be
identified with the strongly coupled 10D $E_8\times E_8$ heterotic
string theory with two orbifold fixed planes (branes). This model
provides a natural setup in which standard model matter fields are
localized on the branes. A schematic representation of this model
can be found in figure \ref{fig:HoravaWitten}.
Witten also showed that this theory can be consistently
compactified so that the universe appears as a 5D spacetime with
two boundary branes \cite{Witten:1996mz}. The six extra dimensions
were compactified in a Calabi-Yau manifold and the action of the
5D effective theory was derived \cite{Lukas:1998yy,Lukas:1998tt}.
These references showed that although the dimensional reduction of
the graviton and the 4-form flux generate a large number of
fields, it is consistent to retain only the 5D graviton and a
scalar field $\phi$. This scalar field is related to the volume of
the Calabi-Yau  manifold as $V_{CY}=e^\phi$. The resulting
dimensionally reduced effective action in the bulk is
\begin{equation}
S_{bulk}=\frac{1}{2\kappa_G^2}\int
d^5x\sqrt{-g_5}\left({}^{(5)}\!\!R-\partial_M\phi\partial^M\phi+V_{bulk}(\phi)\right),
\end{equation}
where $\kappa_G^2$ denotes the 5D gravitational constant and
$V_{bulk}$ is the potential of the bulk scalar field. It is given
by
\begin{equation}
V_{bulk}(\phi)=-\frac{\sigma^2}{3}e^{-2\sqrt{2}\phi},
\end{equation}
where $\sigma$ is the tension of the 3-brane.

The topology of the 5D spacetime is $S^1/\mathbb{Z}_2\times M_4$
where $M_4$ is a smooth manifold. We label the coordinate of the
extra dimension by $y$ and the two endpoints of this space are
located at $y=0$ and $y=r$. We place the positive tension brane at
$y=0$ and the negative tension brane at $y=r$. The branes' actions
are given as
\begin{equation}
S_+=-\frac{\sqrt{2}}{\kappa_G^2}\sigma\int
d^4x\sqrt{-g_4}e^{-\sqrt{2}\phi},\quad
S_-=\frac{\sqrt{2}}{\kappa_G^2}\sigma\int d^4x\sqrt{-\tilde
g_4}e^{-\sqrt{2}\phi},
\end{equation}
where the subscript plus or minus refers to the tension of the
respective brane, so that for instance $S_+$ denotes the action of
the positive tension brane.

The cosmological evolution of this model including matter on the
branes and arbitrary potentials on the branes and in the bulk have
been studied in \cite{Kobayashi:2002pw}. In
\cite{Kobayashi:2002pw}, the authors also obtain the 4D effective
equations that govern the low-energy dynamics of the brane-worlds.
Reference \cite{Lehners:2006pu} showed that this model possesses a
unique global solution that represents the collision of the
branes. They also showed that away from the collision the Riemann
curvature is bounded and they argued that it offers the prospect
of modelling the big bang singularity as a brane collision in an
higher-dimensional spacetime. This scenario is called Ekpyrotic
universe and early proposals can be found in
\cite{Khoury:2001wf,Khoury:2001bz}. Some dynamical solutions have
been found \cite{Chen:2005jp}, but they are unstable. In these
solutions both branes collide with a time-like naked singularity
resulting in the annihilation of the entire spacetime. In the next
chapter, we shall investigate the origin of this instability using
the gradient expansion method. We shall obtain the 4D effective
theory and we will show that the previous exact dynamical
solutions can be reproduced within the 4D theory.

\section{\label{sec:intro6D}The self-tuning 6D brane-world}
\subsection{\label{subsec:EM}The 6D Einstein-Maxwell model}

Co-dimension two brane-worlds in a 6D bulk have recently attracted
much attention as they possess the remarkable property that the
brane's tension does not curve the 4D spacetime but its only
effect is in the extra-dimension, see \cite{Koyama:2007rx} for a
review. This property was called self-tuning and was interpreted
as a promising solution for the cosmological constant problem.

In Chapter \ref{chapter:6D} we will study a 6D supergravity
brane-world model and we shall obtain the low energy effective
theory. Before as an introduction let us see what is the
cosmological constant (CC) problem and how the 6D Einstein-Maxwell
model tries to solve it. The CC problem consists in the huge
difference between the value of the vacuum energy density
predicted by particle physics, $\rho_{vac}\sim(TeV)^4$ and the
observed value of $\rho_{vac}\sim(10^{-3}eV)^4$. There is no
convincing explanation of why the observed value is so small and
why does the CC only started to dominate the energy density of the
universe recently.

In the 6D Einstein-Maxwell model \cite{Carroll:2003db}, the action
is
\begin{equation}
S=\int
d^6x\sqrt{-g}\left(\frac{M^4_6}{2}{}^{(6)}\!\!R-\Lambda_6-\frac{1}{4}F_{ab}F^{ab}\right),
\end{equation}
where $M_6$ is the 6D Planck mass, $\Lambda_6$ is the 6D
cosmological constant and $F_{ab}$ is a Maxwell field that is used
to stabilize the size of the extra-dimensions. ${}^{(6)}\!R$
denotes the 6D Ricci scalar. The ansatz for the line element is
\begin{equation}
ds^2=\eta_{\mu\nu}dx^\mu dx^\nu+\gamma_{ij}dx^idx^j,
\end{equation}
where $\eta_{\mu\nu}$ denotes the Minkowski metric and
$\gamma_{ij}$ denotes the internal metric. The gauge field is
chosen to have non-zero components only in the compact directions
and it takes the form
\begin{equation}
F_{ij}=\sqrt{\gamma}B_0\epsilon_{ij},
\end{equation}
where $B_0$ is a constant, $\gamma$ is the determinant of
$\gamma_{ij}$ and $\epsilon_{ij}$ is the Levi-Civita tensor. A
static and stable solution is \cite{Carroll:2003db},
\begin{equation}
\gamma_{ij}dx^idx^j=a_0^2\left(d\theta^2+\mathrm{sin}^2\theta
d\varphi^2\right),
\end{equation}
where the radius of the 2-sphere $a_0$ and the magnetic field
strength $B_0$ are given in terms of the bulk cosmological
constant as
\begin{equation}
a_0^2=\frac{M^4_6}{2\Lambda_6},\quad B_0^2=2\Lambda_6.
\end{equation}
In order to have a 4D Minkowski spacetime, $B_0$ has to be tuned
to the value of the bulk's CC. A different value of $B_0$ would
induce a de-Sitter or anti-de-Sitter 4D geometry. This is the CC
problem.

\begin{figure}[t]
\centering
 \scalebox{.5}
 {\rotatebox{0}{
    \includegraphics*[0mm,40mmm][150mm,130mm]{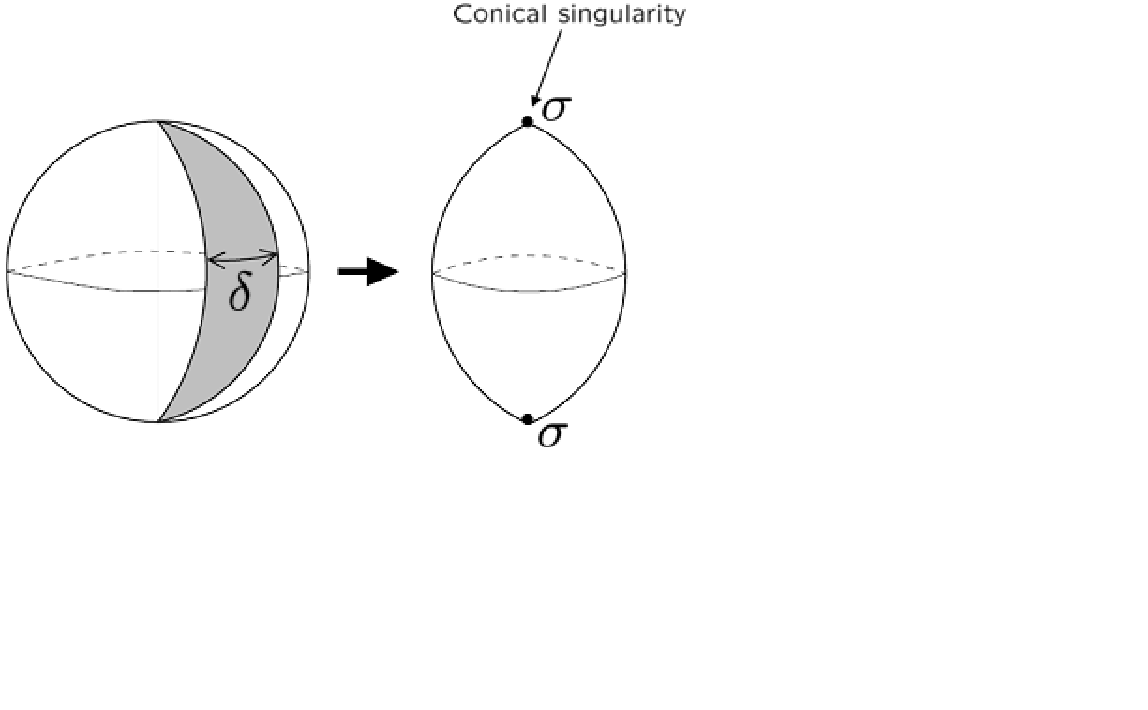}
                 }
 }
\caption{The rugby ball geometry with two branes of tension
$\sigma$ at the conical singularities.}\label{fig:rugbyball}
\end{figure}

Now we can introduce branes with tension at the poles of the
sphere. The branes' action is
\begin{equation}
S_{branes}=-\sigma\int d^4x\sqrt{g},
\end{equation}
where $\sigma$ is the branes' tension. The inclusion of branes
will only modify the extra-dimensions geometry to that of a
``rugby ball"
\begin{equation}
\gamma_{ij}dx^idx^j=a_0^2\left(d\theta^2+\alpha^2\mathrm{sin}^2\theta
d\varphi^2\right),
\end{equation}
where
\begin{equation}
\alpha=1-\frac{\sigma}{2\pi M_6^2}.
\end{equation}
The internal space can de depicted as a sphere with a wedge
(stretching from the north pole to the south pole) taken out and
the open sides glued together, see Figure \ref{fig:rugbyball}. In
other words, the branes introduce a deficit angle
$\delta=2\pi(1-\alpha)$ in the sphere.
Now, we can easily see that the 4D geometry is independent of the
branes' tension $\sigma$. The branes' vacuum energy does not curve
the 4D spacetime as it would in 4D general relativity. This
property was called self-tuning. It does not solve the CC problem
because we still need to tune the value of $B_0$, but this idea
looks promising as it breaks the standard relation between the 4D
vacuum energy and the curvature of the 4D spacetime. Also, if the
self-tuning mechanism worked, it would push the value of the CC to
zero but we would still have to find a way to generate a small CC
to agree with observations.

There are several objections to the idea of self-tuning
\cite{Garriga:2004tq}. The simplest argument is as follows.
Suppose that we have tuned the initial value of $B_0$ to
$B_0^2=2\Lambda_6$ and suppose that the branes' tension were to
change from an initial value to a different final value. This
would imply that the area of the extra-dimensions ($4\pi\alpha$)
would change too, because $\alpha$ is a function of $\sigma$. On
the other hand, the magnetic flux in the internal space has to be
conserved, this necessarily implies that the final magnetic field
strength cannot satisfy $B_{0final}^2=2\Lambda_6$. Therefore the
4D spacetime will no longer be a static Minkowski spacetime but
rather a time-dependent spacetime.
\subsection{\label{subsec:SLED}Supersymmetric large extra-dimensions}

In the previous subsection, we have seen that the failure of the
self-tuning is directly related to the tuning of the magnetic
field strength with the bulk CC, which is necessary to obtain 4D
Minkowski spacetime. The so-called supersymmetric large
extra-dimension (SLED) model was proposed to try to address this
problem. For excellent review articles see \cite{Koyama:2007rx,
Burgess:2004kd, Burgess:2004ib}.

The action of this model is similar to the action of the
Einstein-Maxwell model with an additional field $\phi$, the
dilaton, it reads
\begin{equation}
S=\int
d^6x\sqrt{-{}^{(6)}\!\!g}\left[\frac{M^4_6}{2}\left({}^{(6)}\!\!R-\partial_a\phi\partial^a\phi\right)
-e^\phi\Lambda_6-\frac{1}{4}e^{-\phi}F_{ab}F^{ab}\right].
\end{equation}
Comparing this model with the Einstein-Maxwell model it is easy to
see that there exists a solution with $\phi=\phi_0$, where
$\phi_0$ is a constant. It is also a solution of the
Einstein-Maxwell model with the rescalings
$\Lambda_6\rightarrow\Lambda_6e^{-\phi}$, $B_0^2\rightarrow
B_0^2e^\phi$, provided that $\phi$ is at the minimum of its
potential
\begin{equation}
V'(\phi_0)=-\frac{1}{2}B_0^2e^{-\phi_0}+\Lambda_6e^\phi=0.
\end{equation}
This is the condition to have a flat geometry on the brane
$B_0^2e^{-\phi_0}=2\Lambda_6e^\phi$. In other words, at the
minimum, the vacuum energy is zero and therefore the geometry is
Minkowski, this was called self-tuning.

As before, there are several objections to the idea of
self-tuning, for example, Refs. \cite{Garriga:2004tq,
Vinet:2005dg} derived the 4D effective theory using the metric
ansatz
\begin{equation}
ds^2=g_{\mu\nu}(x)dx^\mu
dx^\nu+M_6^{-2}e^{-2\psi(x)}\left(dr^2+\mathrm{sin}^2rd\theta^2\right),\label{GPmetric}
\end{equation}
and $\phi=\phi(x)$. The 4D effective potential is found to be
\begin{equation}
V(\psi,\phi)=M_6^{-4}e^{\sigma_2}U(\sigma_1), \quad
U(\sigma_1)=\frac{B_0^2}{2\alpha^2}e^{-2\sigma_1}-2M_6^2e^{-\sigma_1}+2\Lambda_6,
\end{equation}
where $\sigma_1=2\psi+\phi$, $\sigma_2=2\psi-\phi$ and $\alpha$ is
related to the branes' tension as $\alpha=1-\frac{\sigma}{2\pi
M_6^2}$. They argued that if we start with a configuration in
which $V(\psi,\phi)=0$, and if we change the branes' tension to a
different final value then $V(\psi,\phi)$ will be different from
zero and $\sigma_2$ will have a runaway potential. The 4D
spacetime becomes non-static.

This argument against self-tuning has raised criticisms
\cite{Vinet:2005dg}. For example, it is argued that the metric
ansatz (\ref{GPmetric}) is restrictive. There is a known class of
static solutions with warping in the bulk that cannot be described
by this ansatz. These solutions might play a role if self-tuning
is to work. Chapter \ref{chapter:6D} is dedicated to the study of
some aspects of this model and we shall derive the low energy
effective theory including warping.

\section{\label{sec:intro10D}Warped compactifications in Type IIB supergravity}

In chapter \ref{chapter:10D}, we will present a method to obtain
the 4D effective theory for warped compactifications including
fluxes and branes in the 10D type IIB supergravity. In this
section we shall introduce the general idea of warped
compactification in type IIB supergravity (sugra). For further
details about type IIB sugra we refer the reader to chapter
\ref{chapter:10D} or to the extensive literature on the subject
\cite{StringT, Quevedo:2002xw, Grana:2005jc} and references
therein.

\begin{figure}[t]
\centering
 \scalebox{.5}
 {\rotatebox{0}{
    \includegraphics*{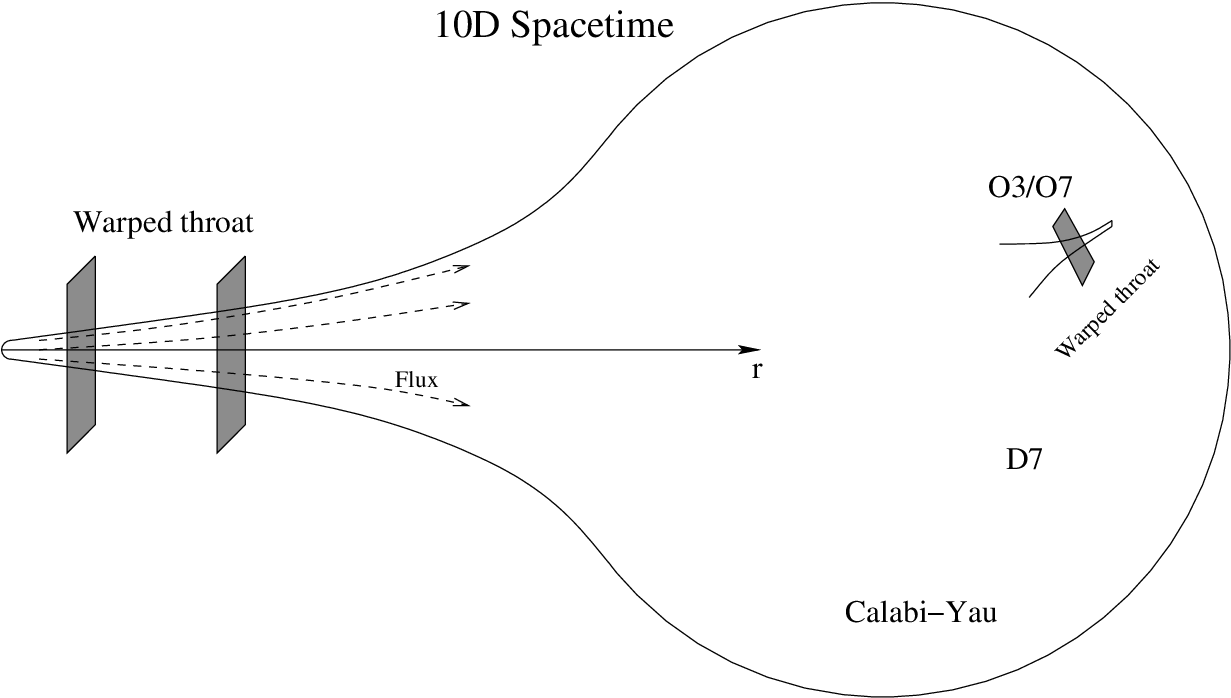}
                 }
 }
\caption{A warped throat compactification.}\label{fig:fig10D}
\end{figure}

The desire of embedding the Randall and Sundrum (RS) solution to
the hierarchy problem in a string theory compactification was
fulfilled with the work of Giddings, Kachru and Polchinski (GKP)
\cite{Giddings:2001yu}. They realized that a warped string
compactification can generate a large hierarchy between the
electroweak scale and the Planck scale in a similar way to the RS
model \cite{Randall:1999ee}. To generate the warping of the
extra-dimensions they needed to use not only positive tension
objects like $D3$ branes but also negative tension objects. Figure
\ref{fig:fig10D} is a schematic representation of their model.

They start with a 10D spacetime, with six of the dimensions
compactified in a Calabi-Yau manifold. In some region of the
internal space, a throat develops. Fluxes and branes curve this
space, giving origin to a warp factor $h(y)$. Their solution for
the metric is
\begin{equation}
ds^2_{10}=h^{-\frac{1}{2}}(y)\eta_{\mu\nu}dx^\mu dx^\nu +
h^{\frac{1}{2}}(y)\gamma_{mn}(y)dy^mdy^n,
\end{equation}
where $\gamma_{mn}$ and $y^m$ denote a Calabi-Yau metric and the
coordinates of the extra-dimensions respectively. $x^\mu$ denotes
the four-dimensional coordinates. The throat region of the
spacetime is given by the solution found by Klebanov and Strassler
\cite{Klebanov:2000hb} and it is well approximated by a 5D
anti-de-Sitter spacetime with the remaining five extra-dimensions
compactified in a small 5-sphere. In this region the RS mechanism
of solving the hierarchy problem works as before.

These warped compactifications also provide models of inflation,
where the inflaton is identified with the position of a brane in
the extra-dimension $r$, where $r$ is a radial direction and it
corresponds to one of the $y^m$. Brane-inflation models will be
discussed in the chapter \ref{chapter:inflationintro}.

\chapter{Moduli instability in warped compactification}\label{chapter:5D}
\chaptermark{Moduli instability in warped compact. ...}
\section{\label{sec:INTRO}Introduction}
The idea of using the degrees of freedom of extra spatial
dimensions to unify gravity and the other interactions has
attracted much interest. Especially, M-Theory has the potential to
achieve this ultimate goal. Ho\v{r}ava and Witten showed that
M-Theory compactified as $S^1/Z_2 \times M^{10}$ reduces to a
known string theory ($E_8\times E_8$ Heterotic String)
\cite{Horava:1995qa}. From a low energy effective theory point of
view, this model has a 5D spacetime (the bulk) with two 4D
boundary hypersurfaces (the branes) \cite{Lukas:1998yy}. The
remaining six spatial dimensions are assumed to be compactified.
All the Standard Model particles are confined to the branes while
gravity can propagate in the bulk. As a consequence of the
compactification of the six spatial dimensions, a 5D effective
scalar field appears in the bulk, which describes the volume of
the compactified 6D space.

A novel aspect of this model is that the fifth dimension is not
homogeneous. The line element for a static solution is given in
the form $ds_5^2=h^{\alpha}(r)dr^2+h^{\beta}(r)ds_4^2(x)$, where
$r$ is the fifth coordinate, $\alpha$, $\beta$ are constants and
$h$ is called the warp factor. In the previous chapter, we showed
how this warped geometry was used by Randall and Sundrum
\cite{Randall:1999vf} to address the hierarchy problem. They have
also showed that it is possible to localize gravity around the
brane with the warped geometry, providing in this way an
alternative to compactification \cite{Randall:1999ee}.

A new class of dynamical solutions that describes an instability
of the warped geometry has been found. Chen \emph{et al.} noticed
that it is possible to obtain a dynamical solution by replacing
the constant modulus in the warp factor $h$ by a linear function
of the 4D coordinates \cite{Chen:2005jp}. This solution describes
an instability of the model as the brane will hit the singularity
in the bulk, where $h=0$. This kind of solution exists also in the
10D type IIB supergravity \cite{Kodama:2005fz}.

In this chapter, we study the moduli instability in a dilatonic
two brane model, where the potentials for the scalar field on the
brane and in the bulk obey the Bogomol'nyi-Prasad-Sommerfield
(BPS) condition \cite{Cvetic:2000pn}. For particular values of
parameters we retrieve either the Ho\v{r}ava-Witten theory or the
Randall-Sundrum model. We identify the origin of the moduli
instability using a 4D effective theory derived in Ref.
\cite{Kobayashi:2002pw} (see Refs. \cite{Brax:2002nt} and
\cite{Kanno:2005zr} for different approaches). This effective
theory is derived by solving the 5D equations of motion using the
gradient expansion method
\cite{Kanno:2002ia,Wiseman:2002nn,Shiromizu:2002qr}, where we
assume that the velocities of the branes are small compared with
the curvature scale of the bulk determined by the warp factor.

Despite the fact that the 4D effective theory is based on a
slow-motion approximation, we will show that the 5D exact
solutions can be reproduced in the 4D effective theory. In order
to understand the relation between 4D solutions in an effective
theory and full 5D solutions, we revisit the gradient expansion
method by employing a new metric ansatz. Using this metric ansatz,
we can clearly see why the moduli instability solution in the 4D
effective theory can be lifted to an exact 5D solution.

We also comment on a claim that the 4D effective theory allows a
much wider class of solutions than the 5D theory
\cite{Kodama:2005cz}. We disagree with that conclusion and we show
that it is based on the restricted form of the 5D metric ansatz
used in Ref. \cite{Kodama:2005cz}. Using our metric ansatz, we
provide a way to lift solutions in the 4D effective theory to 5D
solutions perturbatively in terms of small velocities of the
branes.

The structure of this chapter is as follows. In section
\ref{sec:MODEL1}, the model under consideration is described in
detail. In section \ref{sec:GEM}, we identify the solution in the
4D effective theory that describes the moduli instability. In
section \ref{sec:GEMN}, we revisit the gradient expansion method
to derive the 4D effective theory. We propose a new metric ansatz
which is useful to relate 4D solutions in the effective theory to
5D solutions. Using this formalism we explain why the 4D solution
for the moduli instability can be lifted to an exact 5D solution.
In section \ref{sec:COMM}, we comment on the arguments against the
4D effective theory. Section \ref{sec:CON} is devoted to
conclusions.

\section{\label{sec:MODEL1}The model}

\begin{figure}
\centering
 \scalebox{.7}
 {\rotatebox{0}{
    \includegraphics*{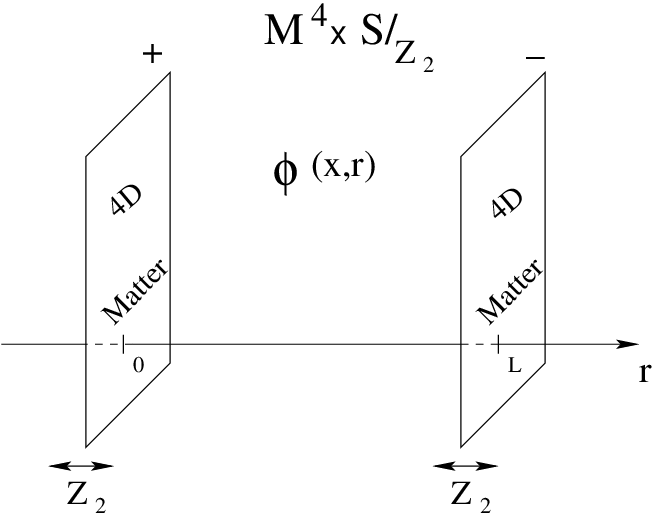}}
 }
\caption{The dilatonic brane-world universe}\label{fig:fig2}
\end{figure}

Our model consists of a 5D spacetime (bulk) filled with a scalar
field. The fifth dimension is a compact space $S^1$ with a $Z_2$
symmetry (i.e. identification $r\rightarrow -r$) and will be
parameterized by the coordinate $r$, $0\leq r\leq L$. The bulk
action takes the form
\formula{Sbulk}{S_{bulk}=\frac{1}{2\kappa_G^2}\int
d^5x\sqrt{-g_5}\left[{}^5\!R-\nabla_M\varphi\nabla^M\varphi+V_{bulk}(\varphi)\right],}
where $\kappa_G$ denotes the 5D gravitational constant. Throughout
this work Latin indices can have values in $\{r,t,x,y,z\}$, while
Greek indices do not include the extra dimension coordinate $r$.
The scalar field potential will be
\formula{Vbulk}{V_{bulk}(\varphi)=-\left(b^2-\frac{2}{3}\right)e^{-2\sqrt{2}b\varphi}\sigma^2,}
where $\sigma$ and $b$ are the remaining parameters of this model.
We can retrieve either the Randall-Sundrum model ($b=0$) or the
Ho\v{r}ava-Witten model ($b=1$) according to the value of $b$. The
orbifold $S^1/Z_2$ has two fixed points, at $r=0$ and $r=L$, and
we can put branes there. The branes' action is
\formula{Sbranes}{S_{branes}=S_++S_-,} where $S_+$ and $S_-$ are
the positive and negative tension brane action, respectively. They
are given by
\formula{Sbrane+-}{S_\pm=\mp\frac{\sqrt{2}}{\kappa_G^2}\int
d^4x\sqrt{-g_4}e^{-\sqrt{2}b\varphi} \sigma.} The action of our
model is \formula{ACTT}{S_{total}=S_{bulk}+S_{branes}.} A
schematic picture of our model is shown in Fig. \ref{fig:fig2}

\section{\label{sec:GEM}Exact solutions for moduli instability
in 4D effective theory}

\subsection{\label{subsec:SOL}The exact 5D solution}

Exact static solutions for the equations of motion that result
from (\ref{ACTT}) were obtained in \cite{Cvetic:2000pn}. An
interesting dynamical solution was found by Chen \emph{et al.}
\cite{Chen:2005jp}. They noticed that if we add a linear function
of time to the warp factor of the solution of
\cite{Cvetic:2000pn}, it would still be a solution of the
equations of motion. Their exact 5D solution reads
\begin{eqnarray}
dS_5^2&=&\left(h\tau-\frac{|r|}{l}\right)^\frac{6b^2-2}{3b^2+1}dr^2+\left(h\tau-\frac{|r|}{l}\right)^{\frac{2}{3b^2+1}}\eta_{\mu\nu}dx^\mu
dx^\nu,\label{ESMSM}
\\
\varphi&=&\frac{3\sqrt{2}b}{3b^2+1}\ln
\left(h\tau-\frac{|r|}{l}\right), \quad
l=\frac{3}{3b^2+1}\frac{\sqrt{2}}{\sigma}, \label{ESMSD}
\end{eqnarray}
where $h$ is an arbitrary constant.

From the above equations we can read the scale factor of the
positive tension brane ($r=0$) as
\formula{sf+}{a^2_+(\tau)=\left(h\tau\right)^\frac{2}{3b^2+1},}
and the scalar field as
\formula{scf+}{\varphi(r=0,\tau)=\frac{3\sqrt{2}b}{3b^2+1}\ln(h\tau).}

Let us choose $h<0$ and $\tau <0$. In this case the proper
distance between branes, the so-called radion, is decreasing
according to
\formula{RAD}{\mathcal{R}=\int_0^L\left(h\tau-\frac{r}{l}\right)^{\frac{3b^2-1}{3b^2+1}}dr=l\frac{3b^2+1}{6b^2}(h\tau)^\frac{6b^2}{3b^2+1}\left[1-\left(1-\frac{L/l}{h\tau}\right)^{\frac{6b^2}{3b^2+1}}\right].}
When $\tau\rightarrow-\infty$, $\mathcal{R}\rightarrow
L\left(h\tau\right)^\frac{3b^2-1}{3b^2+1}$ and for $\tau=L/hl$,
$\mathcal{R}=l(3b^2+1)\left(L/l\right)^\frac{6b^2}{3b^2+1}/6b^2$.
Before the two branes collide, a curvature singularity will appear
in the negative tension brane ($r=L$) at $\tau=L/hl$. This
singularity will move towards the positive tension brane and will
reach it at $\tau=0$. This event represents the total annihilation
of the spacetime.

It is useful to note that if we drop the modulus sign in
(\ref{ESMSM}), the bulk spacetime is a static black brane or black
hole solution depending on $b$, see \cite{Chen:2005jp} for more
details. If $b=1$, there is a timelike curvature singularity at
$h\tau=|r|/l$. From the static bulk point of view, the two branes
are moving in this static bulk and the negative tension brane
first hits the singularity. Even if the bulk spacetime is static,
the existence of the branes, which gives the modulus sign for $r$,
makes the spacetime truly time dependent.

Moduli instability is a serious problem for these types of model.
In this work, we will try to understand it from a 4D effective
theory viewpoint.
\subsection{\label{subsec:ACT}The exact solutions in 4D effective theory}

In \cite{Kobayashi:2002pw}, Kobayashi and Koyama applied the
gradient expansion method to solve perturbatively  the 5D
equations of motion resulting from (\ref{ACTT}). Their $0^{th}$
order solution reads
\begin{eqnarray}
dS_5^2&=&d(t)^2e^{2\sqrt{2}b\phi(t)}dr^2+\mathcal{F}(r,t)^\frac{1}{3b^2}
h_{\mu\nu}(t)dx^\mu dx^\nu,\label{KK0th1}
\\
\varphi(r,t)&=&\frac{1}{\sqrt{2}b}\ln \mathcal{F}(r,t)+\phi(t),
\quad \mathcal{F}(r,t)=1-\sqrt{2}b^2d(t)\sigma|r|. \label{KK0th2}
\end{eqnarray}
The main result of the paper is a set of equations involving only
four dimensional quantities that describe the dynamics of the
unknown functions $d(t)$, $h_{\mu\nu}(t)$ and $\phi(t)$, appearing
in Eqs. (\ref{KK0th1}) and (\ref{KK0th2}). It can be shown that
their dynamical equations can be consistently deduced from the
following action
\formula{EFFACT}{S_{eff}=\frac{l}{2\kappa_G^2}\int
d^4x\sqrt{-h}e^{\sqrt{2}b\phi}\left[\psi
R(h)-\frac{3}{2(1+3b^2)}\frac{1}{1-\psi}\nabla_\alpha\psi\nabla^\alpha\psi-
\psi\nabla_\alpha\phi\nabla^\alpha\phi\right],} where here
$\nabla_\alpha$ means the covariant derivative with respect to
$h_{\mu\nu}$ and $\psi=1-\left(1-d\right)^\frac{3b^2+1}{3b^2}$.
This is the action we obtain if we substitute Eqs. (\ref{KK0th1})
and (\ref{KK0th2}) into (\ref{ACTT}) and integrate over the extra
dimension \cite{Brax:2002nt}.

Performing the conformal transformation
$h_{\mu\nu}=e^{-\sqrt{2}b\phi}f_{\mu\nu}/\psi$ and defining new
scalar fields as
\formula{psi1}{\psi=1-\tanh^2\left[\sqrt{\frac{1+3b^2}{6}}\Psi\right],}
\formula{phih}{\phi=\frac{\Theta}{\sqrt{1+3b^2}}+\lambda\ln\psi,
\quad \lambda\equiv-\frac{3}{\sqrt{2}}\frac{b}{3b^2+1},} we arrive
at the action in the Einstein frame
\formula{EFFACTEF}{S_E=\frac{l}{2\kappa_G^2}\int
d^4x\sqrt{-f}\left[R(f)
-\Theta_{|\alpha}\Theta^{|\alpha}-\Psi_{|\alpha}\Psi^{|\alpha}\right].}

It is clear that the moduli fields have no potential and this is
the origin of the instability. We are interested in cosmological
solutions so we choose $f_{\mu\nu}$ to be a flat FRW metric.  The
Einstein equations resulting from the action (\ref{EFFACTEF}) are
\formula{Gxxtt}{\frac{a''}{a}=-\frac{1}{6}\left({\Psi'}^2+{\Theta'}^2\right),
\quad
\left(\frac{a'}{a}\right)^2=\frac{1}{6}\left({\Psi'}^2+{\Theta'}^2\right).}
and the equations of motion for the scalar fields are
\formula{PsieqThetaeq}{\Psi''+2\frac{a'}{a}\Psi'=0,\quad
\Theta''+2\frac{a'}{a}\Theta'=0,} where $a(\tau)$ is the FRW scale
factor and the prime denotes derivative with respect to the
conformal time $\tau$. Equations (\ref{Gxxtt},\ref{PsieqThetaeq})
can be easily integrated to find \formula{a2}{ \Psi(\tau)=
\sqrt{\frac{6}{3b^2+1}} \alpha \ln a^2+ \gamma,\quad \Theta(\tau)=
\sqrt{1+3 b^2} \beta \ln a^2+ \delta, \quad a^2(\tau)=\xi \tau+
\zeta,} \formula{zzzz}{ \alpha^2 + \frac{(1+3b^2)^2}{6} \beta^2 =
1+3 b^2,} where $\alpha$, $\beta$, $\gamma$, $\delta$, $\zeta$ and
$\xi$ are integration constants. The solutions for the fields in
the original frame can be obtained as
\formula{d1}{d(\tau)=1-\left[\frac{(\mu \tau+\nu)^{2
\alpha}-1}{(\mu\tau+\nu)^{2
\alpha}+1}\right]^\frac{6b^2}{3b^2+1},} \formula{phi1}{\phi(\tau)=
\ln\left[\frac{(\mu\tau+\nu)^{2 \alpha \lambda(b) +\beta}}
{\left[(\mu \tau +\nu)^{2 \alpha}+1\right]^{2\lambda(b)}} \right]+
\varrho,} where $\lambda(b)$ is defined in Eq.~(\ref{phih}) and
integration constants $\mu$, $\nu$, $\varrho$ are redefined from
$\xi$, $\zeta$, $\gamma$ and $\delta$. The radion is calculated as
\formula{radionKK}{\mathcal{R}(\tau)=\frac{1}{\sqrt{2}b^2\sigma}d(\tau)e^{\sqrt{2}b\phi(\tau)},}
and the square of the scale factor on the positive tension brane
is
\formula{sf+KK}{a^2_+(\tau)=\frac{e^{-\sqrt{2}b\phi(\tau)}}{\psi(\tau)}a^2(\tau).}
We have found the remarkable fact that for a particular choice of
the integration constants we can reproduce the exact solution
found by Chen \emph{et al.} described in the previous section. If
we choose integration constants obeying the relations
\formula{relations}{\alpha=\frac{1}{2}, \;\; \beta=-\lambda(b),
\;\; \mu= \frac{2 h l}{L}, \;\; \nu=-1,} we get the same 4D
quantities (scale factor and scalar field on the positive tension
brane) and radion as the Chen \emph{et al.} solution,
Eqs.~(\ref{sf+}, \ref{scf+}, \ref{RAD}).

For this solution, the scale factor in the Einstein frame is given
by
\begin{equation}
a^2(\tau)=\frac{2L}{l}\left(h \tau - \frac{L}{2l} \right).
\end{equation}
If we take $h<0$, $\tau <0$, this corresponds to a collapsing
universe, due to the kinetic energy of the scalar fields. At $\tau
= L/2hl <0$, the universe in the Einstein frame reaches the
Big-bang singularity. However, we should be careful to interpret
this singularity. In fact, in the original 5D theory, $\tau =
L/2hl$ does not correspond to any kind of singularity. In 5D
theory, the negative tension brane hits the singularity at $\tau =
L/hl$ and the positive tension brane hits the singularity at
$\tau=0$. In fact at $\tau = L/2hl$, $\psi =0$ and so the
conformal transformation becomes singular and the Einstein frame
metric loses physical meaning.

\section{\label{sec:GEMN}Gradient expansion method with a new metric ansatz}
In the previous section, we found that the exact solution derived
in \cite{Chen:2005jp} can be reproduced from the $0^{th}$ order of
the perturbative method. In this section, we will revisit the
gradient expansion method which is used to derive the effective
theory to clarify the reason why the exact solution derived in
\cite{Chen:2005jp} can be reproduced within the 4D effective
theory.

\subsection{\label{subsec:5DEQ}5D Equations}
In this subsection we study the 5D equations of motion. From the
action (\ref{ACTT}) we derive the Einstein's equations
\begin{eqnarray}
{}^5G_{AB}&=&\nabla_A\varphi\nabla_B\varphi+\frac{1}{2}g_{AB}\left[-\nabla^C\varphi\nabla_C\varphi+V_{bulk}(\varphi)\right]
\nonumber \\
&&+\sqrt{2}\sigma\left[-\frac{\sqrt{g_4}}{\sqrt{g_5}}g_{\mu\nu}\delta_A^\mu\delta_B^\nu
e^{-\sqrt{2}b\varphi}\delta(r)+\frac{\sqrt{g_4}}{\sqrt{g_5}}g_{\mu\nu}\delta_A^\mu\delta_B^\nu
e^{-\sqrt{2}b\varphi}\delta(r-L)\right]. \label{Einstein1}
\nonumber\\
\end{eqnarray}

Assuming that the 5D line element has the form
\formula{metric1}{dS_5^2=g_{AB}(X)dX^AdX^B=g_{rr}(x,r)dr^2+g_{\mu\nu}(x,r)dx^\mu
dx^\nu,} ($x$ dependence means dependence of all the other
coordinates $\{x,y,z,t\}$, except the extra dimension $r$) we can
extract the junction conditions for the metric tensor from
Einstein's equation,
\formula{JC}{\sqrt{g^{rr}}\left(K_\nu^\mu-\delta_\nu^\mu
K\right){\!}\Bigg|_{\stackrel{\scriptstyle{r=0^+}}{\scriptstyle{r=L^-}}}=\mp\frac{1}{\sqrt{2}}\sigma
e^{-\sqrt{2}b\varphi}\delta_\nu^\mu\Bigg|_{\stackrel{\scriptstyle{r=0^+}}{\scriptstyle{r=L^-}}},}
where the extrinsic curvature $K_{\mu\nu}$ is defined by
$K_{\mu\nu}=-\frac{1}{2}\partial_rg_{\mu\nu}$. The scalar field
equation of motion is
\formula{scalarfeq}{\Box\varphi+\sqrt{2}b\left(b^2-\frac{2}{3}\right)\sigma^2e^{-2\sqrt{2}b\varphi}=-2b\sigma\left[\sqrt{g^{rr}}e^{-\sqrt{2}b\varphi}\delta(r)-\sqrt{g^{rr}}e^{-\sqrt{2}b\varphi}\delta(r-L)\right],}
and so the junction conditions for the scalar field are
\formula{JCSF}{\sqrt{g^{rr}}\partial_r\varphi{\!}\Bigg|_{\stackrel{\scriptstyle{r=0^+}}{\scriptstyle{r=L^-}}}=\mp
b\sigma
e^{-\sqrt{2}b\varphi}\Bigg|_{\stackrel{\scriptstyle{r=0^+}}{\scriptstyle{r=L^-}}}.}

In order to proceed we shall assume that the 5D line element has
further symmetries, described by
\formula{metric2}{dS_5^2=e^{2\sqrt{2}b\tilde{\varphi}(x,r)}H^{\frac{6b^2-2}{3b^2+1}}(x,r)dr^2+H^{\frac{2}{3b^2+1}}(x,r)\tilde{g}_{\mu\nu}dx^\mu
dx^\nu,\quad H(x,r)=C(x)-\frac{r}{l}.} In this section, we will
assume that the position of the second brane is $r=L=l$.  For the
scalar field we will assume the form
\formula{sfd}{\varphi(x,r)=\frac{3\sqrt{2}b}{3b^2+1}\ln
H+\tilde{\varphi}(x,r).} This metric ansatz is inspired by the
time dependent solution (\ref{ESMSM},\ref{ESMSD}) of Chen \emph{et
al.} \cite{Chen:2005jp}. Their solution was found by replacing the
modulus parameter in the static solution by a linear function of
time. In the same manner, we introduce an $x$ dependence in $H$
through the modulus parameter $C(x)$ in a covariant way . We also
introduce the function $\tilde{\varphi}$ for the scalar field
moduli. In order to satisfy the junction conditions (\ref{JCSF})
we must have the exponential factor in the $g_{rr}$ metric
component. The tensor $\tilde{g}_{\mu\nu}$ is left completely
general.

After some mathematical manipulations of the Einstein equations we
obtain
\begin{eqnarray}
&&\frac{1}{l}e^{-2\sqrt{2}b\tilde{\varphi}}H^{\frac{-9b^2+1}{3b^2+1}}\left(\frac{3b^2-5}{3b^2+1}\tilde{K}^\mu_\nu-\frac{1}{3b^2+1}\delta^\mu_\nu\tilde{K}\right)
\nonumber \\
&&-\sqrt{2}be^{-2\sqrt{2}b\tilde{\varphi}}H^{\frac{-3b^2+1}{3b^2+1}}\tilde{\varphi}_{,r}\left(\frac{1/l}{3b^2+1}H^{\frac{-6b^2}{3b^2+1}}\delta^\mu_\nu+H^{\frac{-3b^2+1}{3b^2+1}}\tilde{K}^\mu_\nu\right)
\nonumber \\
&&-H^{\frac{-3b^2-3}{3b^2+1}}\left(C_{|\nu}^{|\mu}+\frac{1}{3b^2+1}C_{|\alpha}^{|\alpha}\delta_\nu^\mu\right)
-H^{-\frac{2}{3b^2+1}}\left[\sqrt{2}b\tilde{\varphi}_{|\nu}^{|\mu}+\left(2b^2+1\right)\tilde{\varphi}^{|\mu}\tilde{\varphi}_{|\nu}\right]
\nonumber \\
&&-\frac{\sqrt{2}b}{3b^2+1}H^{\frac{-3b^2-3}{3b^2+1}}\left[\left(3b^2+1\right)\left(\tilde{\varphi}^{|\mu}C_{|\nu}+\tilde{\varphi}_{|\nu}C^{|\mu}\right)+\tilde{\varphi}^{|\alpha}C_{|\alpha}\delta_\nu^\mu\right]
\nonumber \\
&&+e^{-2\sqrt{2}b\tilde{\varphi}}H^{\frac{-6b^2+2}{3b^2+1}}\left(\tilde{K}^\mu_{\nu,r}-\tilde{K}\tilde{K}^\mu_\nu\right)+H^{-\frac{2}{3b^2+1}}R_\nu^\mu(\tilde{g})=0
,\label{munueq1}
\end{eqnarray}
\begin{eqnarray}
&&\frac{1}{l}\frac{3b^2-3}{3b^2+1}e^{-2\sqrt{2}b\tilde{\varphi}}H^{\frac{-9b^2+1}{3b^2+1}}\tilde{K}
+e^{-2\sqrt{2}b\tilde{\varphi}}H^{\frac{-6b^2+2}{3b^2+1}}\left(\tilde{K}_{,r}-\tilde{K}^{\alpha\beta}\tilde{K}_{\alpha\beta}\right)
\nonumber \\
&&+e^{-2\sqrt{2}b\tilde{\varphi}}H^{\frac{-6b^2+2}{3b^2+1}}\tilde{\varphi}_{,r}\left(\frac{1}{l}\frac{2\sqrt{2}b}{3b^2+1}H^{-1}-\sqrt{2}b\tilde{K}-\tilde{\varphi}_{,r}\right)
\nonumber \\
&&-H^{\frac{-3b^2-3}{3b^2+1}}\left(\frac{3b^2-1}{3b^2+1}C_{|\alpha}^{|\alpha}+\frac{6\sqrt{2}b^3}{3b^2+1}\tilde{\varphi}^{|\alpha}C_{|\alpha}\right)
\nonumber \\
&&-\sqrt{2}bH^{-\frac{2}{3b^2+1}}\left(\tilde{\varphi}_{|\alpha}^{|\alpha}+\sqrt{2}b\tilde{\varphi}^{|\alpha}\tilde{\varphi}_{|\alpha}\right)=0
,\label{rreq1}
\end{eqnarray}
\begin{eqnarray}
&&\tilde{K}^{\alpha}_{\beta|\alpha}-\tilde{K}_{|\beta}+\sqrt{2}b\left(\tilde{\varphi}_{|\beta}\tilde{K}-\tilde{\varphi}_{|\alpha}\tilde{K}^{\alpha}_{\beta}\right)
+H^{-1}\left(\frac{3b^2-2}{3b^2+1}C_{|\beta}\tilde{K}-\frac{3b^2-5}{3b^2+1}C_{|\alpha}\tilde{K}^{\alpha}_{\beta}\right)
\nonumber \\
&&=-\tilde{\varphi}_{,r}\left(\frac{3\sqrt{2}b}{3b^2+1}H^{-1}C_{|\beta}+\tilde{\varphi}_{|\beta}\right)
,\label{rbetaeq1}
\end{eqnarray}
where $\tilde{K}_{\mu\nu}$ is defined like
$\tilde{K}_{\mu\nu}=-\frac{1}{2}\partial_r{\tilde{g}_{\mu\nu}}$,
$\tilde{K}$ is the trace of $\tilde{K}_{\mu\nu}$,
$\scriptstyle{|}_{\scriptstyle{\alpha}}$ denotes covariant
derivative with respect to $\tilde{g}_{\mu\nu}$ and
$R_{\mu\nu}(\tilde{g})$ is the Ricci tensor of
$\tilde{g}_{\mu\nu}$. Equation (\ref{scalarfeq}) transforms into
\begin{eqnarray}
&&\!\!\!\!\!\!\!\!\!e^{-2\sqrt{2}b\tilde{\varphi}}H^{\frac{-3b^2+5}{3b^2+1}}\bigg[\tilde{\varphi}_{,rr}-\sqrt{2}b\left(\tilde{\varphi}_{,r}\right)^2-\tilde{\varphi}_{,r}\tilde{K}
+\frac{1}{l}\frac{3\sqrt{2}b}{3b^2+1}H^{-1}\tilde{K}
+\frac{1}{l}\frac{9b^2-5}{3b^2+1}H^{-1}\tilde{\varphi}_{,r}\bigg]
\nonumber \\
&&\!\!\!\!\!\!\!\!\!+H\tilde{\varphi}_{|\alpha}^{|\alpha}+\frac{3\sqrt{2}b}{3b^2+1}C_{|\alpha}^{|\alpha}+\sqrt{2}bH\tilde{\varphi}_{|\alpha}\tilde{\varphi}^{|\alpha}
+\frac{9b^2+1}{3b^2+1}\tilde{\varphi}_{|\alpha}C^{|\alpha}=0.
\label{sfeq1}
\end{eqnarray}
With this particular ansatz the junction conditions
(\ref{JC},\ref{JCSF}) are significantly simplified and read
\formula{jcKuv}{\left[\tilde{K}_\nu^\mu\right]{\!}\Bigg|_{\stackrel{\scriptstyle{r=0^+}}{\scriptstyle{r=l^-}}}=0,}
\formula{jcsf}{\left[\tilde{\varphi}_{,r}\right]{\!}\Bigg|_{\stackrel{\scriptstyle{r=0^+}}{\scriptstyle{r=l^-}}}=0.}

It is impossible to solve these equations in general, so in the
next section, we will solve them using the gradient expansion
method up to first order in the perturbations.
\subsection{\label{subsec:PM}The gradient expansion}
\subsubsection{\label{subsubsec:APPR}The approximation}

In order to use perturbation theory to solve a system of
differential equations we need to identify the characteristic
scale of the different terms involved in the equations and then
see if there is a small parameter.

The derivatives along the extra dimension of the conformal metric
$\tilde{g}_{\mu\nu}$ as well as the derivative of
$\tilde{\varphi}$ are of order $1/l$. Typically, we take $l$ to be
of order of a few tenths of a millimeter. If the characteristic
brane's curvature length scale is $D$ then $\partial^2_x\,\tilde
g_{\mu\nu}\sim1/D^2\tilde g_{\mu\nu}$. We assume that variations
along the branes' coordinates are small in comparison with $1/l$.
This implies that the radion changes slowly or that the speed of
the branes is small. More precisely, our small parameters will be
\formula{approx}{l^2\partial^2_x(\ldots)\ll1 \;\;\;\;\;\mbox{and}
\;\;\;\;\;\left(l\partial_x\ldots\right)^2\ll1,} where $\ldots$
represents the conformal metric functions, $C$ or
$\tilde{\varphi}$.

As in the usual perturbation method, we expand the unknown
functions in a series
\formula{expanm}{\tilde{g}_{\mu\nu}(r,x)=\stackrel{(0)}{\tilde{g}}_{\mu\nu}(r,x)+
\stackrel{(1)}{\tilde{g}}_{\mu\nu}(r,x)+\cdots,}
\formula{expansf}{\tilde{\varphi}(r,x)=\stackrel{(0)}{\tilde{\varphi}}(r,x)+
\stackrel{(1)}{\tilde{\varphi}}(r,x)+\cdots.} We impose the
boundary conditions at the position of the positive tension brane
\formula{bcm}{\stackrel{(n)}{\tilde{g}}_{\mu\nu}(r=0,x)=0,\;\mbox{for
all}\;n>0,}
\formula{bcsf}{\stackrel{(n)}{\tilde{\varphi}}(r=0,x)=0,\;\mbox{for
all}\;n>0.} Other quantities are naturally expanded as
\formula{oexpan}{\tilde{K}_{\mu\nu}=\stackrel{(0)}{\tilde{K}}_{\mu\nu}+\stackrel{(1)}{\tilde{K}}_{\mu\nu}+\cdots.}

In practical terms, at zeroth order in the gradient expansion
method we ignore all derivatives along the branes' coordinates.
These terms will only enter the first order equations. At zeroth
order, the 5D partial differential equations of motion reduce to
simpler ordinary differential equations on the extra dimension
coordinate. The gradient expansion method has also been used in
the 4D cosmological context
\cite{Tomita:1975kj,Salopek:1992kk,Comer:1994np,Deruelle:1994iz,Soda:1995fz}.

\subsubsection{\label{subsubsec:BACK}$0^{th}$ Order (Background geometry)}

The $0^{th}$ order system can be easily integrated with respect to
the extra dimension coordinate $r$ to get the particular solution
\formula{0sol}{\stackrel{(0)}{\tilde{g}_{\mu\nu}}(r,x)=\stackrel{(0)}{\tilde{g}_{\mu\nu}}(x),}
\formula{0solsf}{\stackrel{(0)}{\tilde{\varphi}}(r,x)=\stackrel{(0)}{\tilde{\varphi}}(x).}
This solution clearly satisfies the $0^{th}$order junction
conditions.

\subsubsection{\label{subsubsec:1ST}$1^{st}$ Order}

At $1^{st}$ order the evolution equations are
\begin{eqnarray}
  &&\!\!\!\!\!\!\!\!\!\!\!\!\!\!\!\!\!\!\frac{1}{l}e^{-2\sqrt{2}b\stackrel{(0)}{\tilde{\varphi}}}H^{\frac{-9b^2+1}{3b^2+1}}\left(\frac{3b^2-5}{3b^2+1}\stackrel{(1)}{\tilde{K}^\mu_\nu}-\frac{1}{3b^2+1}\delta^\mu_\nu\stackrel{(1)}{\tilde{K}}\right)
+e^{-2\sqrt{2}b\stackrel{(0)}{\tilde{\varphi}}}H^{\frac{-6b^2+2}{3b^2+1}}\stackrel{(1)}{\tilde{K}^\mu_{\nu,r}}
  \nonumber \\
  &&\!\!\!\!\!\!\!\!\!\!\!\!\!\!\!\!\!\!-H^{\frac{-3b^2-3}{3b^2+1}}\left(C_{|\nu}^{|\mu}+\frac{1}{3b^2+1}C_{|\alpha}^{|\alpha}\delta_\nu^\mu\right)
-H^{-\frac{2}{3b^2+1}}\left[\sqrt{2}b\stackrel{(0)}{\tilde{\varphi}_{|\nu}^{|\mu}}+\left(2b^2+1\right)\stackrel{(0)}{\tilde{\varphi}^{|\mu}}\stackrel{(0)}{\tilde{\varphi}_{|\nu}}\right]
  \nonumber \\
  &&\!\!\!\!\!\!\!\!\!\!\!\!\!\!\!\!\!\!-\frac{\sqrt{2}b}{3b^2+1}H^{\frac{-3b^2-3}{3b^2+1}}\left[\left(3b^2+1\right)\left(\stackrel{(0)}{\tilde{\varphi}^{|\mu}}C_{|\nu}+\stackrel{(0)}{\tilde{\varphi}_{|\nu}}C^{|\mu}\right)+\stackrel{(0)}{\tilde{\varphi}^{|\alpha}}C_{|\alpha}\delta_\nu^\mu\right]
\nonumber \\
  &&\!\!\!\!\!\!\!\!\!\!\!\!\!\!\!\!\!\!-\frac{\sqrt{2}b}{l}\frac{1}{3b^2+1}e^{-2\sqrt{2}b\stackrel{(0)}{\tilde{\varphi}}}H^{\frac{-9b^2+1}{3b^2+1}}\stackrel{(1)}{\tilde{\varphi}_{,r}}\delta^\mu_\nu
  +H^{-\frac{2}{3b^2+1}}R_\nu^\mu\left(\stackrel{(0)}{\tilde{g}}\right)=0
,\label{1storder1}
\end{eqnarray}
\begin{eqnarray}
  &&\!\!\!\!\!\!\!\!\!\!\!\!\!\!\!\!\!\!\frac{1}{l}\frac{3b^2-3}{3b^2+1}e^{-2\sqrt{2}b\stackrel{(0)}{\tilde{\varphi}}}H^{\frac{-9b^2+1}{3b^2+1}}\stackrel{(1)}{\tilde{K}}
+e^{-2\sqrt{2}b\stackrel{(0)}{\tilde{\varphi}}}H^{\frac{-6b^2+2}{3b^2+1}}\stackrel{(1)}{\tilde{K}_{,r}}
  \nonumber \\
  &&\!\!\!\!\!\!\!\!\!\!\!\!\!\!\!\!\!\!-H^{\frac{-3b^2-3}{3b^2+1}}\left(\frac{3b^2-1}{3b^2+1}C_{|\alpha}^{|\alpha}+\frac{6\sqrt{2}b^3}{3b^2+1}\stackrel{(1)}{\tilde{\varphi}^{|\alpha}}C_{|\alpha}\right)
\nonumber \\
  &&\!\!\!\!\!\!\!\!\!\!\!\!\!\!\!\!\!\!+\frac{1}{l}\frac{2\sqrt{2}b}{3b^2+1}e^{-2\sqrt{2}b\stackrel{(0)}{\tilde{\varphi}}}H^{\frac{-9b^2+1}{3b^2+1}}\stackrel{(1)}{\tilde{\varphi}_{,r}}
  -\sqrt{2}bH^{-\frac{2}{3b^2+1}}\left(\stackrel{(0)}{\tilde{\varphi}_{|\alpha}^{|\alpha}}+\sqrt{2}b\stackrel{(0)}{\tilde{\varphi}^{|\alpha}}\stackrel{(0)}{\tilde{\varphi}_{|\alpha}}\right)=0
,\label{1storder2}
\end{eqnarray}
\begin{eqnarray}
  &&\!\!\!\!\!\!\!\!\!\!\!\!\!\!\!\!\!\!e^{-2\sqrt{2}b\stackrel{(0)}{\tilde{\varphi}}}H^{\frac{-3b^2+5}{3b^2+1}}\left[\stackrel{(1)}{\tilde{\varphi}_{,rr}}+\frac{1}{l}\frac{1}{3b^2+1}H^{-1}\left(3\sqrt{2}b\stackrel{(1)}{\tilde{K}}+\left(9b^2-5\right)\stackrel{(1)}{\tilde{\varphi}_{,r}}\right)\right]
\nonumber \\
&&\!\!\!\!\!\!\!\!\!\!\!\!\!\!\!\!\!\!+H\stackrel{(0)}{\tilde{\varphi}_{|\alpha}^{|\alpha}}+\frac{3\sqrt{2}b}{3b^2+1}C_{|\alpha}^{|\alpha}+\sqrt{2}bH\stackrel{(0)}{\tilde{\varphi}_{|\alpha}}\stackrel{(0)}{\tilde{\varphi}^{|\alpha}}
+\frac{9b^2+1}{3b^2+1}\stackrel{(0)}{\tilde{\varphi}_{|\alpha}}C^{|\alpha}=0
.\label{1storder3}
\end{eqnarray}
The junction conditions at this order are
\formula{jc1}{\left[\stackrel{(1)}{\tilde{K}_\nu^\mu}\right]{\!}\Bigg|_{\stackrel{\scriptstyle{r=0^+}}{\scriptstyle{r=l^-}}}=0,}
\formula{jc1sf}{\left[\stackrel{(1)}{\tilde{\varphi}_{,r}}\right]{\!}\Bigg|_{\stackrel{\scriptstyle{r=0^+}}{\scriptstyle{r=l^-}}}=0.}
In the preceding equations all the indices are raised with the
zeroth order metric. Combining the trace of equation
(\ref{1storder1}) with  equation (\ref{1storder2}) we obtain
\begin{eqnarray}
\frac{1}{l}e^{-2\sqrt{2}b\stackrel{(0)}{\tilde{\varphi}}}\stackrel{(1)}{\tilde{K}}
&=&\frac{3b^2+1}{6}\left[R\left(\stackrel{(0)}{\tilde{g}}\right)-\stackrel{(0)}{\tilde{\varphi}_{|\alpha}}\stackrel{(0)}{\tilde{\varphi}^{|\alpha}}\right]H^{\frac{9b^2-3}{3b^2+1}}
-\frac{\sqrt{2}b}{l}e^{-2\sqrt{2}b\stackrel{(0)}{\tilde{\varphi}}}\stackrel{(1)}{\tilde{\varphi}_{,r}}\nonumber
\\&&-\left(C_{|\alpha}^{|\alpha}+\sqrt{2}bC_{|\alpha}\stackrel{(0)}{\tilde{\varphi}^{|\alpha}}\right)H^{\frac{6b^2-4}{3b^2+1}}
.\label{K1}
\end{eqnarray}
Imposing the junction conditions (\ref{jc1},\ref{jc1sf}) we get
\formula{riccis}{\tilde{R}\left(\stackrel{(0)}{\tilde{g}}\right)=\stackrel{(0)}{\tilde{\varphi}_{|\alpha}}\stackrel{(0)}{\tilde{\varphi}^{|\alpha}},}
and the equation of motion for the 4D effective scalar field
\formula{aa}{C_{|\alpha}^{|\alpha}+\sqrt{2}bC_{|\alpha}\stackrel{(0)}{\tilde{\varphi}^{|\alpha}}=0.}
Equation (\ref{K1}) now reads
\formula{K10}{\stackrel{(1)}{\tilde{K}}=-\sqrt{2}b\stackrel{(1)}{\tilde{\varphi}_{,r}}.}

Using the decomposition of $\stackrel{(1)}{\tilde{K_\nu^\mu}}$ in
\formula{deco}{\stackrel{(1)}{\tilde{K_\nu^\mu}}=\stackrel{(1)}{\tilde{\Sigma_\nu^\mu}}
+\frac{1}{4}\delta_\nu^\mu\stackrel{(1)}{\tilde{K}},} equation
(\ref{1storder1}) can be easily integrated to find
\begin{eqnarray}
&&\!\!\!\!\!\!\!\!\!\!\!\!\!\!\!\!\!\!\!\!\!\!\!\!\!\!e^{-2\sqrt{2}b\stackrel{(0)}{\tilde{\varphi}}}H^{-\frac{3b^2-5}{3b^2+1}}\stackrel{(1)}{\tilde{\Sigma_\nu^\mu}}=
\Bigg[\left(C_{|\nu}^{|\mu}+\sqrt{2}b\left(C_{|\nu}\stackrel{(0)}{\tilde{\varphi}^{|\mu}}+C^{|\mu}\stackrel{(0)}{\tilde{\varphi}_{|\nu}}\right)\right)r
\nonumber
\\&&\!\!\!\!\!\!\!\!\!\!\!\!\!\!\!\!\!\!\!\!\!\!\!\!\!\!-\left(R_\nu^\mu\left(\stackrel{(0)}{\tilde{g}}\right)-\sqrt{2}b\stackrel{(0)}{\tilde{\varphi}_{|\nu}^{|\mu}}-\left(2b^2+1\right)\stackrel{(0)}{\tilde{\varphi}_{|\nu}}\stackrel{(0)}{\tilde{\varphi}^{|\mu}}\right)\left(Cr-\frac{r^2}{2l}\right)\Bigg]_{traceless}
+\chi_\nu^\mu(x) , \label{sigma1}
\end{eqnarray}
where $\chi_\nu^\mu(x)$ is an integration constant and the
subscript $[\;\;]_{traceless}$ means the traceless part of the
quantity between square brackets. In terms of
$\tilde{\Sigma_\nu^\mu}$, the junction conditions (\ref{jc1}) are
\formula{jcsigma1}{\left[\stackrel{(1)}{\tilde{\Sigma}_\nu^\mu}\right]{\!}\Bigg|_{\stackrel{\scriptstyle{r=0^+}}{\scriptstyle{r=l^-}}}=0.}

From the previous junction conditions (\ref{jcsigma1}) we can
obtain the 4D effective equations of motion
\begin{eqnarray}
\!\!\!\!\!\!\!\!\!\!\!\!\!\!\!\!\!\!\!\tilde{R}_\nu^\mu\left(\stackrel{(0)}{\tilde{g}}\right)&=&\left(C-\frac{1}{2}\right)^{-1}\left[C_{|\nu}^{|\mu}+\frac{1}{4}\delta_\nu^\mu
C_{|\alpha}^{|\alpha}+\sqrt{2}b\left(C_{|\nu}\stackrel{(0)}{\tilde{\varphi}^{|\mu}}+C_{|\mu}\stackrel{(0)}{\tilde{\varphi}^{|\nu}}\right)\right]
\nonumber\\
&&+\sqrt{2}b\left(\stackrel{(0)}{\tilde{\varphi}_{|\nu}^{|\mu}}-\frac{1}{4}\delta_\nu^\mu\stackrel{(0)}{\tilde{\varphi}_{|\alpha}^{|\alpha}}\right)
+\left(2b^2+1\right)\stackrel{(0)}{\tilde{\varphi}_{|\nu}}\stackrel{(0)}{\tilde{\varphi}^{|\mu}}-\frac{b^2}{2}\delta_\nu^\mu\stackrel{(0)}{\tilde{\varphi}_{|\alpha}}\stackrel{(0)}{\tilde{\varphi}^{|\alpha}}
,\label{4deffectiveeq}
\end{eqnarray}
and \formula{chi}{\chi_\nu^\mu(x)=0.}

Combining equation (\ref{K10}) with the scalar field equation
(\ref{1storder3}) and integrating it once with respect to the
extra dimension, we get (after using the previous equations of
motion to simplify the result)
\formula{drsf}{e^{-2\sqrt{2}b\stackrel{(0)}{\tilde{\varphi}}}H^{\frac{-3b^2+5}{3b^2+1}}\stackrel{(1)}{\tilde{\varphi}_{,r}}=-\left(Cr-\frac{r^2}{2l}\right)\left(\stackrel{(0)}{\tilde{\varphi}_{|\alpha}^{|\alpha}}+\sqrt{2}b\stackrel{(0)}{\tilde{\varphi}_{|\alpha}}\stackrel{(0)}{\tilde{\varphi}^{|\alpha}}\right)-\stackrel{(0)}{\tilde{\varphi}^{|\alpha}}C_{|\alpha}r+\Xi(x),}
where $\Xi(x)$ is just an integration constant.

The junction conditions (\ref{jc1sf}) give \formula{Xi}{\Xi(x)=0}
and the equation of motion for the second 4D effective scalar
field
\formula{2eq}{\stackrel{(0)}{\tilde{\varphi}_{|\alpha}^{|\alpha}}+\sqrt{2}b\stackrel{(0)}{\tilde{\varphi}_{|\alpha}}\stackrel{(0)}{\tilde{\varphi}^{|\alpha}}=-\left(C-\frac{1}{2}\right)^{-1}\stackrel{(0)}{\tilde{\varphi}^{|\alpha}}C_{|\alpha}.}

\subsection{\label{subsec:ACT2}The 4D Effective theory}
The 4D effective equations of motion are summarized as
\begin{eqnarray}
\!\!\!\!\!\!\!\!\!\!\!\!\!\!\!\!\!\!\!\tilde{R}_\nu^\mu\left(\stackrel{(0)}{\tilde{g}}\right)&=&\left(C-\frac{1}{2}\right)^{-1}\left[C_{|\nu}^{|\mu}+\frac{1}{4}\delta_\nu^\mu
C_{|\alpha}^{|\alpha}+\sqrt{2}b\left(C_{|\nu}\stackrel{(0)}{\tilde{\varphi}^{|\mu}}+C_{|\mu}\stackrel{(0)}{\tilde{\varphi}^{|\nu}}\right)\right]
\nonumber\\
&&+\sqrt{2}b\left(\stackrel{(0)}{\tilde{\varphi}_{|\nu}^{|\mu}}-\frac{1}{4}\delta_\nu^\mu\stackrel{(0)}{\tilde{\varphi}_{|\alpha}^{|\alpha}}\right)
+\left(2b^2+1\right)\stackrel{(0)}{\tilde{\varphi}_{|\nu}}\stackrel{(0)}{\tilde{\varphi}^{|\mu}}-\frac{b^2}{2}\delta_\nu^\mu\stackrel{(0)}{\tilde{\varphi}_{|\alpha}}\stackrel{(0)}{\tilde{\varphi}^{|\alpha}}
,\label{4deffectiveeqA}
\end{eqnarray}
\formula{sfR}{C_{|\alpha}^{|\alpha}+\sqrt{2}bC_{|\alpha}\stackrel{(0)}{\tilde{\varphi}^{|\alpha}}=0,}
\formula{2eqC}{\stackrel{(0)}{\tilde{\varphi}_{|\alpha}^{|\alpha}}+\sqrt{2}b\stackrel{(0)}{\tilde{\varphi}_{|\alpha}}\stackrel{(0)}{\tilde{\varphi}^{|\alpha}}=-\left(C-\frac{1}{2}\right)^{-1}\stackrel{(0)}{\tilde{\varphi}^{|\alpha}}C_{|\alpha},}
and they can be deduced from the following action
\formula{action1}{S_{eff}=\frac{l}{\kappa_G^2}\int
d^4x\sqrt{-\stackrel{(0)}{\tilde{g}}}\left(C-\frac{1}{2}\right)
e^{\sqrt{2}b\stackrel{(0)}{\tilde{\varphi}}}\left[R(\stackrel{(0)}{\tilde{g}})-\stackrel{(0)}{\tilde{\varphi}_{|\alpha}}\stackrel{(0)}{\tilde{\varphi}^{|\alpha}}\right],}
where $\scriptstyle{|}_{\scriptstyle{\alpha}}$ denotes covariant
derivative with respect to $\stackrel{(0)}{\tilde{g}_{\mu\nu}}$.
We should note that this effective action can be derived by
substituting in (\ref{ACTT}) the 5D solutions up to the first
order and integrating it over the fifth dimension
\cite{Kanno:2005zr}.

As a consistency check, we see that if we perform the conformal
transformation
\formula{1}{h_{\mu\nu}(x)=C^{\frac{2}{3b^2+1}}(x)\stackrel{(0)}{\tilde{g}_{\mu\nu}}(x),}
the previous action reduces to (\ref{EFFACT}) if the effective
scalar fields of the two theories are related through
\formula{2}{\psi=2C^{-2}\left(C-\frac{1}{2}\right),}
\formula{3}{\phi(x)=\frac{3\sqrt{2}b}{3b^2+1}\ln C
+\stackrel{(0)}{\tilde{\varphi}}(x).} The check consists in seeing
that these relations are exactly the ones required so that the two
observables (the proper distance between branes and the scalar
field on the positive tension brane) agree in both approaches.

\subsection{\label{subsec:SOL2}The 5D exact solution}

It is straightforward to find a cosmological solution of this 4D
effective theory. For example we can easily find the following
particular solution
\begin{equation}
\stackrel{(0)}{\tilde{g}_{\mu\nu}}(x)=\eta_{\mu\nu},\quad
C(x)=ht,\quad \stackrel{(0)}{\tilde{\varphi}}(x)=0,
\label{4dsolution}
\end{equation}
where $h$ is an integration constant.

Now we are ready to address the question why the above solution
can be lifted to an exact 5D solution. Let us start by calculating
the next order correction ${}^{(1)}{\tilde{g}_{\mu\nu}}$ and
${}^{(1)}{\tilde{\varphi}}$. Eq. (\ref{drsf}) and the boundary
conditions (\ref{bcsf}) give ${}^{(1)}{\tilde{\varphi}}=0$, if we
take as $0^{th}$ order solution Eqs. (\ref{4dsolution}). We can
construct ${}^{(1)}{\tilde{K}_{\mu\nu}}$ from Eqs. (\ref{K10}) and
(\ref{sigma1}). For the $0^{th}$ order solution (\ref{4dsolution})
this gives ${}^{(1)}{\tilde{K}_{\mu\nu}}=0$. After imposing the
boundary conditions (\ref{bcm}), we obtain that the next order
correction vanishes, ${}^{(1)}{\tilde{g}_{\mu\nu}(r,x)}=0$. For
solution (\ref{4dsolution}) it turns out that all the corrections
vanish and the $0^{th}$ order solution is an exact solution of the
non-perturbed 5D Eqs. (\ref{munueq1}-\ref{jcsf}).

For other solutions of the 4D effective theory, higher order
corrections will not vanish and therefore they should be taken
into account in the reconstruction of the 5D metric. Using the
gradient expansion method, we can reconstruct the 5D solution
perturbatively. We should emphasize that the choice of the
$0^{th}$ order metric is quite important in order to reconstruct
5D solutions efficiently. Our metric ansatz has the advantage that
it is possible to recover the exact solution of Chen \emph{et al.}
(\ref{4dsolution}) at $0^{th}$ order. Indeed, if we had started
with an ansatz like Eqs. (\ref{KK0th1},\ref{KK0th2}) we would need
an infinite number of higher order terms to obtain the exact 5D
solution.

\section{\label{sec:COMM}Validity of 4D effective theory}
In this section we will make comments on a work by Kodama and
Uzawa \cite{Kodama:2005cz}. Let us start by briefly describing
their arguments. After deriving the 4D effective theory for warped
compactification of the 5D Ho\v{r}ava-Witten model (they also
extend their analysis to 10D IIB supergravity and obtain the same
conclusions), the authors show that the 4D effective theory allows
a wider class of solutions than the fundamental higher dimensional
theory. Therefore we should be careful in using this effective
theory approach, because we may find 4D solutions that do not
satisfy the equations of motion once lifted back to 5D.

The authors assume a metric ansatz of the form
\formula{met}{dS_5^2=h(x,r)dr^2+h^{\frac{1}{2}}(x,r)\tilde{g}_{\mu\nu}(x)dx^\mu
dx^\nu,} where the warp factor has the form $h(x,r)=C(x)-r/L$.
This corresponds to taking $\tilde{K}_{\mu \nu}=0$,
$\tilde{\varphi}=0$ and $b=1$ (for the Ho\v{r}ava-Witten case) in
our work. Then Eq. (\ref{munueq1}) reduces to
\begin{equation}
-H(r,x)^{-\frac{3}{2}}\left(C(x)_{|\nu}^{|\mu}+\frac{1}{4}C(x)_{|\alpha}^{|\alpha}
\delta_\nu^\mu\right)
+H(r,x)^{-\frac{1}{2}}R_\nu^\mu(\tilde{g}(x))=0.
\end{equation}
In order to satisfy this equation for all values of $r$, we should
have \formula{con}{R_{\mu\nu}(\tilde{g})=0,\quad C_{|\mu \nu}=0.}

They obtain the 4D effective action, by integrating the fifth
dimension, as \formula{actionHK}{S_{eff} \propto \int
d^4x\sqrt{-\tilde{g}}\left(C(x)-\frac{1}{2}\right) R(\tilde{g}),}
which agrees with our effective action (\ref{action1}). As we have
shown, this theory admits solutions with
$R_{\mu\nu}(\tilde{g})\neq 0$ (see Eq.~(\ref{4deffectiveeqA})),
which do not obey the constraint (\ref{con}) obtained from the 5D
equations of motion.

However, it is clear from our analysis that their metric ansatz is
too restrictive.  If we consider a more general metric as our
metric ansatz, we see that the 5D Einstein equations contain more
terms given by $\tilde{K}_{\mu \nu}$. With the inclusion of these
new terms, the 5D equations do not necessarily imply (\ref{con}).
Of course, the non-vanishing $\tilde{K}_{\mu\nu}$ changes the
metric (\ref{met}) and one could argue that the resultant 4D
effective action would be also changed. However, it is shown that
even if we include the first order corrections
${}^{(1)}{\tilde{g}_{\mu\nu}}(x)$ to the metric, the resultant 4D
effective action derived by integrating out the fifth dimension
does not change \cite{Kanno:2005zr}. Therefore, for 4D solutions
that do not satisfy (\ref{con}), we should include the corrections
to the metric (\ref{met}). We have provided this correction
perturbatively. Using Eqs. (\ref{K1}) and (\ref{sigma1}), we can
reconstruct the correction to the metric,
${}^{(1)}{\tilde{g}_{\mu\nu}}(x)$, which is necessary to satisfy
the 5D equation of motion.

We should emphasize that the validity of the 4D effective theory
is based on the conditions (\ref{approx}). If the 4D effective
theory admits a solution that violates the conditions
(\ref{approx}), then there is no guarantee that the 4D solution
can be lifted up to the 5D solution consistently. We should check
the validity of the 4D effective theory by calculating the higher
order corrections to ensure that the higher order corrections can
be neglected consistently.
\section{\label{sec:CON}Conclusion}
In this chapter, we have studied the moduli instability in a two
brane model with a bulk scalar field found by Chen \emph{et al.}.
This model can be viewed as a generalization of the
Ho\v{r}ava-Witten theory and the Randall-Sundrum model. The scalar
field potentials in the bulk and on the branes are tuned in order
to satisfy the BPS condition.

We used a low energy effective theory, which is derived by
assuming that variations along the brane coordinates of the metric
are small compared with variations along the dimension
perpendicular to the brane. The effective theory is a bi-scalar
tensor theory where one of the scalar fields arises from the bulk
scalar field (dilaton) and the other arises from the degree of
freedom of the distance between branes (radion). In the Einstein
frame, the theory consists of two massless scalar fields, and the
lack of potentials for these moduli fields was shown to be
responsible for the instability.

We found that the exact solution derived in \cite{Chen:2005jp} can
be reproduced from the $0^{th}$ order of the perturbative method,
despite the fact that slow-motion approximations are used. We
revisited the gradient expansion method which is used to derive
the effective theory, in order to understand why the exact
solution derived in \cite{Chen:2005jp} can be reproduced within
the 4D effective theory. We proposed a new metric ansatz which is
useful to see the relation between the solutions in the effective
theory and the full solutions for 5D equation of motion. Using
this metric ansatz, it is transparent why the moduli instability
solution can be lifted to a full 5D solution. We have also shown
that not all solutions in the 4D effective theory can be lifted to
exact 5D solutions. For these solutions, the solutions in the
effective theory receive higher order corrections in velocities of
the branes and we need to find 5D solutions perturbatively.

Finally, we comment on the arguments against the 4D effective
theory. Ref. \cite{Kodama:2005cz} claims that the 4D effective
theory allows a much wider class of solutions than the 5D theory.
We argued that this conclusion comes from a too restricted metric
ansatz used in Ref. \cite{Kodama:2005cz}. Using a more general
metric ansatz, we provided a way to reconstruct the full 5D
solutions from the solutions in the 4D effective theory.

The gradient expansion method can be applied to other warped
compactifications such as the ones in supergravity models. In
fact, there have been debates on the validity of the metric ansatz
commonly used to derive the 4D effective theory in 10D type IIB
sugra. Our 10D generalization of the $0^{th}$ order metric ansatz
agrees with that proposed in Ref. \cite{Giddings:2005ff} and the
method presented in this chapter will provide a consistent way to
reduce the 10D theory to the 4D effective theory based on this
metric ansatz. This will be the topic of chapter
\ref{chapter:10D}. Before that, in the next chapter, we will
obtain the low energy effective theory in a 6D supergravity model
using the gradient expansion method.

\chapter{Low energy effective theory on a regularized brane in 6D supergravity}\label{chapter:6D}
\chaptermark{Low energy effective theory in 6D SUGRA}

\section{Introduction}
Recently, much attention has been paid to six-dimensional
supergravity \cite{Gibbons:2003di,Burgess:2004dh,Tolley:2005nu,
Aghababaie:2003ar, Lee:2006ge,Burgess:2006ds}. The most intriguing
property of six-dimensional supergravity is that the
four-dimensional spacetime is always Minkowski even in the
presence of branes with tension. A 3-brane with tension induces
only a deficit angle in the six-dimensional spacetime and the
tension does not curve the four-dimensional spacetime within the
brane. This feature is called self-tuning and it may solve the
cosmological constant problem
\cite{Koyama:2007rx,Carroll:2003db,Chen:2000at,Navarro:2003vw,Aghababaie:2003wz}.
This is the basis of the supersymmetric large extra-dimension
(SLED) proposal \cite{Burgess:2007ui}.

There have been several objections to the idea of self-tuning
\cite{Garriga:2004tq,Vinet:2005dg,Navarro:2003bf,Nilles:2003km,Lee:2003wg,Kehagias:2004fb,Vinet:2004bk}.
The self-tuning relies on the classical scaling property of the
model. The six-dimensional equations of motion are invariant under
the constant rescaling $g_{MN} \to e^{\omega} g_{MN}$ and
$e^{\phi} \to e^{\phi -\omega}$, where $g_{MN}$ denotes the
six-dimensional metric and $\phi$ is the dilaton field. Then there
is a modulus associated with this scaling property.
Ref.~\cite{Garriga:2004tq} derived an effective potential for this
modulus. This modulus is shown to have an exponential potential.
Then there must be a fine-tuning of parameters to ensure that the
potential vanishes in order to have a static solution. This is the
reason why the static solution always has vanishing cosmological
constant. However, if this fine-tuning is broken, the modulus
acquires a runaway potential and the four-dimensional spacetime
becomes non-static. Non-static solutions in six-dimensional
supergravity have been derived and they are supposed to correspond
to the response of the bulk geometry to a change of tension of
branes
 \cite{Cline:2003ak, Tolley:2006ht, Kobayashi:2007hf, Copeland:2007ur}.

However, it is difficult to deal with an arbitrary change of
tension with a brane described by a pure conical singularity. This
is because if we put matter on the brane other than a cosmological
constant, the metric diverges at the position of the brane.
Recently, it was suggested that we can regularize the brane by
resolving it by a codimension one cylindrical 4-brane
\cite{Peloso:2006cq, Papantonopoulos:2006dv, Himmetoglu:2006nw,
Kobayashi:2007kv}. These types of models may be regarded as a
variation of Kaluza-Klein/hybrid brane world
\cite{Louko:2001ik,Carter:2006uk,Kobayashi:2007qe,Kanno:2007wj,Appleby:2007ct}.
Once the brane becomes a codimension one object, it is possible to
put arbitrary matter on the brane without having the divergence of
the metric. Then it becomes possible to study the effect of the
change of tension on the four-dimensional geometry on the brane.

There is another interesting issue of whether it is possible to
recover conventional cosmology at low energies in six-dimensional
models. Recent works have shown that it is impossible to recover
sensible cosmology if one derives cosmological solutions by
considering a motion of branes in a given static bulk spacetime
\cite{Papantonopoulos:2007fk, Minamitsuji:2007fx}. It was
concluded that the time-dependence of the bulk spacetime should be
taken into account.

In this chapter, we derive a four-dimensional effective theory for
the modulus in six-dimensional supergravity with resolved 4-branes
by extending the analysis of Ref.~\cite{Fujii:2007fi} which
studied the low energy effective theory in the Einstein-Maxwell
theory \cite{Gibbons:1986wg,Mukohyama:2005yw}. Arbitrary matter
and potentials for the dilaton on 4-branes are allowed to exist.
We use the gradient expansion technique to solve the
six-dimensional geometry assuming that the deviation from the
static solution is small \cite{Kanno:2002iaa, Shiromizu:2002qr}.
The gradient expansion method has been applied to various types of
brane-worlds
\cite{Shiromizu:2003dr,Onda:2003sj,Arroja:2006zz,Koyama:2006ni,Kobayashi:2006jw,Koyama:2005nq}.
Using this method, it is possible to solve the non-trivial
dependence of the bulk geometry on the four-dimensional
coordinates. By solving the effective four-dimensional equations,
we can derive the time-dependent solutions and compare them with
the exact six-dimensional time dependent solutions found in the
literature \cite{Tolley:2006ht, Kobayashi:2007hf,
Copeland:2007ur}. It is also possible to study whether we can
reproduce sensible cosmology at low energies or not. We also study
the possibility to stabilize the modulus using the potentials for
the dilaton on the branes along the line of
Ref.~\cite{Burgess:2007vi}.

We should mention that we concentrate our attention on a classical
dynamics in this paper and do not address the quantum corrections.
In fact the important feature of SLED is that it could also
provide a way to address the stability to quantum corrections to
the cosmological constant \cite{Burgess:2007ui}. This is because
in the 6D model, the Kaluza-Klein (KK) mass scale is of the order
of a $\mu m^{-1}$ and so is precisely at the energy scale of the
cosmological constant. The combination of the bulk supersymmetry
and scaling solution can maintain the quantum corrections to be of
order $ \mu m^{-4}$ and hence the order of the cosmological
constant.

The chapter is organized as follows. In section \ref{sec:BASIC},
basic equations are summarized. In section \ref{sec:GE}, we solve
the six-dimensional equations of motion using the gradient
expansion method. In section \ref{sec:branes}, the effective
theory on the regularized branes is derived by imposing junction
conditions. Then we derive time dependent cosmological solutions
in the effective theory and compare them with the exact
six-dimensional solutions. The possible way to stabilize the
modulus is discussed. Section \ref{sec:conclusion6D} is devoted to
conclusions.

\section{Basic equations}\label{sec:BASIC}

The relevant part of the supergravity action we consider is
\begin{eqnarray}
S =\int d^6x\sqrt{-g}\left[\frac{M^4}{2}{}^{(6)}\!\!R
-\frac{M^4}{2}\left(\partial \phi\right)^2
-\frac{1}{4}F^2e^{-\phi}-\frac{M^4}{2L_I^2} e^\phi \right],
\end{eqnarray}
where $\phi$ is the dilaton, $M$ is the fundamental scale of
gravity, $F^2:=F_{MN}F^{MN}$,
$(\partial\phi)^2:=g^{MN}\partial_M\phi\partial_N\phi$ and
$F_{MN}=\partial_MA_N-\partial_NA_M$ is the field strength of the
gauge field $A_M$. For the moment we are interested in solving the
6D bulk equations of motion. In Sec. \ref{sec:branes} we will add
two 4-branes (at positions $y=y_\pm$) and $L_I$ denotes the
different bulk curvature scales on either sides of the branes, see
Fig. \ref{fig:fig}. We start with the axisymmetric metric ansatz
\begin{eqnarray}
g_{MN}dx^Mdx^N &=&L_I^2
e^{2\lambda(x)}\frac{dy^2}{f(y)}+\ell^2e^{2[\psi(y,x)-\lambda(x)]}f(y)d
\theta^2+ 2\ell b_{\mu}(y, x)d\theta dx^{\mu} \nonumber
\\&&+a^2(y)\bar{h}_{\mu\nu}(y, x)dx^\mu dx^\nu, \label{aximetric}
\end{eqnarray}
where capital Latin indices numerate the 6D coordinates while the
Greek indices are restricted to the 4D coordinates.

The evolution equations along the $y$-direction are given by
\begin{eqnarray}
n^y \py K_{\hat\mu}^{\;\hat\nu}+\hat K K_{\hat\mu}^{\;\hat\nu}
&=&{}^{5}\!R_{\hat\mu}^{\;\hat\nu}-e^{-\lambda(x)}{}^5\!D_{\hat
\mu}{}^5 \!D^{\hat \nu}e^{\lambda (x)} -\partial_{\hat \mu}
\phi\partial^{\hat \nu} \phi -\frac{1}{4L_I^2} e^{\phi}
\delta_{\hat\mu}^{\;\hat\nu}
\nonumber\\&&-\frac{1}{M^4}\left(F_{\hat\mu M}F^{\hat\nu
M}-\frac{1}{8}\delta_{\hat\mu}^{\;\hat\nu}F^2\right)e^{-\phi},
\end{eqnarray}
where $n^y=e^{-\lambda}{\sqrt {f}}/L_I$, $K_{\hat\mu}^{\;\hat\nu}$
is the extrinsic curvature of $y=$ constant hypersurfaces, $\hat
K$ is its 5D trace, ${}^{5}\!R_{\hat\mu}^{\;\hat\nu}$ is the 5D
Ricci tensor and ${}^{5}\! D_{\hat\mu}$ is the covariant
derivative with respect to the 5D metric. Here, $\hat\mu = \mu$
and $\theta$. The Hamiltonian constraint is
\begin{eqnarray}
{}^{5}\!R+K_{\hat\mu}^{\;\hat\nu}K_{\hat\nu}^{\;\hat\mu}-\hat K^2
=-\frac{2}{M^4}\left(F_{yM}F^{yM}\!-\frac{1}{4}F^2\right)e^{-\phi}
-2(n^y \partial_y \phi)^2+(\partial \phi)^2\!+\!\frac{1}{L_I^2}
e^\phi,\!\!\!\!\!\!\!\!\!\!\!\!\!\!\!\!\!\nonumber\\
\end{eqnarray}
and the momentum constraints are
\begin{eqnarray}
{}^{5}\!
D_{\hat\nu}\left(K_{\hat\mu}^{\;\hat\nu}-\delta_{\hat\mu}^{\;\hat\nu}\hat
K\right) =\frac{1}{M^4}F_{\hat\mu M}F^{yM}n_ye^{-\phi} +D_{\hat
\mu} \phi \;n^y \partial_y \phi ,
\end{eqnarray}
where $n_y=e^{\lambda} L_I/{\sqrt {f}}$.

The Maxwell equations are given by
\begin{eqnarray}
\nabla_M\left(e^{-\phi}F^{MN}\right) =0,
\end{eqnarray}
where $\nabla_M$ is the covariant derivative with respect to the
6D metric. The dilaton equation of motion is
\begin{eqnarray}
\nabla_M\nabla^M \phi+\frac{1}{4M^4}
F^2e^{-\phi}-\frac{1}{2L_I^2}e^\phi=0.
\end{eqnarray}

\section{Gradient expansion approach}\label{sec:GE}

In this section we will use the gradient expansion method
\cite{Kanno:2002iaa, Shiromizu:2002qr} to solve the 6D bulk
equations. We assume that the length scale $\ell$ is of the same
order of $L_I$.  The small expansion parameter is the ratio of the
bulk curvature scale to the 4D intrinsic curvature scale,
\begin{eqnarray*}
\varepsilon = \ell^2|R|.
\end{eqnarray*}
See section \ref{sec:LE} and subsection \ref{subsec:PM} for more
details on the gradient expansion method. We expand the various
quantities as
\begin{eqnarray}
\bar{h}_{\mu\nu}=h_{\mu\nu}(x)+\varepsilon h^{(1)}_{\mu\nu}(y,
x)+\cdots, \quad \psi = \psi^{(0)}+\varepsilon\psi^{(1)}+\cdots,
\nonumber\\
\phi = \phi^{(0)}+\varepsilon\phi^{(1)}+\cdots, \quad
F_{y\theta}=\stac{(0)}{F_{y\theta}}+\varepsilon\!\stac{(1)}{F_{y\theta}}+\cdots,
\nonumber\\
K_{\mu}^{\;\nu}=\stac{(0)}{K_{\mu}^{\;\nu}}+\varepsilon
\!\stac{(1)}{K_{\mu}^{\;\nu}}+\cdots, \quad
K_{\theta}^{\;\theta}=\stac{(0)}{K_{\theta}^{\;\theta}}+\varepsilon
\!\stac{(1)}{K_{\theta}^{\;\theta}}+\cdots.
\end{eqnarray}
As to the other quantities, we follow~\cite{Fujii:2007fi} and
first assume
\begin{eqnarray}
b_{\mu}=\varepsilon^{1/2}b^{(1/2)}_{\mu}+\cdots, \quad
K_{\theta}^{\;\nu}=\varepsilon^{1/2}\!\stac{(1/2)}{K_{\theta}^{\;\nu}}+\cdots,
\nonumber\\
F^{\mu y}=\varepsilon^{1/2}\stac{(1/2)}{F^{\mu y}}+\cdots, \quad
F_{\mu\nu}=\varepsilon\!\stac{(1)}{F_{\mu\nu}}+\cdots,
\end{eqnarray}
and then will show that all the ${\cal O}(\varepsilon^{1/2})$
quantities in fact vanish. Since $\partial_{\mu}A_{\theta}\sim
\varepsilon^{1/2}\py A_{\theta}$, we have
$F_{\mu\theta}=\varepsilon^{1/2}\stac{(1/2)}{F^{\mu
\theta}}+\cdots$. We will show that this ${\cal
O}(\varepsilon^{1/2})$ term in $F_{\mu\theta}$ also vanishes. See
appendix \ref{app:epsilon12} for more details on the ${\cal
O}(\varepsilon^{1/2})$ quantities. The bulk energy-momentum tensor
contains terms like $F_{\mu\lambda}F^{\nu\lambda}$ but these do
not contribute to the low energy effective theory as they are
higher order in the gradient expansion. The 5D Ricci tensor is
given by
\begin{eqnarray}
{}^{5}\!R_{\mu}^{\;\nu} &=&\varepsilon \frac{1}{a^2} \left(
R_{\mu}^{\;\nu}[h] -\cD_\mu \cD^\nu \tilde\psi -\cD_\mu \tilde\psi
\cD^\nu \tilde\psi  \right)+\cdots,
\\
{}^{5}\!R_\theta^{\;\theta} &=&-\varepsilon \frac{1}{a^2} \left(
{\cal D}_\lambda {\cal D}^\lambda \tilde\psi + \cD_\lambda
\tilde\psi \cD^\lambda  \tilde\psi  \right)+ \cdots,
\end{eqnarray}
and ${}^5\! R_{\theta}^{\;\mu}={\cal O}(\varepsilon^{3/2})$, where
$\tilde\psi:=\psi^{(0)}-\lambda$. $R_{\mu}^{\;\nu}[h]$ and ${\cal
D}_\mu$ are respectively the Ricci tensor and the covariant
derivative constructed from $h_{\mu\nu}(x)$.

\subsection{Zeroth order equations}

The $\theta$ component of the Maxwell equations at zeroth order
reads
\begin{eqnarray}
\partial_y \Bigl(a^4 e^{-\phi^{(0)}+\psi^{(0)}}\stac{(0)}{F^{y \theta}}\Bigr)=0,
\end{eqnarray}
while the equation of motion for the dilaton at zeroth order is
given by
\begin{eqnarray}
\frac{1}{a^4}\py \Bigl(a^4f e^{\psi^{(0)}} \py \phi^{(0)}
\Bigr)+\frac{1}{2}
\left(\frac{1}{M^2\ell}\stac{(0)}{F_{y\theta}}\right)^2e^{2\lambda-\phi^{(0)}-\psi^{(0)
}} -\frac{1}{2}e^{2\lambda+\phi^{(0)}+\psi^{(0)}}=0.
\end{eqnarray}

The $(\mu\nu)$ and $(\theta\theta)$ components of the evolution
equations are given respectively by
\begin{eqnarray}
&&f \left[ \py \left(\frac{\py a}{a} \right) +\left(4 \frac{\py
a}{a}+\frac{\py f}{f}+\py\psi^{(0)}  \right) \frac{\py
a}{a}\right] \nonumber\\&&\qquad\qquad\qquad
=\frac{1}{4}\left(\frac{1}{M^2\ell}\stac{(0)}{F_{y\theta}}\right)^2e^{2\lambda-\phi^{(0)}-2\psi^{(0)}}
-\frac{1}{4}e^{2\lambda+\phi^{(0)}},
\\
&&f \left[\py \left(\frac{\py f}{2f}+\py \psi^{(0)} \right) +
\left(4 \frac{\py a}{a}+\frac{\py f}{f}+\py \psi^{(0)}   \right)
\left(\frac{\py f}{2f}+\py \psi^{(0)}\right)    \right]
\nonumber\\&&\qquad\qquad\qquad
=-\frac{3}{4}\left(\frac{1}{M^2\ell}\stac{(0)}{F_{y\theta}}\right)^2e^{2\lambda-\phi^{(0)}-2\psi^{(0)}}
-\frac{1}{4}e^{2\lambda+\phi^{(0)}},
\end{eqnarray}
and the Hamiltonian constraint becomes
\begin{eqnarray}
4f\bigg[3\left(\frac{\py
a}{a}\right)^2&&\!\!\!\!\!\!\!\!\!+\frac{\py a}{a}\left(\frac{\py
f}{f}+2\py\psi^{(0)}\right)\bigg] =\nonumber\\&&
\left(\frac{1}{M^2\ell}\stac{(0)}{F_{y\theta}}\right)^2e^{2\lambda-\phi^{(0)}-2\psi^{(0)}}
+ f\bigl(\py \phi^{(0)}\bigr)^2-e^{2\lambda+\phi^{(0)}}.
\end{eqnarray}

The solutions for the above equations are obtained as
\begin{eqnarray}
a(y)=\sqrt{y},\qquad
f(y)=\frac{1}{4}\left(-y+\frac{\mu}{y}-\frac{q^2}{y^3} \right),
\qquad \lambda(x) = \frac{1}{2}\Phi(x) ,
\end{eqnarray}
and
\begin{eqnarray}
\psi^{(0)}(y, x)&=&\Phi(x)+\sigma(x), \qquad \phi^{(0)}(y, x)=-\ln
y-\Phi(x), \nonumber\\ \stac{(0)}{F}_{y\theta}&=& M^2\ell\frac{
q}{a^4}e^{\phi^{(0)}+\psi^{(0)}} =M^2\ell
\frac{q}{y^3}e^{\sigma(x)},
\end{eqnarray}
where $\mu$ and $q$ are integration constants. The momentum
constraint implies $\partial_{\mu}\sigma = 0$, and therefore
$\sigma=$ constant. This immediately leads to
$\stac{(1/2)}{F_{\mu\theta}}=0$ and hence $F_{\mu\theta}={\cal
O}(\varepsilon^{3/2})$. In the following, we put $\sigma=0$
without loss of generality. The 6D metric at the zeroth order is
given by
\begin{equation}
g_{MN} dx^M dx^N = e^{\Phi(x)} \left[ L_I^2 \frac{dy^2}{f} +
\ell^2 f d \theta^2 \right]  +a^2(y) h_{\mu \nu}(x) dx^{\mu}
dx^{\nu}.
\end{equation}
Then we can see that $\Phi(x)$ is associated with the scaling
symmetry $g_{MN} \to e^{\omega} g_{MN}$ and $e^{\phi} \to e^{\phi
- \omega}$. In fact, we will find that a solution for $h_{\mu
\nu}$ is given by $h_{\mu \nu} = e^{\Phi} \eta_{\mu \nu}$ if the
brane preserves the scaling symmetry, where $\eta_{\mu \nu}$
denotes the 4D Minkowski metric.

\subsection{First order equations}
At first order, the $(\mu\nu)$ component of the evolution
equations is given by
\begin{eqnarray}
&&\!\!\!\!\!\!\!\!\!\frac{\sqrt{f}}{L_I}e^{-\Phi/2}\left[ \py
\!\stac{(1)}{K_{\mu}^{\;\nu}}+ \left(\frac{2}{y}+\frac{\py
f}{2f}\right)\!\stac{(1)}{K_{\mu}^{\;\nu}} +\frac{1}{2y}\Bigl(
\stac{(1)}{K_{\lambda}^{\;\lambda}}+\stac{(1)}{K_{\theta}^{\;\theta}}
\Bigr)\delta_{\mu}^{\;\nu} \right] \nonumber\\&&=\frac{1}{y}\left(
R_{\mu}^{\;\nu}-{\cal D}_\mu {\cal D}^\nu \Phi-\frac{3}{2}{\cal
D}_\mu \Phi {\cal D}^\nu \Phi\right)
-\frac{1}{4L_I^2}e^{\phi^{(0)}} \phi^{(1)} \delta_\mu^{\;\nu}
+\frac{1}{4} \cF\delta_{\mu}^{\;\nu}, \label{ev1st}
\end{eqnarray}
where
\begin{eqnarray}
\cF:=
\frac{1}{M^4}\Bigl(\stac{(0)}{F_{y\theta}}\stac{(1)}{F^{y\theta}}
+\stac{(1)}{F_{y\theta}}\stac{(0)}{F^{y\theta}}\Bigr)
e^{-\phi^{(0)}}-\frac{1}{M^4} \stac{(0)}{F}_{y \theta}
\stac{(0)}{F^{y\theta}}e^{-\phi^{(0)}}\phi^{(1)}.
\end{eqnarray}
The 4D Ricci tensor $R_{\mu}^{\;\nu}$ does not depend on $y$
because it is computed from $h_{\mu\nu}$ which is a function of
$x^{\mu}$ only and the index is raised by $h_{\mu \nu}$.

The 4D traceless part of Eq.~(\ref{ev1st}) is found to be
\begin{eqnarray}
\py\left(y^2\sqrt{f}\;\mathbb{K}_{\mu}^{\;\nu}\right) =
e^{\Phi/2}yL_I\mathbb{R}_{\mu}^{\;\nu}, \label{tl}
\end{eqnarray}
where we defined
$\mathbb{K}_{\mu}^{\;\nu}:=\stac{(1)}{K_{\mu}^{\;\nu}}-(1/4)\delta_{\mu}^{\;\nu}\!\!\stac{(1)}{K_{\lambda}^{\;\lambda}}$
and
\begin{eqnarray}
\mathbb{R}_{\mu}^{\;\nu}:=R_{\mu}^{\;\nu}-\frac{1}{4}
\delta_{\mu}^{\;\nu}R-\left({\cal D}_\mu {\cal D}^\nu \Phi
-\frac{1}{4}\delta_{\mu}^{\;\nu} {\cal D}^2 \Phi
\right)-\frac{3}{2}\left[ {\cal D}_\mu \Phi {\cal D}^\nu \Phi
-\frac{1}{4}\delta_{\mu}^{\;\nu} ({\cal D} \Phi)^2
\right],\nonumber\\
\end{eqnarray}
where $\cD^2\Phi:=h^{\mu\nu}\cD_{\mu}\cD_{\nu}\Phi$ and
$(\cD\Phi)^2:=h^{\mu\nu}\cD_{\mu}\Phi\cD_{\nu}\Phi$. The general
solution to the above equation is given by
\begin{eqnarray}
\mathbb{K}_{\mu}^{\;\nu}=\frac{e^{\Phi/2}}{2\sqrt{f}}L_I\mathbb{R}_{\mu}^{\;\nu}
+\frac{1}{y^2\sqrt{f}}\mathbb{C}_{\mu}^{\;\nu}(x),\label{general-sol}
\end{eqnarray}
where the traceless tensor $\mathbb{C}_{\mu}^{\;\nu}(x)$ is the
integration ``constant'' to be fixed by the boundary conditions.

The 4D trace part of the evolution equations is
\begin{eqnarray}
\frac{\sqrt{f}}{L_I}e^{- \Phi/2} &&\!\!\!\!\!\!\!\!\!\left[ \py
\!\stac{(1)}{K_{\lambda}^{\;\lambda}}+\left(\frac{4}{y}+\frac{\py
f}{2f}\right)
\!\stac{(1)}{K_{\lambda}^{\;\lambda}}+\frac{2}{y}\stac{(1)}{K_{\theta}^{\;\theta}}
\right]\nonumber\\&&=\frac{1}{y}\left[ R-{\cal D}^2
\Phi-\frac{3}{2}({\cal D}\Phi)^2 \right] +
\cF-\frac{1}{L_I^2}\frac{e^{-\Phi}}{y}\phi^{(1)} , \label{tp1st}
\end{eqnarray}
and the $(\theta\theta)$ component of the evolution equations is
\begin{eqnarray}
\frac{\sqrt{f}}{L_I}e^{-\Phi/2}&&\!\!\!\!\!\!\!\!\!\left[ \py
\!\stac{(1)}{K_{\theta}^{\;\theta}}+ \left(\frac{2}{y}+\frac{\py
f}{f}\right)\!\stac{(1)}{K_{\theta}^{\;\theta}} +\frac{\py
f}{2f}\stac{(1)}{K_{\lambda}^{\;\lambda}} \right]
\nonumber\\&&=-\frac{1}{2y}\left[{\cal D}^2 \Phi+({\cal
D}\Phi)^2\right] -\frac{3}{4} \cF-\frac{1}{4L_I^2
}\frac{e^{-\Phi}}{y}\phi^{(1)}. \label{thth1st}
\end{eqnarray}
The Hamiltonian constraint at first order reduces to
\begin{eqnarray}
&&\!\!\!\!\!\!\!\!\!\!\!\!\!\!\!\!\!\!\!\!\!\!\!\!\frac{1}{y}\left[R-{\cal
D}^2\Phi-\frac{3}{2}({\cal D}\Phi)^2 \right]+ \cF
\nonumber\\&&\!\!\!\!\!\!\!\!\!\!\!\!\!\!\!\!\!\!=2\frac{\sqrt{f}}{L_I}e^{-\Phi/2}\left[
\left(\frac{3}{2y}+\frac{\py
f}{2f}\right)\stac{(1)}{K_{\lambda}^{\;\lambda}}+
\frac{2}{y}\stac{(1)}{K_{\theta}^{\;\theta}}\right]+\frac{1}{L_I^2}\frac{e^{-\Phi}}{y}\phi^{(1)}
+ \frac{2f}{L_I^2}\frac{e^{-\Phi}}{y}\py\phi^{(1)}. \label{ham1st}
\end{eqnarray}
The dilaton equation of motion at first order reads
\begin{eqnarray}
&&\frac{f}{L_I^2}e^{-\Phi}\left[\py^2\phi^{(1)}+\left(\frac{2}{y}+\frac{\py
f}{f}\right)\py \phi^{(1)}\right]
-\frac{\sqrt{f}}{L_I}\frac{e^{-\Phi/2}}{y}\left(
\stac{(1)}{K_{\lambda}^{\;\lambda}}+\stac{(1)}{K_{\theta}^{\;\theta}}
\right) \nonumber\\&&\qquad
-\frac{1}{y}\left[\cD^2\Phi+(\cD\Phi)^2\right]
-\frac{1}{2L_I^2}\frac{e^{-\Phi}}{y}\phi^{(1)}+\frac{1}{2}\cF=0.\label{dil1st}
\end{eqnarray}

Now we define convenient quantities
\begin{eqnarray}
\cJ:=n^y\py\phi^{(1)}+\frac{1}{2}\stac{(1)}{K_{\lambda}^{\;\lambda}}
\end{eqnarray}
and
\begin{eqnarray}
\cK:&=&\frac{3}{4}
\stac{(1)}{K_{\lambda}^{\;\lambda}}+\stac{(1)}{K_{\theta}^{\;\theta}}
+\frac{\sqrt{f}}{L_I}e^{-\Phi/2}\left(\frac{\py
f}{2f}-\frac{1}{2y}\right)\psi^{(1)}\nonumber\\&&+
\frac{y}{M^4\ell^2L_I\sqrt{f}}\stac{(0)}{F_{y\theta}}
e^{-\Phi/2}A_{\theta}^{(1)}
-\frac{\sqrt{f}}{L_I}\frac{e^{-\Phi/2}}{y}\phi^{(1)}.
\end{eqnarray}
The evolution equations for these variables can be derived using
Eqs.~(\ref{tp1st})--(\ref{dil1st}). With some manipulation one
arrives at
\begin{eqnarray}
\py\left(y^2\sqrt{f}\cJ\right)&=& \frac{1}{2}e^{\Phi/2}yL_I\left[
 R+\cD^2\Phi+\frac{1}{2}(\cD\Phi)^2
\right],\label{evJ}\\
\py\left(y^2\sqrt{f}\cK\right)&=&\frac{1}{4}e^{\Phi/2}yL_I\left[
 R-3\cD^2\Phi-\frac{7}{2}(\cD\Phi)^2
\right].\label{evK}
\end{eqnarray}
The two equations have the same structure as that of
Eq.~(\ref{tl}). The general solution for each evolution equation
contains one integration ``constant'' which will be determined by
the boundary conditions.

In terms of the above variables, the momentum constraint equations
are simplified to
\begin{eqnarray}
 \cD_{\nu}\left(e^{\Phi/2}\mathbb{K}_{\mu}^{\;\nu}\right)
- \cD_{\mu}\left(e^{\Phi/2}\cK\right)
+e^{\Phi/2}\cJ\cD_{\mu}\Phi=0.\label{momKJ}
\end{eqnarray}

\section{Junction conditions and effective theory on a regularized brane}\label{sec:branes}

\begin{figure}
\scalebox{.5}
 { \includegraphics*{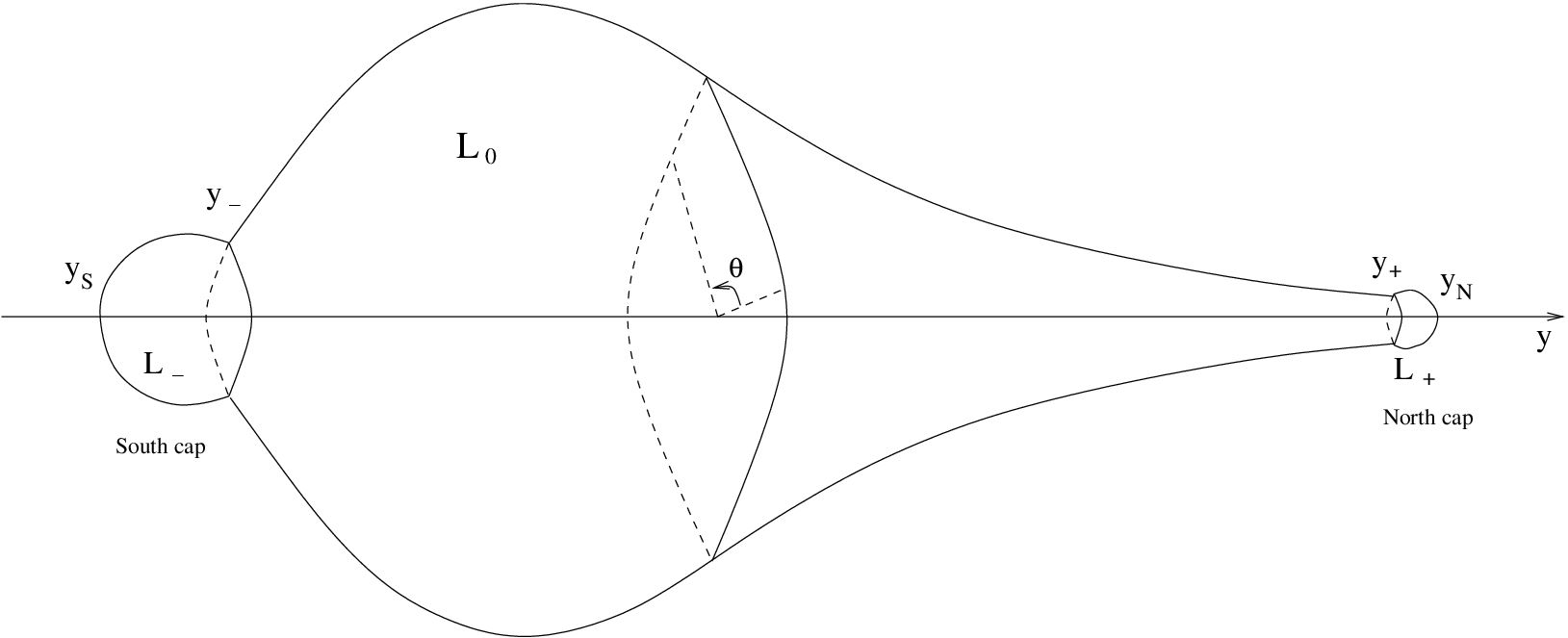}
  }
\caption{Schematic representation of the bulk spacetime with two
regularized caps.}\label{fig:fig}
\end{figure}

Our choice of parameters $\mu$, $q$ implies that $f(y)$ vanishes
at $y_N$ and $y_S$. These points are conical singularities that
are sourced by 3-branes. In order to accommodate usual matter on
the branes we need to resolve these singularities. We will use the
regularization scheme of~\cite{Peloso:2006cq, Burgess:2007vi}. The
conical branes are replaced with cylindrical codimension-one
branes at $y=y_\pm$ and their interiors are filled with regular
caps. See figure \ref{fig:fig} for a sketch of the model. The
geometry of the caps and the central bulk is described by the 6D
solutions found in the previous section, with different curvature
scales $L_+$ ($L_-$) for the north (south) cap and $L_0$ for the
central bulk.

The action of each brane is taken to be
\begin{eqnarray}
S_{{\rm brane}}&=&-\int d^5x\sqrt{-q}\left[
V(\phi)+\frac{1}{2}U(\phi)(\partial_{\hat\mu}\Sigma-\e A_{\hat\mu}
) (\partial^{\hat\mu}\Sigma-\e A^{\hat\mu}
)\right]\nonumber\\&&+\int d^5x\sqrt{-q}\,{\cal L}_{{\rm m}},
\end{eqnarray}
where $q_{\hat\mu\hat\nu}$ is the induced metric on the 4-brane,
$V(\phi)$ and $U(\phi)$ are the couplings to the dilaton, and
${\cal L}_{{\rm m}}$ is the Lagrangian of usual matter localized
on the brane. At this stage we assume that the brane matter ${\cal
L}_{{\rm m}}$ does not couple to the dilaton field. We introduce a
Stueckelberg field $\Sigma$, which is obtained by integrating out
the massive radial mode of a brane Higgs field. The equation of
motion for $\Sigma$ gives the gradient expansion form of the
solution as~\cite{Fujii:2007fi}
\begin{eqnarray}
\Sigma(\theta, x)=n\theta+c^{(0)}(x)+\varepsilon
c^{(1)}(x)+\cdots,
\end{eqnarray}
where $n$ must be an integer because of the periodicity
$\theta\simeq\theta+2\pi$.

The jump conditions for the Maxwell field are
\begin{eqnarray}
\left[\left[ n^MF_{MN}e^{-\phi}\right]\right]=-\e
U(\partial_N\Sigma-\e A_N),\label{gen-max}
\end{eqnarray}
while for the dilaton field we have
\begin{eqnarray}
\left[\left[ n^M\partial_M\phi
\right]\right]=\frac{1}{M^4}\left[\frac{dV}{d\phi}
+\frac{1}{2}\frac{dU}{d\phi} (\partial_{\hat \lambda}\Sigma -\e
A_{\hat \lambda})(\partial^{\hat \lambda}\Sigma-\e A^{\hat
\lambda}) \right],\label{gen-dil}
\end{eqnarray}
where $[[ F ]]_{y_{{\rm b}}}:=\lim_{\epsilon\to
0}\left(F|_{y_{{\rm b}}+\epsilon}-F|_{y_{{\rm
b}}-\epsilon}\right)$. Here and hereafter in this section all the
quantities are evaluated at the position of the brane under
consideration. The Israel conditions are given by
\begin{eqnarray}
\left[\left[
K_{\hat\mu}^{\;\hat\nu}-\delta_{\hat\mu}^{\;\hat\nu}\hat
K\right]\right] =-\frac{1}{M^4}T_{\hat \mu ({\rm tot})}^{\;\hat
\nu}\label{gen-is}
\end{eqnarray}
where
\begin{eqnarray}
T_{\hat \mu ({\rm tot})}^{\;\hat \nu}&=&U \left[ (\partial_{\hat
\mu}\Sigma -\e A_{\hat \mu})(\partial^{\hat \nu}\Sigma-\e A^{\hat
\nu}) -\frac{1}{2}\delta_{\hat \mu}^{\;\hat \nu} (\partial_{\hat
\lambda}\Sigma -\e A_{\hat \lambda})(\partial^{\hat
\lambda}\Sigma-\e A^{\hat \lambda}) \right]\nonumber\\&&-V
\delta_{\hat \mu}^{\;\hat \nu} +T_{\hat \mu}^{\;\hat \nu},
\end{eqnarray}
and $T_{\hat \mu}^{\;\hat \nu}$ represents the matter
energy-momentum tensor.

\subsection{Zeroth order}
At zeroth order in the gradient expansion the junction
conditions~(\ref{gen-max})--(\ref{gen-is}) are written as
\begin{eqnarray}
\!\!\!\!\!\!\!\!\!\!\!\!\!\!\!\!\!\!\!\!\!\!
\!\!\!\!\!\!\!\!\!\!\!\!\!\!\!\!\!\!\!\!\!\!\!\!\!\!\!\!\text{Maxwell:}&&\;\;
\left[\left[ \frac{\sqrt{f}}{L_I}\!\stac{(0)}{F}_{y \theta}
y\,e^{\Phi/2}\right]\right]=-\e U^{(0)} \left(n-\e
A_\theta^{(0)}\right),\label{maxj1}
\end{eqnarray}
\begin{eqnarray}
\text{Dilaton:}&&\;\; \left[\left[
\frac{\sqrt{f}}{L_I}\frac{1}{y}e^{-\Phi/2} \right]\right]
=-\frac{1}{M^4}\left[
\frac{dV^{(0)}}{d\phi^{(0)}}+\frac{1}{2}\frac{dU^{(0)}}{d\phi^{(0)}}\frac{e^{-\Phi}}{\ell^2f}
\left(n-\e
A_{\theta}^{(0)}\right)^2\right],\nonumber\\\label{dilj1}
\end{eqnarray}
\begin{eqnarray}
\!\!\!\!\!\!\!\!\!\!\!\!\!\!\!\!\!\!\!\!\text{Israel\;}(\mu\nu):\;\;\;\;\;\;
\Bigg[\Bigg[ \frac{{\sqrt {f}}}{L_I}
&&\!\!\!\!\!\!\!\!\!\left(\frac{3}{2y}+\frac{\py f}{2 f}
\right)e^{-\Phi/2} \Bigg] \Bigg]\nonumber\\&&=-\frac{1}{M^4}\left[
V^{(0)} +\frac{1}{2}U^{(0)} \frac{e^{-\Phi}}{\ell^2f} \left(n-\e
A_{\theta}^{(0)}\right)^2\right],\label{Imn1}
\end{eqnarray}
\begin{eqnarray}
\text{Israel\;}(\theta\theta):&&\;\; \left[\left[ \frac{{\sqrt
{f}}}{L_I} \frac{2}{y} e^{-\Phi/2}\right] \right]=
-\frac{1}{M^4}\left[ V^{(0)} -\frac{1}{2}U^{(0)}
\frac{e^{-\Phi}}{\ell^2f} \left(n-\e
A_{\theta}^{(0)}\right)^2\right].\nonumber\\\label{Ithth1}
\end{eqnarray}
The above conditions relate several parameters with each other,
and the detail of the parameter counting of the configuration is
found in Ref.~\cite{Burgess:2007vi}. In particular, the dilaton
jump condition~(\ref{dilj1}) and the Israel
condition~(\ref{Ithth1}) imply
\begin{eqnarray}
\frac{V^{(0)}}{2}-\frac{dV^{(0)}}{d\phi^{(0)}}-\frac{1}{2}\frac{e^{-\Phi}}{\ell^2
f} \left(\frac{U^{(0)}}{2}+\frac{dU^{(0)}}{d\phi^{(0)}}\right)
\left(n-\e A_{\theta}^{(0)}\right)^2=0.\label{lambda3}
\end{eqnarray}

The classical scaling symmetry is preserved by the special choice
of the potentials~\cite{Aghababaie:2003ar, Burgess:2007vi}
\begin{eqnarray}
V(\phi)=v e^{\phi/2}, \qquad U(\phi)=u e^{-\phi/2}.
\label{scalepotential}
\end{eqnarray}
With these potentials the junction
conditions~(\ref{maxj1})--(\ref{Ithth1}) put no constraints on
$\Phi(x)$ and Eq.~(\ref{lambda3}) is trivially satisfied. In this
case the first order analysis will provide the equation of motion
for $\Phi(x) $, as will be seen in the next subsection. In the
following, we assume that at the zeroth order, the potentials are
given by (\ref{scalepotential}), that is,
$U^{(0)}(\phi^{(0)})=u^{(0)}e^{-\phi^{(0)}/2}$ and
$V^{(0)}(\phi^{(0)})=v^{(0)}e^{\phi^{(0)}/2}$. Then we expand the
potentials as follows:
\begin{eqnarray}
V(\phi) &=& V^{(0)}(\phi^{(0)}) + \varepsilon
\left( V^{(1)}(\phi^{(0)}) + \frac{d V^{(0)}}{d \phi^{(0)}} \phi^{(1)} \right), \\
U(\phi) &=& U^{(0)}(\phi^{(0)}) + \varepsilon \left(
U^{(1)}(\phi^{(0)}) + \frac{d U^{(0)}}{d \phi^{(0)}} \phi^{(1)}
\right),
\end{eqnarray}
where $V^{(1)}(\phi^{(0)})$ and $U^{(1)}(\phi^{(0)})$ stand for
the deviations from the zeroth order potentials.

\subsection{First order}

The 4D traceless part of the Israel conditions at first order is
given by
\begin{eqnarray}
\left[\left[\mathbb{K}_{\mu}^{\;\nu}\right]\right]=-\frac{1}{M^4}\mathbb{T}_{\mu}^{\;\nu},
\label{ktless}
\end{eqnarray}
where
$\mathbb{T}_{\mu}^{\;\nu}:=T_{\mu}^{\;\nu}-(1/4)\delta_{\mu}^{\;\nu}T_{\lambda}^{\;\lambda}$.
The 4D trace part of the Israel conditions reduces to
\begin{eqnarray}
\Biggl[\Biggl[\frac{3}{4}
\stac{(1)}{K_{\lambda}^{\;\lambda}}+\stac{(1)}{K_{\theta}^{\;\theta}}
\Biggr]\Biggr] \!\!&=&\!\! \frac{1}{4M^4}T_{\lambda}^{\;\lambda} -
\frac{1}{M^4} \Delta V +\frac{U^{(0)}}{M^4}\Delta
\nonumber\\&&\!\!\!\!-\frac{1}{M^4}\left[
\frac{dV^{(0)}}{d\phi^{(0)}} +\frac{1}{2}
\frac{dU^{(0)}}{d\phi^{(0)}} \frac{e^{-\Phi}}{\ell^2f} \left(n-\e
A_{\theta}^{(0)}\right)^2 \right]\phi^{(1)},\label{tr1}
\end{eqnarray}
where we defined
\begin{equation}
\Delta V = V^{(1)}(\phi^{(0)}) + \frac{1}{2} U^{(1)}(\phi^{(0)})
\frac{e^{-\Phi}}{\ell^2 f} \left (n-\e A_{\theta}^{(0)} \right)^2,
\end{equation}
and
\begin{eqnarray}
\Delta:=\frac{e^{-\Phi}}{\ell^2 f}\left(n-\e
A_{\theta}^{(0)}\right) \left[ \e A_{\theta}^{(1)}+ \left(n-\e
A_{\theta}^{(0)}\right)\psi^{(1)} \right].
\end{eqnarray}

Using the zeroth order junction conditions, Eq.~(\ref{tr1}) simply
gives
\begin{eqnarray}
[[\cK]] =\frac{1}{4M^4}T_{\lambda}^{\;\lambda}-\frac{1}{M^4}
\Delta V.\label{ktr}
\end{eqnarray}
The $(\theta\theta)$ component of the Israel conditions is
\begin{eqnarray}
\Bigl[\Bigl[\stac{(1)}{K_{\lambda}^{\;\lambda}}\Bigr]\Bigr]&=&\frac{T_{\theta}^{\;\theta}}{M^4}
-\frac{U^{(0)}}{M^4}\Delta-\frac{1}{M^4}\left[
 \frac{dV^{(0)}}{d\phi^{(0)}} -\frac{1}{2} \frac{dU^{(0)}}{d\phi^{(0)}}
\frac{e^{-\Phi}}{\ell^2 f}\left(n-\e A_{\theta}^{(0)}\right)^2
\right]\phi^{(1)} \nonumber\\&&-\frac{1}{M^4}\Delta V+
U^{(1)}(\phi^{(0)}) \frac{e^{-\Phi}}{M^4\ell^2 f} \left (n-\e
A_{\theta}^{(0)} \right)^2,
\end{eqnarray}
and the dilaton jump condition is
\begin{eqnarray}
\left[\left[n^y\py\phi^{(1)}\right]\right] &=&
-\frac{1}{M^4}\frac{dU^{(0)}}{d\phi^{(0)}}\Delta+\frac{1}{M^4}\left[
 \frac{d^2V^{(0)}}{d\phi^{(0) 2}} +\frac{1}{2} \frac{d^2 U^{(0)}}{d \phi^{(0) 2}}
\frac{e^{-\Phi}}{\ell^2 f}\left(n-\e A_{\theta}^{(0)}\right)^2
\right]\phi^{(1)}\nonumber\\&&-\frac{1}{M^4}\frac{d}{d \Phi}
(\Delta V)-\frac{1}{2}U^{(1)}(\phi^{(0)})
\frac{e^{-\Phi}}{M^4\ell^2 f} \left (n-\e A_{\theta}^{(0)}
\right)^2.
\end{eqnarray}

Using the fact that the zeroth order potential have the scale
invariant forms (\ref{scalepotential}), the above two conditions
are combined to give
\begin{eqnarray}
\left[\left[\cJ\right]\right] =
\frac{1}{2M^4}T_{\theta}^{\;\theta} -\frac{1}{M^4} \frac{d}{d
\Phi} (\Delta V) - \frac{1}{2 M^4} \Delta V.
\end{eqnarray}
Therefore, the momentum constraints become
\begin{equation}
\cD_{\nu}\left(e^{\Phi/2}T_{\mu}^{\;\nu} -e^{\Phi/2} \Delta V
\delta^{\;\nu}_{\mu}\right) =\left( \frac{1}{2}
T_{\theta}^{\;\theta} -\frac{d}{d\Phi} (\Delta V) - \frac{1}{2}
\Delta V \right) e^{\Phi/2}\cD_{\mu}\Phi.
\end{equation}
In terms of the energy-momentum tensor integrated along the
$\theta$-direction,
\begin{eqnarray}
\overline{T}_{\hat\mu}^{\;\hat\nu}:=2\pi\ell
\sqrt{f}e^{\Phi/2}T_{\hat\mu}^{\;\hat\nu},
\end{eqnarray}
this can be rewritten as
\begin{eqnarray}
\cD_{\nu} \overline{T}_{\mu}^{\;\nu}
=\frac{1}{2}\overline{T}_{\theta}^{\;\theta}  \cD_{\mu}\Phi.
\end{eqnarray}

To fix the integration constants completely, we need the boundary
conditions at the north and south poles. Near a pole with the
coordinate $y=y_p$, where $p=\{N,S\}$, we have $f\sim y-y_p$. In
order for the evolution equations~(\ref{tl}), (\ref{evJ}), and
(\ref{evK}) to be regular at the poles, we require
\begin{eqnarray}
\mathbb{K}_{\mu}^{\;\nu},\;\cK,\;\cJ \lesssim |y-y_p|^{1/2}\to 0.
\end{eqnarray}
Now we can determine all the integration constants included in the
general solutions for $\mathbb{K}_{\mu}^{\;\nu}$, $\cK$ and $\cJ$.
Since the structure of the evolution equations and boundary
conditions are identical for these three variables, we summarize
the procedure to fix the integration constants in
appendix~\ref{app:solve}, and here we focus on the resulting
effective theory on the brane.

Using Eqs.~(\ref{ktless}) and~(\ref{ktr}) together with the
solution for $\mathbb{K}_{\mu}^{\;\nu}$ and $\cK$ in terms of $R$
and $\Phi$, we end up with the effective equations
\begin{eqnarray}
e^{\Phi}\bigg(&&\!\!\!\!\!\!\!\!\!R_{\mu}^{\;\nu}[q^+]-\frac{1}{2}\delta_{\mu}^{\;\nu}
R[q^+] - \Phi^{;\nu}_{;\mu}+
\delta_{\mu}^{\;\nu}\Phi^{;\lambda}_{;\lambda} - \frac{3}{2}
\Phi_{;\mu} \Phi^{;\nu}+\frac{5}{4} \delta_{\mu}^{\;\nu}
\Phi_{;\lambda} \Phi^{;\lambda} \bigg) \nonumber\\&&=\kappa_+^2
\left(\overline{T}_{\mu}^{+\nu} - \overline{\Delta V}^+
\delta^{\nu}_{\mu} \right) +\frac{a_-^2 }{a_+^2}\kappa_-^2 \left(
\overline{T}_{\mu}^{-\nu} - \overline{\Delta
V}^-\delta^{\nu}_{\mu} \right), \label{effective}
\end{eqnarray}
where the 4D gravitational couplings are defined as
\begin{eqnarray}
\kappa^2_{\pm}:=\frac{a_\pm^2}{2\pi\ell_*^2M^4}, \quad \text{with}
\quad \ell_*^2=\ell\int_{y_S}^{y_N}L_I ydy,
\end{eqnarray}
$;$ denotes a covariant derivative with respect to the induced
metric $q^+_{\mu \nu} =a_+^2 h_{\mu \nu}$, $R_{\mu \nu}[q^+]$ is
Ricci tensor computed from $q^+_{\mu \nu}$ and the potential
integrated along the $\theta$-direction is defined as
\begin{equation}
\overline{\Delta V} = 2 \pi \ell \sqrt{f} e^{\Phi/2} \Delta V.
\end{equation}

The first order equations for $\cJ$ give the equation of motion
for $\Phi$:
\begin{eqnarray}
\left(e^{\Phi} \right)^{;\mu}_{;\mu} &=&\frac{\kappa^2_+}{4}\left(
\overline{T}_{\lambda}^{+\lambda}-\overline{T}_{\theta}^{+\theta}
+ 2 \frac{d}{d \Phi} (\overline{\Delta V}^+) - 4 \overline{\Delta
V}^+ \right)
\nonumber\\&&+\frac{a_-^2}{a_+^2}\frac{\kappa^2_-}{4}\left(
\overline{T}_{\lambda}^{-\lambda}-\overline{T}_{\theta}^{-\theta}+
 2 \frac{d}{d \Phi} (\overline{\Delta V}^-) - 4 \overline{\Delta V}^-
\right).
\end{eqnarray}

For simplicity let us ignore the matter energy-momentum tensor and
the potential on the south brane:
$\overline{T}^{-\hat\nu}_{\hat\mu}=\overline{\Delta V}^-=0$. In
the absence of the $(\theta\theta)$ component of the energy
momentum tensor on the north brane, the 4D effective equations can
be deduced from the action
\begin{eqnarray}
S_{{\rm eff}}=\int d^4x\sqrt{-q^+}\left[\frac{e^\Phi}{2
\kappa_+^2} \left(R[q^+]-\omega_{{\rm BD}} \Phi_{;\mu}\Phi^{;
\mu}\right)- \overline{\Delta V}^+ + \overline{{\cal L}}_{{\rm
m}}^+\right],
\end{eqnarray}
with the Brans-Dicke parameter $\omega_{{\rm BD}}=1/2$ (see also
appendix B of Ref.~\cite{Kobayashi:2007hf}).

\subsection{The exact time-dependent solutions in the 4D effective theory}\label{sec:4Det}
We now consider cosmological solutions in the 4D effective theory
and compare them with the known solutions to the full 6D field
equations~\cite{Tolley:2006ht, Kobayashi:2007hf, Copeland:2007ur}.

Let us assume $\overline{T}_{\mu}^{\pm \nu}=\overline{\Delta
V}^-=0$ and $\overline{T}_{\theta}^{\pm\theta}=0$. We consider the
case where the first order potential is scale invariant form. Then
$\Delta V^+ \propto e^{-\Phi/2}$ and $\overline{\Delta V}^+ =$
const. $:=\Lambda/\kappa_+^2$. We go to the Einstein frame defined
by
\begin{equation}
\tilde h_{\mu\nu}=e^{\Phi} q^+_{\mu\nu},
\end{equation}
and then the equations of motion become
\begin{eqnarray}
\tilde{R}_\mu^{\;\nu}[\tilde
h]-\frac{1}{2}\delta_\mu^{\;\nu}\tilde{R}[\tilde h] &=&-\Lambda
e^{-2\Phi}\delta_\mu^{\;\nu}+2
\Phi_{|\mu}\Phi^{|\nu}-\Phi_{|\lambda}\Phi^{|\lambda}\delta_\mu^{\;\nu},
\\
\Phi_{|\mu}^{\;|\mu}&=&-\Lambda e^{-2\Phi},
\end{eqnarray}
where $|$ stands for the covariant derivative with respect to the
Einstein frame metric $\tilde h_{\mu\nu}$.

Taking $\tilde h_{\mu\nu}$ to be a flat Friedman-Robertson-Walker
metric, $\tilde
h_{\mu\nu}dx^{\mu}dx^{\nu}=A^2(\tau)(-d\tau^2+d\mathbf{x}^2)$, the
equations of motion reduce to
\begin{equation}
\frac{A''}{A}=\frac{2}{3}\Lambda A^2e^{-2\Phi}-\frac{1}{3}\Phi'^2,
\quad \left(\frac{A'}{A}\right)^2=\frac{1}{3}\Lambda
A^2e^{-2\Phi}+\frac{1}{3}\Phi'^2,
\end{equation}
and
\begin{equation}
\Phi''+2\frac{A'}{A}\Phi'=\Lambda A^2e^{-2\Phi},
\end{equation}
where ${}':=d/d\tau$. For $\Lambda=0$ a solution of the above
equations is
\begin{equation}
A^2(\tau)=A_1\tau+A_2, \quad \Phi(\tau)=\pm\sqrt{3}\ln A(\tau) +
A_3,\label{power}
\end{equation}
where $A_i\;(i =1,2,3)$ are integration constants. For
$\Lambda\neq0$ a solution is
\begin{equation}
A(\tau)=e^{C \tau }, \quad \Phi(\tau)=\ln
A(\tau),\label{exponential}
\end{equation}
where $C^2=\Lambda/2$. This indicates that $q^+_{\mu \nu} =
e^{\Phi} \eta_{\mu \nu}$ which is expected from the scaling
symmetry.

The brane scale factor $a_{{\rm b}}$ and the ``radion'' $\Psi$ are
given by
\begin{equation}
a_{{\rm b}} = e^{-\Phi/2}A, \quad \Psi=e^{\Phi}.
\end{equation}
The solution (\ref{power}) gives the same 4D observables (the
brane scale factor and radion) as an exact 6D solution found by
Copeland and Seto (equation~(70) of~\cite{Copeland:2007ur}). In
the same way the solution (\ref{exponential}) reproduces the 4D
quantities of their equation (76). (The latter solution was first
found by Tolley \emph{et al.}~\cite{Tolley:2006ht}.) Thus we show
that the solutions to the full 6D equations are reproduced by our
4D effective theory on a regularized brane with the scale
invariant potential and with or without additional ``tension''
$\Lambda$.

At the zeroth order, the amplitudes $v^{(0)}$ and $u^{(0)}$ of the
potentials are fine-tuned. If we change the amplitude of the scale
invariant potentials, which is equivalent to adding a cosmological
constant in the 4D effective theory, we get a runaway potential
for $\Phi$ and the 4D spacetime becomes non-static.

\subsection{Breaking the scale invariance}
Finally, let us consider the case where the first order potential
breaks the scale invariance. As the BD parameter is given by
$1/2$, this model violates the constraints coming from the solar
system experiments unless the BD scalar $\Phi$ is stabilized. It
is suggested that the potential on a brane can naturally stabilize
the modulus. For example, if we consider potentials
$V_1(\phi^{(0)}) = v^{(1)} e^{s \phi^{(0)}}$ and $U_1(\phi^{(0)})
= u^{(1)} e^{t \phi^{(0)}}$, the effective potential is given by
\begin{equation}
\overline{\Delta V} = 2 \pi \ell \sqrt{f} e^{\Phi/2} \left(
\frac{v^{(1)}}{y_+^s} e^{-s \Phi} + \frac{u^{(1)}}{2\ell^2fy_+^t}
e^{-(t+1) \Phi} (n- \e A^{(0)}_{\theta})^2 \right).
\end{equation}
As we saw in the previous subsection, if we take $s=1/2$ and
$t=-1/2$, $\overline{\Delta V}$ is independent of $\Phi$. However,
in general, it is possible to have a potential with a minimum by
choosing $s, t, v^{(1)}$ and $u^{(1)}$ appropriately
\cite{Burgess:2007vi}. Then $\Phi$ can be stabilized and general
relativity (GR) is recovered.

\section{Conclusions}\label{sec:conclusion6D}
In this chapter, we derived the low energy effective theory in the
six-dimensional supergravity with resolved 4-branes. The gradient
expansion method is used to solve the bulk geometry. The resultant
effective theory is a Brans-Dicke theory with the Brans-Dicke
parameter given by $\omega_{\rm BD}=1/2$. If we choose the dilaton
potentials on the branes so that they keep the scaling symmetry in
the bulk and if we tune their amplitudes then there is no
potential in the effective theory and the modulus is massless.
Thus the static four-dimensional spacetime has vanishing
cosmological constant. It is also possible to obtain
time-dependent solutions due to the dynamics of the modulus field
and we showed that they are identified with the six-dimensional
exact time dependent solutions found in \cite{Copeland:2007ur}.

Even if the potentials preserve the scaling symmetry, it was found
that there appears an effective cosmological constant in the
four-dimensional effective theory by changing the amplitude of the
potentials. Then in the Einstein frame, the modulus field acquires
an exponential potential and the static solution is no longer
allowed. Again, we showed that the cosmological solutions obtained
in the effective theory can be identified with the six-dimensional
exact time dependent solutions found in \cite{Tolley:2006ht,
Kobayashi:2007hf, Copeland:2007ur}.

Our effective theory allows us to discuss cosmology with arbitrary
matter on the brane. As the BD parameter is given by $1/2$, it is
impossible to reproduce realistic cosmology without stabilizing
the modulus field. As it was suggested by
Ref.~\cite{Burgess:2007vi}, it is easy to generate a potential for
the modulus $\Phi$ with a minimum by breaking the scaling symmetry
from the dilaton potentials on the branes. Then it is possible to
reproduce GR at low energies. However, once we stabilize the
modulus, the cosmological constant on a brane curves the
four-dimensional spacetime in the same way as in GR.

Our result would indicate that it is possible to reproduce
sensible cosmology in this six-dimensional supergravity model at
low energies but it would be difficult to address the classical
part of cosmological constant problem in this set-up. However, we
should mention that our effective theory is valid only up to the
energy scale determined by inverse of the size of
extra-dimensions. This condition is roughly given by $H \ell_* <1$
where $H$ is the Hubble parameter. If we consider scales smaller
than $\ell_*$ gravity becomes six-dimensional and from the
table-top experiments, $\ell_*$ is smaller than a few $\mu$m. Then
for $H > 10^{-2}$ eV, our universe becomes six-dimensional and it
is impossible to use the four-dimensional effective theory. In
order to address the behaviour of the universe at high energies,
we should deal with time-dependent solutions directly in
six-dimensional spacetime. This remains an open question.

Another issue that we did not touch in this chapter is the quantum
part of the cosmological constant problem, that is the stability
to quantum corrections. As mentioned in the introduction, there
are reasons for believing that the combination of the bulk
supersymmetry and scaling symmetry can maintain the quantum
corrections to be of the order of the cosmological constant
\cite{Burgess:2007ui}. In order to address this issue, it is
necessary to perform 6D calculations which take into account the
loops of the KK modes.

Finally, we briefly comment on the limit where the codimension one
branes are shrunk to codimension two objects. Our effective theory
shows no pathological behaviour in this limit as long as the
four-dimensional energy-momentum tensor integrated along the
$\theta$-direction remains finite. However, in this limit, the
first order extrinsic curvature $\!\stac{(1)}K_{\mu \nu}$ diverges
and then the first order correction to the four-dimensional metric
diverges. Then it is not clear whether there is a physical meaning
in this limit. This is related to a deep issue of whether it is
possible to put ordinary matter on codimension 2 objects
\cite{Kaloper:2006ek, deRham:2007pz} and we also leave this
problem as an open question.

\chapter{On the 4D effective theory in warped compactifications
with fluxes and branes}\label{chapter:10D}
\chaptermark{The 4D effective theory in type IIB SUGRA}

\section{\label{sec:INTRO2}Introduction}
In this chapter, using the gradient expansion method, we shall
derive the 4D effective theory for warped compactifications with
fluxes and branes in the 10D type IIB supergravity.

Warped string compactifications with fluxes and branes have
provided a novel approach to long standing problems of particle
physics and cosmology. Inspired by the earlier work by Randall and
Sundrum \cite{Randall:1999ee}, Giddings \emph{et al} (GKP)
\cite{Giddings:2001yu} showed that, in type IIB string theory, it
is possible to realize the warped compactifications that can
accommodate the large hierarchy between the electroweak scale and
the Planck scale. In the GKP model, the warping of the
extra-dimensions is generated by the fluxes and the presence of
D-branes and orientifold planes. All moduli fields are stabilized
due to fluxes except for the universal K\"ahler modulus. The
warped compactifications also provide a promising background for
cosmological inflation. An inflaton can be identified with the
internal coordinate of a mobile D3 brane moving in a warped throat
region \cite{Kachru:2003sx}.

So far, most works are essentially based on effective 4D theories
derived by a dimensional reduction. However, it is a non-trivial
problem to derive the 4D effective theories including the warping
and branes. In fact there have been debates on the validity of the
derivation of the 4D effective theories. In a conventional
Kaluza-Klein theory, the effective theory can be derived by
assuming a factorizable ansatz for the higher-dimensional fields.
This approach has been applied even in the presence of the
warping. For example, the 10D metric ansatz of the form
\begin{equation}
ds_{10}^2 = h^{-1/2}(y) e^{-6 u(x)} g_{\mu \nu}(x) dx^{\mu}
dx^{\nu} + h^{1/2}(y) e^{2 u(x)} \gamma_{mn} dy^m dy^n,
\label{wrong}
\end{equation}
is often used where $x$ denotes the coordinates of the 4D
non-compact spacetime and $y$ denotes the coordinates of the 6D
compact manifold. The function $u(x)$ is identified with the
universal K\"ahler modulus. However, it has been criticized that
the higher-dimensional dynamics do not satisfy this ansatz
\cite{deAlwis:2003sn,deAlwis:2004qh}. Thus it is required to
derive the 4D effective theory by starting from a correct ansatz
and consistently solving the 10D equations of motion.

These attempts were initiated in
Refs.~\cite{Kodama:2005fz,Giddings:2005ff,Kodama:2005cz}.
Ref~\cite{Kodama:2005fz} derived an exact time-dependent 10D
solution that describes the instability of the warped
compactification due to the non-stabilized K\"ahler modulus. It
was shown that this dynamical solution cannot be described by the
metric ansatz (\ref{wrong}). Ref.~\cite{Giddings:2005ff} derived
the potential for the moduli fields by consistently solving the
10D equations of motion using the metric ansatz that is consistent
with the dynamical solutions in the 10D theory. It was found that
the universal K\"ahler modulus is not a simple scaling of the
internal metric as is assumed in Eq.~(\ref{wrong}). A similar
consistent ansatz was proposed in Ref~\cite{Kodama:2005cz} and the
4D effective theory without potentials was derived.

In this chapter, we build on these works to present a systematic
way to derive the 4D effective theory by consistently solving the
10D equations of motion of type IIB supergravity. We exploit the
so-called gradient expansion method, which has been shown in the
previous chapters to be a powerful method to derive the 4D
effective theory in the context of 5D/6D brane world models
\cite{Kanno:2002ia,Wiseman:2002nn,Shiromizu:2002qr,Arroja:2006zz,Arroja:2007ss}.
Our method also provides a scheme to study the gravitational
backreaction to the warped geometry due to the moduli dynamics,
branes and fluxes.

\section{\label{sec:AAE}10D equations of motion}
Let us start by describing the type IIB supergravity based on
Ref.~\cite{Giddings:2001yu}. In the Einstein frame, the bosonic
part of the action for the type IIB supergravity is given by
\begin{eqnarray}
S_{IIB} &=& \frac{1}{2\kappa_{10}^2}\int
d^{10}x\sqrt{-g_{10}}\left({}^{(10)}\!\!R-\frac{\partial_M\tau\partial^M\bar{\tau}}{2\left(\mbox{Im}
\tau\right)^2}-\frac{G_{(3)}\cdot\bar{G}_{(3)}}{12\mbox{Im}\tau}-\frac{\tilde{F}_{(5)}^2}{4\cdot5!}\right) \nonumber\\
&&-\frac{i}{8\kappa_{10}^2} \int\frac{C_{(4)}\wedge
G_{(3)}\wedge\bar{G}_{(3)}}{\mbox{Im}\tau}, \label{10DACTION}
\end{eqnarray}
where the combined 3-form flux is $G_{(3)}=F_{(3)}-\tau H_{(3)}$
with $F_{(3)}=dC_{(2)}$, $H_{(3)}=dB_{(2)}$ and
\begin{equation}
\tau=C_{(0)}+ie^{-\phi},
\end{equation}
where $\phi$ is the so-called dilaton, $C_{(j)}$ is the
Ramond-Ramond potential of rank $(j)$ and $B_{(2)}$ is the NS-NS
potential. The bar denotes the complex conjugate. The 5-form field
is given by
\begin{equation}
\tilde{F}_{(5)}=F_{(5)}-\frac{1}{2}C_{(2)}\wedge
H_{(3)}+\frac{1}{2}B_{(2)}\wedge F_{(3)},
\end{equation}
with $F_{(5)}=dC_{(4)}$. The total action in our model is
\begin{equation}
S=S_{IIB}+S_{loc},\label{10DACTIONTOTAL}
\end{equation}
where the term $S_{loc}$ is the action for the localized sources
such as D3-branes and O3 planes;
\begin{equation}
S_{loc}=\sum_j \left(-\mu_3\int d^4x\sqrt{-g_j}+T_3\int
d^4xC_4\right),
\end{equation}
where the integrals are calculated over the 4D non-compact space
at the point $j$ in the compact space and $g_j$ is the determinant
of the induced metric on a brane at the point $j$ ($\mu_3$ is
positive/negative for D3-branes/O3 planes).

We can derive the equations of motion for the fields from the
action (\ref{10DACTION}). The trace reversed Einstein equations
are given by
\begin{eqnarray}
R_{AB}&=&\frac{\mbox{Re}\left(\partial_A\bar{\tau}\partial_B\tau\right)}{2\left(\mbox{Im}\tau\right)^2}+\frac{1}{4\mbox{Im}\tau}\left[\mbox{Re}\left(G_{ACD}\bar{G}_B^{\;\;\;CD}\right)-\frac{1}{12}G_{CDE}\bar{G}^{CDE}{}^{(10)}g_{AB}\right]\nonumber\\
&&+\frac{1}{96}\tilde{F}_{AP_1...P_4}\tilde{F}_B^{\;\;P_1...P_4}
+\kappa_{10}^2\left(T_{AB}^{loc}-\frac{1}{8}{}^{(10)}g_{AB}T^{loc}\right),\label{10DEE}
\end{eqnarray}
where $T_{AB}^{loc}$ is the energy-momentum tensor for the
localized sources defined by
\begin{equation}
T_{AB}^{loc}=-\frac{2}{\sqrt{-{}^{(10)}g}}\frac{\delta
S_{loc}}{\delta {}^{(10)}g^{AB}},
\end{equation}
and ${}^{(10)}g_{AB}$ is the metric for the 10D spacetime.

Following Refs.~\cite{Kodama:2005cz,Giddings:2005ff}, we take the
10D metric as
\begin{equation}
dS_{10}^2= h^{-\frac{1}{2}}(x,y)g_{\mu\nu}(x,y)dx^\mu dx^\nu +
h^{\frac{1}{2}}(x,y)\gamma_{mn}(y)dy^mdy^n, \label{metric}
\end{equation}
with
\begin{equation}
h(x,y)=h_1(y)+h_0(x),
\end{equation}
where upper case Latin indices run from 0 to 9, Greek indices run
from 0 to 3 (non-compact dimensions) and lower case Latin indices
run from 4 to 9 (compact dimensions). In the following, all
indices are raised by $g_{\mu \nu}$ and $\gamma_{m n}$. The
function $h_0(x)$ is the so-called universal K\"ahler modulus. It
should be emphasized that this is not a simple scaling of the
internal metric.

Using our metric ansatz, we can calculate the 10D Ricci tensor.
The mixed component is calculated as
\begin{equation}
R_{\mu p}=- g^{\alpha\beta} K_{\mu\beta
p|\alpha}+K_{p,\mu}-\frac{1}{2}h^{-1}h_{,\mu} K_p,
\end{equation}
where we defined
\begin{equation}
K_{\mu\nu p}\equiv-\frac{1}{2} g_{\mu\nu,p}, \quad K_p\equiv
g^{\mu\nu} K_{\mu\nu p},
\end{equation}
and $|$ denotes the covariant derivative with respect to
$g_{\mu\nu}$, that is,
\begin{equation}
K_{\mu\delta p|\alpha}\equiv K_{\mu\delta
p,\alpha}-{}^{(4)}\Gamma_{\mu\alpha}^\sigma K_{\sigma\delta p}-
{}^{(4)}\Gamma_{\delta\alpha}^\sigma K_{\mu\sigma p},
\end{equation}
where ${}^{(4)} \Gamma_{\mu\alpha}^\sigma$ is the Christoffel
symbol constructed from $g_{\mu \nu}$. The non-compact components
are given by
\begin{eqnarray}
R_{\mu\nu}&=&{}^{(4)}R_{\mu\nu}(g)-h^{-1}h_{|\mu\nu}+\frac{1}{4}h^{-1}
g_{\mu\nu}h_{|\alpha}^{|\alpha}+ \frac{1}{4}h^{-2}
g_{\mu\nu}h_{;a}^{;a}-\frac{1}{4}h^{-3}h_{;a}h^{;a} g_{\mu\nu}
\nonumber\\ &&+h^{-1} K_{\mu\nu
b}^{\;\;\;;b}-\frac{1}{4}h^{-2}h^{;b} K_b g_{\mu\nu}-h^{-1}
K_{\mu\nu}^{\;\;b} K_b + 2h^{-1} K_{\nu\;b}^{\;\delta}
K_{\mu\delta}^{\;\;\;b},
\end{eqnarray}
where $;$ denotes covariant derivative with respect to
$\gamma_{ab}$, that is,
\begin{equation}
K_{\mu\nu b;a}\equiv K_{\mu\nu b,a}-{}^{(6)}\Gamma_{ab}^c
K_{\mu\nu c},
\end{equation}
where ${}^{(6)}\Gamma_{ab}^c$ is the Christoffel symbol
constructed from $\gamma_{ab}$. The compact components of the
Ricci tensor are
\begin{eqnarray}
R_{ab}&=&{}^{(6)} R_{ab}(\gamma)+\frac{1}{4}h^{-2}h_{;d}h^{;d}
\gamma_{ab}-\frac{1}{4}h^{-1}h_{;d}^{;d} \gamma_{ab}
-\frac{1}{4}h_{|\delta}^{|\delta}
\gamma_{ab}-\frac{1}{2}h^{-2}h_{;a}h_{;b}
\nonumber\\
&&+K_{a;b}-\frac{1}{2}h^{-1}\left(h_{,b} K_a+h_{,a}
K_b\right)+\frac{1}{4}h^{-1}h^{;c} K_c \gamma_{ab}
-K^{\alpha\beta}_{\;\;\;a} K_{\alpha\beta b}.
\end{eqnarray}
Note that the self duality of the 5-form field must be imposed by
hand
\begin{equation}
\tilde{F}_{(5)}=\ast \tilde{F}_{(5)}.\label{F5SD}
\end{equation}
We shall take the self-dual 5-form field in the form
\begin{equation}
\tilde{F}_{(5)}=\left(1+\ast\right)\sqrt{-g}d_y\alpha(x,y) \wedge
dx^0\wedge dx^1\wedge dx^2\wedge dx^3,\label{F5}
\end{equation}
Then the self duality condition (\ref{F5SD}) is automatically
satisfied and the only non-zero components of $\tilde{F}_{(5)}$
are
\begin{equation}
\tilde{F}_{(5)a\mu_0\cdots\mu_3}=\alpha_{;a} {}^{(4)}
\varepsilon_{\mu_0\cdots\mu_3},
\end{equation}
\begin{equation}
\tilde{F}_{(5)n_1\cdots n_5}=-h^2 \alpha^{;c} {}^{(6)}
\varepsilon_{cn_1\cdots n_5},
\end{equation}
where ${}^{(4)}\varepsilon_{\mu_0\cdots\mu_3}$ denotes the
Levi-Civita tensor with respect to $g_{\mu\nu}$ and
${}^{(6)}\varepsilon_{cn_1\cdots n_5}$ denotes the Levi-Civita
tensor with respect to $\gamma_{mn}$.

\section{\label{sec:GEM2}Gradient expansion method}
In this section, we will use the gradient expansion method to
solve the 10D Einstein equations (\ref{10DEE}).
\subsection{\label{subsec:Approx}Gradient expansion}
The gradient expansion is based on the assumption that $x$
derivatives are suppressed compared with $y$ derivatives
\begin{equation}
\partial_x^2 \ll \partial_y^2.
\end{equation}
See section \ref{sec:LE} and subsection \ref{subsec:PM} for more
details on the gradient expansion method. Using this assumption,
we can reduce the partial differential equations with respect to
$x$ and $y$ to a set of ordinary differential equations with
respect to $x$.
We expand the metric as
\begin{equation}
g_{\mu\nu}(x,y)=\stackrel{(0)}{g_{\mu\nu}}(x,y)+\stackrel{(1)}{g_{\mu\nu}}(x,y)+\cdots,
\end{equation}
where the first order quantities are of the order
$\partial_x^2/\partial_y^2$. Accordingly, $K_{\mu \nu p}$ is also
expanded as
\begin{equation}
K_{\mu\nu p}=\stackrel{(0)}{K_{\mu\nu p}}+\stackrel{(1)}{K_{\mu\nu
p}}+\cdots.
\end{equation}

\subsection{\label{subsec:zero}First order equations}
We assume that at zeroth-order, the axion/dilaton is constant and
the 3-form field only has non-zero components in the compact
space, i.e.
\begin{equation}
G_{(3)}=\frac{1}{3!}G_{abc}(y)dy^a\wedge dy^b\wedge dy^c.
\end{equation}
Furthermore, we assume that the zeroth-order metric is independent
of the compact coordinates
\begin{equation}
\stackrel{(0)}{g_{\mu\nu}}(x,y)=\stackrel{(0)}{g_{\mu\nu}}(x).\label{0metric}
\end{equation}

The Bianchi identity/equation of motion for the 5-form field
becomes
\begin{equation}
\alpha_{;c}^{;c} + 2 h^{-1} h_{;c} \alpha^{;c}
=ih^{-2}\frac{G_{pqr}\ast_6\bar{G}^{pqr}}{12\mbox{Im}\tau}+2\kappa_{10}^2h^{-2}
T_3\rho_3^{loc},\label{F5Bi2}
\end{equation}
at zeroth order where $\ast_6$ is the Hodge dual with respect to
$g_{ab}$ and we defined a rescaled D3 charge density
$\rho_3^{loc}$ which does not depend on $h$. We define $\omega$ as
\begin{equation}
\omega = \alpha -h^{-1}.
\end{equation}
Then the Bianchi identity is rewritten as
\begin{equation}
-h_{;c}^{;c} = - h^2 \omega^{;c}_{;c} - 2 h h^{;c} \omega_{;c} +i
\frac{G_{pqr}\ast_6\bar{G}^{pqr}}{12\mbox{Im}\tau} +2\kappa_{10}^2
T_3\rho_3^{loc}.
\end{equation}
The non-compact components of the Einstein equations (\ref{10DEE})
can be rewritten as
\begin{eqnarray}
\!\!\!\!\!\!{}^{(4)}R_{\mu\nu} && -
h^{-1}\left(h_{|\mu\nu}-\frac{1}{4}\stackrel{(0)}{g_{\mu\nu}}h_{|\alpha}^{|\alpha}\right)
+h^{-1}\stackrel{(1)}{K_{\mu\nu
b}^{\;\;\;;b}}-\frac{1}{4}h^{-2}h^{;b}\stackrel{(1)}{K_b}\stackrel{(0)}{g_{\mu\nu}}
+ \frac{1}{4} \omega^{;c}_{;c} \stackrel{(0)}{g_{\mu \nu}}
\nonumber\\ &&
=-\frac{h^{-2}}{96\mbox{Im}\tau}\stackrel{(0)}{g_{\mu\nu}}\left|iG_{(3)}-\ast_6G_{(3)}\right|^2+
\frac{\kappa_{10}^2}{2}h^{-2}\stackrel{(0)}{g_{\mu\nu}}\left(T_3\rho_3^{loc}-\mu_3(y)\right)
,\label{1stmunu}
\end{eqnarray}
where ${}^{(4)} R_{\mu\nu}$ denotes the Ricci tensor constructed
from $g_{\mu \nu}$, $\mu_3(y)= \mu_3 \delta(y-y_i)/\sqrt{\gamma}$
which is independent of $h$ and we dropped the non-linear term in
$\omega$. In the same way, the compact equations can be rewritten
as
\begin{eqnarray}
\!\!\!\!\!\!{}^{(6)}R_{ab} &-& \frac{1}{4}h_{|\delta}^{|\delta}
\gamma_{ab}
+\stackrel{(1)}{K_{a;b}}-\frac{1}{2}h^{-1}\left(h_{,b}\stackrel{(1)}{K_a}+h_{,a}\stackrel{(1)}{K_b}\right)+\frac{1}{4}h^{-1}h^{;c}\stackrel{(1)}{K_c}\gamma_{ab} \nonumber\\
&=&  \frac{1}{4} h \omega^{;c}_{;c} \gamma_{ab}+ \frac{1}{2}
\left(h_{;a} \omega_{;b} +h_{;b} \omega_{;a}   \right) \nonumber\\
&&
+\frac{h^{-1}}{4\mbox{Im}\tau}\left[\mbox{Re}\left(G_{acd}\bar{G}_b^{cd}\right)-\frac{1}{12}G_{cde}\bar{G}^{cde}\gamma_{ab}-\frac{i}{12}
G_{pqr}\ast_6\bar{G}^{pqr}\gamma_{ab}\right] \nonumber\\
&& +\kappa_{10}^2\left(T_{ab}^{loc}-\frac{1}{8} h^{\frac{1}{2}}
\gamma_{ab}T^{loc}-\frac{1}{2}h^{-1} T_3 \rho_3^{loc}
\gamma_{ab}\right), \label{1stab}
\end{eqnarray}
where ${}^{(6)} R_{ab}$ denotes the Ricci tensor constructed from
$\gamma_{a b}$.

The GKP solution is obtained by taking
\begin{equation}
\ast_6G_{(3)}=iG_{(3)} \label{ISD}, \quad \omega=0,
\end{equation}
with local sources that satisfy $\mu_3(y)=T_3 \rho_3^{loc}$. With
these conditions, it is straightforward to show that the 10D
equations motion are satisfied. Then the warp factor is determined
by
\begin{equation}
-h^{;a}_{;a} = \frac{1}{12 \mbox{Im} \tau} G_{pqr} \bar{G}^{pqr} +
2 \kappa_{10}^2 T_3 \rho_3^{loc}. \label{h1}
\end{equation}
Our strategy is to assume that $\omega$, ${}^{(6)} R_{ab}$ and the
right hand sides of Eqs.~(\ref{1stmunu}) and (\ref{1stab}) are
first order in the gradient expansion and take into account these
contributions as a source for the 4D dynamics of $g_{\mu \nu}(x)$
and $h_0(x)$.

\section{4D Effective equations}
In this section we solve the 10D Einstein equations to get the 4D
effective Einstein equations. Hereafter we omit the superscript
$(i)$ that denotes the order of the gradient expansion. A key
point is the consistency condition that implies that the
integration of the total derivative term over the 6D internal
dimension vanishes
\begin{equation}
\int  d^6 y \sqrt{\gamma} v_{;c} =0,
\end{equation}
for an arbitrary $v$. This condition provides the boundary
conditions for the 10D gravitational fields and yields the 4D
effective equations.

First let us take the trace of Eq.~(\ref{1stmunu});
\begin{equation}
{}^{(4)}R+\left(h^{-1} K_c  + \omega_{;c}  \right)^{;c}
=-\frac{h^{-2}}{24\mbox{Im}\tau}\left|iG_{(3)}-\ast_6G_{(3)}\right|^2
+2\kappa_{10}^2h^{-2}\left(T_3\rho_3^{loc}-\mu_3(y)\right).
\label{tracemunu}
\end{equation}
Then integrating Eq.~(\ref{tracemunu}) over the internal manifold,
we get
\begin{eqnarray}
{}^{(4)} R &=& -\frac{1}{24V_{(6)}\mbox{Im}\tau}\int d^6y\sqrt{\gamma}h^{-2}\left|iG_{(3)}-\ast_6G_{(3)}\right|^2 \nonumber\\
&& +2\frac{\kappa_{10}^2}{V_{(6)}}\int
d^6y\sqrt{\gamma}h^{-2}\left(T_3\rho_3^{loc}-\mu_3(y)\right),
\end{eqnarray}
where $V_{(6)}$ is the volume of the internal space
\begin{equation}
V_{(6)}\equiv \int d^6y\sqrt{\gamma}.
\end{equation}

On the other hand, by combining Eq.~(\ref{1stmunu}) with
Eq.~(\ref{tracemunu}), we obtain the traceless part of the
equation
\begin{equation}
{}^{(4)}R_{\mu\nu}-\frac{1}{4} g_{\mu\nu} {}^{(4)}R-
h^{-1}\left(h_{|\mu\nu}-\frac{1}{4} g_{\mu\nu}
h_{|\delta}^{|\delta} \right)+ h^{-1} \left( K_{\mu\nu b}
-\frac{1}{4} g_{\mu \nu} K_b \right)^{;b}=0. \label{Sigma2}
\end{equation}
Integrating this over the compact space we obtain
\begin{equation}
{}^{(4)} R_{\mu\nu}-\frac{1}{4}g_{\mu\nu}
{}^{(4)}R=H^{-1}\left(h_{|\mu\nu}-\frac{1}{4} g _{\mu\nu}
h_{|\delta}^{|\delta}\right), \label{Meom}
\end{equation}
where the function $H(x)$ is defined as
\begin{equation}
H(x)\equiv h_0(x)+C,
\end{equation}
where
\begin{equation}
C \equiv \frac{1}{V_{(6)}}\int d^6y\sqrt{\gamma}h_1(y).
\end{equation}
Here $h_1(y)$ is obtained by solving Eq.~(\ref{h1}). Finally, we
need an equation that determines the dynamics of $h_0(x)$. Let us
calculate the trace of Eq.~(\ref{1stab})
\begin{eqnarray}
{}^{(6)} R &-& \frac{3}{2}h_{|\delta}^{|\delta}
+K_a^{;a}+\frac{1}{2}h^{-1}h^{;a} K_a = \frac{3}{2} h \left(
\omega^{;c}_{;c} + \frac{2}{3} h^{-1} h_{;c} \omega^{;c}
\right) \nonumber\\
&&+\frac{h^{-1}}{16\mbox{Im}\tau}\left|iG_{(3)}-\ast_6G_{(3)}\right|^2
+3\kappa_{10}^2h^{-1}\left(\mu_3(y)-T_3\rho_3^{loc}\right).
\label{bp}
\end{eqnarray}
Combining Eq.~(\ref{bp}) with Eq.~(\ref{tracemunu}) we get
\begin{eqnarray}
{}^{(6)}R + \frac{1}{2} h {}^{(4)} R -\frac{3}{2} h^{|\mu}_{|\mu}
+ \left( \frac{3}{2} K_{c} -h \omega_{;c} \right)^{;c}
\!\!\!&=&\!\!\! \frac{h^{-1}}{24\mbox{Im}\tau}\left|iG_{(3)}-\ast_6G_{(3)}\right|^2 \nonumber\\
&&\!\!\!\!\!\!+2
\kappa_{10}^2h^{-1}\left(\mu_3-T_3\rho_3^{loc}\right).
\label{aequation}
\end{eqnarray}
Then integrating this over the compact space we obtain the
equation of motion for $H(x)$
\begin{eqnarray}
H_{|\delta}^{|\delta}&=&\frac{H}{3}{}^{(4)} R +
\frac{2}{3V_{(6)}}\int d^6y\sqrt{\gamma} {}^{(6)} R \nonumber\\
&&-\frac{1}{36 V_{(6)}\mbox{Im}\tau}\int d^6y\sqrt{\gamma} h^{-1}\left|iG_{(3)}-\ast_6G_{(3)}\right|^2  \nonumber\\
&&-\frac{4}{3}\frac{\kappa_{10}^2}{V_{(6)}}\int d^6y\sqrt{\gamma}
h^{-1}\left(\mu_3(y)-T_3\rho_3^{loc}\right).
\end{eqnarray}
\section{\label{sec:4D}The 4D effective theory}
The effective 4D equations are summarized as
\begin{equation}
{}^{(4)} G_{\mu\nu} = H^{-1} \left(H_{|\mu \nu} -  g_{\mu\nu}
H^{|\delta}_{|\delta} -V g_{\mu \nu} \right),
\end{equation}
\begin{equation}
H_{|\delta}^{|\delta} =-\frac{4}{3} V + \frac{2}{3} H \frac{d
V}{dH},
\end{equation}
where the potential $V(H)$ is given by
\begin{eqnarray}
\!\!\!\!\!\!V(H)&=&-\frac{1}{2V_{(6)}}\int d^6y\sqrt{\gamma}
{}^{(6)} R
+\frac{1}{48 V_{(6)}\mbox{Im}\tau}\int d^6y\sqrt{\gamma}h^{-1} \left|iG_{(3)}-\ast_6G_{(3)}\right|^2 \nonumber\\
&&+\frac{\kappa_{10}^2}{V_{(6)}}\int d^6y\sqrt{\gamma}
h^{-1}\left(\mu_3(y)-T_3 \rho_3^{loc}\right).
\end{eqnarray}
They can be deduced from the following 4D effective action
\begin{equation}
S_{eff}=\frac{1}{2 \kappa_4^2} \int d^4x\sqrt{-g
}\left[H{}^{(4)}R(g) -2 V(H)\right], \label{action2}
\end{equation}
where $\kappa_4^2 = \kappa_{10}^2 /V_{(6)}$, which can be
determined by integrating the 10D action over the six internal
dimensions.

Performing the conformal transformation
$g_{\mu\nu}=H^{-1}f_{\mu\nu}$ we can write the previous 4D action
in the 4D Einstein frame as
\begin{equation}
S_E=\frac{1}{2 \kappa_4^2} \int
d^4x\sqrt{-f}\left[R(f_{\mu\nu})-\frac{3}{2}\left(\nabla \ln
H\right)^2-2V(H)H^{-2}\right],
\end{equation}
where now $\nabla$ denotes the covariant derivative with respect
to the Einstein frame metric $f_{\mu\nu}$. The 4D effective
equations of motion in the Einstein frame are given by
\begin{equation}
R_{\mu\nu}(f)=\frac{3}{2}\left(\nabla_\mu\ln
H\right)\left(\nabla_\nu\ln H\right) + V(H)H^{-2}f_{\mu\nu},
\end{equation}
\begin{equation}
\nabla_\alpha\nabla^\alpha\ln
H=-\frac{4}{3}V(H)H^{-2}+\frac{2}{3}\frac{dV}{dH}H^{-1}.
\end{equation}
By defining $\rho(x) = i H(x)$, the kinetic term can be rewritten
into a familiar form
\begin{equation}
S_{E, kin}= \frac{1}{2 \kappa_4^2} \int d^4 x \sqrt{-f} \left[R -
6 \frac{\partial_{\mu} \rho \partial^{\mu} \bar{\rho}} {|\rho -
\bar{\rho}|^2} \right].
\end{equation}
We would find the same result for the kinetic term for the
universal K\"ahler modulus even if the wrong metric ansatz
Eq.~(\ref{wrong}) was used to perform a dimensional reduction
where $\rho(x) = i e^{4 u(x)}$. In a region where the warping is
negligible $h_1(y) \ll h_0(x)$, $H(x)$ can be identified as $e^{4
u(x)}$, which is the simple scaling of the internal metric.
However, in a region where the warping is not negligible, the
original 10D dynamics is completely different between
(\ref{wrong}) and (\ref{metric}) \cite{Kodama:2005cz}.

The potential associated with the 3-form agrees with the result
obtained in Ref.~\cite{Giddings:2005ff}. It should be emphasized
that this potential is positive-definite. Nevertheless, the 3-form
contribution to the 4D Ricci scalar ${}^{(4)} R$ is negative
definite \cite{deAlwis:2003sn,deAlwis:2004qh} in accordance with
the no-go theorem for getting de Sitter spacetime
\cite{Maldacena:2000mw}. The resolution is the kinetic term of the
modulus $H(x)$, that is, ${}^{(4)}R$ is not directly related to
$V$ \cite{Giddings:2005ff}. In fact the 4D Ricci scalar is related
to the potential as
\begin{equation}
{}^{(4)}R = H^{-1} \left(4V + 3 H^{|\delta}_{|\delta} \right) =2
\frac{dV}{dH}.
\end{equation}
At the minimum of the potential we have ${}^{(4)} R = V= dV/dH=0$,
but, if we move away from the minimum and the modulus $H(x)$ is
moving, the potential cannot be read off from ${}^{(4)}R$.

We also notice that D3 branes with $\mu_3(y)= T_3 \rho_3^{loc}$ do
not give any gravitational energy in the 4D effective theory. On
the other hand, anti-D3 branes with $\mu_3(y)= -T_3 \rho_3^{loc}$
give a potential energy. This was used to realize de Sitter vacuum
\cite{Kachru:2003aw}.

\section{Discussions}
In this chapter, we presented a systematic way to derive the 4D
effective theory by consistently solving the type IIB supergravity
equations of motion in warped compactifications with fluxes and
branes. We used the gradient expansion method to solve the 10D
equations of motion. The consistency condition that the
integration of the total derivative terms over the internal 6D
space vanishes gives the boundary conditions for the 10D
gravitational fields. These boundary conditions give the 4D
effective equations. Once the solutions for the 4D effective
equations are obtained, we can determine the backreaction to the
10D geometry by solving Eqs.~(\ref{tracemunu}),(\ref{Sigma2}) and
(\ref{aequation}).

We did not introduce the stabilization mechanism for the modulus
$H(x)$. It is essential to stabilize this universal K\"ahler
modulus to get a viable phenomenology \cite{Kachru:2003aw}.
Usually, non-perturbative effects are assumed to give a potential
to this moduli. The non-perturbative effects will modify the 10D
dynamics and the 4D effective potential have to be consistent with
this 10D dynamics. Most works so far introduce the
non-perturbative potentials for the modulus field directly in the
4D effective theory and it is not clear this is consistent with
the original 10D dynamics. It is desirable to derive the potential
for the moduli fields by consistently solving the 10D Einstein
equation with non-perturbative corrections. In addition, a mobile
D3 brane plays a central role to realize inflation
\cite{Kachru:2003sx}. The D3 brane with $\mu_3= T_3$ probes a
no-scale compactification and the brane can sit at any point of
the compact space with no energy cost. In fact, the D3 brane does
not give any gravitational energy in the 4D effective theory.
However, once the stabilization mechanism is included, this is no
longer true. This is in fact an important effect which generally
yields a potential for the D3 brane that is not enough flat for
slow roll. It was also pointed out that the gravitational
backreaction of the D3 brane is essential to calculate the
corrections to the potential \cite{Baumann:2006th}.

The method presented in this chapter can be extended to include
the non-perturbative effects by introducing an effective 10D
energy-momentum tensor in the 10D Einstein equations. It is also
possible to include a moving D3 brane in our scheme along the line
of Ref.~\cite{Koyama:2005nq,Kanno:2005vq}, which studied the
dynamics of D-branes with self-gravity in a 5D toy model. Then we
can calculate the potential for a mobile D3 brane by taking into
account the stabilization and the backreaction of the D3 brane. In
addition, it is essential to study how the motion of the D3 brane
is coupled to 4D gravity in order to address the dynamics of
inflation. These issues are left as open questions.

\chapter{Introduction to brane inflation}\label{chapter:inflationintro}

So far we have seen how to obtain the 4D effective theories of
gravity from the more fundamental higher-dimensional theories. It
is now important to consider applications of these 4D effective
theories in cosmology. Specifically we will discuss the
application of brane-world models to inflationary cosmology know
as brane inflation.

This chapter is meant to give a brief introduction to the idea of
brane inflation. There are several good reviews of brane inflation
in the literature, see
\cite{Cline:2007pm,Cline:2006hu,McAllister:2007bg}. We shall
discuss the main features of brane inflation, in particular the
so-called DBI-inflation model, where DBI stands for
Dirac-Born-Infeld. The DBI-inflation model is rather interesting
because the particular and unusual form of the kinetic energy of
the inflaton can give rise to novel observational signatures. For
example, it can produce large non-Gaussianities in the
distribution of the primordial curvature perturbation. The next
two chapters will be dedicated to the study of non-Gaussianities
in rather general models of inflation. The models considered will
include the DBI-inflation model as a particular case, which was
our main motivation to study models with non-standard kinetic
energy.

Before introducing the general ideas of brane inflation, in the
next section we shall briefly introduce a much older model of
standard inflation and explain why is inflation required in the
first place.

\section{\label{sec:standardinflation}Standard inflation}

Inflation is a period of accelerated expansion in the very early
universe. During this period the scale factor grew at least thirty
orders of magnitude. This extremely fast and brief expansion is
needed to solve or aleviate the fine tuning problems of the
standard big bang theory. For nice reviews on inflation see for
example \cite{Langlois:2004de,Bassett:2005xm}.

For instance, within the standard big bang model there is no
explanation for why is the cosmic microwave background radiation
(CMBR) so homogeneous and isotropic. In fact, regions of the CMBR
sky separated by more than one degree would have been causally
disconnected at the time when the photons were emitted. This is
called the horizon problem. Inflation solves this problem by
increasing the particle horizon so that all regions of the present
observable universe were in causal contact in the past.

From observations we know that today the ratio of the energy
density to the critical density is within a few percent of unity
\cite{Komatsu:2008hk}. But during the standard big bang theory
this ratio is driven away from one. This implies that in the early
universe this ratio would have to be extremely fine tuned to
unity. This is the flatness problem. The inflationary solution is
simple. Because of the accelerated expansion, the total energy
density is driven towards the critical density to an arbitrary
precision depending on the duration of inflation. Inflation was
originally proposed to solve these and other problems of the
standard big bang theory \cite{Guth:1980zm}.

Later it was realized that inflation can also explain much more.
For example, it can be used to explain the observed fluctuations
in the CMBR. Inflation increases the wavelength of the
fluctuations of any light scalar field. When these cross the
horizon they become classical perturbations that will seed the
primordial perturbations that will eventually grow into large
scale structures, like clusters of galaxies that we see today.
When the inflationary predictions are compared to observations
they agree quite well, providing us strong evidence for inflation.

Accelerated expansion, i.e. $\ddot a >0$, implies that the
equation of state parameter of the fluid is
\begin{equation}
\omega<-\frac{1}{3}.
\end{equation}
To have quasi-exponential expansion we further require that the
Hubble parameter varies slowly, i.e.
\begin{equation}
\epsilon=-\frac{\dot H}{H^2}\ll 1,
\end{equation}
this implies $\omega\approx-1$, that is the cosmological constant
case. $\epsilon$ is called the slow-roll parameter. In order to
inflation to last long enough to solve the horizon and flatness
problems we need the inflaton field to accelerate slowly.
Equivalently, we define another slow-roll parameter like
\begin{equation}
\delta=-\frac{\ddot \phi}{H\dot\phi},
\end{equation}
and we require it to be small. $\phi$ denotes the scalar field
that is sustaining inflation. For a standard kinetic term, single
field inflation model with potential $V(\phi)$ the previous two
conditions are equivalent to require that the slope and the
curvature of the potential are small.

Usually inflation is caused by one (or more) scalar field as just
described. One important open question in inflation is the origin
of these scalar fields. Using a more fundamental theory, like
string theory, brane inflation tries to address this problem by
finding natural candidates for the inflaton. Natural candidates
are the scalar fields that describe the positions of a brane in
the higher-dimensional spacetime. In the next section, we shall
introduce the main ideas of brane inflation.

\section{\label{sec:braneinflation}Brane inflation}

\begin{figure}[t]
\centering
 \scalebox{.7}
 {\rotatebox{0}{
    \includegraphics*{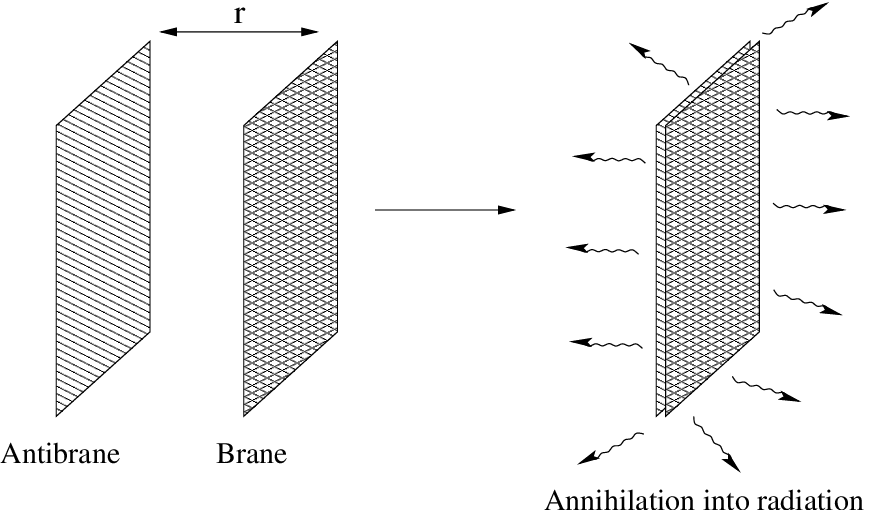}
                 }
 }
\caption{Brane-antibrane inflation and
reheating.}\label{fig:braneantibrane}
\end{figure}

The first brane inflation model was proposed in 1998 by Dvali and
Tye \cite{Dvali:1998pa}. They pointed out that the interaction
energy between a $D3$-brane and its antibrane (a
$\overline{D3}$-brane) in a higher-dimensional spacetime can give
rise to inflation. In this model, inflation ends when the two
branes collide, annihilating each other. The energy released by
the annihilation is used to reheat the universe that then follows
the usual big bang model. Figure \ref{fig:braneantibrane} is a
schematic representation of this scenario.
There are two (or more) branes separated by a distance $r$. This
distance $r$ plays the role of the inflaton in the 4D effective
theory. When the distance between the branes is of order $1/M_s$,
where $M_s$ is the string mass scale, the field becomes tachyonic
and inflation ends. From a 4D viewpoint brane-antibrane inflation
is very similar to hybrid inflation.

For a large separation, the interaction energy between the
$D3/\overline{D3}$ pair is \cite{Kachru:2003sx}
\begin{equation}
V(r)=2T_3\left(1-\frac{1}{2\pi^3}\frac{T_3}{M_{(10)}^8r^4}\right),\label{pot}
\end{equation}
where $T_3$ is the tension of the $D3$-brane, $M_{(10)}^2$ is the
10D Planck mass and it is given by
\begin{equation}
M_{(10)}^8=\frac{M_{Pl}^2}{L^6},
\end{equation}
where $M_{Pl}^2$ is the 4D Planck mass and $L^6$ is the volume of
the compact manifold.

There is a problem with this model/potential. If one defines the
usual slow-roll parameters as
\begin{equation}
\varepsilon=\frac{M_{Pl}^2}{2}\left(\frac{V'}{V}\right)^2, \quad
\eta=M_{Pl}^2\frac{V''}{V},
\end{equation}
where a prime denotes derivative with respect to the canonically
normalized field $\phi$, given by $\phi=\sqrt{T_3}r$, then to have
long lasting inflation we need $\varepsilon,\eta\ll 1$. For the
potential (\ref{pot})
\begin{equation}
\eta\sim -0.3\left(\frac{L}{r}\right)^6,
\end{equation}
and $\eta\ll 1$ is only possible if $r>L$. To have $r>L$ is
impossible in a manifold of size $L$. One possible and natural way
to evade this $\eta$-problem is to consider brane inflation in a
warped geometry \cite{Kachru:2003sx}. We will discuss such models
in the next section.

\section{\label{sec:KKLMMTmodel}The KKLMMT model: slow-moving brane}

In this section, we present the Kachru \emph{et al.} (KKLMMT)
model \cite{Kachru:2003sx} of warped brane inflation. They
consider a $D3$-brane and an anti-$D3$-brane in a 5D
anti-de-Sitter space ($AdS_5$). In \cite{Kachru:2003sx}, the
authors also provide a concrete example of warped brane inflation
in string theory. In that case the $AdS_5$ is replaced by the
Klebanov-Strassler geometry \cite{Klebanov:2000hb}, but the latter
can be well approximated by $AdS_5\times S^5$ in the region far
away from the tip of the geometry and this is a region of interest
to study inflation. Thus, we shall present the $AdS_5$ case
because it captures the essential features despite being simpler.

\begin{figure}[t]
\centering
 \scalebox{.3}
 {\rotatebox{0}{
    \includegraphics*{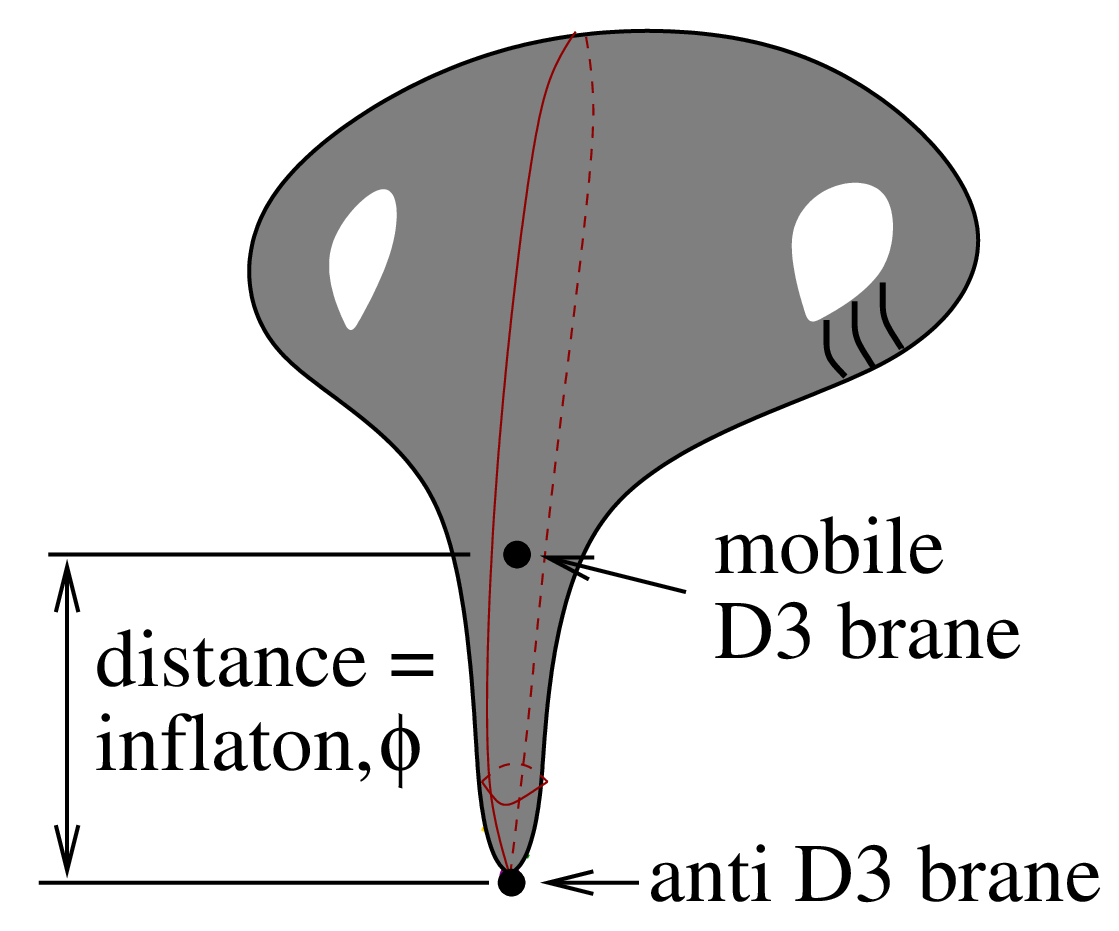}
                 }
 }
\caption{$D3$ and $\overline{D3}$ branes in the KKLMMT model. From
\cite{Cline:2007pm}.}\label{fig:KKLMMT}
\end{figure}

The $\overline{D3}$-brane is fixed at the infrared end of the
geometry (region with small values of the warp factor). The
$D3$-brane is mobile and experiences a small attractive force
towards the $\overline{D3}$-brane as in figure \ref{fig:KKLMMT}.
If the $\overline{D3}$-brane(s) were to be absent then the
$D3$-brane would feel no force in this background as the
electrostatic repulsion coming from the 5-form background flux
exactly cancels the gravitational attraction.

They consider a string theory compactification on $AdS_5\times
S^5$. It is a solution of the 10D supergravity with a 5-form field
strength $F_5$. The metric is
\begin{equation}
ds^2=h^{-\frac{1}{2}}\eta_{\mu\nu}dx^\mu
dx^\nu+h^{\frac{1}{2}}dr^2,\label{AdS5metric}
\end{equation}
where the warp factor $h$ is $h=R^4/r^4$. $R$ is the
characteristic length scale of the $AdS_5$ spacetime and it is
related to the 5-form charge. We choose to truncate the $AdS_5$ to
the region where $r_0<r<r_{max}$. From (\ref{AdS5metric}) one sees
that the region of small $r$ is the bottom of the gravitational
well, this justifies the assumption of placing the
$\overline{D3}$-brane(s) at the infrared end of the geometry. We
denote the position of the $\overline{D3}$-brane by $r_0$. The
background 5-form field is
\begin{equation}
\left(F_5\right)_{rx^\mu}=\frac{4r^3}{R^4},
\end{equation}
and the 4-form gauge potential $C_4$ is
\begin{equation}
\left(C_4\right)_{x^\mu}=\frac{r^4}{R^4}.
\end{equation}
The location of the $D3$-brane is denoted by $r_1$. The motion of
the $D3$-brane is described by the Born-Infeld action plus the
Chern-Simons action as
\begin{equation}
S=-T_3\int
d^4x\sqrt{-g}\frac{r_1^4}{R^4}\sqrt{1-\frac{R^4}{r_1^4}g^{\mu\nu}\partial_\mu
r_1\partial_\nu r_1}+T_3\int
d^4x\left(C_4\right)_{x^\mu},\label{D3braneaction}
\end{equation}
where $g^{\mu\nu}$ is the inverse metric along the non-compact
directions. $T_3$ is the $D3$-brane tension. The action for a
$\overline{D3}$-brane is similar to the previous action with the
sign of the second term changed, because a $\overline{D3}$-brane
has opposite 5-form charge. Now it is easy to see that if the
$D3$-brane is slow-moving and there are no $\overline{D3}$-branes
then the two contributions in (\ref{D3braneaction}) cancel each
other partially to give
\begin{equation}
S=T_3\int d^4x\sqrt{-g}\frac{1}{2}g^{\mu\nu}\partial_\mu
r_1\partial_\mu r_1.
\end{equation}
This means that at low energies the $D3$-brane behaves as a free
field.

Now, if we introduce a single $\overline{D3}$-brane this will
perturb the metric and the $F_5$. The $D3$-brane will acquire a
potential energy as \cite{Kachru:2003sx}
\begin{equation}
V=2T_3\frac{r_0^4}{R^4}\left(1-\frac{1}{N}\frac{r_0^4}{r_1^4}\right),
\end{equation}
where $N$ is the number of $D3$-branes that are present elsewhere
in the manifold and create the warping. The previous expression is
valid in the limit $r_1\gg r_0$. The first term is a constant
potential energy due to the presence of the $\overline{D3}$-brane,
but the effective tension of the antibrane is reduced by a factor
$r_0^4/R^4$. The second term is location dependent and the minus
sign indicates attraction between the branes. It is slow varying
as the inverse fourth power of the radial location of the
$D3$-brane. This term is also suppressed relatively to the first
constant term. Then we can naturally get $\varepsilon, \eta\ll 1$.

This model is not totally free of problems, in particular a
version of the so-called $\eta$-problem discussed in the previous
section is still present due to coupling to gravity, see
\cite{Kachru:2003sx} for details.

\section{\label{sec:DBImodel}The DBI model: ``relativistic" regime}

We have seen in the previous section that if the brane is moving
slowly (``slow-roll" regime) and if the potential is flat enough
then inflation occurs. References
\cite{Silverstein:2003hf,Alishahiha:2004eh,Chen:2004gc,Chen:2005ad}
noted that inflation is also possible in a different regime of the
parameter space. They realized that inflation is possible even if
the potential is steep and the $D3$-brane is ``relativistic". In
their model, the warped geometry slows down the inflaton field
even on a steep potential. They called this model DBI-inflation
because the action that governs the movement of the brane has the
form of the DBI action. It reads
\begin{eqnarray}
S\!\!\!&=&\!\!\!\frac{M^2_{Pl}}{2}\int
d^4x\sqrt{-g}R\nonumber\\&&\!\!\!\!\!\!-\int
d^4x\sqrt{-g}\left(f(\phi)^{-1}\sqrt{1+f(\phi)g^{\mu\nu}\partial_\mu\phi\partial_\nu\phi}-f(\phi)^{-1}+V(\phi)\right),
\end{eqnarray}
where $f(\phi)$ is related to the warp factor and $\phi$ is the
rescaled position of the $D3$-brane in the throat. $V(\phi)$ is
the potential energy that arises from interactions with the
background fields. For an $AdS$-like throat,
$f(\phi)\propto\phi^{-4}$.

There are two different situations that have been studied in the
literature. The ultraviolet (UV) model
\cite{Silverstein:2003hf,Alishahiha:2004eh}, where the inflaton
moves from the UV side of the throat to the tip of the geometry
(the infrared (IR) side), under the potential
\begin{equation}
V(\phi)\simeq\frac{1}{2}m^2\phi^2, \quad
m\gg\frac{M_{Pl}}{\sqrt{\lambda}},
\end{equation}
where $\lambda$ is a parameter that depends on the background
fluxes.

The other model that has been studied is the IR model
\cite{Chen:2004gc,Chen:2005ad}, where the brane moves away from
the tip of the throat towards the Calabi-Yau part of the manifold.
In this case the potential is
\begin{equation}
V(\phi)\simeq V_0-\frac{1}{2}m^2\phi^2, \quad m\sim H.
\end{equation}
In the spatially homogeneous case the Lagrangian is
\begin{equation}
P(X,\phi)=-f(\phi)^{-1}\sqrt{1-2Xf(\phi)}+f(\phi)^{-1}-V(\phi),
\end{equation}
where $X$ is defined as
$X=-1/2\partial^\mu\phi\partial_\mu\phi=\dot\phi^2/2$. The
requirement that the argument of the square root in the Lagrangian
has to be non-negative gives a speed limit for the inflaton as
\begin{equation}
1-f\dot\phi^2\geq 0.
\end{equation}
We can define the speed of sound $c_s$ as
$c_s=\sqrt{1-f\dot\phi^2}$. If the inflaton's speed is close to
saturating the previous bound its speed is said to be
relativistic. In this case, $c_s\ll 1$ and
$\dot\phi\simeq\pm1/\sqrt{f(\phi)}$. The requirement that
$\epsilon\equiv-\frac{\dot H}{H^2}\ll 1$ implies that the
potential energy dominates over the kinetic energy during
inflation despite the fact that the inflaton is moving
relativistically.

In the next chapter, we shall study linear and non-linear
perturbations of general models of inflation that include
DBI-inflation as a particular case. We will be interested in
calculating the higher-order statistics of the curvature
perturbation.

\section{\label{sec:MultiDBI}Multiple field DBI-inflation}

So far in this introductory chapter we have only discussed single
field models of brane inflation. The inflaton field was identified
with the position of a $D3$-brane in a higher-dimensional
spacetime, like an $AdS_5$ throat for example. However, in more
realistic string theory compactifications
\cite{Giddings:2001yu,Kachru:2003sx} the throat is a 6D space that
possesses a radial direction but also five angular coordinates. In
general, one expects a probe $D3$-brane to move in all these
directions. From a 4D effective theory point of view, each of
these positions of the brane in the compact space is described by
a 4D scalar field. Thus, DBI-inflation is naturally a multiple
field inflationary model.

Some cosmological implications of including angular momentum in
DBI brane inflation have been considered in
\cite{Easson:2007dh,Huang:2007hh}. Due to some confusion in recent
literature about the correct action for multi-field DBI-inflation
here we shall outline the derivation of the multi-field DBI action
\cite{Langlois:2008wt,Langlois:2008qf}.

We consider a $D3$-brane with tension $T_3$ moving in a 10D warped
string compactification. The 10D metric is
\begin{equation}
ds^2=H_{AB}dY^AdY^B=h^{-\frac{1}{2}}(y^K)g_{\mu\nu}dx^\mu dx^\nu
+h^{\frac{1}{2}}(y^K)G_{IJ}(y^K)dy^Idy^J,
\end{equation}
where $h$ is the warp factor, $g_{\mu\nu}$ and $x^\mu$ denote the
metric and the coordinates of the 4D non-compact space
respectively. $G_{IJ}$ and $y^I$ are the metric and the
coordinates of the 6D internal space respectively. The 10D
coordinates are denoted by $Y^A=\{x^\mu,y^I\}$. The dynamics of
the $D3$-brane is described by the DBI Lagrangian
\begin{equation}
P=-T_3\sqrt{-\mathrm{det}\gamma_{\mu\nu}},
\end{equation}
where $\mathrm{det}$ denotes the determinant and $\gamma_{\mu\nu}$
is the induced metric on the brane. The induced metric is given by
\begin{equation}
\gamma_{\mu\nu}=H_{AB}\partial_\mu Y^A_{(b)}
\partial_\nu Y^B_{(b)},
\end{equation}
where $Y^A_{(b)}(x^\mu)$ is the brane embedding. They are given by
$Y^A_{(b)}=\left(x^\mu,\varphi^I(x^\mu)\right)$. Finally the
induced metric on the brane can be written as
\begin{equation}
\gamma_{\mu\nu}=h^{-\frac{1}{2}}\left(g_{\mu\nu}+hG_{IJ}\partial_\mu\varphi^I\partial_\nu\varphi^J\right).
\end{equation}
This implies that the DBI Lagrangian is
\begin{equation}
P=-T_3h^{-1}\sqrt{-g}\sqrt{\mathrm{det}\left(\delta_\nu^\mu+hG_{IJ}\partial^\mu\varphi^I\partial_\nu\varphi^J\right)}.
\end{equation}
Introducing the rescaled quantities
\begin{equation}
f=\frac{h}{T_3}, \quad \phi^I=\sqrt{T_3}\varphi^I,
\end{equation}
and a potential $V\left(\phi^I\right)$, the DBI Lagrangian is
\begin{equation}
P=-f(\phi^I)^{-1}\left(\sqrt{\mathcal{D}}-1\right)-V(\phi^I),
\end{equation}
where
\begin{equation}
\mathcal{D}=\mathrm{det}\left(\delta_\nu^\mu+fG_{IJ}\partial^\mu\phi^I\partial_\nu\phi^J\right).
\end{equation}
The potential results from the interaction of the brane with
background fields and has been explicitly calculated in some
particular cases. For the rest of this work the potential is left
unspecified and general.

It is useful to define a ``kinetic matrix" as
\begin{equation}
X^{IJ}=-\frac{1}{2}\partial_\mu\phi^I\partial^\mu\phi^J.
\end{equation}
Noting that
\begin{equation}
\mathrm{det}(A)=-\frac{1}{4!}\varepsilon_{\alpha_1\alpha_2\alpha_3\alpha_4}\varepsilon^{\beta_1\beta_2\beta_3\beta_4}
A_{\beta_1}^{\alpha_1}A_{\beta_2}^{\alpha_2}A_{\beta_3}^{\alpha_3}A_{\beta_4}^{\alpha_4},
\end{equation}
where
\begin{equation}
A_\beta^\alpha=\delta_\beta^\alpha+fB_I^\alpha B_\beta^I, \quad
B_I^\alpha\equiv G_{IJ}\partial^\alpha\phi^J,\quad
B_\beta^I\equiv\partial_\beta\phi^I,
\end{equation}
and after using the identity
\begin{equation}
\varepsilon_{\alpha_1\alpha_2\alpha_3\alpha_4}\varepsilon^{\alpha_1\cdots\alpha_j\beta_{j+1}\cdots\beta_4}
=
-(4-j)!j!\delta_{\alpha_{j+1}}^{[\beta_{j+1}}\cdots\delta_{\alpha_4}^{\beta_4]},
\end{equation}
one can show that
\begin{equation}
\mathcal{D}=1-2fG_{IJ}X^{IJ}+4f^2X_I^{[I}X_J^{J]}-8f^3X_I^{[I}X_J^JX_K^{K]}+16f^4X_I^{[I}X_J^JX_K^KX_L^{L]}.\label{curlyD}
\end{equation}
Because of the antisymmetrization over the field indices the last
term vanishes if there are at most three fields. For two fields
the last two terms of the previous expression vanish. For the
single field case only the first two terms remain and one recovers
the result obtained in single field DBI-inflation presented in the
previous section. It turns out that the non-linear terms in
$X^{IJ}$ of equation (\ref{curlyD}) also vanish for a multi-field
DBI model if one considers homogeneous configurations, i.e.,
configurations in which the fields only depend on time and
$X^{IJ}$ is given by $X^{IJ}=1/2\dot\phi^I\dot\phi^J$. This fact
mislead the authors of
\cite{Easson:2007dh,Huang:2007hh,Langlois:2008mn} to ignore these
terms even in the inhomogeneous case. References
\cite{Langlois:2008wt,Langlois:2008qf} pointed out the importance
of these terms and correctly conclude that they cannot be ignored.
Because of these terms multi-field DBI-inflation is not a
particular case of multi-field K-inflation as previously thought.
In chapter \ref{chapter:bispectrum}, we will study linear and
non-linear perturbations in general multiple field inflationary
model. This general class of models includes K-inflation and
DBI-inflation as particular examples. We shall also discuss the
non-Gaussianity produced in these models, focusing in the three
point functions of the adiabatic and entropy perturbations.

\chapter{Non-Gaussianity from the trispectrum in general single field
inflation}\label{chapter:trispectrum}
\chaptermark{The trispectrum in single field inflation}

\section{\label{sec:INTRO4}Introduction}

The theory of slow-roll inflation generically predicts that the
observed cosmic microwave background radiation (CMBR) anisotropies
are nearly scale invariant and very Gaussian. Indeed, the latest
observations of CMBR by WMAP5 \cite{Komatsu:2008hk} confirm these
expectations. This constitutes one of the biggest achievements of
modern cosmology.

Despite its successes the theory of inflation still has many open
questions. For example, we do not know the origin of the scalar
field whose energy drives inflation, not to mention that we have
never detected directly in the laboratory these kind of particles.
The energy scale at which inflation happened is unknown by many
orders of magnitude. There are many models of inflation that give
similar predictions for the power spectrum of primordial
perturbations, so which one (if any) is the correct one?

For us to move a step forward in our understanding of the very
early universe we have to work on two fronts. First the
observational side. In the next few years, with improved
experiments like the Planck
satellite\footnote{http://www.rssd.esa.int/index.php?project=Planck},
we will measure the CMBR anisotropies to an incredible resolution.
For example the observational bounds on the bispectrum (the three
point correlation function of the primordial curvature
perturbation $\zeta$) will shrink from the present WMAP5 value
$-9<f_{NL}<111$ \cite{Komatsu:2008hk} to $|f_{NL}|\lesssim 5$
\cite{Komatsu:2001rj}, where the parameter $f_{NL}$ parameterizes
the size of the bispectrum. It is because this parameter is
constrained to be small that we say that the CMBR anisotropies are
very Gaussian. The observational bounds on the trispectrum (four
point function) will also tighten significantly from the rather
weak present constraint of $|\tau_{NL}|<10^8$
\cite{Boubekeur:2005fj,Alabidi:2005qi} to the future constraint of
$|\tau_{NL}|\sim 560$ \cite{Kogo:2006kh}, where $\tau_{NL}$
denotes the size of the trispectrum. These previous observational
bounds on the non-linearity parameters are for non-Gaussianity of
the local type. These bounds change depending on the shape of the
wave vectors' configuration \cite{Babich:2004gb}. This is one of
the reasons why it is important to calculate the shape dependence
of the non-Gaussianity.

In face of these expected observational advances, it is then
imperative to push forward our theoretical knowledge of our
theories and calculate more observational consequences of the
different inflationary models to make a comparison with
observations possible. One possible direction to be followed by us
and many others is to calculate higher order statistics (like the
trispectrum) of the primordial curvature perturbation. These
higher order statistics contain much more information about the
inflationary dynamics and if we observe them they will strongly
constrain our models. Because these higher order statistics have a
non-trivial momentum dependence (shape) they will help to
discriminate between models that have a similar power spectrum
(two point function).

Calculations of the bispectrum for a single field inflationary
model were done by Maldacena \cite{Maldacena:2002vr}. He showed
that the primordial bispectrum is too small (of the order of the
slow-roll parameters) to be observed even with Planck. Subsequent
work generalized Maldacena's result to include more fields and
more complicated kinetic terms \cite{Seery:2005wm, Seery:2005gb,
Chen:2006nt}. In \cite{Chen:2006nt}, Chen \emph{et al.} have
calculated the bispectrum for a quite general model of single
field inflation. They showed that for some models even the
next-to-leading order corrections in the slow-roll expansion may
be observed.

In this chapter, we will focus our attention on the calculation of
the trispectrum. In \cite{Seery:2006vu}, Seery \emph{et al.} have
calculated the trispectrum for slow-roll multi-field models (with
standard kinetic terms) and they showed that at horizon crossing
it is too small to be observed. But there are models of single
field inflation, well motivated from more fundamental theories,
that can produce a significant amount of non-Gaussianity, such as
Dirac-Born-Infeld (DBI) inflation
\cite{Alishahiha:2004eh,Huang:2006eh}. In \cite{Huang:2006eh} the
authors have computed the trispectrum for a model where the
inflaton's Lagrangian is a general function of the field's kinetic
energy and the field's value, their result was obtained using a
simple method \cite{Gruzinov:2004jx}, that only gives the correct
leading order answer. In this chapter we will provide the
equations necessary to calculate the next-to-leading order
corrections to the trispectrum. We argue that for some models
these corrections might become equally observable in the future.
In fact, we will calculate the fourth order action in the uniform
curvature gauge that is exact in the slow-roll expansion and
therefore in principle one could calculate all slow-roll
corrections to the trispectrum of the field perturbations.

We will also compute the exact fourth order action for the
curvature perturbation $\zeta$ in the comoving gauge. For a
simpler inflation model (with the standard kinetic term) this was
recently done in \cite{Jarnhus:2007ia}. However
\cite{Jarnhus:2007ia} did not consider second order tensor
perturbations. We will argue that this is an oversimplification
that leads to erroneous results. The reason simply being that at
second order in perturbation theory, scalar degrees of freedom
will source second order tensor perturbations and this will give a
non-zero contribution for the fourth order action and hence the
trispectrum.

There are other reasons why we will perform the calculation in the
comoving gauge. First of all, in doing so we work all the time
with the gauge invariant variable $\zeta$ that is directly related
with the observational quantities. The comoving gauge action can
also be used for other practical purposes. For example, it can be
used to calculate loop effects that can possibly have important
observational consequences. It can also be used to calculate the
trispectrum of models where the potential has a ``feature" (see
\cite{Chen:2006xjb,Chen:2008wn} for an example of such calculation
for the bispectrum). In the vicinity of the sudden potential
``jump" the slow-roll approximation temporarily fails and one
might get an enhancement of the trispectrum. There are well
motivated models of brane inflation where the throat's warp factor
suddenly jumps \cite{Hailu:2006uj}.

This chapter is organized as follows. In the next section, we
introduce the model under consideration. In section
\ref{sec:Perturbations4} we shall study non-linear perturbations.
First, we compute the fourth order action in the comoving gauge
including both scalar and second order tensor degrees of freedom.
After that we compute the fourth order action in the uniform
curvature gauge. In section \ref{sec:trispectrumformalism}, we
present the formalism needed to calculate the trispectrum. In
section \ref{sec:leadingtrispectrum}, we calculate the leading
order trispectrum using the comoving gauge action. We comment on
previous works and on the observability of next-to-leading order
corrections. Section \ref{sec:conclusion} is devoted to
conclusions. 

\section{\label{sec:MODEL}The model}
In this work, we will consider a fairly general class of models
described by the following action
\begin{equation}
S=\frac{1}{2}\int d^4x\sqrt{-g}\left[M^2_{Pl}R+2P(X,\phi)\right],
\label{action4}
\end{equation}
where $\phi$ is the inflaton field, $M_{Pl}$ is the Planck mass
that we will set to unity hereafter, $R$  is the Ricci scalar and
$X\equiv-\frac{1}{2}g^{\mu\nu}\partial_\mu\phi\partial_\nu\phi$ is
the inflaton's kinetic energy. $g_{\mu\nu}$ is the metric tensor.
We label the inflaton's Lagrangian by $P$ and we assume that it is
a well behaved function of two variables, the inflaton field and
its kinetic energy.

This general field Lagrangian includes as particular cases the
common slow-roll inflation model, DBI-inflation
\cite{Silverstein:2003hf,Alishahiha:2004eh} and K-inflation
\cite{ArmendarizPicon:1999rj}.

We are interested in flat, homogeneous and isotropic
Friedman-Robertson-Walker universes described by the line element
\begin{equation}
ds^2=-dt^2+a^2(t)\delta_{ij}dx^idx^j, \label{FRW}
\end{equation}
where $a(t)$ is the scale factor. The Friedman equation and the
continuity equation read
\begin{equation}
3H^2=\rho,
\end{equation}
\begin{equation}
\dot{\rho}=-3H\left(\rho+P\right), \label{continuity4}
\end{equation}
where the Hubble rate is $H=\dot{a}/a$, $\rho$ is the energy of
the inflaton and it is given by
\begin{equation}
\rho=2XP_{,X}-P ,\label{energy}
\end{equation}
where $P_{,X}$ denotes the derivative of $P$ with respect to $X$.

It was shown in \cite{Garriga:1999vw} that for this model the
speed of propagation of scalar perturbations (``speed of sound")
is $c_s$ given by
\begin{equation}
c_s^2=\frac{P_{,X}}{\rho_{,X}}=\frac{P_{,X}}{P_{,X}+2XP_{,XX}}.
\label{sound speed}
\end{equation}
We define the slow variation parameters, analogues of the
slow-roll parameters, as:
\begin{equation}
\epsilon=-\frac{\dot{H}}{H^2}=\frac{XP_{,X}}{H^2}, \label{epsilon}
\end{equation}
\begin{equation}
\eta=\frac{\dot{\epsilon}}{\epsilon H}, \label{eta}
\end{equation}
\begin{equation}
s=\frac{\dot{c_s}}{c_sH}. \label{s}
\end{equation}
We should note that these slow variation parameters are more
general than the usual slow-roll parameters and that the smallness
of these parameters does not imply that the field is rolling
slowly. We assume that the rate of change of the speed of sound is
small (as described by $s$) but $c_s$ is otherwise free to change
between zero and one.

It is convenient to introduce the following parameters that
describe the non-linear dependence of the Lagrangian on the
kinetic energy:
\begin{equation}
\Sigma=XP_{,X}+2X^2P_{,XX}=\frac{H^2\epsilon}{c_s^2},
\label{Sigma}
\end{equation}
\begin{equation}
\lambda=X^2P_{,XX}+\frac{2}{3}X^3P_{,XXX}, \label{lambda}
\end{equation}
\begin{equation}
\Pi=X^3P_{,XXX}+\frac{2}{5}X^4P_{,XXXX}. \label{Pi}
\end{equation}
These parameters are related to the size of the bispectrum and
trispectrum. The power spectrum of the primordial quantum
fluctuation was first derived in \cite{Garriga:1999vw} and reads
\begin{equation}
P_k^\zeta=\frac{1}{36\pi^2}\frac{\rho^2}{\rho+P}=\frac{1}{8\pi^2}\frac{H^2}{c_s\epsilon},
\label{PowerSpectrum}
\end{equation}
where it should be evaluated at the time of horizon crossing
${c_s}_*k=a_*H_*$. The spectral index is
\begin{equation}
n_s-1=\frac{d\ln P_k^\zeta}{d\ln k}=-2\epsilon-\eta-s.
\label{SpectralIndex}
\end{equation}
WMAP observations of the perturbations in the CMBR tell us that
the previous power spectrum is almost scale invariant therefore
implying that the three slow variation parameters should be small
at horizon crossing, roughly of order $10^{-2}$.

\section{\label{sec:Perturbations4}Non-linear
perturbations}

In this section, we will consider perturbations of the background
(\ref{FRW}) beyond linear order. There is a vast literature on
second order perturbations that is important when one is
interested in calculating three point correlation functions, see
for example
\cite{Maldacena:2002vr,Seery:2005wm,Chen:2006nt,Acquaviva:2002ud}.
We are interested in non-Gaussianities that come from the
trispectrum and so we need to use third order perturbation theory.
For that we need to compute the fourth order in the perturbation
action. In this section we will obtain the fourth order action in
two different gauges. As a check on our calculations we will
compute the leading order (in slow roll) trispectrum in both
gauges. We will follow the pioneering approach developed by
Maldacena \cite{Maldacena:2002vr} and used in several subsequent
papers \cite{Seery:2005wm,Seery:2005gb,Huang:2006eh,Seery:2006vu}.

For reasons that will become clear later it is convenient to use
the ADM metric formalism \cite{Arnowitt:1962hi}. The ADM line
element reads
\begin{equation}
ds^2=-N^2dt^2+h_{ij}\left(dx^i+N^idt\right)\left(dx^j+N^jdt\right),
\label{ADMmetricphi4}
\end{equation}
where $N$ is the lapse function, $N^i$ is the shift vector and
$h_{ij}$ is the 3D metric.

The action (\ref{action4}) becomes
\begin{equation}
S=\frac{1}{2}\int dtd^3x\sqrt{h}N\left({}^{(3)}\!R+2P\right)+
\frac{1}{2}\int dtd^3x\sqrt{h}N^{-1}\left(E_{ij}E^{ij}-E^2\right).
\end{equation}
The tensor $E_{ij}$ is defined as
\begin{equation}
E_{ij}=\frac{1}{2}\left(\dot{h}_{ij}-\nabla_iN_j-\nabla_jN_i\right),
\end{equation}
and it is related to the extrinsic curvature by
$K_{ij}=N^{-1}E_{ij}$. $\nabla_i$ is the covariant derivative with
respect to $h_{ij}$ and all contra-variant indices in this section
are raised with $h_{ij}$ unless stated otherwise.

The Hamiltonian and momentum constraints are respectively
\begin{eqnarray}
{}^{(3)}\!R+2P-2\pi^2N^{-2}P_{,X}-N^{-2}\left(E_{ij}E^{ij}-E^2\right)&=&0,\nonumber\\
\nabla_j\left(N^{-1}E_i^j\right)-\nabla_i\left(N^{-1}E\right)&=&\pi
N^{-1}\nabla_i\phi P_{,X},\label{LMphi4}
\end{eqnarray}
where $\pi$ is defined as
\begin{equation}
\pi\equiv \dot{\phi}-N^j\nabla_j\phi.\label{pi}
\end{equation}
We decompose the shift vector $N^i$ into scalar and intrinsic
vector parts as
\begin{equation}
N_i=\tilde{N_i}+\partial_i\psi,
\end{equation}
where $\partial_i\tilde{N^i}=0$, here indices are raised with
$\delta_{ij}$.

Before we consider perturbations around our background let us
count the number of degrees of freedom (dof) that we have. There
are five scalar functions, the field $\phi$, $N$, $\psi$,
$\mbox{det} h$ and $h_{ij}\sim\partial_i\partial_j H$, where $H$
is a scalar function and $\mbox{det} h$ denotes the determinant of
the 3D metric. Also, there are two vector modes $\tilde{N}^i$ and
$h_{ij}\sim\partial_i\chi_j$, where $\chi^j$ is an arbitrary
vector. Both $\tilde{N}^i$ and $\chi^j$ satisfy a divergenceless
condition and so carry four dof. Furthermore, we also have a
transverse and traceless tensor mode $\gamma_{ij}$ that contains
two additional dof. Because our theory is invariant under change
of coordinates we can eliminate some of these dof. For instance, a
spatial reparametrization like
$x^i=\tilde{x}^i+\partial^i\tilde{\epsilon}(\tilde{x},\tilde{t})+\epsilon^i_{(t)}(\tilde{x},\tilde{t})$,
where $\tilde{\epsilon}$ and  $\epsilon^i_{(t)}$ are arbitrary and
$\partial_i\epsilon^i_{(t)}=0$, can be chosen so that it removes
one scalar dof and one vector mode. A time reparametrization would
eliminate another scalar dof. Constraints in the action will
eliminate further two scalar dof and a vector mode. In the end we
are left with one scalar, zero vector and one tensor modes that
correspond to three physical propagating dof.

In the next subsection we shall use two different gauges that
correctly parameterize these dof. Because physical observables are
gauge invariant we know that both gauges have to give the same
result for the trispectrum for instance. It seems then unnecessary
to perform the calculation twice in different gauges. In practice,
we will see that both gauges have advantages and disadvantages and
one is more suitable for some applications than the other.
Furthermore, it provides a good consistency check on the
calculation.

\subsection{\label{subsec:PerturbationsComoving}Non-linear
perturbations in the comoving gauge}

In this subsection, we will compute the fourth order action for
the general model (\ref{action4}) in the comoving gauge. In this
gauge the scalar degree of freedom is the so-called curvature
perturbation $\zeta$ that is also gauge invariant. There are a few
works on this subject using this gauge, see e.g.
\cite{Jarnhus:2007ia}, where the authors have calculated the
fourth order action for a standard kinetic term inflation but they
neglected second order tensor perturbations. We will show that
this is an oversimplification that may lead to an erroneous result
for the four point correlation function.

In the comoving gauge, the inflaton fluctuations vanish and the 3D
metric is perturbed as
\begin{eqnarray}
&&\delta\phi=0,\nonumber\\  &&h_{ij}=a^2e^{2\zeta}\hat{h}_{ij},
\quad
\hat{h}_{ij}=\delta_{ij}+\gamma_{ij}+\frac{1}{2}\gamma_{ik}\gamma_j^k+\cdots\label{zetagauge}
\end{eqnarray}
where $\mbox{det} \hat{h}=1$, $\gamma_{ij}$ is a tensor
perturbation that we assume to be a second order quantity, i. e.
$\gamma_{ij}=\mathcal{O}(\zeta^2)$. It obeys the traceless and
transverse conditions $\gamma_i^i=\partial^i\gamma_{ij}=0$
(indices are raised with $\delta_{ij}$). $\zeta$ is the gauge
invariant scalar perturbation. In (\ref{zetagauge}), we have
ignored the first order tensor perturbations
${}^{(1)}{\gamma_{ij}}_{GW}$. This is because any correlation
function involving this tensor mode will be smaller than a
correlation function involving only scalars, see results of
\cite{Maldacena:2002vr}. In the literature the second order tensor
perturbations are often neglected, however they should be taken
into account. The reason for this is because at second order the
scalars will source the tensor perturbations equation. Later in
this section, we will elaborate further on this point. Higher
order tensor perturbations, like ${}^{(3)}\gamma_{ij}$, do not
contribute to the fourth order action.

We expand $N$ and $N^i$ in power of the perturbation $\zeta$
\begin{eqnarray}
N=1+\alpha_1+\alpha_2+\cdots,\\
\tilde{N_i}=\tilde{N_i}^{(1)}+\tilde{N_i}^{(2)}+\cdots,\\
\psi=\psi_1+\psi_2+\cdots,
\end{eqnarray}
where $\alpha_n$, $\tilde{N_i}^{(n)}$ and $\psi_n$ are of order
$\zeta^n$.

Some useful expressions for the quantities appearing in
(\ref{LMphi4}), valid to all orders in perturbations but for
$\gamma_{ij}=0$:
\begin{eqnarray}
{}^{(3)}\!R=-2a^{-2}e^{-2\zeta}\left(\partial_i\zeta\partial^i\zeta+2\partial_i\partial^i\zeta\right),
\end{eqnarray}
\begin{eqnarray}
E_{ij}E^{ij}-E^2&=&-6\left(H+\dot{\zeta}\right)^2+\frac{4H}{a^2}\left(1+\frac{\dot{\zeta}}{H}\right)e^{-2\zeta}
\left(\partial^2\psi+\partial^k\zeta\partial_k\psi+\partial^k\zeta\tilde{N}_k\right)\nonumber
\\\nonumber&&
+a^{-4}e^{-4\zeta}\Bigg[\frac{1}{2}\partial_i\tilde{N}_j\left(\partial^i\tilde{N}^j+\partial^j\tilde{N}^i\right)+2\partial^i\partial^j\psi\partial_i\tilde{N}_j
+\partial_i\partial_j\psi\partial^i\partial^j\psi
\\\nonumber&&
-\partial^2\psi\partial^2\psi
-2\partial_i\tilde{N}_j\left(\partial^j\zeta\tilde{N}^i+\partial^i\zeta\tilde{N}^j\right)-4\tilde{N}_k\partial_i\zeta\partial^i\partial^k\psi
\\&&
-2\partial_i\tilde{N}_j\left(\partial^j\zeta\partial^i\psi+\partial^i\zeta\partial^j\psi\right)
-4\partial_i\partial_j\psi\partial^j\zeta\partial^i\psi\nonumber
\\&&+2\partial^j\zeta\left(\partial_j\zeta\tilde{N}_i\tilde{N}^i+2\partial_j\zeta\tilde{N}_i\partial^i\psi+\partial_j\zeta\partial_i\psi\partial^i\psi\right)
\Bigg],
\end{eqnarray}
\begin{eqnarray}
&&\nabla_j\left(N^{-1}E_i^j\right)-\nabla_i\left(N^{-1}E\right)=
\nonumber\\ &&
-N^{-2}\Bigg[-2\left(H+\dot{\zeta}\right)\partial_iN
+\frac{a^{-2}e^{-2\zeta}}{2}\bigg(-\partial^jN\left(\partial_iN_j+\partial_jN_i\right)
\nonumber\\
&&\quad\quad\quad\quad+2\partial_jN\left(\partial^j\zeta
N_i+\partial_i\zeta
N^j\right)+2\partial_iN\partial^2\psi\bigg)\Bigg] \nonumber\\&&
+N^{-1}\Bigg[-2\partial_i\dot{\zeta}
+a^{-2}e^{-2\zeta}\bigg(\frac{1}{2}\left(\partial^j\zeta\partial_j\tilde{N}_i-\partial_j\zeta\partial_i\tilde{N}^j\right)
+\partial_j\zeta\partial^j\zeta N_i-\partial_i\zeta\partial^j\zeta
N_j \nonumber\\ &&\quad\quad\quad\quad+\partial_i\partial^j\zeta
N_j+\partial^2\zeta
N_i-\frac{1}{2}\partial^2\tilde{N}_i\bigg)\Bigg].
\end{eqnarray}
In the previous equations, indices in the right-hand side are
raised with $\delta_{ij}$ while indices in the left-hand side are
raised with $h_{ij}$. In the rest of this section indices will be
raised with $\delta_{ij}$.

Now, the strategy is to solve the constraint equations for the
lapse function and shift vector in terms of $\zeta$ and then plug
in the solutions in the expanded action up to fourth order.

At first order in $\zeta$, a particular solution for equations
(\ref{LMphi4}) is \cite{Maldacena:2002vr,Seery:2005wm}:
\begin{equation}
\alpha_1=\frac{\dot{\zeta}}{H}, \quad \tilde{N_i}^{(1)}=0, \quad
\psi_1=-\frac{\zeta}{H}+\chi, \quad
\partial^2\chi=a^2\frac{\epsilon}{c_s^2}\dot{\zeta}.\label{N1order}
\end{equation}
At second order, the constraint equation for the lapse function
gives
\begin{eqnarray}
\frac{4H}{a^2}\partial^2\psi_2&=&-2a^{-2}\partial_i\zeta\left(\partial^i\zeta+2H\partial^i\psi_1\right)
-4\alpha_1\left(a^{-2}\partial_i\partial^i\zeta-2\Sigma\zeta\right)-2\alpha_1^2\left(\Sigma+6\lambda\right)\nonumber\\
&&-a^{-4}\left(\partial^i\partial_k\psi_1\partial_i\partial^k\psi_1-\partial^2\psi_1\partial^2\psi_1\right)
+4\alpha_2\left(\Sigma-3H^2\right),
\end{eqnarray}
and the equation for the shift vector gives
\begin{equation}
2H\partial_i\alpha_2-\frac{1}{2}a^{-2}\partial^2\tilde{N_i}^{\!\!(2)}=
-a^{-2}\left(\partial_k\alpha_1\partial^k\partial_i\psi_1-\partial_i\alpha_1\partial^2\psi_1
+\partial^2\zeta\partial_i\psi_1+\partial_i\partial^k\zeta\partial_k\psi_1\right).
\label{Niconstraint2order}
\end{equation}
Due to the fact that $\tilde{N^i}$ is divergenceless and that any
vector can be separated into a incompressible and irrotational
part one can separate the contributions from $\alpha_2$ and
$\tilde{N_i}^{\!\!(2)}$ in the previous equation. The irrotational
part of Eq. (\ref{Niconstraint2order}) gives
\begin{equation}
2H\alpha_2=\partial^{-2}\partial^iF_i, \label{alpha2}
\end{equation}
and the incompressible part gives
\begin{equation}
\frac{1}{2a^2}\tilde{N_i}^{\!\!(2)}=-\partial^{-2}F_i+
\partial^{-4}\partial_i\partial^kF_k, \label{Ni2}
\end{equation}
where $F_i$ is defined as the right-hand side of equation
(\ref{Niconstraint2order}). The operator $\partial^{-2}$ is
defined by $\partial^{-2}(\partial^2\varphi)=\varphi$ and in
Fourier space it just brings in a factor of $-1/k^2$.

It was shown in \cite{Chen:2006nt} that to compute the effective
action of order $\zeta^n$, within the ADM formalism, one only
needs to use the solution for the Lagrange multipliers $N$ and
$N^i$ up to order $\zeta^{n-2}$. Therefore in order to calculate
the fourth order effective action the knowledge of the Lagrange
multipliers up to second order is required. It is given in
equations (\ref{N1order}), (\ref{alpha2}) and (\ref{Ni2}).

The second order action is
\begin{equation}
S_2=\int
dtd^3x\left[a^3\frac{\epsilon}{c_s^2}\dot{\zeta}^2-a\epsilon\left(\partial\zeta\right)^2\right].
\label{2action}
\end{equation}

The third order action is
\cite{Maldacena:2002vr,Seery:2005wm,Chen:2006nt}
\begin{eqnarray}
S_3\!\!&=&\!\!\int dtd^3x\left[-\epsilon
a\zeta\left(\partial\zeta\right)^2-a^3\left(\Sigma+2\lambda\right)\frac{\dot{\zeta}^3}{H^3}
+\frac{3a^3\epsilon}{c_s^2}\zeta\dot{\zeta}^2\right.\nonumber\\
&&\!\!\!\!\left.+\frac{1}{2a}\left(3\zeta-\frac{\dot{\zeta}}{H}\right)\left(\partial_i\partial_j\psi_1\partial_i\partial_j\psi_1-\partial^2\psi_1\partial^2\psi_1\right)
-\frac{2}{a}\partial_i\psi_1\partial_i\zeta\partial^2\psi_1\right].\label{3action}
\end{eqnarray}

The scalar fourth order action is
\begin{eqnarray}
S_4\!\!&=&\!\!\frac{1}{2}\int dtd^3xa^3\left[-a^{-2}\epsilon
\zeta^2\left(\partial\zeta\right)^2+\alpha_1^4\left(2\Sigma+9\lambda+\frac{10}{3}\Pi\right)
-6\zeta\alpha_1^3\left(\Sigma+2\lambda\right)\right.\nonumber\\
&&\!\!\!\!-2\alpha_2^2\left(\Sigma-3H^2\right)
+a^{-4}\left(\frac{\zeta^2}{2}+\zeta\alpha_1+\alpha_1^2\right)\left(\partial^kN_j^{(1)}\partial_k{N^j}^{(1)}-\partial^jN_j^{(1)}\partial^kN_k^{(1)}\right)\nonumber\\
&&\!\!\!\!-2a^{-4}\left(\zeta+\alpha_1\right)\partial^k{N^j}^{(1)}\left(\partial_kN_j^{(2)}-\delta_{kj}\partial^nN_n^{(2)}-2\partial_j\zeta N_k^{(1)}\right)\nonumber\\
&&\!\!\!\!+a^{-4}\left(-4\partial_kN_j^{(1)}\partial^j\zeta
{N^k}^{(2)}+2N_k^{(1)}\partial_j\zeta {N^k}^{(1)}\partial^j\zeta\right)+9\zeta^2\alpha_1^2\Sigma\nonumber\\
&&\!\!\!\!\left.+\frac{a^{-4}}{2}\partial_k\tilde{N}_j^{(2)}\partial^k\tilde{N}^{j(2)}-2a^{-4}\partial_kN_j^{(2)}\left(\partial^j\zeta
{N^k}^{(1)}+\partial^k\zeta {N^j}^{(1)}\right)\right].
\label{SSSS}
\end{eqnarray}
Here, no slow-roll approximation has been made. The previous
action has to be supplemented with the action containing terms
with one and two tensors
\begin{equation}
S_{\gamma^2}=\frac{1}{8}\int
dtd^3x\left[a^3\dot{\gamma}_{ij}\dot{\gamma}^{ij}-a\partial_k\gamma_{ij}\partial^k\gamma^{ij}\right],\label{GG}
\end{equation}
\begin{eqnarray}
S_{\gamma\zeta^2}=\int
dtd^3x\Bigg[&&\!\!\!\!\!\!\!\!\!\!-2\frac{a}{H}\gamma_{ij}\partial^i\dot{\zeta}\partial^j\zeta-a\gamma_{ij}\partial^i\zeta\partial^j\zeta
-\frac{1}{2}a\left(3\zeta-\frac{\dot{\zeta}}{H}\right)\dot{\gamma}_{ij}\partial^i\partial^j\psi_1
\nonumber\\&&\!\!\!\!\!\!\!\!\!\!+\frac{1}{2}a^{-1}\partial_k\gamma_{ij}\partial^i\partial^j\psi_1\partial^k\psi_1\Bigg].\label{GSS}
\end{eqnarray}

\subsubsection{\label{subsubsec:MSvariable}Canonical variable for quantization $\zeta_n$}

In order to calculate the quantum four point correlation function
we follow the standard procedure in quantum field theory. However
there is an important subtlety here. The gauge invariant quantity
$\zeta$ is not the correct variable to be quantized, because it is
not a canonical field . The canonical field to be quantized is the
field perturbation $\delta\phi$, or a convenient parameterization
$\zeta_n$ defined by
\begin{equation}
\zeta_n=-\frac{H}{\dot{\phi}_0}\delta\phi,\label{zetan}
\end{equation}
where $\phi_0$ is the background value of the field. We will see
that $\zeta$ is related to $\zeta_n$ by a non-linear
transformation so for the power spectrum calculation both
procedures of quantizing $\zeta$ or $\zeta_n$ give the same answer
because the difference between these two variables is a second
order quantity. However, for the calculation of higher order
correlation functions (like the bispectrum or trispectrum)
$\zeta_n$ is the correct variable to be quantized as it is linear
in $\delta\phi$. For the trispectrum, quantizing $\zeta$ gives
different results from quantizing $\zeta_n$. We will find the
relation between $\zeta$ and $\zeta_n$ through the gauge
transformation equations from the uniform curvature gauge
(discussed in detail in the next subsection) to the comoving
gauge. In the uniform curvature gauge the ansatz is
\begin{eqnarray}
&&\phi(x,t)=\phi_0+\delta\phi(x,t),\nonumber\\
&&h_{ij}=a^2\hat{h}_{ij}, \quad
\hat{h}_{ij}=\delta_{ij}+\tilde{\gamma}_{ij}+\frac{1}{2}\tilde{\gamma}_{ik}\tilde{\gamma}_j^k+\cdots
\end{eqnarray}
where $\mbox{det} \hat{h}=1$ and $\tilde{\gamma}_{ij}$ is a tensor
perturbation that we assume to be a second order quantity, i.e.,
$\tilde{\gamma}_{ij}=\mathcal{O}(\delta\phi^2)$. The gauge
transformations are
\begin{eqnarray}
\zeta(\B{x})&=&\zeta_n(\B{x})+F_2(\zeta_n(\B{x}))+F_3(\zeta_n(\B{x})),\label{gte0}\\
\gamma_{ij}&=&\tilde{\gamma}_{ij}(t)+\mu_{ij}\label{gtt},
\end{eqnarray}
where $F_2(\zeta_n),\mu_{ij}=\mathcal{O}(\zeta_n^2)$,
$F_3(\zeta_n)=\mathcal{O}(\zeta_n^3)$ are the terms coming from
the second and third order gauge transformations respectively,
they can be found explicitly in appendix \ref{GT} or in
\cite{Maldacena:2002vr,Jarnhus:2007ia}.

We now need to find the fourth order action for the variable
$\zeta_n$. Schematically, we write the different contributions as
\begin{equation}
S_{\zeta_n}=S_4(\zeta_n)+S_{\gamma^2}(\zeta_n)+S_{\gamma\zeta^2}(\zeta_n)+S_3(F_2(\zeta_n))+S_2(F_2(\zeta_n)).\label{zetanaction}
\end{equation}
The first three terms come from Eqs. (\ref{SSSS}-\ref{GSS}) when
we substitute $\zeta$ with $\zeta_n$ and $\gamma_{ij}$ with Eq.
(\ref{gtt}). Due to the non-linear relation between $\zeta$ and
$\zeta_n$, the third order action for $\zeta$, Eq.
(\ref{3action}), will after the change of variables give a
contribution to the fourth order action like $S_3(F_2(\zeta_n))$.
Similarly, the second order action, Eq. (\ref{2action}), will also
contribute with $S_2(F_2(\zeta_n))$. In principle, one would also
need to compute the third order gauge transformation as the second
order action gives origin to fourth order terms like
$\dot{\zeta_n}\dot{F}_3(\zeta_n)$ (where $F_3(\zeta_n)$ is the
third order piece of the gauge transformation). Fortunately, terms
involving $F_3$ can be shown to be proportional to the first order
equations of motion for $\zeta_n$, therefore when computing the
trispectrum these terms will vanish and we do not need to
calculate the third order gauge transformation explicitly at this
point. It can be easily seen that equations (\ref{3action}),
(\ref{SSSS}), (\ref{GG}), (\ref{GSS}) or their counterparts in
terms of $\zeta_n$, Eq. (\ref{zetanaction}), have terms that are
not slow roll suppressed. However, because in pure de Sitter space
$\zeta$ is a gauge mode we expect the action (\ref{zetanaction})
to be slow-roll suppressed (of order $\epsilon$). One can perform
many integrations by parts to show that the unsuppressed terms of
(\ref{zetanaction}) can be reduced to total derivative terms and
slow-roll suppressed terms given by
\begin{eqnarray}
S_{parts}=\int dt&&\!\!\!\!\!\!\!\!\!\!d^3x\Big\{
-\frac{3\epsilon}{aH^2}\Big[
\frac{1}{4}\left(\partial_j\zeta_n\partial^j\zeta_n\right)^2+\frac{1}{8}\zeta_n^2\partial^2\left(\partial_j\zeta_n\partial^j\zeta_n\right)
\nonumber\\&&
+\partial_j\zeta_n\partial^j\zeta_n\partial^{-2}\partial^l\partial^k\left(\partial_l\zeta_n\partial_k\zeta_n\right)
+2\partial_i\left(\partial^k\zeta_n\partial^i\zeta_n\right)\partial^{-2}\partial_l\left(\partial_k\zeta_n\partial^l\zeta_n\right)
\nonumber\\&&
+\frac{1}{2}\left(\partial^{-2}\partial_j\partial_k\left(\partial^j\zeta_n\partial^k\zeta_n\right)\right)^2
\Big]\nonumber
\\&&-\frac{5\epsilon}{8a^3H^4}\Big[\partial_l\zeta_n\partial^l\zeta_n\partial^k\partial^j\left(\partial_k\zeta_n\partial_j\zeta_n\right)
-
\frac{1}{2}\partial_j\zeta_n\partial^j\zeta_n\partial^2\left(\partial_k\zeta_n\partial^k\zeta_n\right)
\nonumber
\\&&\;\;\;\;\;\;\;\;\;\;\;\;\;\;\;\;\;\;-\frac{1}{2}\partial^k\partial^j\left(\partial_k\zeta_n\partial_j\zeta_n\right)\partial^{-2}\partial^m\partial^l\left(\partial_m\zeta_n\partial_l\zeta_n\right)
\Big]\Big\}.\label{Sparts}
\end{eqnarray}
For us to be able to obtain the previous result it is crucial to
include the contributions from the tensor actions (\ref{GG}) and
(\ref{GSS}), otherwise the trispectrum calculated using the
$\mathcal{O}(\epsilon^0)$ of (\ref{SSSS}) does not vanish, giving
the wrong leading order result. Neglecting tensor perturbations
(sourced by the scalars) for the calculation of the trispectrum is
not consistent and leads to wrong results. This is one of the
results of our work. The contribution from $\tilde{\gamma}_{ij}$
that comes through $\gamma_{ij}$ in (\ref{GG}) and (\ref{GSS})
will result in terms that are already slow-roll suppressed and no
further integrations by parts are need on these terms (see Eq.
(\ref{tildegammaeq}) of next subsection). The final action for
$\zeta_n$ is then
\begin{equation}
{S_4}_{\zeta_n}=S_{\zeta_n}^{(\epsilon)}+S_{parts},\label{zetanactionsup}
\end{equation}
where $S_{\zeta_n}^{(\epsilon)}$ denotes the terms of
(\ref{zetanaction}) that are suppressed by at least one slow-roll
parameter. This final action is slow-roll suppressed as expected
and no slow-roll approximation was made so it is also exact.

\subsection{\label{subsec:PerturbationsUniformeR4}Non-linear
perturbations in the uniform curvature gauge}

In order to calculate the intrinsic four point correlation
function of the field perturbation we need to compute the action
of fourth order in the perturbations. In this subsection we will
obtain the fourth order action in the uniform curvature gauge. In
this gauge, the scalar degree of freedom is the inflaton field
perturbation $\delta\phi(x^\mu)$. There are several works in the
literature where the authors also calculate the trispectrum. In
\cite{Seery:2006vu}, Seery \emph{et al.} calculate the trispectrum
of a multi-field inflation model, however the result is only valid
for fields with standard kinetic energies, i.e.
$P(X_1,\ldots,X_n,\phi_1,\ldots,\phi_n)=X_1+\cdots+X_n-V$, where
$X_n$ is the kinetic energy of $\phi_n$ and $V$ is the potential.
In this chapter we will generalize their result for an arbitrary
function $P(X,\phi)$ but for single field only. Recently, Huang
and Shiu have obtained the fourth order action for the model under
consideration (\ref{action4}). However the result was obtained by
only perturbing the field Lagrangian. This procedure gives the
right result as long as we are interested in the leading order
contribution in the small speed of sound limit and in the
slow-roll approximation, which was their case of interest. In the
present section we will compute the fourth order action that is
valid to all orders in slow roll and in the sound of speed
expansion. To do that we have to perturb the full action
(\ref{action4}) up to fourth order in the field perturbations. The
procedure to obtain the fourth order action in this gauge is
similar to the one used in subsection
\ref{subsec:PerturbationsComoving}.

In this gauge, the inflaton perturbation does not vanish and the
3D metric takes the form
\begin{eqnarray}
&&\phi(x,t)=\phi_0+\delta\phi(x,t),\nonumber\\
&&h_{ij}=a^2\hat{h}_{ij}, \quad
\hat{h}_{ij}=\delta_{ij}+\tilde{\gamma}_{ij}+\frac{1}{2}\tilde{\gamma}_{ik}\tilde{\gamma}_j^k+\cdots\label{deltaphigauge}
\end{eqnarray}
where $\mbox{det} \hat{h}=1$ and $\tilde{\gamma}_{ij}$ is a tensor
perturbation that we assume to be a second order quantity, i.e.,
$\tilde{\gamma}_{ij}=\mathcal{O}(\delta\phi^2)$. It obeys the
traceless and transverse conditions
$\tilde{\gamma}_i^i=\partial^i\tilde{\gamma}_{ij}=0$ (indices are
raised with $\delta_{ij}$). In the literature the second order
tensor perturbations are often neglected, however based on our
results it should be taken into account. We can always use the
gauge freedom at second order to eliminate the trace and the
vector perturbations of $h_{ij}$. The presence of
$\tilde{\gamma}_{ij}$ makes the three dimensional hypersurfaces
non-flat so using the name uniform curvature gauge might be
misleading.

We expand $N$ and $N^i$ in powers of the perturbation
$\delta\phi(x,t)$
\begin{eqnarray}
N=1+\alpha_1+\alpha_2+\cdots,\\
\tilde{N_i}=\tilde{N_i}^{(1)}+\tilde{N_i}^{(2)}+\cdots,\\
\psi=\psi_1+\psi_2+\cdots,
\end{eqnarray}
where $\alpha_n$, $\tilde{N_i}^{(n)}$ and $\psi_n$ are of order
$\delta\phi^n$ and $\phi_0(t)$ is the background value of the
field. At first order in $\delta\phi$, a particular solution for
equations (\ref{LMphi4}) is \cite{Maldacena:2002vr,Seery:2006vu}:
\begin{equation}
\alpha_1=\frac{1}{2H}\dot{\phi_0}\delta\phi P_{,X}, \quad
\tilde{N_i}^{(1)}=0, \quad
\partial^2\psi_1=\frac{a^2\epsilon}{c_s^2}\frac{d}{dt}\left(-\frac{H}{\dot{\phi}}\delta\phi\right)
. \label{N1orderphi}
\end{equation}
At second order, the constraint equation for the lapse function
gives
\begin{eqnarray}
\frac{4H}{a^2}\partial^2\psi_2&=&\frac{1}{a^4}\left(\partial^2\psi_1\partial^2\psi_1-\partial_i\partial_j\psi_1\partial^i\partial^j\psi_1\right)
+6H^2\left(3\alpha_1^2-2\alpha_2\right)\nonumber\\\nonumber
&&+\frac{8H\alpha_1}{a^2}\partial^2\psi_1-\left(3\dot{\phi_0}^2\alpha_1^2-2\dot{\phi_0}^2\alpha_2+\dot{\delta\phi}^2-4\dot{\phi_0}\dot{\delta\phi}\alpha_1\right)P_{,X}
\\&&\nonumber
-\frac{1}{a^2}\left(\partial_i\delta\phi\partial^i\delta\phi-2\dot{\phi_0}\partial_i\delta\phi\partial^i\psi_1\right)P_{,X}\\\nonumber
&&+\delta\phi^2P_{,\phi\phi}+2\dot{\phi_0}\delta\phi\left(\dot{\phi_0}\alpha_1-\dot{\delta\phi}\right)P_{,X\phi}\\\nonumber
&&-\frac{\dot{\phi_0}^2}{a^2}\bigg(-10a^2\dot{\phi_0}\dot{\delta\phi}\alpha_1-2\dot{\phi_0}\partial_i\delta\phi\partial^i\psi_1
+6a^2\dot{\phi_0}^2\alpha_1^2
\nonumber\\&&\nonumber-2\dot{\phi_0}^2a^2\alpha_2+4a^2\dot{\delta\phi}^2-\partial_i\delta\phi\partial^i\delta\phi\bigg)P_{,XX}\\
&&-\dot{\phi_0}^4\left(\dot{\phi_0}^2\alpha_1^2+\dot{\delta\phi}^2-2\dot{\phi_0}\dot{\delta\phi}\alpha_1\right)P_{,XXX}
\nonumber\\&&+2\dot{\phi_0}^3\delta\phi\left(\dot{\phi_0}\alpha_1-\dot{\delta\phi}\right)P_{,XX\phi}-\dot{\phi_0}^2\delta\phi^2P_{,X\phi\phi},
\end{eqnarray}
and the equation for the shift vector gives
\begin{eqnarray}
2H\partial_i\alpha_2-\frac{1}{2}a^{-2}\partial^2\tilde{N_i}^{\!\!(2)}&=&
4\alpha_1H\partial_i\alpha_1+a^{-2}\left(\partial_i\alpha_1\partial^2\psi_1-\partial_k\alpha_1\partial_i\partial^k\psi_1\right)
\nonumber\\
&&+\partial_i\delta\phi\left(\dot{\delta\phi}-\dot{\phi_0}\alpha_1\right)P_{,X}+\partial_i\delta\phi\dot{\phi_0}^2\left(\dot{\delta\phi}-\dot{\phi_0}\alpha_1\right)P_{,XX}
\nonumber\\
&&+\partial_i\delta\phi\dot{\phi_0}\delta\phi P_{,X\phi},
\label{Niconstraint2orderphi}
\end{eqnarray}
where $P_{,\phi}$ means derivative of $P$ with respect to $\phi$.

Due to the fact that $\tilde{N^i}$ is divergenceless and that any
vector can be separated into a incompressible and irrotational
part one can separate the contributions from $\alpha_2$ and
$\tilde{N_i}^{\!\!(2)}$ in the previous equation. The irrotational
part of Eq. (\ref{Niconstraint2orderphi}) gives
\begin{equation}
2H\alpha_2=\partial^{-2}\partial^iF_i, \label{alpha2phi}
\end{equation}
and the incompressible part gives
\begin{equation}
\frac{1}{2a^2}\tilde{N_i}^{\!\!(2)}=-\partial^{-2}F_i+
\partial^{-4}\partial_i\partial^kF_k,
\label{Ni2phi}
\end{equation}
where $F_i$ is defined as the right-hand side of equation
(\ref{Niconstraint2orderphi}).

The scalar fourth order action, where no slow-roll approximation
has been made, is
\begin{equation}
S_4=S_A+S_B,\label{SSSSphi}
\end{equation}
where
\begin{eqnarray}
S_A&\!\!=&\!\!\int dtd^3x\Bigg[ -\frac {a\delta\phi}{2} \Big[
a^{2}\alpha_1\dot{\delta\phi}^2+2\partial_i\delta\phi
\partial^i\psi_1\left(\dot{\delta\phi}-\alpha_1\dot{\phi}_0\right)+a^2\dot{\phi}_0^2\alpha_1^3+\alpha_1(\partial\delta\phi)^2
\nonumber\\
&&\!\!\!\!
+2\dot{\phi}_0\partial^i\delta\phi\left(\tilde{N_i}^{\!\!(2)}
+\partial_i\psi_2\right)\Big] P_{,X\phi} -\frac{1}{8a} \Big[
4\dot{\phi}_0^4a^4\alpha_2\left(\alpha_2-2\alpha_1^2\right)+3\dot{\phi}_0^4a^4\alpha_1^4
\nonumber\\&&\!\!\!\!-18\dot{\phi}_0^2a^4\alpha_1^2\dot{\delta\phi}^2
+8\dot{\phi}_0^2a^2\partial^i\delta\phi
\left(\tilde{N_i}^{\!\!(2)}+
\partial_i\psi_2\right)\left(\dot{\delta\phi}-\dot{\phi}_0\alpha_1\right)
\nonumber\\&&\!\!\!\!-4\dot{\phi}_0\partial_i\delta\phi\partial^i\psi_1\bigg((\partial\delta\phi)^2-3\dot{\phi}_0^2a^2\alpha_1^2
+8\dot{\phi}_0a^2\alpha_1\dot{\delta\phi}-3a^2\dot{\delta\phi}^2\bigg)
+12\dot{\phi}_0a^4 \alpha_1\dot{\delta\phi}^3\nonumber
\\\nonumber
&&\!\!\!\!+4\dot{\phi}_0^3a^4\alpha_1^3 \dot{\delta\phi}
-4\,\dot{\phi}_0^2\partial_i\delta\phi\partial^i\psi_1\partial_k\delta\phi
\partial^k\psi_1-2 \dot{\phi}_0 ^2a^2 \alpha_1^2
(\partial\delta\phi)^2 -4 \alpha_1\dot{\phi}_0\dot{\delta\phi}a^2
(\partial\delta\phi)^2
\\\nonumber
&&\!\!\!\!-a^4\left(\dot{\delta\phi}^2\!-\!a^{-2}(\partial\delta\phi)^2\right)^2\Big]P_{,XX}
\!+\!\frac{\dot{\phi}_0^2a}{12}\Big[6\dot{\phi}_0\partial_i\delta\phi
\partial^i\psi_1\left(2\dot{\phi}_0\alpha_1\dot{\delta\phi}-\dot{\delta\phi}^2\!-\!\dot{\phi}_0^2\alpha_1^2\right)
\\\nonumber
&&\!\!\!\!
+3(\partial\delta\phi)^2\left(2\dot{\phi}_0\alpha_1\dot{\delta\phi}-\dot{\phi}_0^2\alpha_1^2\right)
-16\dot{\phi}_0a^2\alpha_1\dot{\delta\phi}^3
+24\dot{\phi}_0^2a^2\alpha_1^2\dot{\delta\phi}^2-12\dot{\phi}_0^3a^2\alpha_1^3\dot{\delta\phi}
\\\nonumber &&\!\!\!\!+\dot{\phi}_0^4a^2\alpha_1^4
+3a^2\dot{\delta\phi}^2\left(\dot{\delta\phi}^2-a^{-2}(\partial\delta\phi)^2\right)\Big]
P_{,XXX} +\frac{\dot{\phi}_0a\delta\phi}{2}\Big[
a^2\dot{\delta\phi}^3+3\dot{\phi}_0^2a^2\alpha_1^2\dot{\delta\phi}
\nonumber\\&&\!\!\!\!+\left((\partial\delta\phi)^2+2\dot{\phi}_0\partial_i\delta\phi\partial^i\psi_1\right)\left(\dot{\phi}_0\alpha_1-\dot{\delta\phi}\right)
-4\dot{\phi}_0a^2 \alpha_1 \dot{\delta\phi}^2 \Big]
P_{,XX\phi}\nonumber
\\
\nonumber &&\!\!\!\! -\frac{a\delta\phi^2}{8}\Big[
2(\partial\delta\phi)^2-2a^2\dot{\delta\phi}^2+4a^2\alpha_1\dot{\phi}_0\dot{\delta\phi}+2a^2\dot{\phi}_0^2\alpha_1^2+4
\dot{\phi}_0\partial_i\delta\phi\partial^i\psi_1 \Big]
P_{,X\phi\phi}
\\
\nonumber &&\!\!\!\! +\frac{1}{24}\dot{\phi}_0^4a^3 \Big[
6\dot{\phi}_0^2\alpha_1^2\dot{\delta\phi}^2-4\dot{\phi}_0\alpha_1
\dot{\delta\phi}^3-4\dot{\phi}_0^3\alpha_1^3\dot{\delta\phi}+\dot{\phi}_0^4
\alpha_1^4+\dot{\delta\phi}^4 \Big]
 P_{,XXXX}
\\
\nonumber &&\!\!\!\!-\frac{1}{6}\dot{\phi}_0^3\delta\phi a^3
\left(
-\dot{\delta\phi}^3+3\dot{\phi}_0\alpha_1\dot{\delta\phi}^2+\dot{\phi}_0^3\alpha_1^3-3\dot{\phi}_0^2\alpha_1^2\dot{\delta\phi}
\right) P_{,XXX\phi}
\\
 &&\!\!\!\!+\frac{1}{4}\dot{\phi}_0^2 \delta\phi^2a^3
 \left( -2\alpha_1\dot{\phi}_0\dot{\delta\phi}+\dot{\delta\phi}^2+\dot{\phi}_0^2\alpha_1^2
\right)P_{,XX\phi\phi}
\nonumber\\&&\!\!\!\!-\frac{1}{6}\dot{\phi}_0\delta\phi^3a^3
\left(\alpha_1\dot{\phi}_0-\dot{\delta\phi}\right)P_{,X\phi\phi\phi}
\Bigg], \label{SSSSphiA}
\end{eqnarray}
\vspace{-1.5cm}
\begin{eqnarray}
S_B&=&\int dtd^3x\Bigg[\alpha_1^3a^3\delta\phi P_{,\phi}
+\frac{1}{2}a^3\alpha_1^2\delta\phi^2P_{,\phi\phi}+\frac{1}{6}a^3\alpha_1\delta\phi^3P_{,\phi\phi\phi}+\frac{1}{24}a^3\delta\phi^4P_{,4\phi}
\nonumber\\\nonumber
&&-\frac{1}{2a}\bigg[-(\partial_i\delta\phi\partial^i\psi_1)^2-2\alpha_1\dot{\delta\phi}a^2\partial_i\delta\phi\partial^i\psi_1
+\alpha_1^2a^2(\partial\delta\phi)^2\\\nonumber
&&+2a^2\partial_i\delta\phi\left(\tilde{N_i}^{\!\!(2)}+\partial_i\psi_2\right)\left(\dot{\delta\phi}-\alpha_1\dot{\phi}_0\right)
+\dot{\phi}_0^2a^4\alpha_2\left(\alpha_2-2\alpha_1^2\right)\bigg]P_{,X}
\\
&&+\frac{1}{4a}\bigg[2\partial^i\tilde{N^j}^{\!\!(2)}\partial_{(i}\tilde{N_{j)}}^{\!\!(2)}-4\alpha_1\partial_i\partial_k\psi_1\left(\partial^i\partial^k\psi_2
+\partial^i\tilde{N^k}^{\!\!(2)}\right) \nonumber\\
&&+12a^4H^2\alpha_2\left(\alpha_2-2\alpha_1^2\right)+4\alpha_1\partial^2\psi_1\partial^2\psi_2\bigg]
\Bigg]. \label{SSSSphiB}
\end{eqnarray}
The previous actions should be supplemented with the pure tensor
terms and the tensor-scalar coupling terms:
\begin{equation}
S=\frac{1}{8}\int
dtd^3x\left[a^3\dot{\tilde{\gamma}}_{ij}\dot{\tilde{\gamma}}^{ij}-a\partial_k\tilde{\gamma}_{ij}\partial^k\tilde{\gamma}^{ij}\right],
\label{GGphi}
\end{equation}
\begin{equation}
S=\int
dtd^3x\left[aP_{,X}\tilde{\gamma}^{ij}\partial_j\delta\phi\left(\frac{1}{2}\partial_i\delta\phi+\dot{\phi_0}\partial_i\psi_1\right)\right].
\label{GSSphi}
\end{equation}

This constitutes the main result of this subsection. It is a good
check for our calculation to see that the previous action
(\ref{SSSSphi}) reduces in some particular cases to previously
known results present in the literature.

For example, if we restrict our model to the standard inflation
case, i.e., $P(X,\phi)=X-V(\phi)$, where $V(\phi)$  is the
inflaton potential, then all the terms in the scalar action
(\ref{SSSSphiA}) vanish and the only contribution to the fourth
order action comes from (\ref{SSSSphiB}). These terms exactly
reproduce the result of Seery \emph{et al.} \cite{Seery:2006vu},
their equation (36), restricted to single field. However, in the
total fourth order action there are also the tensor contributions
(\ref{GGphi}) and (\ref{GSSphi}). In general, to proceed one has
to calculate the equation of motion for the second order tensor
perturbations $\tilde{\gamma}_{ij}$ from Eqs. (\ref{GGphi}),
(\ref{GSSphi}) to get
\begin{equation}
\tilde{\gamma}_{ij}''+2\frac{a'}{a}\tilde{\gamma}_{ij}'-\partial^2\tilde{\gamma}_{ij}=
\left(2P_{,X}\partial_j\delta\phi\partial_i\delta\phi+4P_{,X}\dot{\phi_0}\partial_j\delta\phi\partial_i\psi_1\right)^{TT},
\label{tildegammaeq}
\end{equation}
where TT means the transverse and traceless parts of the
expression inside the parenthesis (see appendix \ref{TTpart} for
details of how to extract the TT parts of a tensor) and then solve
this equation to obtain $\tilde{\gamma}_{ij}$ as a function of
$\delta\phi$. One can immediately see that at second order the
scalars will source the tensor perturbation equation as it was
previously shown by others \cite{Ananda:2006af,Osano:2006ew}. At
this order in perturbation theory, equation (\ref{tildegammaeq})
should also have a source term quadratic in the first order tensor
perturbations, ${{}^{(1)}\tilde{\gamma}_{ij}}_{GW}$. We neglect
these terms because we expect that any correlation function where
${{}^{(2)}\tilde{\gamma}_{ij}}_{GW}$ enters, which is sourced by
the first order tensor modes squared, must be smaller than a
correlation function with only scalars, see Ref.
\cite{Maldacena:2002vr} for an example. In Fourier space, the
source term of (\ref{tildegammaeq}) is suppressed by $k^2$, where
$k$ is the wave number. Once we have the solution of
$\tilde{\gamma}_{ij}$ in terms of $\delta\phi$ we can plug back
the result in (\ref{GGphi}) and (\ref{GSSphi}) to get the total
fourth order scalar action.

\section[The general formalism to calculate the trispectrum]{\label{sec:trispectrumformalism}The general formalism to calculate the\\ trispectrum}
\subsection{\label{subsec:trispectrumzetan}The trispectrum of $\zeta_n$}

Now we shall give the basic equations needed to calculate the
trispectrum \cite{Maldacena:2002vr,Weinberg:2005vy}. First we need
to solve the second order equation of motion for $\zeta_n$
(obtained from (\ref{2action})). Defining new variables
\begin{equation}
v_k=zu_k, \quad z=\frac{a\sqrt{2\epsilon}}{c_s},
\end{equation}
where the Fourier mode function $u_k$ is given by
\begin{equation}
u_k=\int d^3x\zeta_n(t,\B{x})e^{-i\B{k}\cdot\B{x}},
\end{equation}
the equation of motion for $\zeta_n$ is
\begin{equation}
v_k''+c_s^2k^2v_k-\frac{z''}{z}v_k=0, \label{Mukhanoveq}
\end{equation}
where prime denotes derivative with respect to conformal time
$\tau$. This is also known as the Mukhanov equation. The previous
equation can be solved, at leading order in slow roll and if the
rate of change of the sound speed is small \cite{Chen:2006nt}, to
give
\begin{equation}
u_k\equiv u(\tau,\B{k})=\frac{iH}{\sqrt{4\epsilon
c_sk^3}}\left(1+ikc_s\tau\right)e^{-ikc_s\tau}.\label{modefc}
\end{equation}
We do not need to impose any constraints in the sound speed and it
can be arbitrary. Only its rate of change is assumed to be small.
The next-to-leading order corrections to the previous solutions
are also known and can be found in \cite{Chen:2006nt}. In the
general case, we would have to solve Eq. (\ref{Mukhanoveq})
without assuming slow roll. This can be done numerically.

In order to calculate the $\zeta_n$ correlators we follow the
standard procedure in quantum field theory. $\zeta_n$ is promoted
to an operator that can be expanded in terms of creation and
annihilation operator as
\begin{equation}
\zeta_n(\tau,\B{k})=u(\tau,\B{k})a(\B{k})+u^*(\tau,-\B{k})a^\dag(-\B{k}).
\end{equation}
The standard commutation relation applies
\begin{equation}
\left[a(\B{k_1}),a^\dag(\B{k_2})\right]=(2\pi)^3\delta^{(3)}(\B{k_1}-\B{k_2}).
\end{equation}
The vacuum expectation value of the four point operator in the
interaction picture (at first order) is
\cite{Maldacena:2002vr,Weinberg:2005vy}
\begin{eqnarray}
\langle\Omega|&&\!\!\!\!\!\!\!\!\!\zeta_n(t,\B{k_1})\zeta_n(t,\B{k_2})\zeta_n(t,\B{k_3})\zeta_n(t,\B{k_4})|\Omega\rangle\nonumber\\&&=
-i\int_{t_0}^td\tilde{t}\langle 0|
\left[\zeta_n(t,\B{k_1})\zeta_n(t,\B{k_2})\zeta_n(t,\B{k_3})\zeta_n(t,\B{k_4}),H_I(\tilde{t})\right]|0\rangle,
\label{vev}
\end{eqnarray}
where $t_0$ is some early time during inflation when the inflaton
vacuum fluctuation is deep inside the horizon, $t$ is some time
after horizon exit. $|\Omega\rangle$ is the interacting vacuum
which is different from the free theory vacuum $|0\rangle$. If one
uses conformal time, it is a good approximation to perform the
integration from $-\infty$ to $0$ because $\tau\approx-(aH)^{-1}$.
$H_I$ denotes the interaction hamiltonian and it is given by
$H_I=\pi\dot\zeta_n-L$, where $\pi$ is defined as $\pi=\frac{\partial L}{\partial \dot\zeta_n}$ and $L$ is the lagrangian. In this work, we will only calculate the contribution for the four point function
that comes from a part of the interaction hamiltonian determined by the fourth order lagrangian
$H_I=-L_4$, where $L_4$ is the total lagrangian obtained from the
action (\ref{zetanactionsup}). We should point out that the other terms that we do not consider
here in the fourth order interaction hamiltonian are indeed important to obtain the full leading
order result (see equation (\ref{trizetan})) as was recently shown by \cite{Huang:2006eh}.

Of course in the end we are interested in the four point
correlation function of $\zeta$ and not of $\zeta_n$. At leading
order in slow roll these two correlation functions are equal but
they will differ at next-to-leading order.

\subsection{\label{subsec:trispectrumzeta}The trispectrum of $\zeta$}

In this subsection we calculate the relation between the
trispectrum of $\zeta$, on large scales, and the trispectrum  of
$\zeta_n$ calculated using the formalism of the previous
subsection. This relation also involves lower-order correlation
functions of $\zeta_n$ present in the literature. The variables
$\zeta$ and $\zeta_n$ are related up to third order by
\begin{equation}
\zeta(\B{x})=\zeta_n(\B{x})+F_2(\zeta_n(\B{x}))+F_3(\zeta_n(\B{x})),\label{gte}
\end{equation}
where $F_2(\zeta_n)=\mathcal{O}(\zeta_n^2)$,
$F_3(\zeta_n)=\mathcal{O}(\zeta_n^3)$ are the terms coming from
the second and third order gauge transformations respectively.
$F_2$ can be found in appendix \ref{GT}, it is
\begin{equation}
F_2(\zeta_n)=\left(\frac{\epsilon}{2}+\frac{\ddot{\phi}_0}{2H\dot{\phi}_0}\right)\zeta_n^2
+\frac{1}{H}\zeta_n\dot{\zeta}_n+\beta,
\end{equation}
where $\beta$ is given in Eq. (\ref{beta}). In the large scale
limit (super-horizon scales), we can ignore $\beta$ as it contains
gradient terms. $F_3$ was calculated in \cite{Jarnhus:2007ia} and
reads
\begin{eqnarray}
F_3(\zeta_n)&=&\left(\frac{\dddot{\phi}_0}{3H^2\dot{\phi}_0}+\frac{\epsilon\ddot{\phi}_0}{H\dot{\phi}_0}+
\frac{\epsilon^2}{3}+\frac{\epsilon\eta}{3}\right)\zeta_n^3+\left(\frac{3\ddot{\phi}_0}{2H\dot{\phi}_0}+
2\epsilon\right)\frac{\dot{\zeta}_n\zeta_n^2}{H}+\frac{\zeta_n\dot{\zeta}_n^2}{H^2}
+ \frac{\ddot{\zeta}_n\zeta_n^2}{2H^2} \nonumber\\&&+ f_a(\zeta_n)
+ f_b(\zeta_n,\tilde{\gamma}_{ij}),
\end{eqnarray}
where $f_a$ denotes terms that contain gradients (it can be found
in \cite{Jarnhus:2007ia}). $f_b$ is the part of the third order
gauge transformations that contains $\tilde{\gamma}_{ij}$. The
explicit form of $f_b$ is to the best of our knowledge still
unknown. To find out the explicit dependence of these terms on
$\zeta_n$ one would have to solve the equations of motion for
$\tilde{\gamma}_{ij}$, equation (\ref{tildegammaeq}). We do not do
this in this work. We believe that these terms will vanish in the
large scale limit and therefore do not contribute to our
calculation.

A field redefinition like
$\zeta=\zeta_n+a_1\zeta_n^{(a)}\zeta_n^{(b)}+a_2\zeta_n^{(c)}\zeta_n^{(d)}$,
where $\zeta_n^{(a,b,c,d)}$ denotes one of $\zeta_n,
\dot{\zeta}_n, \ddot{\zeta}_n$, gives after using Wick's theorem a
relation between both trispectrum like
\begin{equation}
\langle\zeta(\B{x}_1)\zeta(\B{x}_2)\zeta(\B{x}_3)\zeta(\B{x}_4)\rangle_c=\{T\}+\{PB\}+\{PPP\}+\mathcal{O}(P^\zeta_k)^4,
\end{equation}
where
\begin{equation}
\{T\}=\langle\zeta_n(\B{x}_1)\zeta_n(\B{x}_2)\zeta_n(\B{x}_3)\zeta_n(\B{x}_4)\rangle,
\end{equation}
\begin{eqnarray}
&&\!\!\!\!\!\!\!\!\!\!\!\!\!\!\{PB\}=\nonumber\\
&&\!\!\!\!\!\!\!\!\!\!\!\!\!\!a_1\Big[\langle\zeta_n^{(a)}(\B{x}_1)\zeta_n(\B{x}_2)\rangle\langle\zeta_n^{(b)}(\B{x}_1)\zeta_n(\B{x}_3)\zeta_n(\B{x}_4)\rangle
+\langle\zeta_n^{(a)}(\B{x}_1)\zeta_n(\B{x}_3)\rangle\langle\zeta_n^{(b)}(\B{x}_1)\zeta_n(\B{x}_2)\zeta_n(\B{x}_4)\rangle
\nonumber\\&&\!\!\!\!\!\!\!\!\!\!\!\!\!\!\!
\quad+\langle\zeta_n^{(a)}(\B{x}_1)\zeta_n(\B{x}_4)\rangle\langle\zeta_n^{(b)}(\B{x}_1)\zeta_n(\B{x}_2)\zeta_n(\B{x}_3)\rangle
+\langle\zeta_n^{(b)}(\B{x}_1)\zeta_n(\B{x}_2)\rangle\langle\zeta_n^{(a)}(\B{x}_1)\zeta_n(\B{x}_3)\zeta_n(\B{x}_4)\rangle
\nonumber\\&&\!\!\!\!\!\!\!\!\!\!\!\!\!\!\!
\quad+\langle\zeta_n^{(b)}(\B{x}_1)\zeta_n(\B{x}_3)\rangle\langle\zeta_n^{(a)}(\B{x}_1)\zeta_n(\B{x}_2)\zeta_n(\B{x}_4)\rangle
+\langle\zeta_n^{(b)}(\B{x}_1)\zeta_n(\B{x}_4)\rangle\langle\zeta_n^{(a)}(\B{x}_1)\zeta_n(\B{x}_2)\zeta_n(\B{x}_3)\rangle
\nonumber\\&&\!\!\!\!\!\!\!\!\!\!\!\!\!\!\! \quad+3\;perm. \Big]
\nonumber\\&\!\!\!\!\!\!+&\!\!\!\!\!a_2\Big[(a\rightarrow
c,b\rightarrow d)+3\;perm.\Big],
\end{eqnarray}
\begin{eqnarray}
\{PPP\}&\!\!\!\!=&\!\!\!\!a_1^2\Big[\langle\zeta_n^{(a)}(\B{x}_1)\zeta_n^{(a)}(\B{x}_2)\rangle\Big(\langle\zeta_n^{(b)}(\B{x}_1)\zeta_n(\B{x}_3)\rangle\langle\zeta_n^{(b)}(\B{x}_2)\zeta_n(\B{x}_4)\rangle
+(\B{x}_3\longleftrightarrow \B{x}_4)\Big) \nonumber\\&& \quad
\!\!\!\!\!\!\!\!+\langle\zeta_n^{(a)}(\B{x}_1)\zeta_n^{(b)}(\B{x}_2)\rangle\Big(\langle\zeta_n^{(b)}(\B{x}_1)\zeta_n(\B{x}_3)\rangle\langle\zeta_n^{(a)}(\B{x}_2)\zeta_n(\B{x}_4)\rangle
+(\B{x}_3\longleftrightarrow \B{x}_4)\Big) \nonumber\\&& \quad
\!\!\!\!\!\!\!\!+\langle\zeta_n^{(b)}(\B{x}_1)\zeta_n^{(a)}(\B{x}_2)\rangle\Big(\langle\zeta_n^{(a)}(\B{x}_1)\zeta_n(\B{x}_3)\rangle\langle\zeta_n^{(b)}(\B{x}_2)\zeta_n(\B{x}_4)\rangle
+(\B{x}_3\longleftrightarrow \B{x}_4)\Big) \nonumber\\&& \quad
\!\!\!\!\!\!\!\!+\langle\zeta_n^{(b)}(\B{x}_1)\zeta_n^{(b)}(\B{x}_2)\rangle\Big(\langle\zeta_n^{(a)}(\B{x}_1)\zeta_n(\B{x}_3)\rangle\langle\zeta_n^{(a)}(\B{x}_2)\zeta_n(\B{x}_4)\rangle
+(\B{x}_3\longleftrightarrow \B{x}_4)\Big) \nonumber\\&& \quad
\!\!\!\!\!\!\!\!+5\;perm. \Big]
\nonumber\\&\!\!\!\!\!\!\!\!\!\!\!\!+&\!\!\!\!\!\!\!\!a_2^2\Big[(a\rightarrow
c,b\rightarrow d)+5\;perm.\Big]
\nonumber\\&\!\!\!\!\!\!\!\!\!\!\!\!+&\!\!\!\!\!\!\!\!2a_1a_2\Big[\langle\zeta_n^{(a)}(\B{x}_1)\zeta_n^{(c)}(\B{x}_2)\rangle\Big(\langle\zeta_n^{(b)}(\B{x}_1)\zeta_n(\B{x}_3)\rangle\langle\zeta_n^{(d)}(\B{x}_2)\zeta_n(\B{x}_4)\rangle
+(\B{x}_3\longleftrightarrow \B{x}_4)\Big) \nonumber\\&& \quad+
\langle\zeta_n^{(a)}(\B{x}_1)\zeta_n^{(d)}(\B{x}_2)\rangle\Big(\langle\zeta_n^{(b)}(\B{x}_1)\zeta_n(\B{x}_3)\rangle\langle\zeta_n^{(c)}(\B{x}_2)\zeta_n(\B{x}_4)\rangle
+(\B{x}_3\longleftrightarrow \B{x}_4)\Big) \nonumber\\&& \quad+
\langle\zeta_n^{(b)}(\B{x}_1)\zeta_n^{(c)}(\B{x}_2)\rangle\Big(\langle\zeta_n^{(a)}(\B{x}_1)\zeta_n(\B{x}_3)\rangle\langle\zeta_n^{(d)}(\B{x}_2)\zeta_n(\B{x}_4)\rangle
+(\B{x}_3\longleftrightarrow \B{x}_4)\Big) \nonumber\\&& \quad+
\langle\zeta_n^{(b)}(\B{x}_1)\zeta_n^{(d)}(\B{x}_2)\rangle\Big(\langle\zeta_n^{(a)}(\B{x}_1)\zeta_n(\B{x}_3)\rangle\langle\zeta_n^{(c)}(\B{x}_2)\zeta_n(\B{x}_4)\rangle
+(\B{x}_3\longleftrightarrow \B{x}_4)\Big) \nonumber\\&& \quad+
5\;perm. \Big],
\end{eqnarray}
where ``perm" means the other permutations of the preceding terms
and $\mathcal{O}(P^\zeta_k)^4$ denotes terms that are suppressed
by successive powers of the power spectrum. $(a\rightarrow
c,b\rightarrow d)$ means terms equal to the immediately preceding
terms with $a,b$ replaced by $c,d$ respectively.
$(\B{x}_3\longleftrightarrow \B{x}_4)$ means a term obtained from
the preceding term with $\B{x}_3$ and $\B{x}_4$ interchanged.

If the field redefinition contains third order pieces like
$\zeta=\zeta_n+b_1\zeta_n^{(a)}\zeta_n^{(b)}\zeta_n^{(c)}$, they
contribute with additional terms as
\begin{eqnarray}
\langle&&\!\!\!\!\!\!\!\!\!\!\!\!\!\!\zeta(\B{x}_1)\zeta(\B{x}_2)\zeta(\B{x}_3)\zeta(\B{x}_4)\rangle_c=\langle\zeta_n(\B{x}_1)\zeta_n(\B{x}_2)\zeta_n(\B{x}_3)\zeta_n(\B{x}_4)\rangle
\nonumber\\&+&\!\!\!\!b_1\Big[\langle\zeta_n^{(a)}(\B{x}_1)\zeta_n(\B{x}_2)\rangle\Big(\langle\zeta_n^{(b)}(\B{x}_1)\zeta_n(\B{x}_3)\rangle\langle\zeta_n^{(c)}(\B{x}_1)\zeta_n(\B{x}_4)\rangle
+(b\longleftrightarrow c)\Big) \nonumber\\&& \quad+
\langle\zeta_n^{(b)}(\B{x}_1)\zeta_n(\B{x}_2)\rangle\Big(\langle\zeta_n^{(a)}(\B{x}_1)\zeta_n(\B{x}_3)\rangle\langle\zeta_n^{(c)}(\B{x}_1)\zeta_n(\B{x}_4)\rangle
+(a\longleftrightarrow c)\Big) \nonumber\\&& \quad+
\langle\zeta_n^{(c)}(\B{x}_1)\zeta_n(\B{x}_2)\rangle\Big(\langle\zeta_n^{(a)}(\B{x}_1)\zeta_n(\B{x}_3)\rangle\langle\zeta_n^{(b)}(\B{x}_1)\zeta_n(\B{x}_4)\rangle
+(a\longleftrightarrow b)\Big) \nonumber\\&& \quad+3\;perm. \Big]
+ \mathcal{O}(P^\zeta_k)^4,
\end{eqnarray}
where for example $(b\longleftrightarrow c)$ means a terms
obtained from the term immediately preceding with $b$ and $c$
interchanged. To the best of our knowledge the expectation values
involving operators containing derivatives of $\zeta_n$ have not
yet been calculated in the literature. However, once the mode
function equation (\ref{Mukhanoveq}) is solved, one has all the
ingredients needed to calculate these expectation values,
including the interaction Hamiltonian.

\section{\label{sec:leadingtrispectrum}Calculation of the leading
order trispectrum}

In this section, we will use the formalism of the previous section
and the fourth order exact interaction Hamiltonian of subsection
\ref{subsec:PerturbationsComoving} to calculate the leading order
trispectrum, under the assumption that the ``slow-roll" parameters
(\ref{epsilon}-\ref{s}) are always small until the end of
inflation.

\subsection{\label{subsec:leadingtrispectrumzetan}The leading order
trispectrum of $\zeta_n$}

To calculate the leading order trispectrum of $\zeta_n$ in slow
roll, we need to evaluate Eq. (\ref{vev}) where $H_I$ is read from
the order $\epsilon$ terms of the action (\ref{zetanactionsup}).
The interaction Hamiltonian we get contains terms with
$\tilde{\gamma}_{ij}$. Fortunately it can be shown that to compute
the leading order trispectrum we do not need to know the solution
for $\tilde{\gamma}_{ij}$ and the knowledge of its equation of
motion (Eq. (\ref{tildegammaeq})) is sufficient. At this order we
use the solution for the mode functions Eq. (\ref{modefc}). The
integrals in Eq. (\ref{vev}) can then be performed analytically to
give
\begin{eqnarray}
\langle&&\!\!\!\!\!\!\!\!\!\!\Omega|\zeta_n(\B{k}_1)\zeta_n(\B{k}_2)\zeta_n(\B{k}_3)\zeta_n(\B{k}_4)|\Omega\rangle
=
(2\pi)^3\delta^3(\B{k}_1+\B{k}_2+\B{k}_3+\B{k}_4)\frac{H^6}{\epsilon^3c_s^3}\frac{1}{\Pi_ik_i^3}\nonumber\\&&
\times\left[\frac{3}{4}\left(10\Pi+3\lambda\right)\frac{c_s^2}{H^2\epsilon}A_1-\frac{1}{2^6}\left(3\lambda-\frac{H^2\epsilon}{c_s^2}+H^2\epsilon\right)
\frac{1}{H^2\epsilon}A_2-\frac{1}{2^8}\frac{c_s^2-1}{c_s^4}A_3\right],\nonumber\\\label{trizetan}
\end{eqnarray}
where the momentum dependent functions $A_i$ are defined as
\begin{eqnarray}
A_1&=&\frac{\Pi_ik_i^2}{K^5},\nonumber\\
A_2&=&\frac{k_1^2k_2^2(\textbf{k}_3\cdot\textbf{k}_4)}{K^3}\left(1+\frac{3(k_3+k_4)}{K}+\frac{12k_3k_4}{K^2}\right)+\textrm{perm.},
\nonumber \\
A_3&=&\frac{(\textbf{k}_1\cdot\textbf{k}_2)(\textbf{k}_3\cdot\textbf{k}_4)}{K}\left[1+\frac{\sum_{i<j}k_ik_j}{K^2}
+\frac{3k_1k_2k_3k_4}{K^3}
\left(\sum_i\frac{1}{k_i}\right)+12\frac{k_1k_2k_3k_4}{K^4}\right]\nonumber\\&&+\textrm{perm.},
\end{eqnarray}
and ``perm.'' refers to the $24$ permutations of the four momenta.
Note that the quantities of Eq. (\ref{trizetan}) are evaluated at
the moment $\tau_*$ at which the total wave number $K=\sum_{i=1}^4
k_i$ exits the horizon, i.e., when $K{c_s}_*=a_*H_*$. This leading
order result agrees with the result of Huang and Shiu
\cite{Huang:2006eh} \footnote{The full leading order
result for the four point function can be found in the revised version of \cite{Huang:2006eh} which takes into account all the contributions for
the fourth order interaction hamiltonian.} that did their calculation
in the uniform curvature gauge and using a simpler method that it
is only valid to calculate the leading order contribution for
models with $c_s\ll 1$. In fact, we can compare our uniform
curvature gauge result (\ref{SSSSphi}) with the result of Huang
and Shiu \cite{Huang:2006eh}. We see that the last terms of the
fourth, sixth and ninth lines of equation (\ref{SSSSphiA}) are
exactly the ones obtained by \cite{Huang:2006eh}, their equation
(15), using the method of just expanding the field Lagrangian as
in \cite{Gruzinov:2004jx,Creminelli:2003iq}. For a model with a
general field Lagrangian these terms are the ones that give the
leading order contribution for the trispectrum, in the small sound
speed limit, equation (\ref{trizetan}). The contribution coming
from the tensor part will be of next-to-leading order in this
case.

For standard kinetic term inflation, $\Pi=\lambda=0$ and $c_s=1$
and Eq. (\ref{trizetan}) vanishes, the leading order is then given
by the next order in slow roll.

\subsection{\label{subsec:leadingtrispectrumzeta}The leading order trispectrum of $\zeta$}

It is well known that if the slow-roll conditions are satisfied
until the end of inflation and we can ignore gradient terms then
the gauge invariant curvature perturbation $\zeta$ remains
constant on super-horizon scales to all order in perturbation
theory. In this subsection, we will see that this fact greatly
simplifies the relation between the trispectrum of $\zeta$ and
$\zeta_n$.

In the large scales limit, Eq. (\ref{gte}) simplifies to give
\begin{equation}
\zeta=\zeta_n+a\zeta_n^2+\frac{1}{H}\zeta_n\dot{\zeta}_n+\mathcal{O}(\zeta_n^3),\label{gtels}
\end{equation}
where $a$ is defined as
\begin{equation}
a=\frac{\epsilon}{2}+\frac{\ddot{\phi}_0}{2H\dot{\phi}_0}.
\end{equation}
Using the fact that $\dot{\zeta}=0$ on super-horizon scales and
the equation resulting from a time derivative of Eq. (\ref{gtels})
one can show that
\begin{equation}
\dot{\zeta}_n=-\dot{a}\zeta_n^2+\mathcal{O}(\zeta_n^3).\label{zetandot}
\end{equation}
This equation has a simple interpretation. The variable $\zeta_n$
is not constant outside the horizon, only the gauge invariant
quantity $\zeta$ is. This is the reason why the term
$\frac{1}{H}\zeta_n\dot{\zeta}_n$ in the second order gauge
transformation cannot be ignored when one is calculating the
trispectrum of $\zeta$. Substituting Eq. (\ref{zetandot}) in Eq.
(\ref{gte}) and taking the large scale limit we get
\begin{equation}
\zeta=\zeta_n+a\zeta_n^2+ b\zeta_n^3+\cdots \label{ct}
\end{equation}
where $\cdots$ means cubic terms that contain at least one time
derivative of $\zeta_n$ and that will only give a contribution to
the five point function. The variable $b$ is defined as
\begin{eqnarray}
b&=&\frac{\dddot{\phi}_0}{3H^2\dot{\phi}_0}+\frac{\epsilon\ddot{\phi}_0}{H\dot{\phi}_0}+
\frac{\epsilon^2}{3}+\frac{\epsilon\eta}{3}-\frac{\dot{a}}{H}\nonumber\\
&=&-\frac{\dddot{\phi}_0}{6H^2\dot{\phi}_0}+\frac{\epsilon\ddot{\phi}_0^2}{2H\dot{\phi}_0}+\frac{\ddot{\phi}_0^2}{2H^2\dot{\phi}_0^2}
+\frac{\epsilon^2}{3}-\frac{\eta\epsilon}{6}.
\end{eqnarray}

We shall now compare (\ref{ct}) with the result given by the
$\delta N$ formalism
\cite{Starobinsky:1986fxa,Sasaki:1995aw,Lyth:2004gb,Lyth:2005fi}.
In the $\delta N$ approach $\zeta$ is expanded in series in terms
of the field perturbation as
\begin{equation}
\zeta=N'\delta\phi+\frac{1}{2}N''\delta\phi^2+\frac{1}{6}N'''\delta\phi^3+\mathcal{O}(\delta\phi)^4,\label{deltaN}
\end{equation}
where $N$ is the number of e-folds and a prime denotes derivative
with respect to $\phi$. Now comparing Eq. (\ref{ct}) with the
previous equation and observing that
$\zeta_n=-\frac{H}{\dot{\phi}_0}\delta\phi$ we expect
\begin{equation}
\frac{N''}{2}=\frac{H^2}{\dot{\phi}_0^2}a, \quad
\frac{N'''}{6}=-\frac{H^3}{\dot{\phi}_0^3}b.
\end{equation}
We verified that this is indeed the case.

Using Wick's theorem one can now relate the connected part of the
four point correlation function of $\zeta$ with the four point
correlation function of $\zeta_n$ calculated in the previous
section \cite{Byrnes:2006vq,Seery:2006js}. This relation also
involves lower order correlation functions of $\zeta_n$, like the
bispectrum
$\langle\zeta_n(\B{x}_1)\zeta_n(\B{x}_2)\zeta_n(\B{x}_3)\rangle$
(the leading and next-to-leading order in slow roll bispectrum was
previously calculated in \cite{Chen:2006nt}). The relation is
\begin{eqnarray}
\langle&&\!\!\!\!\!\!\!\!\!\!\!\zeta(\B{x}_1)\zeta(\B{x}_2)\zeta(\B{x}_3)\zeta(\B{x}_4)\rangle_c=\langle\zeta_n(\B{x}_1)\zeta_n(\B{x}_2)\zeta_n(\B{x}_3)\zeta_n(\B{x}_4)\rangle
\nonumber\\&\!\!\!\!+&\!\!\!\!\!2a\Big[\langle\zeta_n(\B{x}_1)\zeta_n(\B{x}_2)\rangle\langle\zeta_n(\B{x}_1)\zeta_n(\B{x}_3)\zeta_n(\B{x}_4)\rangle
+\langle\zeta_n(\B{x}_1)\zeta_n(\B{x}_3)\rangle\langle\zeta_n(\B{x}_1)\zeta_n(\B{x}_2)\zeta_n(\B{x}_4)\rangle
\nonumber\\&&\quad+\langle\zeta_n(\B{x}_1)\zeta_n(\B{x}_4)\rangle\langle\zeta_n(\B{x}_1)\zeta_n(\B{x}_2)\zeta_n(\B{x}_3)\rangle
+3\;perm \Big]
\nonumber\\&\!\!\!\!+&\!\!\!\!\!4a^2\Big[\langle\zeta_n(\B{x}_1)\zeta_n(\B{x}_2)\rangle\langle\zeta_n(\B{x}_1)\zeta_n(\B{x}_3)\rangle\langle\zeta_n(\B{x}_2)\zeta_n(\B{x}_4)\rangle
\nonumber\\&&\quad+\langle\zeta_n(\B{x}_1)\zeta_n(\B{x}_2)\rangle\langle\zeta_n(\B{x}_1)\zeta_n(\B{x}_4)\rangle\langle\zeta_n(\B{x}_2)\zeta_n(\B{x}_3)\rangle
+ 5\;perm\Big]
\nonumber\\&\!\!\!\!+&\!\!\!\!\!6b\Big[\langle\zeta_n(\B{x}_1)\zeta_n(\B{x}_2)\rangle\langle\zeta_n(\B{x}_1)\zeta_n(\B{x}_3)\rangle\langle\zeta_n(\B{x}_1)\zeta_n(\B{x}_4)\rangle
+ 3\;perm \Big] + \mathcal{O}(P^\zeta_k)^4,
\end{eqnarray}
where ``perm" means the other permutations of the preceding terms
and $\mathcal{O}(P^\zeta_k)^4$ denotes terms that are suppressed
by successive powers of the power spectrum.  Now, one can easily
see that at leading order in slow roll the trispectrum for $\zeta$
and $\zeta_n$ are equal, this is because the constants $a$ and $b$
are slow-roll suppressed. These terms will only contribute to the
next-to-leading order corrections.

\subsection{\label{subsec:nextobleadingcorrections}The next-to-leading order
corrections for the trispectrum}

In subsection \ref{subsec:leadingtrispectrumzetan}, we showed that
for standard kinetic term inflation the leading order result,
equation (\ref{trizetan}), vanishes and in fact in this case the
leading order of the trispectrum of $\zeta_n$ is of order
$\epsilon^{-2}$ (the next-to-leading order is the leading order).
To obtain these leading order contributions it is easier to
perform the calculation using the uniform curvature gauge action
Eq. (\ref{SSSSphiA})-(\ref{GSSphi}). Eq. (\ref{SSSSphiA}) vanishes
exactly for standard kinetic term inflation. The action
(\ref{SSSSphiB}) is exact in the slow-roll approximation but it is
instructive to determine the slow-roll order of the different
terms. One can see that the leading order contribution comes from
terms of order $\mathcal{O}(\epsilon^0)$, as pointed out in
\cite{Seery:2006vu}. If we take $P(X,\phi)=X-V(\phi)$ then the
leading order (in slow roll) source of Eq. (\ref{tildegammaeq})
will be of order $\mathcal{O}(\epsilon^0)$. We therefore do not
expect $\tilde{\gamma}_{ij}$ to be slow-roll suppressed and the
actions (\ref{GGphi}), (\ref{GSSphi}) will contain unsuppressed
terms of the same order as the leading order term of the action
(\ref{SSSSphiB}). These tensor contributions were absent in the
analysis of \cite{Seery:2006vu} and we have shown that they are of
the same order as the fourth order action considered in
\cite{Seery:2006vu}, our Eq. (\ref{SSSSphiB}). It is still an open
question how these new contributions will change the trispectrum
result of Seery \emph{et al.}.

For the general Lagrangian case, the leading order trispectrum was
given in the previous subsection and in \cite{Huang:2006eh}.
Contrary to the method of \cite{Huang:2006eh}, our method of
obtaining the fourth order action (\ref{SSSSphi}) does not rely on
any approximation and therefore the action (\ref{SSSSphi}) is
valid to all orders in slow roll and in the sound speed expansion
and it can be used to study the next-to-leading order corrections.
Depending on the momentum shape of these next-to-leading terms
they might become big enough to be observed in the next generation
of experiments. A similar argument applies for the next-to-leading
order corrections for the bispectrum, as it was shown in
\cite{Chen:2006nt}. For example, for DBI inflation,
\cite{Huang:2006eh} showed that the leading order non-Gaussianity
parameters $\tau_{NL}$ scales like $\tau_{NL}\sim0.1/c_s^4$ (for a
specific momentum configuration) and $f_{NL}\sim1/c_s^2$. They
argue that if $c_s\sim0.1$ then $f_{NL}$ is still inside the value
range allowed by observations but $\tau_{NL}\sim 10^3$ could be
detected with the Planck satellite CMBR experiment. Therefore,
assuming that the slow-roll parameter $\epsilon$ is of order
$\epsilon\sim0.01$ (at horizon crossing) these next-to-leading
order corrections for the trispectrum could possible be observed
with the Planck satellite. A more careful and systematic study of
the momentum dependence of these new terms is required and it is
left for future work.

\section{\label{sec:conclusion}Conclusion}

We have computed the fourth order action for scalar and second
order tensor perturbations in the comoving gauge. Our result is
exact in the slow-roll (SR) expansion but practically it is useful
to study the SR suppression of the different terms. We were able
to show that after many integrations by parts the unsuppressed
terms contained in the previous action can be reduced to total
derivatives terms plus corrections that are SR suppressed. The
resulting action has the correct order in SR. It is suppressed by
$\epsilon$ as it should be, because in pure de Sitter space the
curvature perturbation is a pure gauge mode. An important lesson
from our work is that in order to obtain the correct SR order for
the action, the second order tensor perturbations cannot be
ignored as assumed in previous works \cite{Seery:2006vu} and
\cite{Jarnhus:2007ia}. We found the explicit form of these tensor
perturbations in the comoving gauge by using the gauge
transformations from the uniform curvature gauge. Fortunately, for
a general inflation model like (\ref{action4}), we showed that we
do not need to solve the equations of motion for the tensor
perturbations if we are interested in calculating the leading
order trispectrum. However, to calculate the next-to-leading order
corrections to that result, or to calculate the leading order
trispectrum for standard kinetic term inflation, we do need to
solve explicitly the equations of motion for the tensor
perturbations. This will be left for future work.

Using the comoving gauge action we have calculated the leading
order in SR trispectrum of $\zeta$. We compared our result with
the result of \cite{Huang:2006eh}, obtained using the uniform
curvature gauge, and we found an agreement.

For the uniform curvature gauge action, that is also exact in the
SR expansion, we identified the terms that will contribute to the
next-to-leading order corrections to the trispectrum. We pointed
out that depending on the model and on the momentum configuration,
some of these corrections might be observable with the Planck
satellite. After taking particular limits, the previous action
nicely reduces to previously known results
\cite{Seery:2006vu,Huang:2006eh} with the caveat that the above
mentioned works ignore tensor contributions.

Finally we have obtained the relations between the trispectrum of
$\zeta$ and $\delta\phi$ (on large scales) using the third order
gauge transformations and compared the result with the $\delta N$
formalism.

To conclude, we have provided the necessary equations (fourth
order action and the relation between $\zeta$ and $\delta\phi$) to
calculate the trispectrum for a fairly general model of inflation
that are also valid for models where SR is temporarily
interrupted, i.e., around a ``step" in the inflaton's Lagrangian
\cite{Hailu:2006uj}. In this case, it is impossible to apply the
$\delta N$ approach and it is required to evaluate the n-point
functions numerically \cite{Chen:2006xjb} (see
\cite{Tanaka:2007gh} for a different approach). We leave this more
practical application of our results for future work.

\chapter{Non-Gaussianity from the bispectrum in general multiple field
inflation}\label{chapter:bispectrum}
\chaptermark{The bispectrum in multiple field inflation}

\section{Introduction}
In this chapter, we shall study the non-Gaussianity of the
primordial curvature perturbation produced by a phase of multiple
field inflation. In particular, we shall concentrate on the
bispectrum or three point correlation function\footnote{This
chapter in based on our work \cite{Arroja:2008yy}. While we were
writing up that work, similar results appeared on the arXiv
\cite{Langlois:2008qf}.}.

The simplest single field inflation models predict that the
non-Gaussianity of the cosmic microwave background (CMB)
fluctuations will be very difficult to be detected even in future
experiments such as
Planck\footnote{http://www.rssd.esa.int/index.php?project=Planck}.
The detection of large non-Gaussianity would mean that the
simplest models of inflation are rejected. There are observational
hints \cite{Yadav:2007yy,Jeong:2007mx} of some deviation from
perfect Gaussianity in the WMAP3 data \cite{Spergel:2006hy} but
the WMAP5 data is consistent with Gaussianity
\cite{Komatsu:2008hk,Curto:2008ym}.

There are a few models where the primordial fluctuations generated
during inflation have a large non-Gaussianity. In the single field
case, if the inflaton field has a non-trivial kinetic term, it is
known that the non-Gaussianity can be large. For example, in
$K$-inflation models where the kinetic term of the inflaton field
is generic, the sound speed of the perturbations can be much
smaller than $1$ \cite{ArmendarizPicon:1999rj,Garriga:1999vw}
which leads to large non-Gaussian fluctuations. Additionally,
Dirac-Born-Infeld (DBI) inflation, motivated by string theory, can
also give large non-Gaussian fluctuations
\cite{Silverstein:2003hf,Alishahiha:2004eh,Chen:2004gc,Chen:2005ad}.
The inflaton is identified with the position of a moving D3 brane
whose dynamics is described by the DBI action. Again, due to the
non-trivial form of the kinetic term, the sound speed can be
smaller than $1$ and the non-Gaussianity becomes large
\cite{Chen:2006nt,Chen:2005fe}. In the previous chapter, the third
and fourth order actions for a single inflaton field with a
generic kinetic term have been calculated by properly taking into
account metric perturbations. Three and four point functions have
also been obtained \cite{Chen:2006nt,Huang:2006eh,Arroja:2008ga}.
For the detailed observational consequences of single-field
DBI-inflation see
\cite{Kecskemeti:2006cg,Lidsey:2006ia,Baumann:2006cd,Bean:2007hc,
Lidsey:2007gq,Peiris:2007gz,Kobayashi:2007hm,Lorenz:2007ze}.

Multi-field inflation models where the curvature perturbation is
modified on large scales due to the entropy perturbations have
been also recently studied. In the case of the standard kinetic
term, it is not easy to generate large non-Gaussianity from
multi-field dynamics \cite{Rigopoulos:2005ae, Vernizzi:2006ve,
Battefeld:2006sz, Battefeld:2007en, Yokoyama:2007dw,
Yokoyama:2007uu} (see however Ref.~\cite{Sasaki:2008uc} and the
curvaton scenario e.g. \cite{Lyth:2002my,Gordon:2002gv,:2008ei}).
In the DBI-inflation case, the position of the brane in each
compact direction is described by a scalar field. Then
DBI-inflation is naturally a multi-field model
\cite{Easson:2007dh}. The effect of the entropy perturbations in
the inflationary models based on string theory constructions in a
slightly different context is also considered in
\cite{Lalak:2007vi,Brandenberger:2007ca}. Recently, Huang \emph{et
al.} calculated the bispectrum of the perturbations in multi-field
DBI-inflation {\it with the assumption that} the kinetic term
depends only on $X=-G^{IJ}
\partial_{\mu} \phi^I
\partial^{\mu} \phi^J /2$ where $\phi^J$ are the scalar fields
$(I=1,2,...)$ and $G_{IJ}$ is the metric in the field space, as
occurs in K-inflation \cite{Huang:2007hh}. They found that in
addition to the usual bispectrum of adiabatic perturbations, there
exists a new contribution coming from the entropy perturbations.
Then they showed that the entropy field perturbations propagate
with the speed of light and the contribution from the entropy
perturbations is suppressed. This property can also be confirmed
by the analysis of a more general class of multi-field models
where the kinetic terms are given by arbitrary functions of X
\cite{Langlois:2008mn,Gao:2008dt}. However, Langlois \emph{et al.}
pointed out that their assumption {\it cannot} be justified for
the multi-field DBI-inflation \cite{Langlois:2008wt}. Even though
the action depends only on $X$ in the homogeneous background,
there exist other kind of terms which contribute only to
inhomogeneous perturbations. They find that this dramatically
changes the behaviour of the entropy perturbations. In fact, it
was shown that the entropy perturbations propagate with the same
sound speed as the adiabatic perturbations.
Ref.~\cite{Langlois:2008wt} also calculated the bispectrum by
generalizing the work of Huang {\it et al.} \cite{Huang:2006eh}.

In this chapter, we study a fairly general class of multi-field
inflation models with a general kinetic term which includes
K-inflation and DBI-inflation. We study the sound speeds of the
adiabatic perturbations and entropy perturbations and clarify the
difference between K-inflation and DBI-inflation. Then we
calculate the third order action by properly taking into account
the effect of gravity. We continue to obtain the three point
functions at leading order in slow-roll and in the small sound
speed limit. We can recover the results for K-inflation and
DBI-inflation easily from this general result.

The structure of the chapter is as follows. In section
\ref{sec:MODEL5}, we describe our model and derive equations in
the background. In section \ref{sec:Perturbations}, we study the
perturbations using the ADM formalism. Additionally, the second
and third order actions are derived by properly taking into
account the metric perturbations. Then we decompose the
perturbations into adiabatic and entropy directions and write down
the action in terms of the decomposed fields. In section
\ref{sec:Lperturbations}, we study the sound speed in several
models including K-inflation and DBI-inflation. It is shown that
in general, adiabatic and entropy sound speeds are different and
both can be smaller than 1. In section
\ref{sec:leadingorderTPFmulti}, the third order action at leading
order in slow-roll and in the small sound speed limit is obtained
in terms of the decomposed fields. Then the three point functions
are derived for a generalized model which includes K-inflation and
DBI-inflation as particular cases. Section
\ref{sec:conclusionbispectrum} is devoted to the conclusion.

\section{\label{sec:MODEL5}The model}
We consider a very general class of models described by the
following action
\begin{equation}
S=\frac{1}{2}\int
d^4x\sqrt{-g}\left[M^2_{Pl}R+2P(X^{IJ},\phi^I)\right],
\label{action5}
\end{equation}
where $\phi^I$ are the scalar fields $(I=1,2,...,N)$, $M_{Pl}$ is
the Planck mass that we will set to unity hereafter, $R$  is the
Ricci scalar and
\begin{equation}
X^{IJ}\equiv-\frac{1}{2}g^{\mu\nu}\partial_\mu\phi^I\partial_\nu\phi^J,
\end{equation}
is the kinetic term, $g_{\mu\nu}$ is the metric tensor. We label
the fields' Lagrangian by $P$ and we assume that it is a well
behaved function so that derivatives of $P$ with respect to
$X^{IJ}$ can be defined. Greek indices run from 0 to 3. Lower case
Latin letters $(i,j,...)$ denote spatial indices. Upper case Latin
letters denote field indices. See section \ref{sec:MODEL} for the
single field case with non-standard kinetic term.

The Einstein field equations in this model are
\begin{equation}
G_{\mu\nu}=Pg_{\mu\nu}+P_{,X^{IJ}}\partial_\mu\phi^I\partial_\nu\phi^J\equiv
T_{\mu\nu},
\end{equation}
where $P_{,X^{IJ}}$ denotes the derivative of $P$ with respect to
$X^{IJ}$ \footnote{Strictly speaking, we adopt the symmetrized
derivative, $P_{,X^{IJ}} \equiv \frac12 \left(\frac{\partial
P}{\partial X^{IJ}} + \frac{\partial P}{\partial
X^{JI}}\right)$.}. The generalized Klein-Gordon equation reads
\begin{equation}
g^{\mu\nu}\left(P_{,X^{IJ}}\partial_\nu\phi^I\right)_{;\mu}+P_{,J}=0,
\end{equation}
where $;$ denotes covariant derivative with respect to
$g_{\mu\nu}$ and $P_{,J}$ denotes the derivative of $P$ with
respect to $\phi^J$.

In the background, we are interested in flat, homogeneous and
isotropic\linebreak Friedman-Robertson-Walker universes described
by the line element
\begin{equation}
ds^2=-dt^2+a^2(t)\delta_{ij}dx^idx^j, \label{FRW5}
\end{equation}
where $a(t)$ is the scale factor. The Friedman equation and the
continuity equation read
\begin{equation}
3H^2=E_0, \label{EinsteinEq}
\end{equation}
\begin{equation}
\dot{E_0}=-3H\left(E_0+P_0\right), \label{continuity}
\end{equation}
where the Hubble rate is $H=\dot{a}/a$, $E_0$ is the total energy
of the fields and it is given by
\begin{equation}
E_0=2X_0^{IJ}P_{0,X^{IJ}}-P_0 ,\label{energy5}
\end{equation}
where a subscript zero denotes background quantities,
$X_0^{IJ}=1/2\dot \phi_0^I\dot \phi_0^J$. The equations of motion
for the scalar fields reduce to
\begin{equation}
P_{0,X^{IJ}}\ddot\phi_0^I+\left(3HP_{0,X^{IJ}}+\dot
P_{0,X^{IJ}}\right)\dot \phi_0^I - P_{0,J}=0.\label{KG1}
\end{equation}

\section{\label{sec:Perturbations}Perturbations}
In this section, we will consider perturbations of the background
(\ref{FRW5}) beyond linear order. For this purpose, we will
construct the action at second and third order in the
perturbations and it is convenient to use the ADM metric formalism
\cite{Langlois:2008mn,Maldacena:2002vr,Seery:2005wm,Seery:2005gb,Chen:2006nt,Arnowitt:1962hi,Gao:2008dt}.
The ADM line element reads
\begin{equation}
ds^2=-N^2dt^2+h_{ij}\left(dx^i+N^idt\right)\left(dx^j+N^jdt\right),
\label{ADMmetricphi}
\end{equation}
where $N$ is the lapse function, $N^i$ is the shift vector and
$h_{ij}$ is the 3D metric. The action (\ref{action5}) becomes
\begin{equation}
S=\frac{1}{2}\int
dtd^3x\sqrt{h}N\left({}^{(3)}\!R+2P(X^{IJ},\phi^I)\right)+
\frac{1}{2}\int dtd^3x\sqrt{h}N^{-1}\left(E_{ij}E^{ij}-E^2\right).
\end{equation}
The tensor $E_{ij}$ is defined as
\begin{equation}
E_{ij}=\frac{1}{2}\left(\dot{h}_{ij}-\nabla_iN_j-\nabla_jN_i\right),
\end{equation}
and it is related to the extrinsic curvature by
$K_{ij}=N^{-1}E_{ij}$. $\nabla_i$ is the covariant derivative with
respect to $h_{ij}$. $X^{IJ}$ can be written as
\begin{equation}
X^{IJ}=-\frac{1}{2}h^{ij}\partial_i\phi^I\partial_j\phi^J+\frac{N^{-2}}{2}v^Iv^J\,,
\end{equation}
where $v^I$ is defined as
\begin{equation}
v^I\equiv \dot{\phi}^I-N^j\nabla_j\phi^I.\label{vi}
\end{equation}
The Hamiltonian and momentum constraints are respectively
\begin{eqnarray}
{}^{(3)}\!R+2P-2N^{-2}P_{,X^{IJ}}v^Iv^J-N^{-2}\left(E_{ij}E^{ij}-E^2\right)&=&0,\nonumber\\
\nabla_j\left(N^{-1}E_i^j\right)-\nabla_i\left(N^{-1}E\right)&=&
N^{-1}P_{,X^{IJ}}v^I\nabla_i\phi^J.\nonumber\\\label{LMphi}
\end{eqnarray}
We decompose the shift vector $N^i$ into scalar and intrinsic
vector parts as
\begin{equation}
N_i=\tilde{N_i}+\partial_i\psi,
\end{equation}
where $\partial_i\tilde{N^i}=0$, here and in the rest of the
section indices are raised with $\delta_{ij}$.

\subsection{\label{subsec:PerturbationsUniformeR}Perturbations in the uniform curvature gauge}
In the uniform curvature gauge, the 3D metric takes the form
\begin{eqnarray}
&&h_{ij}=a^2\delta_{ij},\nonumber\\
&&\phi^I(x,t)=\phi_0^I(t)+Q^I(x,t),
\end{eqnarray}
where $Q^I$ denotes the field perturbations. In the following, we
will usually drop the subscript $``0"$ on $\phi_0^I$ and simply
identify $\phi^I$ as the homogeneous background fields unless
otherwise stated.

We expand $N$ and $N^i$ in powers of the perturbation $Q^I$
\begin{eqnarray}
N&=&1+\alpha_1+\alpha_2+\cdots,\\
\tilde{N_i}&=&\tilde{N_i}^{(1)}+\tilde{N_i}^{(2)}+\cdots,\\
\psi&=&\psi_1+\psi_2+\cdots,
\end{eqnarray}
where $\alpha_n$, $\tilde{N_i}^{(n)}$ and $\psi_n$ are of order
$(Q^I)^n$. At first order in $Q^I$, a particular solution for
equations (\ref{LMphi}) is:
\begin{eqnarray}
\label{constraint}
\alpha_1&=&\frac{P_{,X^{IJ}}}{2H} \dot{\phi^I}
Q^J, \quad
\tilde{N_i}^{(1)}=0, \nonumber\\
\partial^2\psi_1&=&\frac{a^2}{2H} \bigg[
-6H^2\alpha_1+P_{,K}Q^K- \dot{\phi}^I \dot{\phi}^J P_{,X^{IJ}K}Q^K
\nonumber\\&&+\left(P_{,X^{IJ}}+ \dot{\phi}^L \dot{\phi}^M
P_{,X^{LM}X^{IJ}}\right) \left(\dot{\phi}^I \dot{\phi}^J
\alpha_1-\dot \phi^J\dot Q^I\right) \bigg]\,.
\end{eqnarray}

The second order action is calculated as
\begin{eqnarray}
S_{(2)}=\int
dtd^3x\frac{a^3}{2}&\!\!\!\!\Bigg[&\!\!\!\!X_1^{IJ}X_1^{LM}P_{,X^{IJ}X^{LM}}
+P_{,X^{IJ}}\dot Q^I\dot Q^J - a^{-2}P_{,X^{IJ}}\partial^i
Q^I\partial_iQ^J
\nonumber\\&&\!\!\!\!\!\!\!\!\!\!\!\!-\frac{3}{2}\dot \phi^J\dot
\phi^LP_{,X^{IJ}}P_{,X^{LM}}Q^IQ^M +\frac{\dot
\phi^JP_{,X^{IJ}}Q^I}{H}\left(P_{,X^{LM}}X_1^{LM}
+P_{,K}Q^K\right) \nonumber\\&&\!\!\!\!\!\!\!\!\!\!\!\!
+2P_{,X^{IJ}K}Q^K \left(-\dot{\phi}^I \dot{\phi}^J \alpha_1+\dot
\phi^I\dot Q^J\right)+P_{,KL}Q^KQ^L \nonumber\\
&&\!\!\!\!\!\!\!\!\!\!\!\!+P_{,X^{IJ}}\left(3 \dot{\phi}^I
\dot{\phi}^J\alpha_1^2- 4\alpha_1\dot\phi^I\dot
Q^J\right)\Bigg]\,,
\end{eqnarray}
where
\begin{equation}
X_1^{IJ}\equiv-\alpha_1 \dot{\phi}^I \dot{\phi}^J
+\dot\phi^{(I}\dot Q^{J)}\,.
\end{equation}
After integrating by parts in the action and employing the
background field equations, the second order action can be finally
written in the rather simple form
\begin{eqnarray}
S_{(2)} &=& \frac12 \int dt d^3 x a^3 \bigl[ (P_{,X^{IJ}} +
P_{,X^{IK} X^{JL}} \dot{\phi}^K \dot{\phi}^L ) \dot{Q}^I \dot{Q}^J
\nonumber\\
&& - \frac{1}{a^2} P_{,X^{IJ}}
\partial_i Q^I \partial^i Q^J
-{\cal{M}}_{IJ} Q^I Q^J + {\cal N}_{IJ} Q^J \dot{Q}^I \bigr] \,,
\label{2nd_order_action}
\end{eqnarray}
with the effective squared mass matrix
\begin{eqnarray}
{\cal{M}}_{IJ} &=& -P_{,IJ} + \frac{X^{LM}}{H}\dot{\phi}^K
(P_{,X^{JK}} P_{,I X^{LM}}+P_{,X^{IK}} P_{,J X^{LM}})
\nonumber\\
&&-\frac{1}{H^2} X^{MN} X^{PQ} P_{,X^{MN} X^{PQ}} P_{,X^{IK}}
P_{,X^{JL}} \dot{\phi}^K \dot{\phi}^L
\nonumber\\
&&-\frac{1}{a^3} \frac{d}{dt} \left[\frac{a^3}{H} P_{,X^{IK}}
P_{,X^{JL}} \dot{\phi}^K \dot{\phi}^L\right]
\,,\\
{\cal{N}}_{IJ} &=& 2\left(P_{,J X^{IK}}- \frac{X^{MN}}{H}
P_{,X^{IK} X^{MN}} P_{,X^{JL}} \dot{\phi}^L\right) \dot{\phi}^K\,.
\end{eqnarray}
In the same way, the third order action is given by
\begin{eqnarray}
S_{(3)}&\!\!\!\!=&\!\!\!\!\int
dtd^3xa^3\Bigg[\left[3H^2\alpha_1^2+\frac{2H}{a^2}\alpha_1\partial^2\psi_1+\frac{1}{2a^4}
\left(\partial^2\psi_1\partial^2\psi_1-\partial_i\partial_j\psi_1\partial^i\partial^j\psi_1\right)\right]\alpha_1
\nonumber
\\
&&\!\!\!\!\!\!\!\!+\bigg[-\frac12 \alpha_1^3 \dot{\phi}^I
\dot{\phi}^J +\alpha_1^2\dot \phi^I\dot Q^J +a^{-2}\alpha_1\dot
\phi^I\partial^i\psi_1\partial_i Q^J-\frac{\alpha_1}{2}\dot Q^I
\dot Q^J\nonumber\\&&-a^{-2}\partial_iQ^J\left(\dot
Q^I\partial^i\psi_1+\frac{1}{2}\alpha_1\partial^iQ^I\right)
\bigg]P_{,X^{IJ}}+\bigg[\alpha_1^2 \dot{\phi}^I \dot{\phi}^J-
\frac{3}{2}\alpha_1\dot \phi^I\dot
Q^J\nonumber\\&&+\frac{1}{2}\dot Q^I \dot Q^J
-a^{-2}\partial_iQ^J\left(\dot
\phi^I\partial^i\psi_1+\frac{1}{2}\partial^iQ^I\right)\bigg]X_1^{LM}P_{,X^{IJ}X^{LM}}
\nonumber
\\
&&\!\!\!\!\!\!\!\!+\bigg[\frac12 \alpha_1^2 \dot{\phi}^I
\dot{\phi}^J -\alpha_1\dot \phi^I\dot Q^J+\frac{1}{2}\dot Q^I \dot
Q^J -a^{-2}\partial_iQ^J\left(\dot
\phi^I\partial^i\psi_1+\frac{1}{2}\partial^iQ^I\right)\bigg]P_{,X^{IJ}K}Q^K\nonumber
\\
&&\!\!\!\!\!\!\!\!+\frac{\alpha_1}{2}P_{,IJ}Q^IQ^J+\frac{1}{6}X_1^{IJ}X_1^{LM}X_1^{QR}P_{,X^{IJ}X^{LM}X^{QR}}+\frac{1}{2}X_1^{IJ}X_1^{LM}P_{,X^{IJ}X^{LM}K}Q^K
\nonumber
\\
&&\!\!\!\!\!\!\!\!+\frac{1}{2}X_1^{IJ}P_{,X^{IJ}LM}Q^LQ^M+\frac{1}{6}P_{,IJK}Q^IQ^JQ^K\Bigg]\,.
\end{eqnarray}

\subsection{Decomposition into adiabatic and entropy perturbations}
We can decompose the perturbations into the instantaneous
adiabatic and entropy perturbations, where the adiabatic direction
corresponds to the direction of the background fields' evolution
while the entropy directions are orthogonal to this
\cite{Gordon:2000hv}. For this purpose, following
\cite{Langlois:2008mn}, we introduce an orthogonal basis
${e_{n}^I} (n=1,2,...,N)$ in the field space. The orthonormal
condition is defined as
\begin{equation}
P_{,X^{IJ}} e_n^I e_m^J = \delta_{nm}\,, \label{orthonormal}
\end{equation}
so that the gradient term $P_{,X^{IJ}} \partial_i Q^I \partial^i
Q^J$ is diagonalized\footnote{It is worth noting that the
normalization given in
\cite{Langlois:2008wt,Langlois:2008qf,Langlois:2008mn} is $G_{IJ}
e^I_n e^J_m = \delta_{nm}$, unless $P_{,X^{IJ}} = G_{IJ}$, that
is, the case with the canonical kinetic terms, our results written
in terms of $Q_n$ are different from the results of
\cite{Langlois:2008wt,Langlois:2008qf,Langlois:2008mn}. But it can
be shown that when written in terms of $v_n$ which are the
canonical variables for the quantum theory, our results are
consistent with the results of
\cite{Langlois:2008wt,Langlois:2008qf,Langlois:2008mn}.}. Here we
assumed that $P_{,X^{IJ}}$ is invertible and it can be used as a
metric in field space. The adiabatic vector is
\begin{equation}
e_1^I = \frac{\dot{\phi}^I}{\sqrt{P_{,X^{JK}} \dot{\phi}^J
\dot{\phi}^K}}\,,
\end{equation}
which satisfies the normalization given by
Eq.~(\ref{orthonormal}). The field perturbations are decomposed on
this basis as
\begin{equation}
Q^I =Q_n e_{n}^I\,.
\end{equation}
We defined the matrix $Z_{mn}$ which describes the time variation
of the basis as
\begin{equation}
\dot{e}^{I}_n = e^I_m Z_{mn}\,,
\end{equation}
which satisfies $Z_{mn} = - Z_{nm}- \dot{P}_{,X^{IJ}}  e^I_m
e^J_n$ as a consequence of $(P_{,X^{IJ}} e_n^I e_m^J)^{\dot{}}
=0$.

In terms of the decomposed fields, the second order action
(\ref{2nd_order_action}) can be rewritten as
\begin{eqnarray}
S_{(2)} &=& \frac12 \int dt d^3 x a^3 \bigl[ {\cal K}_{mn} (D_t
Q_m)(D_t Q_n) - \frac{1}{a^2} \delta_{mn}
\partial_i Q_m \partial^i Q_n
\nonumber\\
&& - {\cal{M}}_{mn} Q_m Q_n + {\cal{N}}_{mn} Q_n (D_t Q_m)
 \bigr]
\,, \label{2nd_order_action_dec}
\end{eqnarray}
where
\begin{eqnarray}
D_t Q_m &\equiv& \dot{Q}_m + Z_{mn} Q_n\,,\\
{\cal K}_{mn} &\equiv& \delta_{mn} + (P_{,X^{MN}} \dot{\phi}^M
\dot{\phi}^N)
P_{,X^{IK} X^{JL}} e_1^{I} e_n^{K} e_1^{J} e_m^{L}\,,\\
{\cal{M}}_{mn} &\equiv& {\cal{M}}_{IJ} e^I_m e^J_n\,,\\
{\cal{N}}_{mn} &\equiv& {\cal{N}}_{IJ} e^I_m e^J_n.
\end{eqnarray}
From the constructions, ${\cal{K}}_{mn}$, ${\cal{M}}_{mn}$ and
${\cal{N}}_{mn}$ are symmetric with respect to $m$ and $n$. The
explicit form of the effective squared mass matrix in this
representation is
\begin{eqnarray}
{\cal{M}}_{mn} &=& -P_{,mn} + \frac1H\ (P_{,X^{KL}} \dot{\phi}^K
\dot{\phi}^L )^{3/2}
(P_{,m X^{MN}} e^M_1 e^N_1)\delta_{n1}\nonumber\\
&&-\frac{1}{4 H^2} (P_{,X^{KL}} \dot{\phi}^K \dot{\phi}^L )^3
(P_{,X^{MN} X^{PQ}} e^M_1 e^N_1 e^P_1 e^Q_1)
\delta_{m1} \delta_{n1}\nonumber\\
&& -\frac{1}{a^3} \frac{d}{dt} \left[ \frac{a^3}{H} (P_{,X^{MN}}
\dot{\phi}^M \dot{\phi}^N) P_{,X^{IK}} P_{,X^{JL}} e^K_1 e^L_1
\right]
e^I_m e^J_n\,, \\
{\cal{N}}_{mn} &\equiv& -\frac1H (P_{,X^{PQ}} \dot{\phi}^P
\dot{\phi}^Q)^2 (P_{,X^{KL}X^{MN}} e^K_m e^L_1 e^M_1 e^N_1)
\delta_{n1}\nonumber\\
&&+2\sqrt{P_{,X^{LM}} \dot{\phi}^L \dot{\phi}^M} (P_{,n X^{IK}}
e^I_m e^K_1)\,,
\end{eqnarray}
where $P_{,mn} \equiv P_{,IJ} e^I_m e^J_n$ and $P_{,n X^{IK}}
\equiv P_{,J X^{IK}} e^J_n$.

The equation of motion is obtained as
\begin{eqnarray}
&&\frac{1}{a^3} \frac{d}{dt} \left[a^3 (2 {\cal{K}}_{mr} D_t Q_m +
{\cal{N}}_{mr} Q_m) \right] -(2 {\cal{K}}_{mn} Z_{nr} +
{\cal{N}}_{mr} ) D_t Q_m
\nonumber\\
&&-\left(2 {\cal{M}}_{mr} + {\cal{N}}_{mn} Z_{mr} \right)
Q_m+\frac{2}{a^2} \partial^2 Q_r=0\,.
\end{eqnarray}

\section{\label{sec:Lperturbations}Linear perturbations}
In this section, we study the linear order perturbations using the
second order action derived in the previous section.

\subsection{K-inflation}
Let us consider K-inflation models where $P(X^{IJ}, \phi^I)$ is a
function of only the trace $X = X^{IJ} G_{IJ}(\phi^K)$ of the
kinetic terms where $G_{IJ}(\phi^K)$ is a metric in the field
space:
\begin{equation}
P(X^{IJ}, \phi^I) = \tilde{P}(X, \phi^I).
\end{equation}
The derivatives of P can be evaluated as
\begin{eqnarray}
P_{,X^{IJ}} &=&  G_{IJ} \tilde{P}_{,X}\,,\\
P_{,I} &=& \frac12 G_{JK,I} \dot{\phi}^J \dot{\phi}^K
\tilde{P}_{,X}
+ \tilde{P}_{,I}\,,\\
P_{,X^{IJ} X^{KL}} &=& G_{IJ} G_{KL} \tilde{P}_{,XX}\,,\\
P_{,X^{IJ} K} &=&  \frac12 G_{LM,K} \dot{\phi}^L \dot{\phi}^M
G_{IJ} \tilde{P}_{,XX} + G_{IJ,K} \tilde{P}_{,X}
+ G_{IJ} \tilde{P}_{,XK}\,,\\
P_{,IJ} &=&  \frac14 G_{KL,I} G_{MN,J} \dot{\phi}^K \dot{\phi}^L
\dot{\phi}^M \dot{\phi}^N \tilde{P}_{,XX} +\frac12 G_{KL,IJ}
\dot{\phi}^K \dot{\phi}^L \tilde{P}_{,X}
\nonumber\\
&&+\frac12 \dot{\phi}^M \dot{\phi}^N(G_{MN,J} \tilde{P}_{,XI} +
G_{MN,I} \tilde{P}_{,XJ} ) + \tilde{P}_{,IJ}\,,
\end{eqnarray}
and the sound speed is defined as
\begin{eqnarray}
c_s^2 \equiv  \frac{\tilde{P}_{,X}} {\tilde{P}_{,X} + 2X
\tilde{P}_{,XX}}\,.
\end{eqnarray}

In terms of the decomposed field, the second order action can be
written as \cite{Langlois:2008mn}
\begin{eqnarray}
S_{(2)} &=& \frac12 \int dt d^3 x a^3 \biggl[ \left\{ \delta_{mn}
+ \left(\frac{1}{c_s^2} -1\right) \delta_{1m} \delta_{1n} \right\}
(D_t Q_m) (D_t Q_n)
\nonumber\\
&&- \frac{1}{a^2} \delta_{mn}
\partial_i Q_m \partial^i Q_n
-{\cal{M}}_{mn} Q_m Q_n + {\cal{N}}_{mn} Q_n (D_t Q_m) \biggr]\,,
\label{2nd_order_action_kinf_dec}
\end{eqnarray}
where we do not show the explicit forms of ${\cal{M}}_{mn}$ and
${\cal{N}}_{mn}$.

The fact that the sound speeds for the entropy perturbations are
unity has been recognized in Ref.~\cite{Langlois:2008mn}. This is
because the non-trivial second derivative of $P$ only affects the
adiabatic perturbations.

\subsection{DBI-inflation \label{subsec:DBI}}
An interesting class of models is the DBI-inflation which
describes the motion of a D3 brane in a higher dimensional
spacetime. The DBI action is given by \cite{Leigh:1989jq}
\begin{equation}
S=-\int d^4 x \frac{1}{f(\phi^K)}\sqrt{-\mbox{det} [g_{\mu \nu} +
f(\phi^K) G_{IJ}(\phi^K) \partial_{\mu} \phi^I
\partial_{\nu} \phi^J]},
\end{equation}
where det denotes a determinant, $G^{IJ}$ is a metric in field
space and $\phi^I$ corresponds to positions of a brane in higher
dimensional spacetime. Recently it was pointed out by
Ref.~\cite{Langlois:2008wt} that multi-field DBI-inflation is {\it
not} included in multi-field K-inflation discussed in the previous
subsection. Indeed, $P(X^{IJ})$ is not a function of $X$, but it
is given by
\begin{equation}
P(X^{IJ}, \phi^I) = \tilde{P}(\tilde{X}, \phi^I), \quad \tilde{X}
= \frac{(1-{\cal D})}{2f}\,,\label{tildePtildeX}
\end{equation}
where
\begin{eqnarray}
{\cal D} &=& \mbox{det}(\delta^I_J - 2 f X^I_J)\nonumber\\
&=& 1- 2 f G_{IJ} X^{IJ} + 4 f^2 X_I^{[I} X_J^{J]} -8 f^3 X_I^{[I}
X_J^{J} X_K^{K]}+ 16 f^4 X_I^{[I} X_J^J X_K^K
X_L^{L]}.\nonumber\\\label{determinant}
\end{eqnarray}
where $[ \; ]$ denotes the anti-symmetric bracket and we used
$G^{IJ}$ to raise the indices. In the background, $\tilde{X}=X$.
However, this does not mean that the full action is a function of
$X$ only. The DBI action takes a specific form of $\tilde{P}$
\begin{equation}
\tilde{P}(\tilde{X}, \phi^I) = -\frac{1}{f}\left( \sqrt{1-2f
\tilde{X}}-1 \right)-V(\phi^I),
\end{equation}
where we allow for a potential $V(\phi^I)$. The sound speed is
defined as
\begin{equation}
c_s^2 \equiv \frac{\tilde{P}_{,\tilde{X}}}
{\tilde{P}_{,\tilde{X}}+ 2X \tilde{P}_{,\tilde{X}
\tilde{X}}}\,.\label{DBIcs}
\end{equation}

The derivatives of $P$ evaluated on the background can be
calculated as
\begin{eqnarray}
P_{,X^{IJ}} &=& \tilde{P}_{,\tilde{X}} \left(
\frac{d \tilde{X}}{ d X^{IJ}} \right)\,, \\
P_{,X^{IJ}X^{KL}} &=& \tilde{P}_{,\tilde{X} \tilde{X}} \left(
\frac{d \tilde{X}}{ d X^{IJ}} \right) \left( \frac{d \tilde{X}}{ d
X^{KL}} \right) + \tilde{P}_{,\tilde{X}} \left(\frac{d^2
\tilde{X}}{d X^{IJ} dX^{KL}} \right)\,,
\end{eqnarray}
where
\begin{eqnarray}
\frac{d\tilde{X}}{ d X^{IJ}} &\!\!\!\!=&\!\!\!\! c_s^2 G_{IJ} +2 f X_{IJ}\,,\\
\frac{d^2 \tilde{X}}{d X^{IJ} dX^{KL}} &\!\!\!\!=&\!\!\!\! -2f
\left( G_{IJ} G_{KL} - \frac{1}{2}  G_{IK}G_{JL}
-\frac{1}{2}G_{IL}G_{JK} \right) + O(X^{IJ})\,.
\end{eqnarray}
Here we do not explicitly write down the higher order terms in
$X^{IJ}$ in the second derivative as they will not contribute to
the final result. In the following, we will omit these terms. We
can also show that
\begin{eqnarray}
P_{,I} &=& \frac12 G_{JK,I} \dot{\phi}^J \dot{\phi}^K
\tilde{P}_{,\tilde{X}}
+ \tilde{P}_{,I}\,,\\
P_{,IJ} &=&  \frac14 G_{KL,I} G_{MN,J} \dot{\phi}^K \dot{\phi}^L
\dot{\phi}^M \dot{\phi}^N \tilde{P}_{,\tilde{X}\tilde{X}} +
\frac12 G_{KL,IJ} \dot{\phi}^K \dot{\phi}^L \tilde{P}_{,\tilde{X}}
\nonumber\\
&&+\frac12 \dot{\phi}^M \dot{\phi}^N (G_{MN,J}
\tilde{P}_{,\tilde{X} I} +  G_{MN,I} \tilde{P}_{,\tilde{X} J} )
+ \tilde{P}_{,IJ}\,,\\
P_{,X^{IJ} K} &=& \bigl((1-2fX)G_{IJ,K} - 2f_{,K} X G_{IJ}
-2f G_{LM,K} X^{LM} G_{IJ}\nonumber\\
&& 2 f_{,K} X_{IJ} + 2f G_{IL,K} X^L _{\;\;J} + 2f G_{JM,K} X_{I}
^{\;\;M}\bigr)\tilde{P}_{,\tilde{X}}
\nonumber\\
&&+(c_s^2 G_{IJ} +2fX_{IJ}) \left( \frac12 G_{LM,K} \dot{\phi}^L
\dot{\phi}^M \tilde{P}_{,\tilde{X}\tilde{X}} +
\tilde{P}_{,\tilde{X} K} \right)\,.
\end{eqnarray}
It is worth noting that even though $P_{,X^{IJ} K}$ seems to be a
bit complicated, we can show that
\begin{eqnarray}
P_{,X^{IJ} K} \dot{\phi}^J = \frac12 G_{LM,K} \dot{\phi}^L
\dot{\phi}^M \dot{\phi}_I \tilde{P}_{,\tilde{X}\tilde{X}} +
\dot{\phi}_I \tilde{P}_{,\tilde{X} K} +G_{IJ,K}\dot{\phi}^J
\tilde{P}_{,\tilde{X}}\,,
\end{eqnarray}
which is just the same form as K-inflation case. We can also show
that
\begin{equation}
P_{,X^{IJ}} \dot{\phi}^I \dot{\phi}^J =2 X
\tilde{P}_{,\tilde{X}}\,.
\end{equation}

The orthonormality conditions for the basis give
\begin{eqnarray}
e_{n}^I e_{m I} &=& \frac{1}{\tilde{P}_{,\tilde{X}} c_s^2}
\delta_{mn}-\frac{1}{\tilde{P}_{,\tilde{X}}} \frac{1-c_s^2}{c_s^2}
\delta_{m1} \delta_{n1}\,.
\end{eqnarray}
Using these results, the second order action can be written in
terms of the decomposed perturbations as
\begin{eqnarray}
S_{(2)} &=& \frac12 \int dt d^3 x a^3 \biggl[ \frac{1}{c_s^2}
\delta_{mn}(D_t Q_m) (D_t Q_n) - \frac{1}{a^2} \delta_{mn}
\partial_i Q_m \partial^i Q_n
\nonumber\\
&&-{\cal{M}}_{mn} Q_m Q_n + {\cal{N}}_{mn} Q_n (D_t Q_m)
\biggr]\,. \label{2nd_order_action_DBIinf_dec}
\end{eqnarray}

Unlike K-inflation models, all field perturbations have the same
sound speeds as was pointed out by Ref.~\cite{Langlois:2008wt}. In
order to understand the difference between K-inflation and
DBI-inflation, we will consider a generalized model where both
cases are included.

\subsection{Generalized case \label{subsec:generalized_model}}
Let us consider models described by
\begin{equation}
P(X^{IJ}, \phi^I) = \tilde{P}(Y, \phi^I)\,,
\end{equation}
where
\begin{equation}
Y=G_{IJ}(\phi^K) X^{IJ} + \frac{b(\phi^K)}{2} (X^2-X^{J}_I
X^I_J)\,.
\end{equation}
The functional form of $Y$ is chosen so that $Y = X \equiv G_{IJ}
X^{IJ}$ in the background as in the DBI-inflation model. This
model includes as particular cases K-inflation model for $b=0$ and
the DBI-inflation for $b=-2f$ and if $\tilde P$ has the DBI form.
This might be surprising as the DBI action contains additional
terms of order $f^2$ and $f^3$ in $\tilde X$ (see equations
(\ref{tildePtildeX}) and (\ref{determinant})), but it turns out
that these terms do not contribute to the second order action and
the leading order third order action.

Following a similar procedure to the previous subsection, the
second order action can be written in terms of the decomposed
perturbations as
\begin{eqnarray}
S_{(2)} &=& \frac12 \int dt d^3 x a^3 \biggl[ \left\{ \delta_{mn}
+ \frac{2X \tilde{P}_{,YY}}{ \tilde{P}_{,Y}} \delta_{1m}
\delta_{1n} +\frac{bX}{1+bX} (\delta_{n1} \delta_{m1}-\delta_{mn})
\right\}\nonumber\\
&& \times(D_t Q_m) (D_t Q_n) - \frac{1}{a^2} \delta_{mn}
\partial_i Q_m \partial^i Q_n
-{\cal{M}}_{mn} Q_m Q_n + {\cal{N}}_{mn} Q_n (D_t Q_m)
\biggr]\,,\nonumber\\ \label{2nd_order_action_general_dec}
\end{eqnarray}

Now we are in a position to explain the difference between
K-inflation and DBI-inflation. As in K-inflation case, the
non-trivial second derivative of $P$ affects only the adiabatic
perturbations. On the other hand, the non-linear terms of $X^{IJ}$
in $Y$ only affects the entropy perturbations. This can be seen
from the fact that the sound speed for adiabatic perturbations
$c_{ad}^{2}$ and for entropy perturbations $c_{en}^{2}$ are given
by
\begin{equation}
c_{ad}^{2} \equiv \frac{\tilde{P}_{,Y}}{\tilde{P}_{,Y} + 2X
\tilde{P}_{,YY}}\,, \quad c_{en}^{2} \equiv 1 + bX\,,
\end{equation}
and they are independently determined by $\tilde{P}_{,YY}$ and
$d^2 Y/ (dX^{IJ} dX^{KL})$ respectively. Thus in general they are
different. Let us derive the condition under which the two sound
speeds are the same, i.e., $c_{ad}^{2}=c_{en}^{2}$. This condition
is given by
\begin{equation}
2X \frac{\tilde{P}_{,YY}}{\tilde{P}_{,Y}}
 = -\frac{bX}{1+bX}.
\end{equation}
Then we find that the DBI action is a solution of this equation
where $b=-2f$ \footnote{We thank Gianmassimo Tasinato for a
discussion on this point. While we were writing up this work,
there appeared a paper in the arXiv \cite{Langlois:2008qf}. It is
argued that the speciality of the DBI action comes from the fact
that it describes the fluctuations of the positions of the brane
in the higher-dimensional spacetime and they propagate at the
speed of light. From the point of view of observers on the brane,
the sound speeds are smaller than $1$ due to the Lorentz factor
$1/c_s^2$.}, although we should note that it has not been proved
that multi-field DBI inflation is the only case where the sound
speeds are the same.

\section{\label{sec:leadingorderTPFmulti}The leading order in slow-roll three point function}
In this section, we will calculate the leading order in slow-roll
third order action for the generalized model of the previous
subsection and then we shall calculate the leading order three
point function for both adiabatic and entropy directions. Finally
we will obtain the three point function of the comoving curvature
perturbation.

\subsection{\label{subsec:approximations}Approximations: slow-roll}
In order to control the calculations and to obtain analytical
results we need to make use of some approximations. We will use
the slow-roll approximation, where we define a set of parameters
and assume that these parameters are always small until the end of
inflation. We define the slow-roll parameters as

\begin{equation}
\epsilon \equiv -\frac{\dot H}{H^2}=\frac{X\tilde
P_{,Y}}{H^2},\quad \eta \equiv \frac{\dot \epsilon}{\epsilon H},
\end{equation}
\begin{equation}
\chi_{ad} \equiv \frac{\dot c_{ad}}{c_{ad}H},\quad \chi_{en}
\equiv \frac{\dot c_{en}}{c_{en}H}.
\end{equation}
It is important to note that these slow-roll parameters are more
general than the usual slow-roll parameters and that their
smallness does not necessarily imply that the fields are rolling
slowly. Assuming that the parameters $\chi_{ad}$ and $\chi_{en}$
are small implies that the rates of change of the adiabatic and
entropy sound speeds are small, but the sound speeds themselves
can have any value and we assume that they are between zero and
one.

It is convenient to define a parameter that describes the
non-linear dependence of the Lagrangian on the kinetic term as
\begin{equation}
\lambda \equiv \frac{2}{3}X^3\tilde P_{,YYY}+X^2\tilde P_{,YY}.
\end{equation}
We will also assume that the rate of change of this new parameter
is small, as given by
\begin{equation}
l \equiv \frac{\dot \lambda}{\lambda H}.
\end{equation}
At the end of this section, we will show that the size of the
leading order three point function of the fields is fully
determined by five parameters evaluated at horizon crossing:
$\epsilon$, $\lambda$, $H$ and both sound speeds.

It turns out that the equations of motion for both adiabatic and
entropy perturbations at first order form a coupled system of
second order linear differential equations, see appendix
\ref{app:eomfluctuations} for details. In general, the coupling
(denoted by $\xi$ in equation (\ref{coupling})) between adiabatic
and entropy modes cannot be neglected but in this work we will
study the simpler decoupled case, where we assume that $\xi$ is
small when the scales of interest cross outside the sound
horizons, i.e., we will assume that
$\xi\sim\mathcal{O}(\epsilon)$. With these approximations the
adiabatic and entropy modes are decoupled and the system of
equations of motion can be solved analytically. For simplicity, we
will also assume that the mass term present in the entropy
equation of motion is small, i.e., $\mu_s^2/H^2\ll 1$ (refer to
appendix \ref{app:eomfluctuations} for more details). When
calculating the leading order three point functions, we assume
that the quantities related to the time derivatives of the basis
vectors given by $Z_{mn}$ are also slow-roll suppressed. Finally,
the calculation of the three point functions in the next
subsections is valid in the limit of small sound speeds. Our
results will also include sub-leading terms of $\mathcal{O}(1)$
but these terms will in general (for small sound speeds) receive
corrections coming from terms of the order of $\epsilon/c_s^2$,
that we have neglected.

\subsection{Third order action at leading order}
At leading order in the previous approximations, we can neglect
terms containing $\alpha_1$, $\psi_1$ and derivatives of $P$ with
respect to the fields. Then the third order action for the general
model (\ref{action5}) is calculated as
\begin{eqnarray}
S_{(3)} &=& \frac{1}{2} \int dx^3 dt a^3 \left[ P_{,X^{IK}X^{JL}}
\dot{\phi}^{(I} \dot{Q}^{K)} \dot{Q}^{J} \dot{Q}^L -\frac{1}{a^2}
P_{,X^{IK}X^{JL}} \dot{\phi}^{(I} \dot{Q}^{K)}
\partial_i Q^J \partial^i Q^L
\right. \nonumber\\
&& \left. +\frac{1}{3} P_{,X^{IK}X^{JL}X^{MN}} \dot{\phi}^{(I}
\dot{Q}^{K)} \dot{\phi}^{(J} \dot{Q}^{L)} \dot{\phi}^{(M}
\dot{Q}^{N)} \right]\,.
\end{eqnarray}
After decomposition into the new adiabatic/entropy basis the third
order action can be written as
\begin{equation}
S_{(3)} =\int dx^3 dt a^3 \left[\frac{1}{2} \Xi_{nml} \dot Q_n
\dot Q_m \dot Q_l -\frac{1}{2a^2} \Upsilon_{nml} \dot
Q_n(\partial_i Q_m)(\partial^i Q_l)
 \right]\,,\label{leadingorderaction}
\end{equation}
where we define the coefficients $\Xi_{nml}$ and $\Upsilon_{nml}$
as
\begin{eqnarray}
\Xi_{nml} &=& P_{,X^{IK}X^{JL}} \sqrt{P_{,X^{MN}} \dot{\phi}^M
\dot{\phi}^N}
e_1^{(I} e_{(n}^{K)} e_m^J e_{l)}^L \nonumber\\
&& +\frac{1}{3} P_{,X^{IK}X^{JL}X^{MN}} (P_{,X^{PQ}} \dot{\phi}^P
\dot{\phi}^Q)^{3/2}
e_1^{(I}e_n^{K)} e_1^{(J} e_m^{L)} e_1^{(M} e_{l}^{N)}\,,\\
\Upsilon_{nml} &=& P_{,X^{IK}X^{JL}} \sqrt{P_{,X^{MN}}
\dot{\phi}^M \dot{\phi}^N} e_1^{(I} e_n^{K)} e_m^J e_l^L\,.
\end{eqnarray}
We shall now give some useful formulae of the previous quantities
for the different inflationary models considered in this work.

\subsubsection{K-inflation}
For K-inflation model we have
\begin{eqnarray}
\Xi_{nml} &=& (2X \tilde{P}_{,X})^{-\frac{1}{2}} \left( \frac{2X
\tilde{P}_{,XX}}{\tilde{P}_{,X}} \delta_{1 (n} \delta_{ml)} +
\frac{4}{3}\frac{X^2 \tilde{P}_{,XXX}}{\tilde{P}_{,X}}
\delta_{n1}\delta_{m1}\delta_{l1} \right)
, \\
\Upsilon_{nml} &=&(2X \tilde{P}_{,X})^{-\frac{1}{2}} \frac{2X
\tilde{P}_{,XX}}{\tilde{P}_{,X}} \delta_{n1} \delta_{m l}\,.
\end{eqnarray}

\subsubsection{DBI-inflation}
For the DBI-inflation scenario they are given by
\begin{eqnarray}
\Xi_{nml} &=& (2X \tilde{P}_{,\tilde{X}})^{-\frac{1}{2}}
\frac{1-c_s^2}{c_s^4}
\delta_{1 (n} \delta_{ml)}\,, \\
\Upsilon_{nml} &=& (2X \tilde{P}_{,\tilde{X}})^{-\frac{1}{2}}
\left( \frac{1-c_s^2}{c_s^2} \delta_{n1} \delta_{m l} -2
\frac{1-c_s^2}{c_s^2} \left( \delta_{n1} \delta_{ml} - \delta_{n
(m} \delta_{l)1} \right) \right)\,,
\end{eqnarray}
where $c_s^2$ should be understood as the sound speed defined in
Eq.~(\ref{DBIcs}).

\subsubsection{Generalized case}
For the generalized case of subsection
\ref{subsec:generalized_model},  $\Xi_{nml}$ and $\Upsilon_{nml}$
can be written as
\begin{eqnarray}
\Xi_{nml} &\!\!\!=&\!\!\! (2X \tilde{P}_{,Y})^{-\frac{1}{2}}
\Bigg[\left( \frac{4}{3}\frac{X^2
\tilde{P}_{,YYY}}{\tilde{P}_{,Y}} - \frac{(1-c_{ad}^2)(1-
c_{en}^2)}{c_{ad}^2 c_{en}^2} \right)
\delta_{n1}\delta_{m1}\delta_{l1}\nonumber\\&&\quad\quad\quad\quad\quad+\frac{(1-c_{ad}^2)}{c_{ad}^2
c_{en}^2} \delta_{1 (n} \delta_{ml)} \Bigg]
, \\
\Upsilon_{nml} &\!\!\!=&\!\!\! (2X \tilde{P}_{,Y})^{-\frac{1}{2}}
\left( \frac{1-c_{ad}^2}{c_{ad}^2} \delta_{n1} \delta_{m l} -
\frac{2 ( 1- c_{en}^2) }{c_{en}^2} \left( \delta_{n1} \delta_{ml}
- \delta_{n (m} \delta_{l)1} \right) \right),
\end{eqnarray}
and it is obvious that the DBI-inflation is a specific case of the
general model with $c_{ad}^2 = c_{en}^2 = c_s^2$.

\subsection{The three point functions of the fields}
In this subsection, we derive the three point functions of the
adiabatic and entropy fields in the generalized case and at
leading order in slow-roll and in the small sound speeds limit. We
consider the two-field case with the adiabatic field $\sigma$ and
the entropy field $s$.

The perturbations are promoted to quantum operators as
\begin{equation}
Q_n(\tau,\mathbf{x})=\frac{1}{(2\pi)^3}\int
d^3\mathbf{k}Q_n(\tau,\mathbf{k})e^{i\mathbf{k}\cdot\mathbf{x}},
\end{equation}
where
\begin{equation}
Q_n(\tau,\mathbf{k})=u_n(\tau,\mathbf{k})a_n(\mathbf{k})+u^*_n(\tau,-\mathbf{k})a^\dag_n(-\mathbf{k}).
\end{equation}
$a_n(\mathbf{k})$ and $a^\dag_n(-\mathbf{k})$ are the annihilation
and creation operator respectively, that satisfy the usual
commutation relations
\begin{eqnarray}
\left[a_n(\mathbf{k_1}),a^\dag_m(\mathbf{k_2})\right]&=&(2\pi)^3\delta^{(3)}(\mathbf{k_1}-\mathbf{k_2})\delta_{nm},
\nonumber\\
\left[a_n(\mathbf{k_1}),a_m(\mathbf{k_2})\right]&=&\left[a_n^\dag(\mathbf{k_1}),a_m^\dag(\mathbf{k_2})\right]=0.
\end{eqnarray}
At leading order the solution for the mode functions is (see
appendix \ref{app:eomfluctuations} for details)
\begin{equation}
u_n(\tau,\mathbf{k})=A_n\frac{1}{k^{3/2}}\left(1+ikc_n\tau\right)e^{-ikc_n\tau},
\end{equation}
where $c_n$ stands for either the adiabatic or the entropy sound
speeds.

The two point correlation function is
\begin{equation}
\langle
0|Q_n(\tau=0,\mathbf{k_1})Q_m(\tau=0,\mathbf{k_2})|0\rangle=(2\pi)^3\delta^{(3)}(\mathbf{k_1}+\mathbf{k_2})\mathcal{P}_{Q_n}\frac{2\pi^2}{k_1^3}\delta_{nm},
\end{equation}
where the power spectrum is defined as
\begin{equation}
\mathcal{P}_{Q_n}=\frac{|A_n|^2}{2\pi^2}, \quad
|A_\sigma|^2=\frac{H^2}{2c_{ad}}, \quad
|A_s|^2=\frac{H^2}{2c_{en}},
\end{equation}
and it should be evaluated at the time of horizon crossing
${c_n}_* k_1=a_*H_*$

The vacuum expectation value of the three point operator in the
interaction picture (at first order) is
\cite{Maldacena:2002vr,Weinberg:2005vy}
\begin{eqnarray}
\langle&&\!\!\!\!\!\!\!\!\!\!\Omega|Q_l(t,\mathbf{k_1})Q_m(t,\mathbf{k_2})Q_n(t,\mathbf{k_3})|\Omega\rangle\nonumber\\
&&=-i\int^t_{t_0}d\tilde t \langle 0
|\left[Q_l(t,\mathbf{k_1})Q_m(t,\mathbf{k_2})Q_n(t,\mathbf{k_3}),H_I(\tilde
t)\right]|0\rangle,
\end{eqnarray}
where $t_0$ is some early time during inflation when the field's
vacuum fluctuation are deep inside the horizons, $t$ is some time
after horizon exit. $|\Omega\rangle$ is the interacting vacuum
which is different from the free theory vacuum $|0\rangle$. If one
uses conformal time, it is a good approximation to perform the
integration from $-\infty$ to $0$ because $\tau\approx-(aH)^{-1}$.
$H_I$ denotes the interaction Hamiltonian and it is given by
$H_I=-L_3$, where $L_3$ is the Lagrangian obtained from the action
(\ref{leadingorderaction}).

At this order, the only non-zero three point functions are
\begin{eqnarray}
\langle&&\!\!\!\!\!\!\!\!\!\!\Omega|Q_\sigma(0,\mathbf{k_1})Q_\sigma(0,\mathbf{k_2})Q_\sigma(0,\mathbf{k_3})|\Omega\rangle=
(2\pi)^3\delta^{(3)}(\mathbf{k_1}+\mathbf{k_2}+\mathbf{k_3})\frac{2c_{ad}|A_\sigma|^6}{H}
\frac{1}{\Pi_{i=1}^3k_i^3}\frac{1}{K}
\nonumber\\&&\!\!\!\!\!\!\!\!\!\!\!\!
\times\bigg[6c_{ad}^2(C_3+C_4)\frac{k_1^2k_2^2k_3^2}{K^2}
-C_1k_1^2\mathbf{k_2}\cdot\mathbf{k_3}\left(1+\frac{k_2+k_3}{K}+2\frac{k_2k_3}{K^2}\right)
+2\,\,\mathrm{cyclic\, terms} \bigg],\nonumber\\
\end{eqnarray}
\begin{eqnarray}
\langle&&\!\!\!\!\!\!\!\!\!\!\Omega|Q_\sigma(0,\mathbf{k_1})Q_s(0,\mathbf{k_2})Q_s(0,\mathbf{k_3})|\Omega\rangle=
(2\pi)^3\delta^{(3)}(\mathbf{k_1}+\mathbf{k_2}+\mathbf{k_3})\frac{|A_\sigma|^2|A_s|^4}{H}
\frac{1}{\Pi_{i=1}^3k_i^3}\frac{1}{\tilde K} \nonumber\\&&\!\!
\times\bigg[
C_2c_{en}^2k_3^2\mathbf{k_1}\cdot\mathbf{k_2}\left(1+\frac{c_{ad}k_1+c_{en}k_2}{\tilde
K }+\frac{2c_{ad}c_{en}k_1k_2}{\tilde
K^2}\right)+(k_2\leftrightarrow k_3) \nonumber\\&&\!\!
+4C_3c_{ad}^2c_{en}^4\frac{k_1^2k_2^2k_3^2}{\tilde K^2}
-2(C_1+C_2)c_{ad}^2k_1^2\mathbf{k_2}\cdot\mathbf{k_3}\left(1+c_{en}\frac{k_2+k_3}{\tilde
K}+2c_{en}^2\frac{k_2k_3}{\tilde K^2}\right) \bigg],\nonumber\\
\label{tpfmix}
\end{eqnarray}
where $K=k_1+k_2+k_3$, $\tilde K=c_{ad}k_1+c_{en}(k_2+k_3)$,
$\mathrm{cyclic\, terms}$ means cyclic permutations of the three
wave vectors and $(k_2\leftrightarrow k_3)$ denotes a term like
the preceding one but with $k_2$ and $k_3$ interchanged. The pure
adiabatic three point function is evaluated at the moment $\tau_*$
at which the total wave number $K$ exits the horizon, i.e., when
$K{c_{ad}}_*=a_*H_*$. Because of the different propagation speeds,
the adiabatic and entropy modes become classical at different
times, however at leading order we assume that the background
dependent coefficients of (\ref{tpfmix}) do not vary with time and
so they can also be evaluated at the moment $\tau_*$.

The different constants $C_N$ are given by
\begin{eqnarray}
C_1&=&(2H^2\epsilon)^{-\frac{1}{2}} \frac{1-c_{ad}^2}{c_{ad}^2},
\quad C_2=-2(2H^2\epsilon)^{-\frac{1}{2}}
\frac{1-c_{en}^2}{c_{en}^2},
\nonumber\\
C_3&=&(2H^2\epsilon)^{-\frac{1}{2}}
\frac{1-c_{ad}^2}{c_{ad}^2c_{en}^2}, \quad
C_4=(2H^2\epsilon)^{-\frac{1}{2}}\left(
\frac{2\lambda}{H^2\epsilon} -
\frac{1-c_{ad}^2}{c_{ad}^2c_{en}^2}\right).
\end{eqnarray}

\subsection{The three point function of the comoving curvature perturbation}
In this subsection, we calculate the leading order in slow-roll
three point function of the comoving curvature perturbation in
terms of three point function of the fields obtained in the
previous subsection.

During the inflationary era the comoving curvature perturbation is
defined as
\begin{equation}
\mathcal{R} = - \frac{H}{E_0+P_0} \delta q
\end{equation}
in the uniform curvature gauge where $\delta q_{,i} = \delta
T^{0}_{i}$. Using the constraint equation $\delta q = - 2 H
\alpha_1$ (\ref{constraint}) and Eq.~(\ref{continuity}), we get
\begin{equation}
\mathcal{R}=\frac{H}{\dot\sigma}\frac{Q_\sigma}{\sqrt{\tilde{P}_{,Y}}}.
\end{equation}
It is convenient to define the entropy perturbation $\mathcal{S}$
as
\begin{equation}
\mathcal{S}=\frac{H}{\dot\sigma}\frac{Q_s}{\sqrt{\tilde{P}_{,Y}}}\sqrt{\frac{c_{en}}{c_{ad}}},
\end{equation}
so that
$\mathcal{P}_{\mathcal{S}_*}\simeq\mathcal{P}_{\mathcal{R}_*}$,
where the subscript $*$ means that the quantity should be
evaluated at horizon crossing.

In this work we will ignore the possibility that the entropy
perturbations during inflation can lead to primordial entropy
perturbations that could be observable in the CMB. But we shall
consider the effect of entropy perturbations on the final
curvature perturbation. We will follow the analysis of Wands
\emph{et al.} \cite{Wands:2002bn} (see also
\cite{Gordon:2000hv,Wands:2000dp}), where it has been shown that
even on large scales the curvature perturbation can change in time
because of the presence of entropy perturbations. The way the
entropy perturbations are converted to curvature perturbations is
model dependent but it was shown that this model dependence can be
parameterized by a transfer coefficient $T_{\mathcal{RS}}$
\cite{Wands:2002bn} as
\begin{equation}
\mathcal{R}=\mathcal{R}_*+T_\mathcal{RS}\mathcal{S}_*=\mathcal{A}_\sigma
Q_{\sigma *}+\mathcal{A}_sQ_{s*},
\end{equation}
with
\begin{equation}
\mathcal{A}_\sigma=\left(\frac{H}{\dot\sigma\sqrt{\tilde{P}_{,Y}}}\right)_*,\quad
\mathcal{A}_s=T_\mathcal{RS}\left(\frac{H}{\dot\sigma\sqrt{\tilde{P}_{,Y}}}\sqrt{\frac{c_{en}}{c_{ad}}}\right)_*.
\end{equation}

Using the previous expressions we can now relate the three point
function of the curvature perturbation to the three point
functions of the fields obtained in the previous subsection. The
three point function of the curvature perturbation is given by
\begin{eqnarray}
\langle\mathcal{R}(\mathbf{k_1})\mathcal{R}(\mathbf{k_2})\mathcal{R}(\mathbf{k_3})\rangle
&\!\!=&\!\!\mathcal{A}^3_\sigma\langle
Q_\sigma(\mathbf{k_1})Q_\sigma(\mathbf{k_2})Q_\sigma(\mathbf{k_3})\rangle \nonumber\\
&&\!\!\!\!+\mathcal{A}_\sigma\mathcal{A}_s^2\left(\langle
Q_\sigma(\mathbf{k_1})Q_s(\mathbf{k_2})Q_s(\mathbf{k_3})\rangle+2\,perms\right).
\end{eqnarray}

For the DBI-inflation case the previous equation can be simplified
and the total momentum dependence of the three point function of
the comoving curvature perturbation is the same as in single field
DBI \cite{Langlois:2008wt}. For our general model this is no
longer the case, i.e., the different terms of the previous
equation have different momentum dependence. Once again one can
see that DBI-inflation is a very particular case and more
importantly it provides a distinct signature that enables us to
distinguish it from other more general models.

\section{\label{sec:conclusionbispectrum}Conclusion}
In this chapter, we studied the non-Gaussianity from the
bispectrum in general multi-field inflation models with a generic
kinetic term. Our model is fairly general including K-inflation
and DBI-inflation as special cases. We derived the second and
third order actions for the perturbations including the effect of
gravity. The second order action is written in terms of adiabatic
and entropy perturbations. It was shown that the sound speeds for
these perturbations are in general different. In K-inflation the
entropy perturbations propagate at the speed of light.
DBI-inflation is a special case where the sound speeds for the
entropy and adiabatic perturbations are the same.

Then we derive the three point function in the small sound speed
limit at leading order in slow-roll expansion. In these
approximations there exists a three point function between
adiabatic perturbations $Q_{\sigma}$ and entropy perturbations
$Q_s$, $\langle
Q_{\sigma}(\mathbf{k_1})Q_{s}(\mathbf{k_2})Q_{s}(\mathbf{k_3})
\rangle$, in addition to the pure adiabatic three point function.
This mixed contribution has a different momentum dependence if the
sound speeds for the entropy and adiabatic perturbations are
different. This provides a possibility to distinguish between
multi-field models and single field models. Unfortunately, in
multi-field DBI case, the sound speeds for the entropy and
adiabatic perturbations are the same, and the mixed contribution
only changes the amplitude of the three point function. This could
help to ease the constraints on DBI-inflation as is discussed in
Ref.~\cite{Langlois:2008wt}.

In order to calculate the effect of entropy perturbations on the
curvature perturbation at recombination, we need to specify a
model that describes how the entropy perturbations are converted
to the curvature perturbations. In addition, even during
inflation, if the trajectory in field space changes non-trivially,
the entropy perturbations can be converted to the curvature
perturbation. In this work, we modelled this transition by the
function $T_\mathcal{RS}$. It would be interesting to study this
mixing in specific string theory motivated models.

\chapter{Conclusions and Discussions}

Brane-worlds in higher-dimensional spacetimes have been proposed
to explain long standing and crucial problems in physics, such as
the hierarchy problem or the cosmological constant problem.
Although it was shown that some of these issues were not (yet)
completely solved by the brane-world proposals, these
higher-dimensional models provide a way to tackle these difficult
problems from a new perspective, which might suggest new
solutions.

To tackle these unsolved questions, brane-world models always
introduce new physics in the higher-dimensional theory. These new
ingredients will always in one way or another give new
contributions to the 4D effective theories. In most cases, these
new 4D effective degrees of freedom will have observational
consequences on our universe.

In this thesis, we gave a step towards calculating the 4D
observational predictions implied by several higher-dimensional
theories. These predictions are confronted to observations of our
universe, for example, the cosmic microwave background and table
top experiments on Newton's law at short distances. It is
extremely encouraging that we are getting strong observational
constraints on these higher-dimensional models which aim to
provide self-consistent microphysical descriptions of the early
and late time universe.

The main goal of this thesis was the study of the 4D effective
theories of several higher-dimensional proposals and their
observational consequences. In order to achieve it we used a
powerful technique called the gradient expansion method.

For example, in chapter \ref{chapter:5D} we applied this method to
the 5D dilatonic two brane model that reduces to the Randall and
Sundrum model and to the Ho\v{r}ava-Witten theory for particular
choices of our parameters. We were able to show that exact
solutions of the 5D theory, that describe an instability of the
model, can be reproduced within the 4D effective theory obtained
with the gradient expansion method. We have looked at this
instability from a 4D effective theory perspective and we have
shown that it arises because in the Einstein frame the theory
consists of two massless scalar fields, the dilaton and the radion
(a degree of freedom that describes the distance between the
branes). We have argued against some claims that the 4D effective
theory allows a wider class of solutions than the 5D theory and we
have provided a way to obtain the full 5D solutions from the
solutions of the 4D theory.

In chapter \ref{chapter:6D}, we derived the low energy effective
theory in 6D supergravity with resolved co-dimension one branes.
We showed that the 4D effective theory is a Brans-Dicke theory
with the Brans-Dicke parameter equal to one half. We have
calculated the 4D effective potential for the Brans-Dicke field.
Then it was easy to see that one can tune the potentials on the
branes so that this modulus is massless and the static 4D
spacetime has a vanishing cosmological constant. We have obtained
dynamical solutions due to the evolution of the modulus field and
we showed that these solutions can be identified with exact 6D
solutions. If we do not tune the potentials then, in the Einstein
frame, the Brans-Dicke field has an exponential potential and
static solutions are not allowed. Once more we found that these
solutions can be lifted back to 6D exact solutions, showing that
the gradient expansion method is a powerful method. Because the
Brans-Dicke parameter is one half, this theory is immediately
ruled out by solar system constraints on this parameter. However,
we showed that a 4D potential with a minimum can be engineered and
4D realistic cosmology can be obtained. The drawback is that once
the modulus is stabilized, the brane cosmological constant will
curve the 4D spacetime in the same way as in general relativity
and the self-tuning mechanism will not work. There are still some
hopes that this model will provide a quantum solution for the
cosmological constant problem, see \cite{Burgess:2007ui}, but that
discussion is outside the scope of this work.

The final higher-dimensional theory that we studied in this thesis
was the 10D type IIB supergravity. In chapter \ref{chapter:10D},
we presented a systematic way to derive the 4D effective theory in
a warped compactification including fluxes and branes. We have
obtained the 4D effective potential for the volume modulus but
this potential does not stabilize the volume. If one wants to have
a viable model, the volume modulus has to be stabilized. This is
usually done by introducing non-perturbative effects
\cite{Kachru:2003aw} in the 4D effective theory. It would be
desirable to obtain the 4D effective theory from the 10D theory
including 10D non-perturbative effects. This is an area of intense
research at the moment and it is still an open question. This
problem is related to the question of whether there are explicit
working models of slow-roll warped D-brane inflation or not. As
discussed in the main text, the position of a mobile $D3$ brane in
these compactifications can be identified with the inflaton. At
first sight, the potential generated in the compactification seems
to be flat enough to produce long lasting slow-roll inflation.
However, it was argued in \cite{Kachru:2003sx} that volume
stabilization by non-perturbative effects introduces potentially
fatal corrections to the inflaton's potential that will halt
inflation.

Only recently some of these corrections have been calculated
\cite{Baumann:2006th,Burgess:2006cb,Baumann:2007np,Baumann:2007ah}.
It turns out that in the explicit calculated examples
\cite{Burgess:2006cb,Baumann:2007ah} these corrections do not
cancel and the inflaton's potential is always too steep to sustain
inflation. There is an explicit model \cite{Baumann:2007np} where
slow-roll inflation can occur but the model has to be very fine
tuned. The problem of constructing a fully explicit model of brane
inflation in string theory still remains challenging due to the
lack of understanding of all of the corrections to the 4D
effective potential \cite{Baumann:2007np}. It is our hope that the
gradient expansion method will help because it can be extended by
including moduli stabilization effects as an effective 10D
energy-momentum tensor and it can also allow for a moving
$D3$-brane coupled to gravity. Then one can calculate the
potential for the $D3$-brane in a stabilized compactification
including the backreaction of the brane.

To study observational consequences of brane inflation, in
chapters \ref{chapter:trispectrum} and \ref{chapter:bispectrum},
we took a more phenomenological approach. We started with the 4D
effective description of the movement of a D-brane given by the
DBI action and we computed several observational predictions for
our inflationary theory.

In chapter \ref{chapter:trispectrum}, we computed the leading
order in slow-roll trispectrum for a general single field model of
inflation that contains the DBI-inflation model as a particular
case. To achieve this, we had to develop third order perturbation
theory to obtain the fourth order interaction Hamiltonian
including scalar and second order tensor perturbations. We pointed
out that in general second order tensor perturbations cannot be
ignored in the calculation. This fact will have important
implications for existing results of the trispectrum for standard
kinetic term inflation \cite{Seery:2006vu,Jarnhus:2007ia}. Because
our interaction Hamiltonians are exact in the slow-roll expansion,
one can also in principle study the next order slow-roll
corrections. This is left for future work.

As discussed in chapter \ref{chapter:trispectrum}, for single
field DBI-inflation, the size of the bispectrum ($f_{NL}$) is of
order of $1/c_s^2$ and the size of the trispectrum ($\tau_{NL}$)
is of order of $1/c_s^4$. So if the sound speed is $c_s\sim 0.1$
then $f_{NL}$ of order of one hundred is still allowed by
observations and $\tau_{NL}$ will be in the detection range of the
Planck satellite. With the present observational constraints (and
with the expected improvements in the near future) we are already
significantly constraining models of inflation constructed from
higher-dimensional theories, such as DBI-inflation. Recently,
several works
\cite{Baumann:2006cd,Lidsey:2007gq,Peiris:2007gz,Alabidi:2008ej}
showed that single field ultraviolet DBI-inflation is under severe
pressure from observations. Reference \cite{Lidsey:2007gq} in
particular showed that there exists a microphysical upper bound on
the tensor to scalar ratio ($r\lesssim10^{-7}$) that is
incompatible with a lower bound ($r>0.1(1-n_s)$) that applies to
models that generate a red spectral index (as favored by
observations) and a large detectable non-Gaussianity. The infrared
model of DBI-inflation is not excluded by observations
\cite{Bean:2007eh}.

DBI-inflation is naturally a multiple field inflationary model. So
in chapter \ref{chapter:bispectrum}, we studied the
non-Gaussianity in general multi-field inflation models that have
DBI brane inflation as a particular example. Specifically, we
derived the three point function of the curvature perturbation for
a generalized model in the small sound speed limit at leading
order in slow-roll. At this order, we showed that in addition to
the pure three point function of adiabatic perturbations there
exists a mixed three point function of adiabatic and entropy
perturbations. This mixed contribution has a different momentum
dependence if the sound speeds for the entropy and adiabatic
perturbations are different. In multi-field DBI inflation, because
the sound speeds are equal, this mixed contribution only changes
the amplitude of the final three point function, and the momentum
dependence is the same as in the DBI single field case. It has
been shown \cite{Langlois:2008wt} that this mixed term will in
fact reduce the size of the bispectrum and this could help to ease
the observational constraints on the ultraviolet model of
DBI-inflation. To find out how big this reduction is, one has to
know the details of how the entropy perturbations are converted to
the curvature perturbations. This point was not answered in this
thesis and will be left for future projects. An important outcome
of chapter \ref{chapter:bispectrum} is the fact that we now have a
theoretical prediction for the bispectrum for quite general models
of multi-field inflation, such as multi-field DBI-inflation and
K-inflation.

We shall conclude by saying that the future looks promising
because we have been able to calculate more and more observable
predictions from our fundamental theories and observations are
becoming better and more diversified. From the meeting of these
two we are seeing that some models are being ruled out, some still
prevail. More importantly, we are progressing in our knowledge of
the universe.

\bibliographystyle{JHEP}
    \bibliography{bibliography}

\appendix
\chapter{${\cal O}(\varepsilon^{1/2})$ quantities}\label{app:epsilon12}

In this appendix, we analyze the equations of motion present in
chapter \ref{chapter:6D} that are of order $\varepsilon^{1/2}$ in
the gradient expansion. We will show that all the ${\cal
O}(\varepsilon^{1/2})$ quantities vanish.

The $\mu$ component of the ${\cal O}(\varepsilon^{1/2})$ Maxwell
equations reads
\begin{eqnarray}
\py\left( e^{-\phi^{(0)}}a^4 \stac{(1/2)}{F^{y\mu}} \right)=0,
\end{eqnarray}
and thus we have
\begin{eqnarray}
\stac{(1/2)}{F^{ \mu y}} =M^2\frac{C_1^{\mu}(x)}{y^4}.
\end{eqnarray}
The ${\cal O}(\varepsilon^{1/2})$ evolution equation reduces to
\begin{eqnarray}
\frac{e^{-\Phi/2}}{L_I}\frac{1}{y^2}\py\left(y^2\sqrt{f}\stac{(1/2)}{K_{\theta}^{\;\nu}}\right)
&=&-\frac{1}{M^4}\stac{(0)}{F_{\theta y}}\stac{(1/2)}{F^{\nu y}}
\nonumber\\
&=&\ell q\frac{C_1^{\nu}}{y^6},
\end{eqnarray}
which can be integrated to give
\begin{eqnarray}
\stac{(1/2)}{K_{\theta}^{\;\nu}} =
\frac{e^{\Phi/2}}{y^2\sqrt{f}}\left[ -\frac{L_I\ell
q}{3}\frac{C_1^{\nu}(x)}{y^3}+C_2^{\nu}(x)\right].
\end{eqnarray}

The ${\cal O}(\varepsilon)$ evolution equations contain terms like
$\stac{(1/2)}{F_{\mu y}}\stac{(1/2)}{F^{\nu y}}\propto
h_{\mu\lambda}C_1^{\lambda}C_1^{\nu}/f(y)$. In the cap regions, we
thus require $C_1^{\nu}=0$ to avoid the singular behaviour at the
poles. Further, the regularity of
$\stac{(1/2)}{K_{\theta}^{\;\nu}}$ at the poles imposes
$C_2^{\nu}=0$ in the cap regions. To fix the integration constants
in the central bulk, we use the Maxwell jump conditions and Israel
conditions at each brane:
\begin{eqnarray}
\left[\left[n_y\stac{(1/2)}{F^{y\mu
}}e^{-\phi^{(0)}}\right]\right] &=&-\e U\left(
\partial^{\mu}\Sigma-\e A^{\mu}\right)^{(1/2)},
\\
\left[\left[\stac{(1/2)}{K_{\theta}^{\;\nu}}\right]\right]&=&-\frac{U}{M^4}\left(n-\e
A_{\theta}^{(0)}\right) \left(
\partial^{\mu}\Sigma-\e A^{\mu}\right)^{(1/2)}.
\end{eqnarray}
Combining these two equations and noting that
$\stac{(1/2)}{F^{y\mu }}=0=\stac{(1/2)}{K_{\theta}^{\;\nu}}$ in
each cap, we obtain two linear algebraic equations for the bulk
values of $C_1^{\nu}$ and $C_2^{\nu}$:
\begin{eqnarray}
\left.\left[\stac{(1/2)}{K_{\theta}^{\;\nu}}-\frac{1}{\e
M^4}\left(n-\e A_{\theta}^{(0)}\right) n_y\stac{(1/2)}{F^{y\mu
}}e^{-\phi^{(0)}}\right]\right|_{y_{\pm}\mp\epsilon}=0.
\end{eqnarray}
Therefore, $C_1^{\nu}=C_2^{\nu}=0$ in the bulk. Now we also see
that
\begin{eqnarray}
\left(\partial^{\mu}\Sigma-\e A^{\mu}\right)^{(1/2)}=0 \quad
\text{on the branes},
\end{eqnarray}
and $b_{\mu}={\cal O}(\varepsilon^{3/2})$.

\chapter{Solving the bulk evolution equations}\label{app:solve}

The structure of the bulk evolution equations and boundary
conditions of chapter \ref{chapter:6D} are identical for the three
variables $\mathbb{K}_{\mu}^{\;\nu}$, $\cK$ and $\cJ$ (denoted
here by the generic variable $K$). In this appendix, we summarize
the procedure to solve these equations and to find the integration
constants.

All of the key evolution equations in the main text have the form
of
\begin{eqnarray}
\py\left[y^2\sqrt{f} K(y, x) \right] = e^{\Phi(x)/2}y L_IR(x),
\end{eqnarray}
subject to the boundary conditions
\begin{eqnarray}
K(y_N, x)=K(y_S, x)=0,
\\
\left.[[K]]\right|_{y=y_{\pm}}=T^{\pm}(x).\label{app_junk}
\end{eqnarray}
In the south cap we have the solution
\begin{eqnarray}
K=\frac{y^2-y_S^2}{2y^2\sqrt{f}}e^{\Phi/2}L_-R.
\end{eqnarray}
In the bulk the solution can be written as
\begin{eqnarray}
K=\frac{e^{\Phi/2}\left[y^2L_0R+C(x)\right]}{2y^2\sqrt{f}},\label{Kb}
\end{eqnarray}
where the integration constant $C(x)$ is determined by the
condition~(\ref{app_junk}) as
\begin{eqnarray}
C(x)=
\left[-y_-^2L_0+(y_-^2-y_S^2)L_-\right]R+2y_-^2\sqrt{f_-}e^{-\Phi/2}T^-.\label{app:c=}
\end{eqnarray}
In the north cap we have the solution
\begin{eqnarray}
K=\frac{y^2-y_N^2}{2y^2\sqrt{f}}e^{\Phi/2}L_+R,
\end{eqnarray}
and the boundary condition~(\ref{app_junk}) requires
\begin{eqnarray}
\left[(y_N^2-y_+^2)L_++y_+^2L_0\right]R+C =
-2y_+^2\sqrt{f_+}e^{-\Phi/2}T^+.\label{app:eq}
\end{eqnarray}
In the above we defined $f_{\pm}:=f(y_{\pm})$. Substituting
Eq.~(\ref{app:c=}) into Eq.~(\ref{app:eq}) we obtain
\begin{eqnarray}
\left(\int_{y_S}^{y_N}L_Iydy\right)\cdot R =
-\sum_{i=\pm}y_i^2\sqrt{f_i}e^{-\Phi/2}T^i.
\end{eqnarray}

\chapter{\label{GT}Gauge transformations up to second order}

In this appendix, we will find the change of variables that one
needs to perform to go from the uniform curvature gauge
(\ref{deltaphigauge}) to the comoving gauge (\ref{zetagauge}). A
similar result can be found in \cite{Maldacena:2002vr}.
Alternative but equivalent methods to find the gauge
transformations up to second order and beyond can be found in
\cite{Bardeen:1980kt,Kodama:1985bj,Bruni:1996im,Malik:2008yp}.

In order to go from the gauge (\ref{deltaphigauge}) where the
field fluctuation is not zero to the gauge (\ref{zetagauge}) where
$\delta\phi=0$ we need a change of variables that satisfy
$\phi(t+T(t))+\delta\phi(t+T(t))=\phi(t)$.

At first order in perturbation theory we only need to do a time
reparametrization. Let $t$ and $\tilde{t}$ be the time coordinates
in the gauges (\ref{zetagauge}) and (\ref{deltaphigauge})
respectively. The time reparametrization is $\tilde{t}=t+T$. At
first order
\begin{eqnarray}
T=-\frac{\delta\phi}{\dot{\phi_0}}=\frac{\zeta}{H}, \quad
\zeta=-\frac{H}{\dot{\phi_0}}\delta\phi.
\end{eqnarray}
At second order the time reparametrization is
\begin{equation}
T=-\frac{\delta\phi}{\dot{\phi_0}}-\frac{\ddot{\phi_0}\delta\phi^2}{2\dot{\phi_0}^3}+\frac{\dot{\delta\phi}\delta\phi}{\dot{\phi}^2}.
\end{equation}
At this order we also need to perform a spatial reparametrization
given by $\tilde{x}^i=x^i+\epsilon^i(x,t)$, where $\epsilon^i$ is
of second order in the perturbations. The metric in the gauge
(\ref{zetagauge}) becomes
\begin{eqnarray}
h_{ij}&=&-\frac{\partial T}{\partial x^i}\frac{\partial
T}{\partial x^j}+N_j^{(1)}\frac{\partial T}{\partial
x^i}+N_i^{(1)}\frac{\partial T}{\partial
x^j}+a^2T\dot{\tilde{\gamma}}_{ij}+a^2\left(\frac{\partial
\epsilon_j}{\partial x^i}+\frac{\partial \epsilon_i}{\partial
x^j}\right)
\nonumber\\&&+a^2e^{2HT+\dot{H}T^2}\left(\delta_{ij}+\tilde{\gamma}_{ij}(t)+\frac{1}{2}\tilde{\gamma}_{ik}\tilde{\gamma}^k_j\right),
\label{CG}
\end{eqnarray}
where $N_i^{(1)}$ is the first order shift vector in the gauge
(\ref{deltaphigauge}). If the vector $\epsilon^i$ obeys the
equation
\begin{equation}
a^{-2}\delta h_{ij}+\frac{\partial \epsilon_j}{\partial
x^i}+\frac{\partial \epsilon_i}{\partial
x^j}=2\beta\delta_{ij}+\mu_{ij},\label{CG2}
\end{equation}
with $\mu_{ij}$ being a transverse and traceless tensor and
$\delta h_{ij}$ being defined as the first four terms of Eq.
(\ref{CG}), then the gauge transformation equations are given by
\begin{eqnarray}
\zeta&=&HT+\frac{\dot{H}T^2}{2}+\beta,\nonumber\\
\gamma_{ij}&=&\tilde{\gamma}_{ij}(t)+\mu_{ij}.
\end{eqnarray}
To obtain the quantities $\beta$ and $\mu_{ij}$ it proves to be
useful to decompose $\epsilon^i$ in
$\epsilon^i=\partial^i\tilde{\epsilon}+\epsilon_t^i$ with
$\partial_i\epsilon_t^i=0$. After a few mathematical manipulations
of equation (\ref{CG2}) one can obtain
\begin{equation}
\beta=\frac{a^{-2}}{4}\left(\delta
h_i^i-\partial^{-2}\partial^i\partial^j\delta
h_{ij}\right),\label{beta}
\end{equation}
\begin{eqnarray}
\mu_{ij}&=&a^{-2}\bigg(\delta h_{ij}-\frac{1}{2}\delta_{ij}\delta
h_k^k-\partial^{-2}\partial_i\partial^k\delta
h_{kj}-\partial^{-2}\partial_j\partial^k\delta h_{ki}
\nonumber\\&&+\frac{1}{2}\delta_{ij}\partial^{-2}\partial^l\partial^k\delta
h_{lk}+\frac{1}{2}\partial^{-2}\partial_i\partial_j\delta h_k^k
+\frac{1}{2}\partial^{-4}\partial_i\partial_j\partial^l\partial^k\delta
h_{lk} \bigg),
\end{eqnarray}
where $\delta h_{ij}$ can be written explicitly as
\begin{equation}
\delta
h_{ij}=-\frac{1}{H^2}\partial_i\zeta_n\partial_j\zeta_n+\frac{1}{H}\left(\partial_i\zeta_n\partial_j\psi_1+\partial_j\zeta_n\partial_i\psi_1\right)
+\frac{a^2}{H}\zeta_n\dot{\tilde{\gamma}}_{ij},
\end{equation}
where $\psi_1$ is from the uniform curvature gauge and we have
used the variable $\zeta_n$ introduced before, Eq. (\ref{zetan}).
As $\tilde{\gamma}_{ij}$ is of second order now, the term
$\zeta_n\dot{\tilde{\gamma}}_{ij}$ is of third order. We kept it
in Eq. (\ref{CG}) for the sake of comparison with the result of
\cite{Maldacena:2002vr}. For $\zeta$ we have
\begin{equation}
\zeta=\zeta_n+\frac{\epsilon}{2}\zeta_n^2+\frac{\ddot{\phi}_0}{2\dot{\phi}_0H}\zeta_n^2+\frac{1}{H}\zeta_n\dot{\zeta_n}+\beta.
\label{GTzeta}
\end{equation}

\chapter{\label{TTpart}Extraction of TT part of a tensor}

In this appendix, we provide a way to extract the transverse and
traceless (TT) part of a tensor . This will be used to obtain the
TT part of the source term of equation (\ref{tildegammaeq}).

Let $T_{ij}$ to be a given 3D symmetric tensor, as it is the case
for the source of equation (\ref{tildegammaeq}). Then it can be
decomposed into a trace part and a traceless part as
\begin{equation}
T_{ij}=\frac{T}{3}\delta_{ij}+\tilde{T}_{ij}.
\end{equation}
The traceless part (5 degrees of freedom) can be written like
\begin{equation}
\tilde{T}_{ij}=D_{ij}\chi+\partial_i\chi_j+\partial_j\chi_i+\chi_{ij},
\end{equation}
with
$D_{ij}\equiv\partial_i\partial_j-\frac{1}{3}\delta_{ij}\partial^2$,
$\partial^i\chi_i=0$ and $\partial^i\chi_{ij}=0=\chi_i^i$, where
indices are raised by $\delta_{ij}$. The equation
$\partial^iT_{ij}=\frac{1}{3}\partial_jT+\frac{2}{3}\partial_j\partial^2\chi+\partial^2\chi_j$
can be solved using a similar method as the one we used to solve
the second order momentum constraint previously. We then find
\begin{equation}
\chi=\frac{3}{2}\partial^{-4}\partial^jF_j, \quad
\chi_j=\partial^{-2}F_j-\partial_j\partial^i\partial^{-4}F_i,
\end{equation}
where $F_i\equiv\partial^jT_{ij}-\frac{1}{3}\partial_iT$. And
\begin{equation}
\chi_{ij}=T_{ij}-\frac{T}{3}\delta_{ij}-D_{ij}\chi-\partial_i\chi_j-\partial_j\chi_i.
\label{TT}
\end{equation}
In conclusion, given a tensor $T_{ij}$, Eq. (\ref{TT}) defines its
transverse and traceless part.

Let us see how this works at the action level for the particular
case of the tensor perturbations described in the main text. In
the action (\ref{GSSphi}), the source for $\tilde{\gamma}_{ij}$ is
of the form
\begin{equation}
S=\int dtd^3x\tilde{\gamma}^{ij}T_{ij}.
\end{equation}
with $T_{ij}$ being quadratic in $\delta\phi$. We can see that
because $\tilde{\gamma}_{ij}$ is transverse and traceless we are
allowed to replace $T_{ij}$ in the previous action with
$\chi_{ij}$ defined in (\ref{TT}). If we calculate the equations
of motion by varying the resulting action we get as a source
$\chi_{ij}$ and not simply $T_{ij}$ (see Eq.
(\ref{tildegammaeq})), ensuring that both sides of the equations
of motion are transverse and traceless.

\chapter{Equations of motion for the fluctuations}\label{app:eomfluctuations}

Here, we derive the equations of motion for linear perturbations
for the generalized model introduced in
\ref{subsec:generalized_model}. In terms of the field space
``covariant quantities'' \cite{Sasaki:1995aw} which are given by
\begin{eqnarray}
&&{\cal{D}}_t \dot{\phi}^I  \equiv
\ddot{\phi}^I + \Gamma^I _{JK} \dot{\phi}^J  \dot{\phi}^K\,,\\
&&{\cal{D}}_t Q^I  \equiv
\dot{Q}^I + \Gamma^I _{JK} \dot{\phi}^J  Q^K\,,\\
&&{\cal{D}}_I {\cal{D}}_J \tilde{P} \equiv
\tilde{P}_{,IJ} - \Gamma^K _{IJ} \tilde{P}_{,K}\,,\\
&&{\cal{R}}^I _{\;KLJ} \equiv \Gamma^I _{\;KJ,L}- \Gamma^I
_{\;KL,J} + \Gamma^I _{\;LM}\Gamma^M_{\;JK} -\Gamma^I _{\;JM}
\Gamma^M_{\;LK}\,,
\end{eqnarray}
($\Gamma^I _{JK}$ denotes the Christoffel symbols associated with
the field space metric $G_{IJ}$), the second order action can be
expressed as
\begin{eqnarray}
S_{(2)} &\!\!\!\!=&\!\!\!\! \frac12 \int dt d^3 x a^3 \biggl[
\left( \tilde{P}_{,Y}  G_{IJ} + \tilde{P}_{,YY} \dot{\phi}_I
\dot{\phi}_J \right) {\cal{D}}_t Q^I {\cal{D}}_t Q^J
\nonumber\\
&&\!\!\!\!\!\!\!\!- \frac{1}{a^2}  \tilde{P}_{,Y} \left[(1+bX)
G_{IJ} - b X_{IJ}\right]
\partial_i Q^I \partial^i Q^J
-\bar{{\cal{M}}}_{IJ} Q^I Q^J + 2 \tilde{P}_{,YJ} \dot{\phi}_I Q^J
{\cal{D}}_t Q^I
\biggr]\,,\nonumber\\
\end{eqnarray}
with the effective squared mass matrix
\begin{eqnarray}
\bar{{\cal{M}}}_{IJ} &=&- {\cal{D}}_I {\cal{D}}_J \tilde{P} -
\tilde{P}_{,Y} {\cal{R}}_{IKLJ} \dot{\phi}^K \dot{\phi}^L +
\frac{X \tilde{P}_{,Y}}{H} (\tilde{P}_{,YJ} \dot{\phi}_I
+ \tilde{P}_{,YI} \dot{\phi}_J)\nonumber\\
&&+\frac{X \tilde{P}^3 _{,Y}}{2 H^2}(1-\frac{1}{c_{ad}^2})
\dot{\phi}_I \dot{\phi}_J- \frac{1}{a^3} {\cal{D}}_t
\left[\frac{a^3}{2H} \tilde{P}^2 _{,Y}
\left(1+\frac{1}{c_{ad}^2}\right)\dot{\phi}_I \dot{\phi}_J
\right]\,.
\end{eqnarray}

It is worth noting that except for the coefficients of the kinetic
term and the gradient term, this action is the same as K-inflation
case and DBI-inflation case which are derived in
\cite{Langlois:2008mn} and \cite{Langlois:2008wt}, respectively.

From now on we will derive the equations of motion for the
fluctuations. For simplicity, let us now restrict our attention to
the two field case ($I = 1,2$). Then, the perturbations can be
decomposed into $Q^I = Q_\sigma e^I_\sigma + Q_s e^I_s$, where
$e^I_\sigma = e^I_1$ and $e^I_s$ is the unit vector orthogonal to
$e^I_\sigma$. As in standard inflation, it is more convenient to
use conformal time $\tau = \int dt /a(t)$ and define the
canonically normalized fields\footnote{Since the definitions of
$Q_{\sigma}$ and $Q_s$ in this work are different from the
definitions in
\cite{Langlois:2008mn,Langlois:2008wt,Langlois:2008qf} the
relations between $Q_\sigma$ and $v_\sigma$, and $Q_\sigma$ and
$v_s$ are also different from the analogous relations in those
papers.}
\begin{eqnarray}
v_\sigma \equiv \frac{a}{c_{ad}} Q_\sigma\,,\;\;\;\;\; v_s \equiv
\frac{a}{c_{en}}Q_s\,.
\end{eqnarray}

From the similar calculations with K-inflation and DBI-inflation
cases analyzed by \cite{Langlois:2008mn} and
\cite{Langlois:2008wt}, respectively, we find the equations of
motion for $v_\sigma$ and $v_s$ as
\begin{eqnarray}
v_\sigma '' - \xi v_s' + \left( c_{ad}^2 k^2 -\frac{z''}{z}
\right) v_\sigma - \frac{(z \xi)'}{z}v_s &=&0\,,
\label{eom_adiabatic}\\
v_s'' + \xi v_\sigma ' + \left(c_{en}^2 k^2
-\frac{\alpha''}{\alpha} + a^2 \mu_s^2\right) v_s -\frac{z'}{z}
\xi v_\sigma &=& 0 \label{eom_entropy}\,,
\end{eqnarray}
where the primes denote the derivative with respect to $\tau$ and
\begin{eqnarray}
\xi &\equiv& \frac{a}{\dot{\sigma} \tilde{P}_{,Y} c_{ad}} \left[
(1+ c_{ad}^2) \tilde{P}_{,s} -c_{ad}^2 \dot{\sigma}^2
\tilde{P}_{,Ys}\right]\,,\label{coupling}\\
\mu_s^2 &\equiv& -\frac{\tilde{P}_{,ss}}{\tilde{P}_{,Y}} + \frac12
\dot{\sigma}^2 \tilde{R} - \frac{1}{2 c_{ad}^2 X}
\frac{\tilde{P}_{,s}^2}{\tilde{P}_{,Y}^2}
+2 \frac{\tilde{P}_{,Ys} \tilde{P}_{,s}}{\tilde{P}_{,Y}^2}\,,\\
z&\equiv& \frac{a \dot{\sigma}}{ c_{ad} H}
\sqrt{\tilde{P}_{,Y}}\,, \;\;\;\;\;\alpha\equiv a
\sqrt{\tilde{P}_{,Y}}\,,
\end{eqnarray}
with
\begin{eqnarray}
\dot{\sigma} &\equiv& \sqrt{2X}\,,\;\;\;\;\; \tilde{P}_{,s} \equiv
\tilde{P}_{,I} e^I_s \sqrt{\tilde{P}_{,Y}} c_{en}\,,
\nonumber\\\tilde{P}_{,Ys} &\equiv& \tilde{P}_{,YI} e^I_s
\sqrt{\tilde{P}_{,Y}} c_{en}\,, \;\;\;\;\;\tilde{P}_{,ss} \equiv
({\cal{D}}_I {\cal{D}}_J \tilde{P}) e^I_s e^J_s \tilde{P}_{,Y}
c_{en}^2 \,,
\end{eqnarray}
and $\tilde{R}$ denotes the Riemann scalar curvature of the field
space.

If we assume that the effect of the coupling $\xi$ can be
neglected when the scales of interest cross the sound horizons the
two degrees of freedom are decoupled and the system can be easily
quantized. If we further assume the slow-roll approximations, the
time evolution of $H$, $c_{ad}$, and $\dot{\sigma}$ is small with
respect to that of the scale factor and the relations $z''/z
\simeq 2/\tau^2$ and $\alpha''/\alpha \simeq 2/\tau^2$ hold (see
subsection \ref{subsec:approximations} for these approximations).
The solutions of (\ref{eom_adiabatic}) and (\ref{eom_entropy})
with the Bunch-Davies vacuum initial conditions are thus given by
\begin{eqnarray}
v_{\sigma k} &\simeq& \frac{1}{\sqrt{2 k c_{ad}}} e^{-i k c_{ad}
\tau} \left( 1-\frac{i}{k c_{ad} \tau}\right)\,,
\\
v_{s k} &\simeq& \frac{1}{\sqrt{ 2 k c_{en}}} e^{-i k c_{en} \tau}
\left(1-\frac{i}{k c_{en} \tau}\right)\,,
\end{eqnarray}
when $\mu_s^2/H^2$ is negligible for the entropy mode.

Therefore, the power spectra for $Q_\sigma$ and $Q_s$ are obtained
as
\begin{eqnarray}
{\cal{P}}_{Q_\sigma} \simeq \frac{H^2}{4 \pi^2 c_{ad}}\,,
\;\;\;\;\; {\cal{P}}_{Q_s} \simeq \frac{H^2}{4 \pi^2 c_{en}}\,,
\end{eqnarray}
which are evaluated at sound horizon crossing. Here for adiabatic
perturbations, the sound horizon is determined by $c_{ad}$ and for
entropy perturbations, it is determined by $c_{en}$. The ratio of
the power spectra for the adiabatic and entropy modes is thus
given by ${\cal{P}}_{Q_s}/{\cal{P}}_{Q_{\sigma}} = c_{ad}/c_{en}$.



\end{document}